%% file: qsosample_arXiv.tex
\documentclass[preprint2]{aastex6}

\newcommand{\lya}{\mbox{${\rm Ly}\alpha$}}

\newcommand{\Civ}{\ion{C}{4}}
\newcommand{\Ciii}{\ion{C}{3}}
\newcommand{\Feii}{\ion{Fe}{2}}
\newcommand{\Mgii}{\ion{Mg}{2}}

\usepackage{graphicx,amsmath,amssymb,epsfig}
\usepackage[T1]{fontenc}
\usepackage{multirow}
\usepackage{color}

\slugcomment{Accepted for publication in ApJ Supplement Series}

\shorttitle{The eHAQ Survey: Dusty Absorbers toward Red Quasars}
\shortauthors{Krogager et al.}

\begin{document}


\title{The extended High A(V) Quasar Survey:\\ Searching for dusty absorbers toward mid-infrared selected quasars}

\author{
J.-K. Krogager\altaffilmark{1,2}
J. P. U. Fynbo\altaffilmark{2},
K. E. Heintz\altaffilmark{2},
S. Geier\altaffilmark{3},
C. Ledoux\altaffilmark{4},\\
P. M\o ller\altaffilmark{5},
P. Noterdaeme\altaffilmark{1},
B. P. Venemans\altaffilmark{6}, \&
M. Vestergaard\altaffilmark{2,7}
}

\altaffiltext{1}{Institut d'Astrophysique de Paris, CNRS-UPMC, UMR7095, 98bis bd Arago, 75014 Paris, France\\ krogager@iap.fr}
\altaffiltext{2}{Dark Cosmology Centre, Niels Bohr Institute, University of
Copenhagen, Juliane Maries Vej 30, 2100 Copenhagen \O, Denmark}
\altaffiltext{3}{Instituto de Astrof\'isica de Canarias (IAC), 38205 La Laguna, Tenerife, Spain}
\altaffiltext{4}{European Southern Observatory, Alonso de C\'ordova 3107, Vitacura, Casilla 19001, Santiago 19, Chile}
\altaffiltext{5}{European Southern Observatory, Karl-Schwarzschildstrasse 2, 85748 Garching bei M\"unchen, Germany}
\altaffiltext{6}{Max-Planck Institute for Astronomy, K{\"o}nigstuhl 17, 69117 Heidelberg, Germany}
\altaffiltext{7}{Steward Observatory and Department of Astronomy, University of Arizona, 933 N Cherry Avenue, Tucson, AZ 85721, USA}

\begin{abstract}

\noindent
We present the results of a new spectroscopic survey for dusty intervening absorption systems, particularly damped Ly$\alpha$ absorbers (DLAs), towards reddened quasars. The candidate quasars are selected from mid-infrared photometry from the {\it Wide-field Infrared Survey Explorer} combined with optical and near-infrared photometry. Out of 1073 candidates, we secure low-resolution spectra for 108 using the Nordic Optical Telescope on La Palma, Spain. Based on the spectra, we are able to classify 100 of the 108 targets as quasars. A large fraction (50\%) is observed to have broad absorption lines (BALs). Moreover, we find 6 quasars with strange breaks in their spectra, which are not consistent with regular dust reddening.
Using template fitting we infer the amount of reddening along each line of sight ranging from $A(V)\approx0.1$~mag to $1.2$~mag (assuming an SMC extinction curve). In four cases, the reddening is consistent with dust exhibiting the 2175~\AA\ feature caused by an intervening absorber, and for two of these, a \Mgii\ absorption system is observed at the best-fit absorption redshift. In the rest of the cases, the reddening is most likely intrinsic to the quasar. We observe no evidence for dusty DLAs in this survey. However, the large fraction of BAL quasars hampers the detection of absorption systems. Out of the 50 non-BAL quasars only 28 have sufficiently high redshift to detect Ly$\alpha$ in absorption.

\end{abstract}

\keywords{galaxies: active --- quasars: general --- quasars: absorption lines}

\section{Introduction}
The study of dust-reddened quasars, in terms of their number density and amount of dust extinction, provide important clues to the study of evolutionary scenarios for quasars (and active galactic nuclei in general, e.g., \citealt[][and references therein]{Hopkins2005}). In most scenarios quasars are thought to start out as heavily dust obscured phenomena before gradually clearing out the dust (either by strong outflows and/or dust sublimation) leaving behind a population of unobscured, ``typical'' quasars (typical in the sense that they make up the bulk of known quasars; e.g., \citealt{Sanders1988a, Sanders1988b, Hopkins2006}).

In addition to the intrinsic variations in the quasar environments, any intervening dust in galaxies along the line of sight (including our own Milky Way) causes the background quasar to appear dust-reddened as well \citep*{FallPei1989, Fall1989, Pei1991}. Such intervening galaxies are typically associated with the so-called damped Lyman-$\alpha$ absorbers (DLAs, see \citealt*{Wolfe2005}). By studying the absorption in the spectrum of the background quasar, we can recover important quantities about the absorbing gas: e.g., its metallicity, ionization state, dust depletion, and in some cases even the density and temperature \citep[e.g.,][]{Howk2005, Srianand2005, Jorgenson2010, Kanekar2014, Cooke2015, Neeleman2015, Noterdaeme2015a}.
From looking at depletion in large samples of DLAs it has been shown that the more chemically evolved DLAs (i.e., with high metallicity) are more likely to harbour dust \citep{Ledoux2003, DeCia2013}. Thus, the metal-rich and dusty DLAs hold valuable information about the most chemically enriched systems and allow us to study in great detail the molecular absorption (since molecules are primarily formed on the surface of dust grains \citealt{Jenkins1997}) as well as the cold gas phase of the absorbing gas \citep[e.g.][]{Srianand2005, Noterdaeme2007}.

The vast majority of quasars (and thereby also DLAs; \citealt{Prochaska2005, Prochaska2009, Noterdaeme2009b, Noterdaeme2012b}) have been identified  in optical, large-area sky surveys, most notably the Sloan Digital Sky Survey (SDSS, \citealt{York2000}, and its extension, the Baryon Oscillation Spectroscopic Survey [BOSS], \citealt{Dawson2013}) and the 2dF QSO redshift survey \citep{Croom2004}.
The power of selecting quasars in the optical is proven by the sheer number of quasars identified in the SDSS catalog. This optical selection mainly relies on the fact that quasars have very blue optical colours. However, the technique falls short for high redshifts where stars and quasars are no longer easily discernible in the optical color-space. Also, if the quasar has intrinsic spectral differences relative to the ``typical'' quasar population, e.g., dust reddening, weak emission lines or broad absorption lines (BALs), the quasar will appear redder in the observed optical emission, making the distinction between stars and quasars even more difficult \citep{Richards2002, Richards2003}. Furthermore, a dust-rich absorption system along the line of sight can cause a significant reddening effect on the background quasar \citep[e.g.,][]{Fall1989, Noterdaeme2009b, Noterdaeme2012a, Jiang2010, Kaplan2010, Fynbo11, Wang2012, Krogager2016}. All of these effects may cause the background quasar to change appearance to the point where optical searches become blind to these reddened objects.
This introduces a bias in the parent quasar sample from which DLAs are selected. As a result, this bias propagates to a bias in the DLA samples, since the quasars behind chemically evolved and dusty galaxies are missed more often than quasars behind dust-poor galaxies.

The mere existence of this bias and its potential impact on the DLA population has been studied in great detail in the literature \citep{FallPei1989, Fall1989, Pei1991, Ellison2001a, Murphy2004, Vladilo2005, Vladilo2008, Jorgenson2006, Frank2010, Pontzen2009, Kaplan10, Khare2012, Murphy2016}. A large part of these studies have been based on optically selected samples, which are not sensitive to this hidden population of dusty DLAs. However, the surveys by \citet{Ellison2001a} and \citet{Jorgenson2006} were carried out at radio wavelength since the radio emission from quasars is not affected by dust. While they find that dust obscuration is a small effect (for the incidence rate of DLAs) they also conclude that larger radio samples are needed to firmly conclude on the importance of a dust bias \citep[see also the analyses of][]{Ellison2005, Ellison2008}.

In order to gauge the population of reddened quasars, a substantial number of surveys have been undertaken using various selection methods \citep[e.g.,][]{Benn1998, Warren2000, Gregg2002, Richards2003, Hopkins2004, Polletta2006, Lacy2007, Maddox2008, Maddox2012, Glikman2007, Glikman2012, Glikman2013, Urrutia2009, Banerji2012, Hainline2014}, and more recently, wide-field, infrared facilities (such as UKIRT Infrared Deep Sky Survey [UKIDSS, \citealt{UKIDSS}], VISTA Kilo-Degree Infrared Galaxy Survey [VIKING, \citealt{VIKING}], and {\it Wide-field Infrared Survey Explorer} [{\it WISE}\,, \citealt{WISE}]\,) allow quasar candidates to be selected in great numbers based on the shape of their spectral energy distributions (SEDs) instead of using just one single characteristic \citep[e.g.,][]{Warren2000, Warren2007, Peth2011, Maddox2012}. While such efforts are very effective for photometric classification \citep{Richards2015, Peters2015}, the spectroscopic follow-up and classification is much more time-consuming. The BOSS programme has provided a massive improvement in terms of a more unbiased quasar selection due to the development of complex multi-wavelength and multi-epoch selection algorithms \citep[and references therein]{Ross2012}, but the red quasar population is still quite sparsely sampled.\\

In contrast to most other surveys for red quasars, which in various ways focus on a complete census of the quasar population, we initiated our spectroscopic survey (High A(V) Quasar [HAQ] Survey) with the primary focus on finding dust-rich foreground absorption systems. In the first surveys, we used near-infrared and optical photometry alone to select point sources that are consistent with ``typical'' quasars reddened by a foreground dusty absorber. The initial strategy is laid out by \citet[hereafter Paper I]{Fynbo2013a} and a refined set of those criteria were used for a much larger survey presented by \citet[hereafter Paper II]{Krogager2015}. In accordance with previous works analysing the DLA dust bias, we find in the HAQ survey that the incidence rate of dusty absorbers is low, i.e., the dusty foreground absorbers do not lead to a strong bias in terms of the $N$(\ion{H}{1}) distribution; however, since these dusty DLAs are preferentially metal-rich \citep[e.g.,][]{Fynbo2011,Krogager2016} the missing dusty DLAs are important for the overall metal distribution \citep[see also][]{Pontzen2009, Khare2012}. Furthermore, the dusty DLAs are important for a complete census of the cold and molecular gas phases in absorption-selected galaxies \citep{Ledoux2015}. 

In this work, we present the extended HAQ (eHAQ) survey designed to probe quasars with more reddening relative to Paper II (this part of color space was initially included in Paper I, but this led to a large fraction of stellar and galactic contamination) and to eliminate $z\la 1.5$ quasars in order to be able to detect the Ly$\alpha$ absorption line in our observed spectra. The new selection criteria presented in the following allow us to select a purer sample of reddened quasar candidates by basing the sample selection on {\it WISE} mid-infrared photometry. Furthermore, we are able to effectively remove low-redshift quasars as desired. In the following, we present the results of our spectroscopic follow-up campaign of 108 candidates from our new criteria.

Section~\ref{selection} deals with the photometric data used in our work, our selection criteria, and our proposed method to remove stellar contamination. In Sect.~\ref{observations} and Sect.~\ref{results}, we describe our spectroscopic observations together with our classifications and analyses of the sample.
In Sect.~\ref{discussion}, we discuss the implications of our work.
Throughout this work, we will assume a flat $\Lambda$CDM cosmology with $H_0=67.9\, \mathrm{km s}^{-1}\mathrm{Mpc}^{-1}$, $\Omega_{\Lambda}=0.69$ and
$\Omega_{\mathrm{M}} = 0.31$ (Planck Collaboration 2014)\nocite{Planck2014}.

\section{Photometric Data and Target Selection}
\label{selection}

The photometric catalog was compiled by cross-matching the SDSS data release 8 ($u$, $g$, $r$, $i$, and $z$ bands at 0.36, 0.47, 0.62, 0.75, and 0.90~$\mu$m, respectively), the UKIDSS data release 10 ($Y$, $J$, $H$, and $K_s$ bands at 1.03, 1.25, 1.64, and 2.21~$\mu$m, respectively), and {\it WISE} AllSKY data release ($W_1$, $W_2$, $W_3$, and $W_4$ at 3.4, 4.6, 12, and 22 $\mu$m, respectively). In the following, we will refer to the $K_s$ band of UKIDSS simply as the $K$ band.

We matched this optical/infrared catalog to the Faint Images of the Radio Sky at Twenty-cm (FIRST, data release 14Dec17) to compare radio properties of our parent sample and the SDSS/BOSS targets. We stress that the radio data were only included as ancillary data and were not part of the sample selection criteria. The survey provides radio coverage at 1.4~GHz with an average sensitivity of $\sim0.15$~mJy (rms).

The selection of candidate quasars was based primarily on mid-infrared photometry from {\it WISE}. We required the candidates to be detected with a signal-to-noise ratio of at least 5 in the first three {\it WISE} bands (i.e., ${\rm SNR_{W1-3}} \geq 5$).
Moreover, we require that the candidates are detected (at more than 3 sigma significance) in $g$, $r$, $i$, $J$, and $K$, and that they are point sources in both UKIDSS and SDSS (specifically, we use the catalog flags: {\tt mergedclass = -1 \& sdsstype = 6}). We limited our follow-up to candidates that had not already been observed spectroscopically either by us in Paper I and II or in SDSS or BOSS. The declination of the candidates was restricted to be in the range from $-4$\degr\ to $+17$\degr\ in order to allow follow up from facilities in the southern hemisphere while still accessible to the Nordic Optical Telescope.

\begin{figure}
	\centering
	\includegraphics[width=0.48\textwidth]{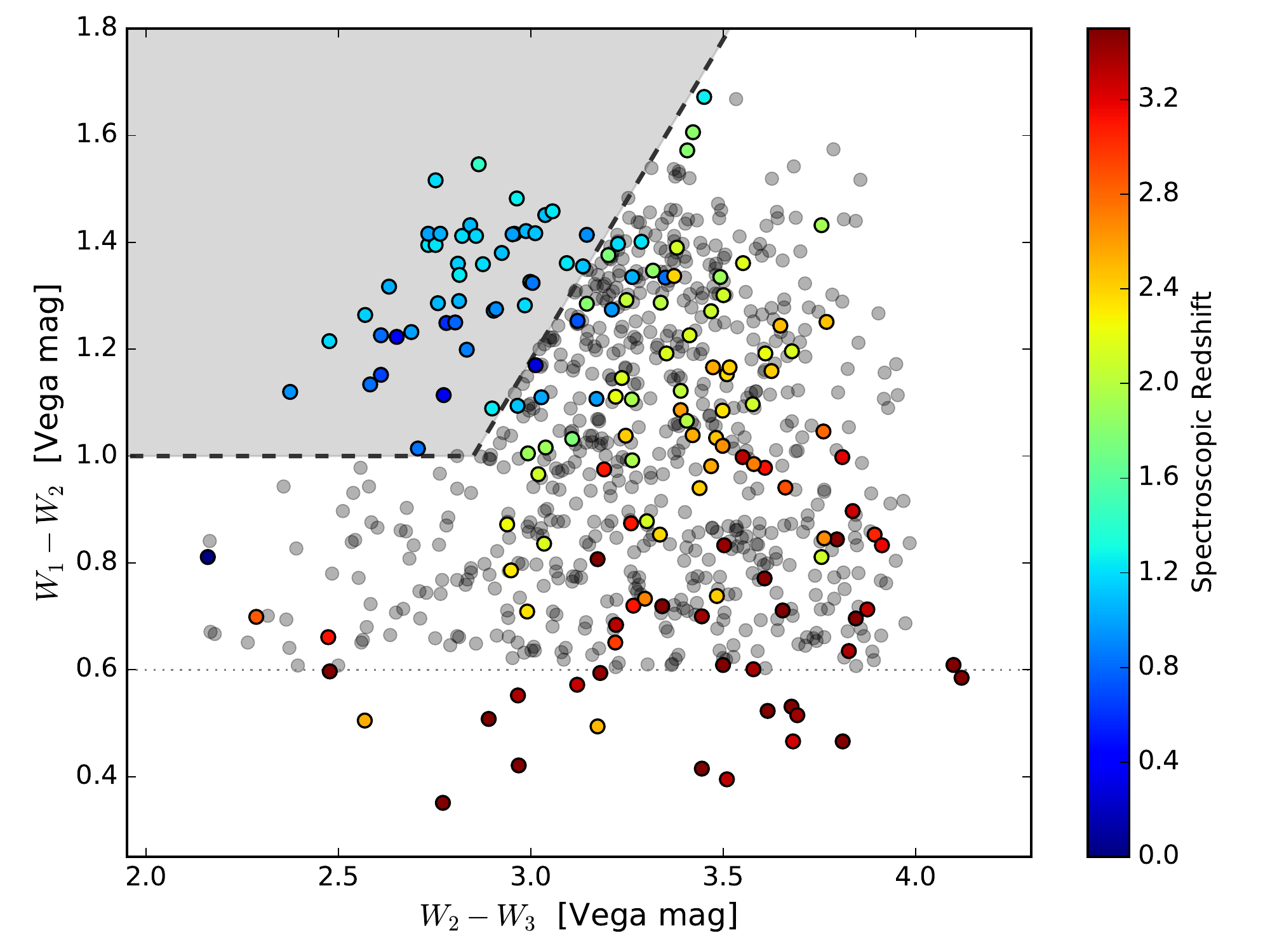}
	\caption{{\it WISE} color-color plots for this work and Paper II. Each colored point corresponds to a quasar from HAQ (Paper II) and its color indicates the spectroscopic redshift. Grey points are candidates for this work. The grey region in the upper left corner indicates the exclusion zone for low-redshift quasars. The dotted line marks the lower limit on $W_1-W_2$ color in order to limit contamination from stars and galaxies.
	}
	\label{fig:WISE_selection}
\end{figure}

Based on experience from the previous surveys (Paper II), we have created a selection cut in the $W_1-W_2$ vs $W_2-W_3$ color-color space in order to select candidates with redshifts larger than $z>1.5$. From the distribution of redshifts of the quasars identified in Paper II, we were able to define a cut in the {\it WISE} color space to exclude low-redshift quasars, see Fig.~\ref{fig:WISE_selection}.
We imposed the following color cuts in {\it WISE}\,:\\

\hspace{-1mm}${\rm for}\ W_2 - W_3 < 2.85:\ W_1 - W_2 < 1~,$

\vspace{1mm}
\hspace{-1mm}${\rm for}\ W_2 - W_3 \geq 2.85:\ W_1 - W_2 < 1.2\times(W_2-W_3)-2.42,$

\vspace{1mm}
\hspace{-1mm}$W_1 - W_2 > 0.6$\ \ and\ \ $2 < W_2 - W_3 < 4$~.\\

The lower boundary on the $W_1 - W_2$ color was imposed to limit contamination from stars and galaxies while still recovering a part of the high-redshift quasars. This was motivated by the results of \citet{Stern2012}, see also discussion by \citet{Richards2015}.

Since we wish to target quasar candidates with potential dusty foreground absorbers, we include two optical color cuts to ensure that the candidates are optically red (consistent with dust reddening)\,:
$g-r > 0.5$ and $r-i > 0.4$ (i.e., the same as in Paper I and II).\\

In the first observing run, we limited our selection to $J_{\rm AB}<19$. In the subsequent runs we relaxed this magnitude cut, however, only a small number of observed candidates have $J_{\rm AB}>19$ (5 out of 108 observed).

\subsection{Stellar rejection}

\begin{figure}
	\centering
	\includegraphics[width=0.48\textwidth]{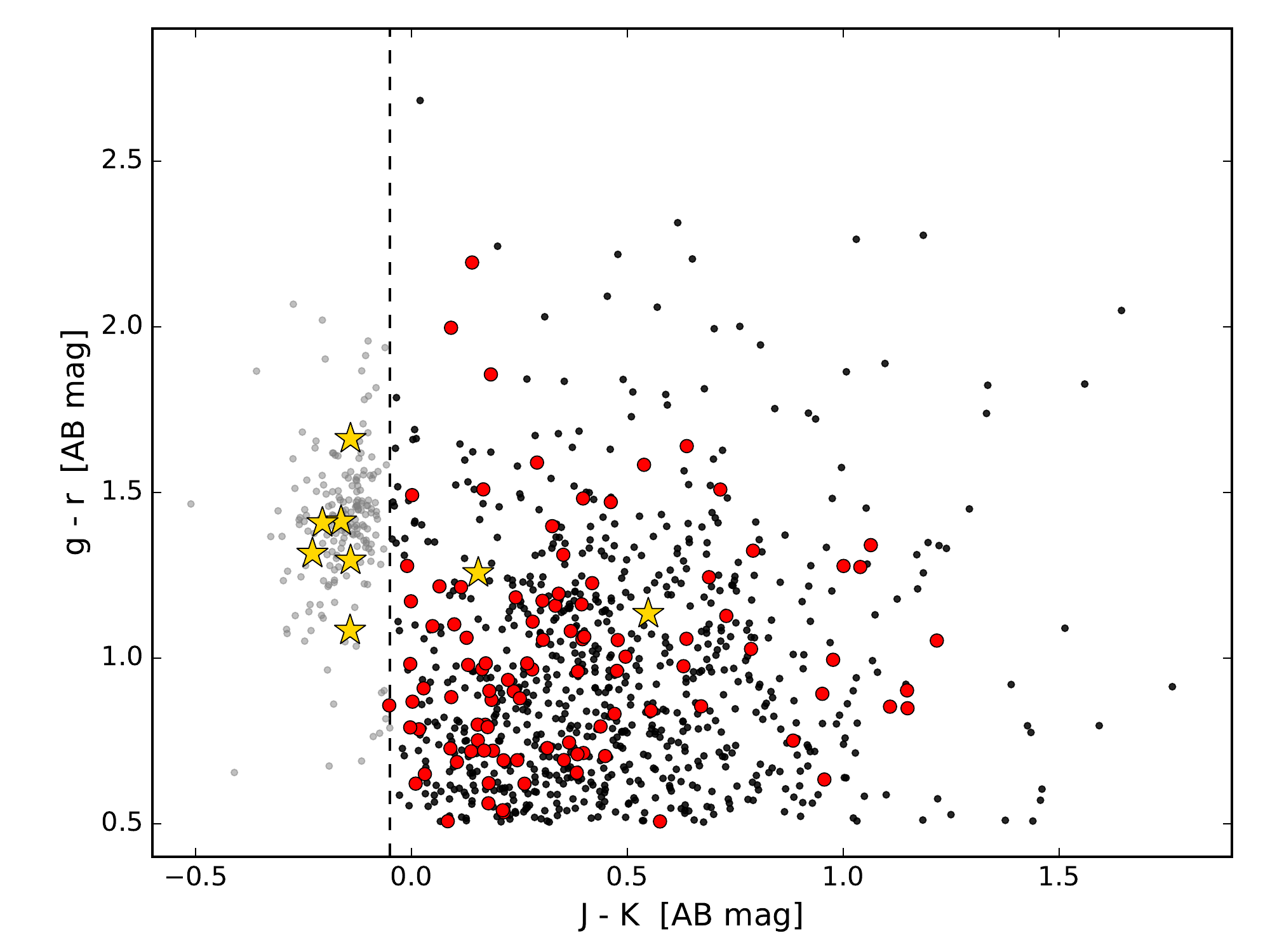}
	\caption{Color-color plot showing optical $g-r$ color vs. near-infrared $J-K$ color for the eHAQ candidates. The small points show the full parent sample where quasars and stars (based on photometry) are shown by black and grey points, respectively. The imposed photometric criterion ($J-K>-0.05$) to reject stars is shown by the vertical dashed line. Large points indicate the targets that were spectroscopically observed: quasars (red circles) and stars (yellow stars). }
	\label{fig:grJK}
\end{figure}

During the first observing run (P50-802), we identified a significant fraction of contamination from stars. The majority of these contaminants were observed to have $J-K<0$. We therefore imposed an additional near-infrared color cut ($J - K > -0.05$) for the subsequent spectroscopic observations in order to remove stellar contaminants. This is illustrated in Fig.~\ref{fig:grJK}. All targets selected after P50-802 have been corrected for the stellar contaminants using the near-infrared cut.\\

\section{Spectroscopic Observations}
\label{observations}

During a series of observing runs in 2015 and 2016, a total of 108 candidate quasars were observed with the Nordic Optical Telescope (NOT)\footnote{Located at the Observatorio del Roque de los Muchachos on La Palma, Spain.} using the Andaluc\'ia faint object spectrograph and camera (ALFOSC).

For all the ALFOSC observations, we binned the CCD pixels by a factor of 2 along the wavelength axis and we aligned the slit with the parallactic angle. The candidates were, as in Paper II, observed using grism \#4, which covers the wavelength range from about 3200~\AA\ to 9100~\AA\ at a resolution of about $R \sim 300$ with a slit width of 1.3 arcsec. Redwards of about 7500 \AA\ the spectra are strongly affected by fringing, which was alleviated by dithering along the slit. In order to prevent second order contamination from wavelengths shorter than 3500~{\AA}, a blocking filter was used for the observations with grism \#4. We used filter no. 94 which blocks out wavelengths shorter than $3560$~{\AA}. Given the red nature of the targets (by selection) and the reduced intensity of the second order spectrum the contamination from the remaining flux which is not blocked by the filter (from $3560-4500$~\AA) will be low (however, it may still be an issue as observed in \citealt{Heintz2016a}). The overlap from the second order starts at $7120$~{\AA}, which is furthermore the spectral range affected by strong fringing. Moreover, we use the overall spectral shape from the broad-band photometry to make sure that we are not affected by significant second order contamination. Based on this consistency check, we find the second order contamination to be negligible.
One candidate (eHAQ2359+1354) was observed with grism \#6 alone to investigate tentative signs of a damped Ly$\alpha$ absorber seen in the spectrum obtained by BOSS. Grism \#6 covers wavelengths from 3200 to 5810 {\AA} at a resolution of about $R \sim 500$ with the 1.0 arcsec slit.

Four additional candidates were observed with the intermediate dispersion spectrograph (IDS) situated at the Cassegrain focus of the Isaac Newton Telescope (INT). For spectra obtained with the IDS, we used the grating `R400R', which covers wavelengths from 5000 to 9500 {\AA} at a resolution of roughly $R \sim 2250$.

Details about the spectroscopic observations are provided in Table~\ref{tab:overview} in Appendix~\ref{appendix:tables}. In the table, we also indicate overlap with existing spectroscopy from BOSS. The overlap was caused by a mistake in the cross-referencing for the target selection.\\

\subsection{Data Reduction}
The spectra were reduced using a combination of IRAF\footnote{IRAF is distributed by the National Optical Astronomy Observatory, which is operated by the Association of Universities for Research in Astronomy (AURA) under cooperative agreement with the National Science Foundation.} and MIDAS\footnote{ESO-MIDAS is a copyright protected software product of the European Southern Observatory. The software is available under the GNU General Public License.} tasks for low-resolution spectroscopy. Cosmic rays were rejected using the software written by \citet{vanDokkum2001}.
In case of photometric observing conditions the spectrophotometric standard star observed on the same night as the science spectra was used for the flux calibration. Otherwise, we used a standard response curve to calibrate the spectra. The spectra and photometry were corrected for Galactic extinction using the extinction maps from \citet{Schlafly2011}. In order to improve the absolute flux-calibration, we scaled the spectra to the $r$-band photometry from SDSS.

For the four targets observed with the INT, we were not able to recover a robust flux calibration. The shape of the spectra does not match the observed photometry from SDSS. However, the wavelength calibration is robust. We therefore only use the spectra to measure the spectroscopic redshifts.\\

\begin{figure*}
	\plotone{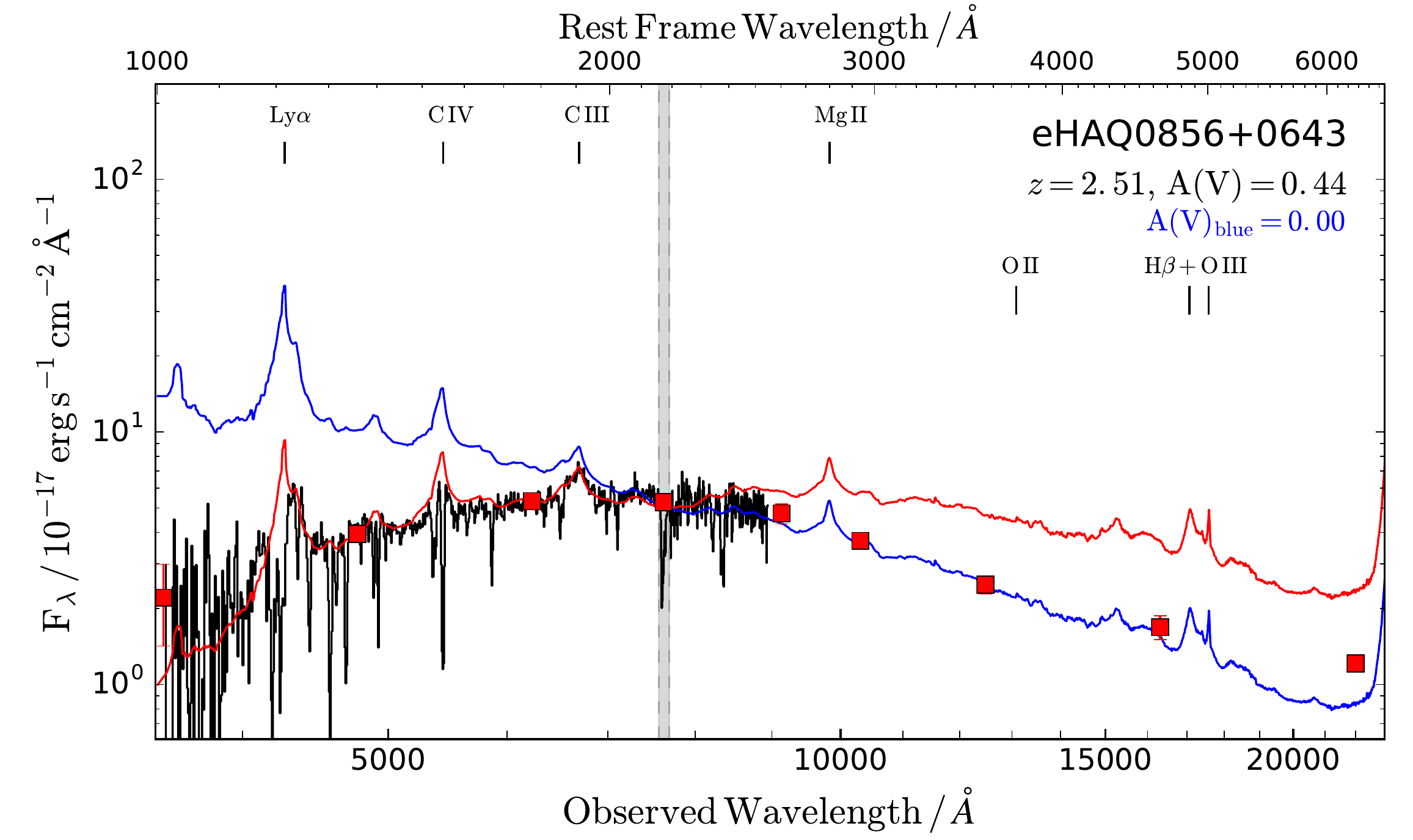}
	\caption{Sample figure of the observed quasar spectra. The entire figure set is
	available online. The observed spectrum is shown as a solid black curve.
	For spectra observed with the INT, the spectra are binned by a factor of 8
	for visual clarity. These cases are marked by the label `{\sc int}' in the lower
	left corner of the figure.
	The filled, red squares indicate the SDSS and UKIDSS photometric
	data points.
	In the upper right corner we provide the estimated emission redshift
	and rest-frame $V$-band extinction. The red curve shows the redshifted
	composite quasar template of \citet{Selsing2016} reddened by the best-fit
	dust model (indicated in the upper right corner).
	The dust reddening applied in the rest-frame of
	the quasar, A(V), and absorber, A(V)$_{\rm abs}$, assumes the SMC and
	LMC extinction laws, respectively.
	In cases where the template is also shown in blue, we were not able to
	fit both optical and infrared data simultaneously; The blue template
	then indicates the best fit to the infrared data only, and the
	corresponding A(V) is given as ${\rm A(V)_{blue}}$. Note that the spectra
	have not been corrected for telluric absorption (marked with a grey band
	at $\sim7600$~\AA). A compected version of the full set of figures
	is shown in Appendix~\ref{appendix:figureset}.
	\label{fig:spectra}}
\end{figure*}

\section{Results}
\label{results}
For all but two candidates, we were able to securely classify the object as either quasar or star based on the available data. 
We identify 8 stars (one of which is tentative) and 100 quasars (one of which is tentative). Individual notes for each target are given in Appendix~\ref{appendix:notes}. Below we describe the stellar and quasar spectra in more detail.

\subsection{Quasars}
\label{res:quasars}

The spectra of the 100 quasars are shown in Fig.~\ref{fig:spectra} (and in Appendix~\ref{appendix:figureset}) together with the SDSS and UKIDSS photometry.
For every quasar we measure the redshift from the characteristic broad emission lines by template matching. In a few cases where the broad emission lines are heavily suppressed (due to BAL or intrinsically weak emission lines), we use absorption features to estimate the quasar's systemic redshift. In order to infer the reddening of each quasar, we fit the spectrum with the quasar template from \citet{Selsing2016} following the same approach as described in Paper II assuming the extinction law for the Small Magellanic Cloud (SMC, $R_V = 2.74$) by \citet{Gordon2003}. We use the SMC extinction curve since we see no evidence for the 2175~\AA\ dust bump in the observed spectra.
Moreover, we note that using the steeper extinction curve derived by \citet{Zafar2015} leads to consistent fits only with a lower best-fit value of A(V) due to the lower value of $R_V=2.41$ for this extinction curve; In terms of $E(B-V)$ the best-fit values are identical. Since the two curves only differ significantly blueward of {\Ciii}, it is only possible to distinguish the two curves in spectra of quasars at redshifts larger than $z\gtrsim1.8$. However, here the intrinsic variations in emission lines (and BAL features) prevents us from significantly distinguishing one from the other due to the limited resolution and signal-to-noise ratio of our data. One exception is the photometric fits performed in Sect.~\ref{WLQ:phot_fit} since for these quasars the weak emission lines do not cause large variations in the SED.
Only in a few cases do we see evidence for a steeper extinction curve and apply the extinction curve from Zafar et al. (marked with $^b$ in Table~\ref{tab:sample}).
For consistency with other works and with Paper I and II, we use the SMC extinction curve in this work, unless stated otherwise.
As noted in Paper I and II (as well as other works; e.g., \citealt{Urrutia2009}), the A(V) values derived from template fitting are mostly indicative since the main source of uncertainty comes from the assumption that all quasars follow the same template. We find an average {\it statistical} uncertainty of 0.01--0.02 mag for the signal-to-noise ratio of our data (see also Sect. 4.1.2 in Paper II). By varying the intrinsic slope of the quasar template before fitting, we recover an estimate of the systematic uncertainty due to these intrinsic variations of around 0.02~mag. Moreover, the best-fit A(V) is correlated with the assumed value of $R_V$ for the SMC or Zafar et al. (2015) curves. For the \citet{Zafar2015} extinction curve ($R_V=2.41$), the corresponding extinction, $A(V)_{\rm Z15}$, is related to the best-fit extinction given in this work, $A(V)_{\rm SMC}$, by the following expression:
$$ A(V)_{\rm Z15} = \frac{R_{V_{\rm Z15}}}{R_{V_{\rm SMC}}} \, A(V)_{\rm SMC} = 0.88 \, A(V)_{\rm SMC}\ .$$

The inferred quasar redshift and A(V) are given in the upper right corner of Fig.~\ref{fig:spectra}. For the target eHAQ2359+1354, we use the available BOSS spectrum to fit the extinction, since the Grism~\#6 spectrum does not allow us to constrain the reddening. In some cases, we are not able to fit both the optical spectrum and the NIR photometry simultaneously and we therefore provide an A(V) value for both the optical and NIR fits. These {\it peculiar} cases are presented later in more detail. The spectroscopic redshift, the best-fit A(V) (assuming SMC-type dust), and our spectral classification is given in Table~\ref{tab:sample} in Appendix~\ref{appendix:tables}. We caution that the BAL classification does not follow a rigorous scheme, it is merely judged by eye in the spectra. 

\begin{figure*}
\centering
	\includegraphics[width=0.6\textwidth]{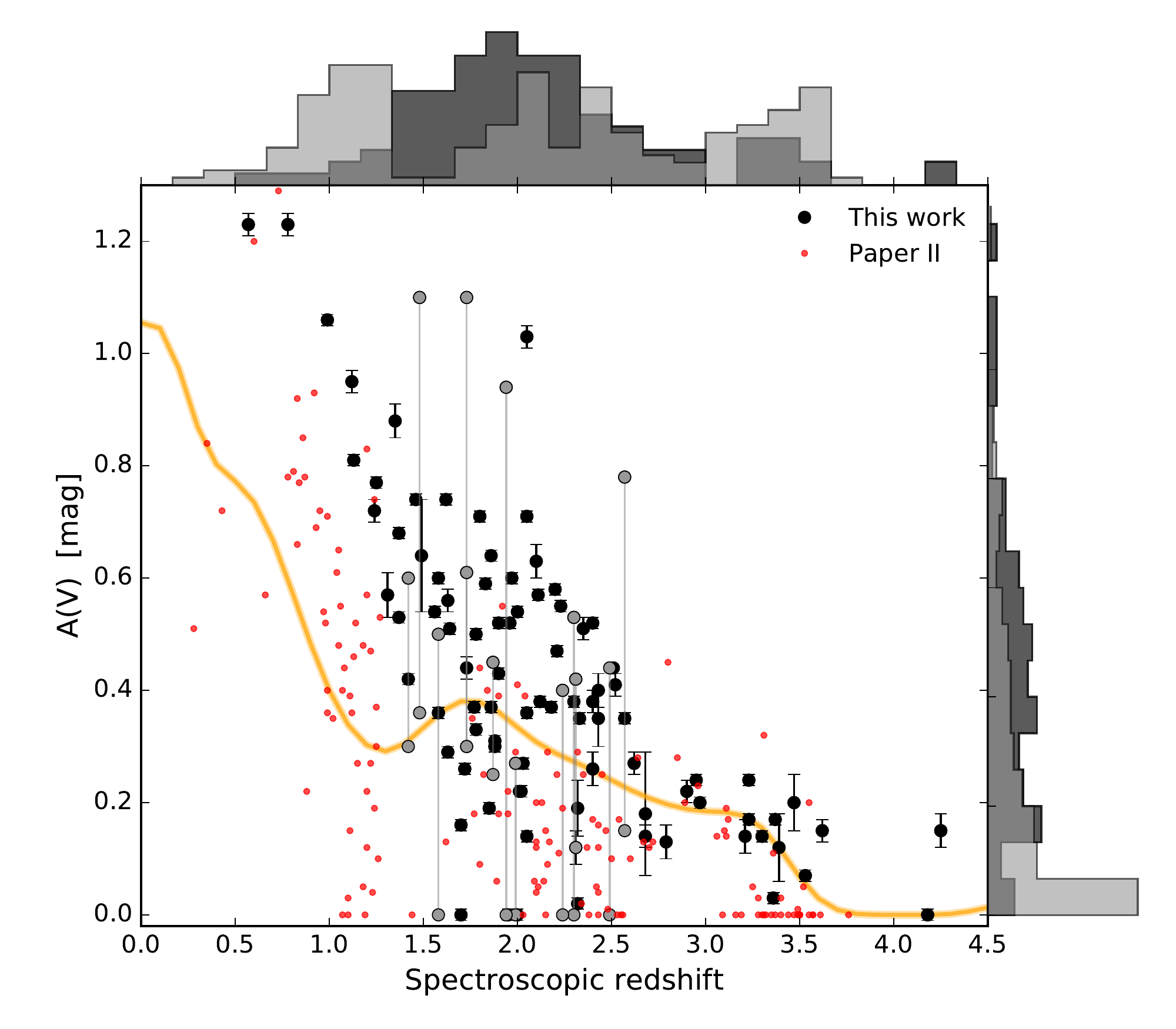}
	\caption{Distribution of redshift vs. the amount of $V$-band extinction,
	A(V), from the spectral fitting.
	The large, filled circles show the sample from this work.
	The errorbar (typically 0.01~mag) is smaller than the point in most
	cases. The small, red points are taken from Paper II. The points that
	are connected with a grey line refer to the peculiar quasars in
	Table~\ref{tab:sample} for which a range of A(V) is given,
	see Sect.~\ref{res:quasars} for details.
	For clarity, the corresponding points are shown in grey.
	The orange curve shows the value of A(V)
	assuming SMC-type extinction for which a quasar at the given
	redshift has an observed color of $g-r=0.5$~mag.
	Since one of our criteria is $g-r>0.5$, points below the curve
	will therefore enter our sample due to another mechanism
	than pure dust-reddening, e.g., strong BAL features or
	weak emission lines. The histograms on the top show the redshift
	probability density functions (PDFs) of this work (dark grey) and
	of Paper II (light grey overlapping).
	Similarly, the histograms on the right show the PDFs of A(V).}
	\label{fig:Av_z}
\end{figure*}

Figure~\ref{fig:Av_z} shows the distribution of redshifts and visual extinction, A(V), for the current quasar sample. For comparison, we also show the sample from Paper II. It is clear that the inclusion of {\it WISE} photometry has enabled us to probe higher amounts of reddening and reject low-redshift quasars while still having a very low level of stellar contamination in our candidate sample.

\subsubsection{Intervening dust}
\label{res:intervening}
Since our main focus for this survey is the intervening dusty systems, we search for absorption systems in the spectra. Table~\ref{tab:absorbers} lists the spectroscopic absorption redshift for all the identified absorption systems in this sample. As in Paper II, we model each spectrum using a combined dust model with a freely variable dusty absorber and an amount of extinction fixed in the quasar rest-frame. The details of this fitting procedure are described in Paper II.
Four quasars are selected by our statistical algorithm as having an intervening absorption system hosting LMC-type dust, see individual fits in Appendix~\ref{appendix:absorber-dust}. These four statistically identified absorption systems are given in Table~\ref{tab:absorbers} together with the spectroscopically identified intervening absorption systems. Two of the statistically identified systems match the spectroscopically identified redshift well within the 1-$\sigma$ uncertainty of the fit. The other two show no absorption at the best-fit redshift. We identify no absorbing systems with SMC-type reddening, which is to be expected due to the degeneracy between dust in the quasar itself (assumed to be SMC type) and dust in the absorber (see Paper II). Two plausible DLAs are identified (eHAQ0930+0148 and eHAQ2359+1354) with intermediate amounts of SMC-type dust reddening along the line of sight. However, we observe no evidence for the dust reddening being caused by the absorber. Due to the poor spectral resolution it is not possible to securely obtain the column densities of \ion{H}{1} from \lya.

\begin{deluxetable*}{llcccl}
\tablecaption{Intervening Absorption Systems.
\label{tab:absorbers}}
\tablehead{ \colhead{Target} &
\colhead{$z_{\rm spec}$\tablenotemark{$a$}} &
\colhead{$z_{\rm abs}$\tablenotemark{$b$}} &
\colhead{${\rm A(V)_{abs}}$\tablenotemark{$b$}} &
\colhead{${\rm A(V)_{QSO}}$} &
\colhead{Notes}
}
\startdata
eHAQ0019+0657  & 1.671	 & \nodata       & \nodata & 0.17 & \ion{Mg}{2} absorber  \\
eHAQ0104+0912  & 1.487	 & $1.52\pm0.03$ & 0.20	   & 0.16 & \ion{Mg}{2} absorber  \\
eHAQ0111+0641  & 2.027	 & $2.04\pm0.09$ & 0.21	   & 0.15 & \ion{Fe}{2} absorption  \\
eHAQ0113+0804  & 1.589	 & \nodata       & \nodata & 0.60 & \ion{Mg}{2} absorber  \\
eHAQ0321+0523  & \nodata & $2.02\pm0.02$ & 0.23    & 0.15 & {\it No absorption lines in spectrum}  \\
eHAQ0835+0127  & 1.717	 & \nodata       & \nodata & 0.14 & \ion{Fe}{2} absorption, tentative  \\
eHAQ0930+0148  & 2.720	 & \nodata       & \nodata & 0.24 & \ion{Si}{2} absorption, Ly$\alpha$  \\
eHAQ1109+1058  & 1.667	 & \nodata       & \nodata & 0.43 & \ion{Mg}{2} absorber  \\
eHAQ1455+0705  & 1.650	 & $1.22\pm0.01$ & 0.42	   & 0.23 & \ion{Mg}{2} absorber  \\
eHAQ2359+1354  & 2.248\tablenotemark{$c$}   & \nodata       & \nodata & 0.13 & \ion{Mg}{2} absorber, Ly$\alpha$\\
\enddata
\tablenotetext{a}{Spectroscopically identified absorption redshift.}
\tablenotetext{b}{Best-fit parameters from intervening dust model. Only cases for which dust in an intervening absorber was preferred are shown. ${\rm A(V)_{abs}}$ assumes LMC type dust in the rest-frame of the absorber.}
\tablenotetext{c}{Based on the spectrum from BOSS.}
\end{deluxetable*}

\subsubsection{Peculiar Quasars}
\label{res:peculiar}
As mentioned above, some quasars are not well described by the quasar template assuming SMC-type reddening. The optical spectrum, in most cases, requires a higher amount of A(V) than what is needed to fit the NIR photometry. This is indicated in Table~\ref{tab:sample} as a range in A(V). We have tried to apply the steeper extinction curve inferred by \citet{Zafar2015}, but this does not provide a good fit either (except for two cases, marked by a $b$ in Table~\ref{tab:sample}). The extinction curve needed to fit these peculiar objects is even steeper than the curve by Zafar et al. and probably exhibits a break in order to reconcile the data.

\begin{figure*}
	\includegraphics[width=0.48\textwidth]{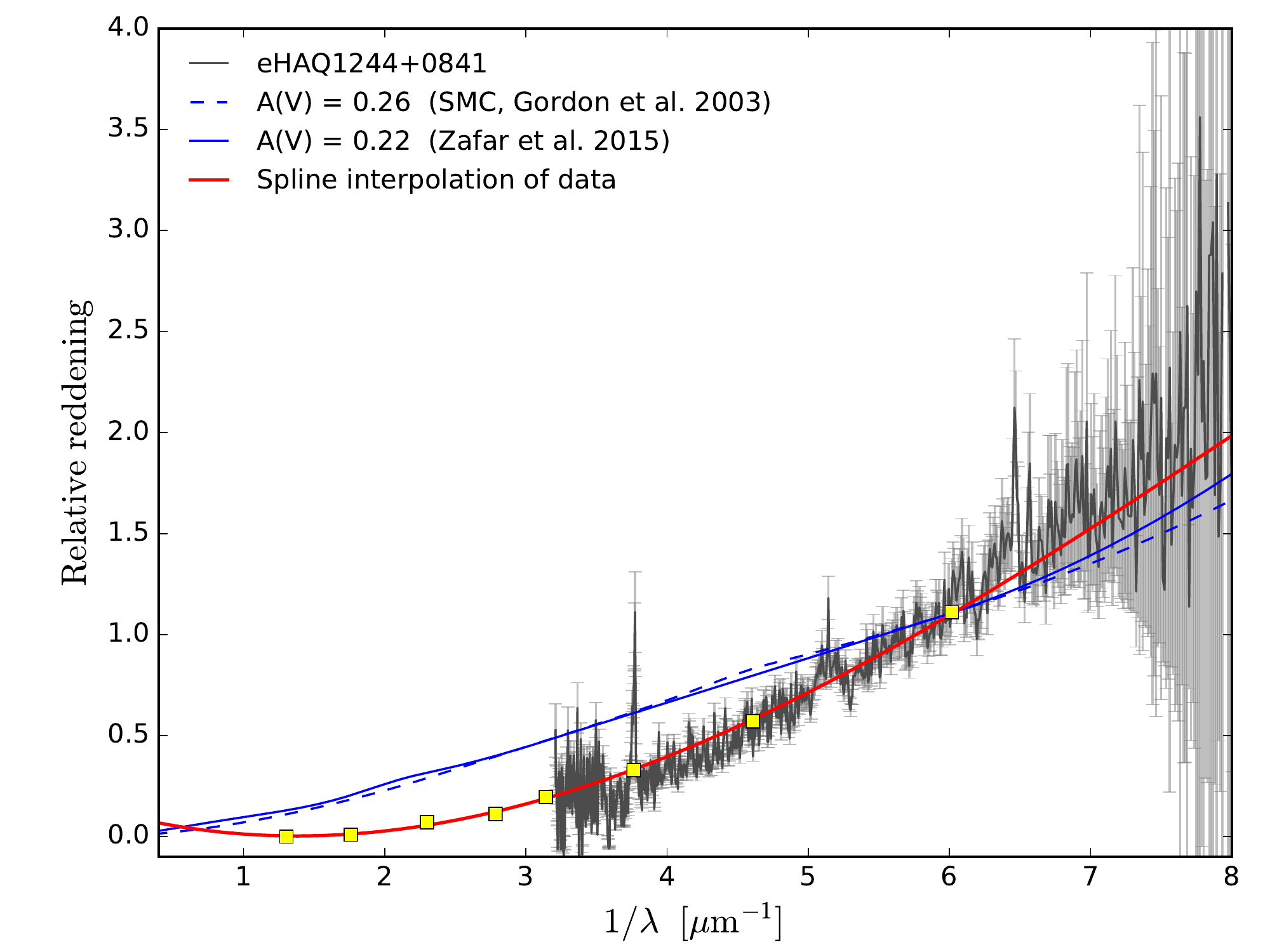}
	\includegraphics[width=0.48\textwidth]{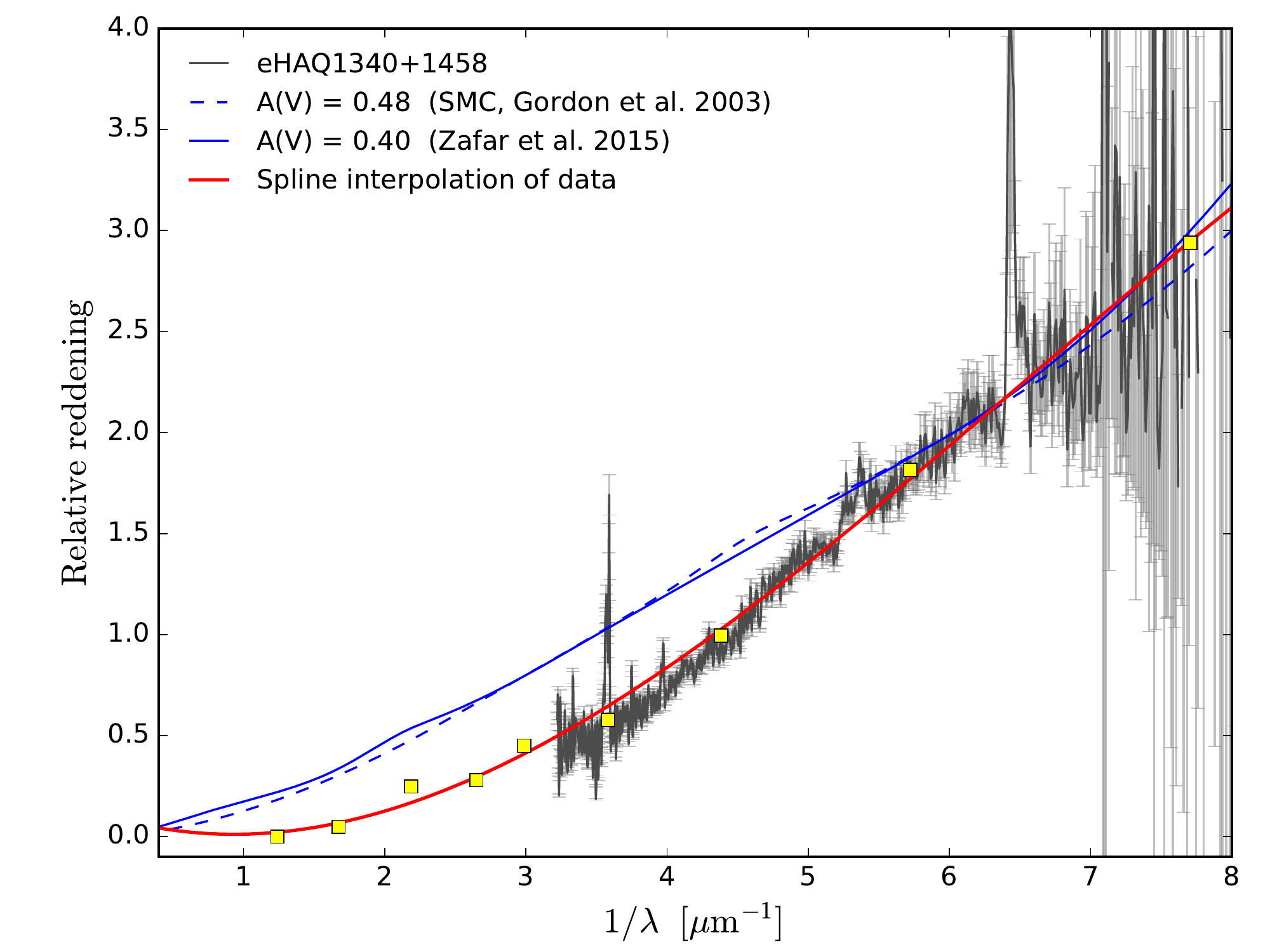}
	\caption{Reddening curves for two quasars: eHAQ1244+0841 ({\it left})
	and eHAQ1340+1458 ({\it right}) shown in the rest-frame of the quasars.
	The spectral data are shown as grey lines
	with errorbars and the broad-band photometry is shown as yellow squares.
	Since we do not know the intrinsic brightness of the quasar, the data have
	been normalized, arbitrarily, to the $K$-band.
	The data thus only indicate the relative reddening and not the total extinction.
	The blue curves show two dust models with varying amounts of reddening to
	best match the data; however, it is clear that they do not match
	the curvature of the data. For comparison, we highlight the
	shape of the reddening curve inferred from the broad-band photometry by a spline
	interpolation (shown in red). The prominent curvature around
	$\lambda^{-1}=3.5~\mu$m$^{-1}$ gives rise to the so-called
	{\it 3000~\AA\ break} in the observed spectra reddened by this type of extinction.}
	\label{fig:steep_extinction}
\end{figure*}

For two of these peculiar quasars (where we do not observe strong absorption), we attempt to recover the reddening curve in order to study the sight-lines in more detail, see Fig.~\ref{fig:steep_extinction}. We stress that since we do not know the absolute intrinsic normalization for these lines-of-sight, we only infer the relative reddening curve and not the absolute extinction.

First we normalize the quasar template \citep{Selsing2016} to the $K$-band and then we calculate the relative reddening, $k_{\lambda}$, as 
\vspace{-1mm}
$$k_{\lambda} = -2.5\, \log (D_{\lambda} / T_{\lambda})\ \ ,$$

where $D_{\lambda}$ refers to the observed data and $T_{\lambda}$ refers to the flux predicted from the template. For photometric data, we convolve the template with the corresponding filter transmission curves to obtain synthetic photometry. 
The resulting curves are shown in Fig.~\ref{fig:steep_extinction}, where we compare with the reddening curves of SMC-type dust \citep{Gordon2003} and quasar-type dust \citep{Zafar2015}, both arbitrarily normalized to illustrate the differences in slope. It is clear that the two known reddening laws provide an equally inadequate description of the data. Both curves can be made to fit part of the data, by correctly normalizing the reddening curve; however, this can only match either the infrared or the optical data -- not both simultaneously. The extinction curve needed to match the observed, peculiar spectra therefore needs a different curvature relative to the SMC-type curve or the curve by \citet{Zafar2015}.\\

We also identify some targets with very strong BAL features, which makes it impossible to determine the intrinsic continuum (e.g., eHAQ0104+0756, eHAQ0044+1321, eHAQ0913+0910). Similarly, we identify a few weak line quasars for which the quasar template provides a bad description since the template was compiled from luminous quasars with regular emission lines. The reddening estimates in these cases are therefore not representative.

Another type of objects denoted as {\it peculiar} are the quasars for which the template seems to provide a bad match over a small spectral range. This is typically caused by intrinsic variations in the pseudo-continuum caused by a large blend of broad iron emission lines from \ion{Fe}{2} and \ion{Fe}{3} and can to some degree be accounted for by including the emission template from \citet*{Vestergaard2001}.

For all these {\it peculiar} cases we mark the quasar by an $a$ in Table~\ref{tab:sample}. The reddening measurement should in these cases be considered rough estimates which will need further investigation over a larger wavelength baseline. For individual comments about the quasars, we refer the reader to the Appendix~\ref{appendix:notes}.\\[1cm]

\subsubsection{Photometric fitting of weak emission-line quasars}
\label{WLQ:phot_fit}
For two quasars (eHAQ0839+0556 and eHAQ1340+0151) classified as {\it peculiar}, we fit the photometry alone, disregarding the spectrum, because the emission lines are very weak and barely visible.
Since the spectrum exhibits no emission features, we use the continuum template of \citet{Richards2006} and apply reddening using the extinction curve of \citet{Zafar2015}, as the slightly different curvature of this extinction curve compared to the SMC curve provides a better fit to the rest-frame UV data. Since both of these quasars appear to have little reddening in the NIR photometry but a high amount of reddening in the optical spectrum, we fit the spectra allowing the intrinsic power-law slope of the quasar and the steepness ($R_V$) of the extinction curve to vary. We parametetrize the change in the power-law slope by the offset, $\Delta \beta$, relative to the intrinsic power-law slope of the template, $\alpha_{\lambda}=-1.7$ (measured from the template by Richards et al.  in the rest-frame wavelength range from $0.1$ to $1.0$~$\mu$m), where $f_{\lambda} \propto \lambda^{\alpha_{\lambda}}$.

\begin{figure*}
	\includegraphics[width=0.5\textwidth]{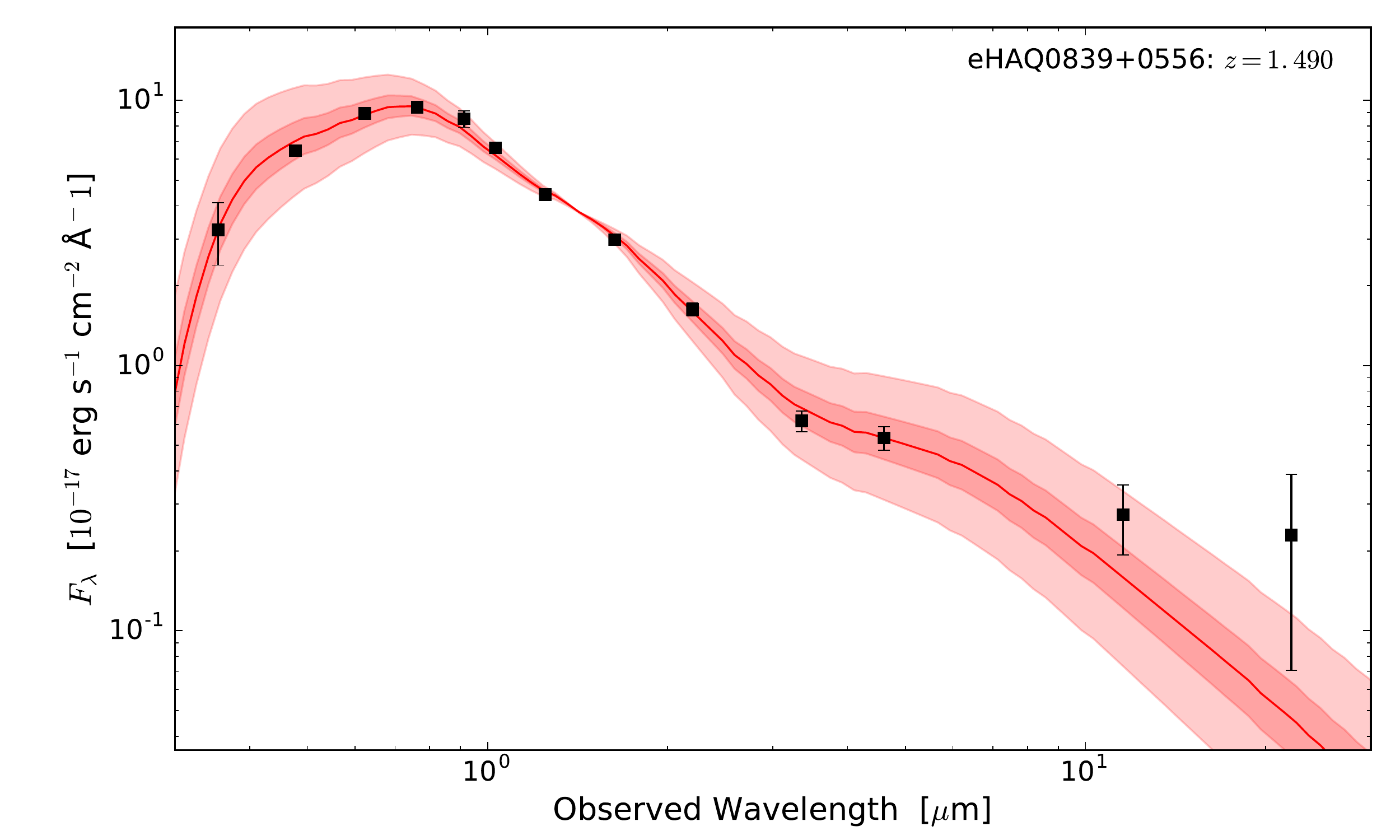}
	\includegraphics[width=0.5\textwidth]{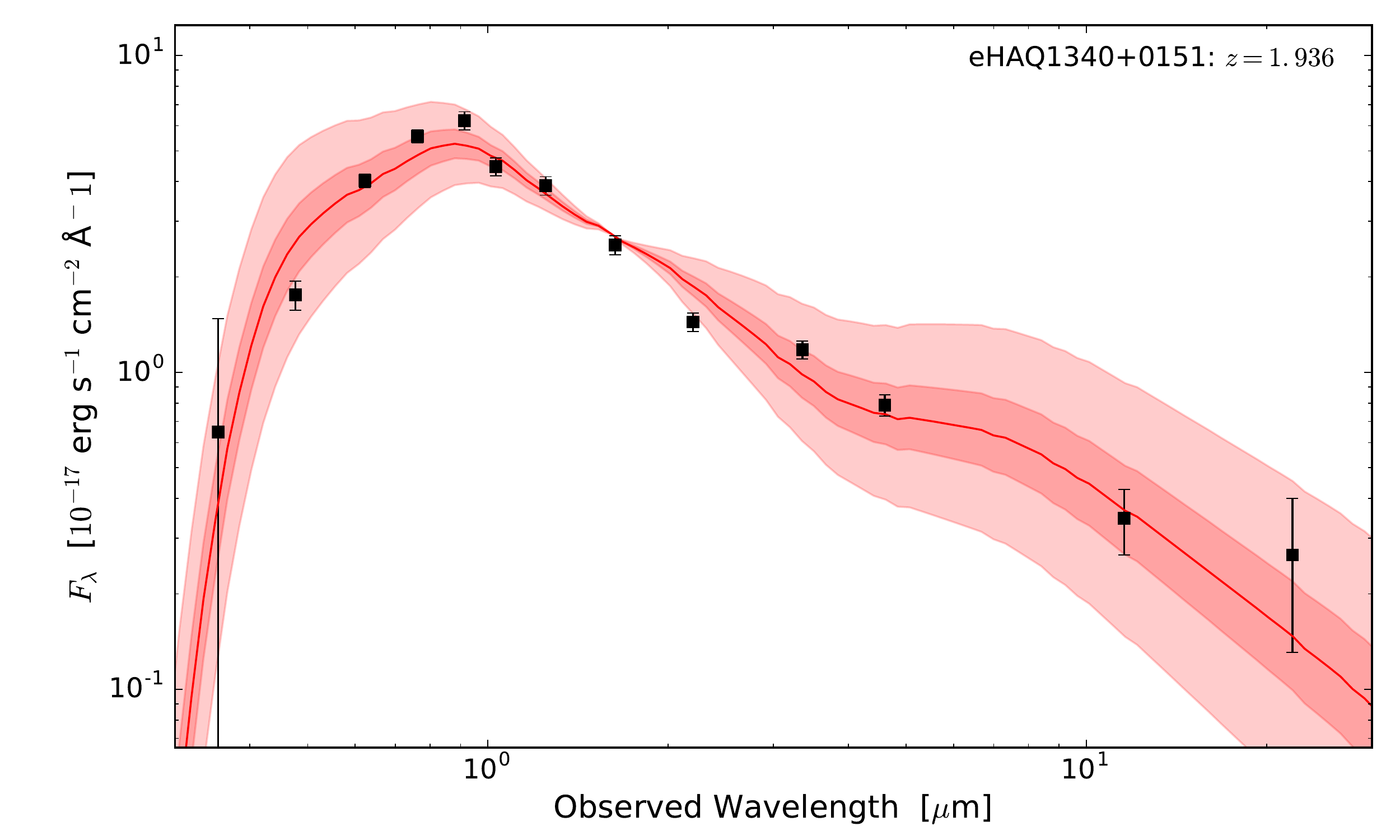}
	\caption{Spectral energy distribution of the weak line quasar eHAQ0839+0556 ({\it left}) and the target, eHAQ1340+0151, tentatively identified as a weak line quasar ({\it right}). Broad-band photometric data are shown as black squares with errorbars. The best-fit model is shown by the red curve surrounded by the 1 and 3 $\sigma$ confidence intervals shown by dark and light red shaded areas, respectively.}
	\label{fig:phot_fit}
\end{figure*}

Since all these parameters are highly degenerate, we use a Markov-chain Monte Carlo routine\footnote{We use the Python code {\tt emcee} written by \citet{emcee}} to constrain the parameters using realistic priors for the power-law slope and $R_V$. For the quasar power-law slope, we use a Gaussian prior centered around $\Delta\beta=0$ with a 1-$\sigma$ width of $0.2$ \citep{Krawczyk2015, Selsing2016}. Similarly, for $R_V$ we use a Gaussian prior centered on $R_V=2.4$ with a 1-$\sigma$ width of $0.3$ \citep{Zafar2015}. Since the spectral identification for eHAQ1340+0151 is merely tentative, we allow the redshift to vary.
The resulting best fits are shown in Figures~\ref{fig:phot_fit}. In both cases, the model performs very well and demonstrates that these weak line quasars can be described by a continuum-only model. The best-fit parameters for both quasars are listed in Table~\ref{tab:mcmc_fit}. The best-fit relative slopes ($\Delta\beta$) both result in a much steeper intrinsic quasar template needed to fit the data, however, these values of $\Delta\beta$ are heavily degenerate with the shape parameter of the extinction curve, $R_V$, and should be considered indicative only. We also caution the reader, that while the continuum-only model seems to provide a good description of the data, the template by \citet{Richards2006} was compiled from a sample of regular Type-1 quasars and hence might not capture the nature of this type of weak line quasars adequately.

\newpage

\begin{deluxetable*}{lccccc}
	\tabletypesize{\small}
	\tablecaption{SED modeling of weak line quasars
	\label{tab:mcmc_fit}}
	\tablehead{ \colhead{Target} & \colhead{$z$} &
				\colhead{$\Delta\beta$\tablenotemark{a}} &
				\colhead{$A(V)$} & \colhead{$R_V$} &
				\colhead{$\log(s)$\tablenotemark{b}}}
	\startdata
	eHAQ0839+0556	& $1.49$ & $-0.88 \pm 0.09$ & $0.64 \pm 0.10$ & $2.04 \pm 0.25$ & $-0.75 \pm 0.04$ \\
	eHAQ1340+0151	& $1.82 \pm 0.11$ & $-0.49 \pm 0.11$ & $0.85 \pm 0.11$ & $2.66 \pm 0.28$ & $-0.82 \pm 0.05$ 
	\enddata
	\tablenotetext{a}{Offset in power-law index relative to the intrinsic quasar
	template. For $\Delta\beta < 0$, the resulting template spectrum is steeper.}
	\tablenotetext{b}{The nuisance parameter, $s$, is an arbitrary scale
	factor in order to match the template to the absolute flux scale of the data.}
\end{deluxetable*}

\subsection{Stellar Contaminants}
\label{res:stars}

The stellar spectra are easily recognizable, except for a few cases (see below). We classify the stellar spectra using an automated algorithm that fits a library of stellar templates \citep{Pickles1998} to the spectra. The best-fit spectral type for each target is given in Appendix~\ref{appendix:stars}. The individual stellar spectra and the broad-band photometry are also shown in Appendix~\ref{appendix:stars}.

For one target (`eHAQ1132+1243') the classification is rather uncertain, due to the lack of strong spectral features. The target is classified as a metal-rich K4 giant. An alternative explanation could be that the target is a quasar at redshift $z=2.26$ with no strong emission lines. This is roughly consistent with the photometric redshift from quasar template fitting: $z_{\rm phot}=3.6 \pm 0.9$. Moreover, the quasar template from \citet{Richards2006} provides a better fit to the {\it WISE} photometry, and the quasar nature would more easily explain the apparent jump between the SDSS and UKIDSS photometry as intrinsic variability between the different epochs of observation in the two surveys. However, more data is needed to securely identify this target.

As mentioned in Sect.~\ref{selection}, we note that all the stellar contaminants have $J-K<0$, except for the two targets classified as K giants, including the just mentioned insecure identification.

\newpage

\section{Discussion}
\label{discussion}
In this work, we present the results of a new spectroscopic survey, the eHAQ survey, targeting reddened quasars with the aim of finding dust-rich absorption systems.
By selecting point sources from a cross-matched catalog of SDSS, UKIDSS and {\it WISE}, we have compiled a sample of 1073 quasar candidates based on their MIR properties. Motivated by results from the previous HAQ survey (Paper II) that did not include {\it WISE}-based selection, we incorporate a criterion to remove low-redshift quasars from our sample. As seen in Fig.~\ref{fig:Av_z}, the redshift distribution is efficiently truncated for $z<1.5$ compared to the distribution of redshifts from Paper II.

Of the 108 targets that we observed, we identify 100 quasars and 8 stars. However, all (except one) of these 8 stars were observed in the first observing run after which we added a criterion to remove stellar contamination ($J-K>-0.05$ for quasars, see Sect.~\ref{selection}).
After correcting the sample for stellar contamination we are able to remove 6 of the 8 stars while retaining all the quasars. We therefore have an efficiency of 98\% for selecting quasars.
Applying the correction for stellar contamination to the parent sample of 1073 candidates yields 880 high-confidence candidate quasars, of which 232 (26\%) are already observed by the SDSS and BOSS surveys.

\subsection{Peculiar Quasar Properties}
Similar to our previous surveys presented in Paper I and II, we observe cases where previously parametrized extinction curves \citep{Gordon2003, Zafar2015} do not reproduce the observed reddening \citep[see also][]{Hall2002, Meusinger2016}. In this work, we identify 14 such cases (Sect.~\ref{res:peculiar}).
We illustrate the relative reddening inferred from the data in two cases, and we compare two known reddening laws to the data. While both reddening laws can be made to match the data for a small wavelength range, neither of them can reproduce the full SED.
We therefore conclude that not only is the needed reddening law steeper than SMC [which merely results in a lower absolute extinction A(V)], but the curvature must also be altered in order to reproduce the apparent breaks in the observed spectra around $\lambda_{\rm rest}\approx3000$~\AA\ ($x\equiv 1/\lambda \approx3.5~\mu$m$^{-1}$; Fig.~\ref{fig:steep_extinction}) -- the so-called {\it `3\,000 \AA\ break'} \citep{Meusinger2016}.

Some part of the mismatch might be ascribed to differences between the assumed template spectrum and the actual intrinsic nature of the quasars in question, e.g., a different slope of the underlying power-law of the quasar spectrum. However, a change in the quasar power-law slope would not significantly change the curvature but only the slope of the inferred reddening laws. Similarly, a significant change in the broad emission lines or broad absorption features would show up as strong wiggles in the observed reddening curves instead of the smooth curve that we observe. The small features observed around $x\approx 5~\mu$m$^{-1}$ and $x\approx 6~\mu$m$^{-1}$ in Fig.~\ref{fig:steep_extinction} are indeed due to variations in the emission around \ion{C}{3} and \ion{C}{4}, but they are very weak and within the expected variance in these regions \citep[see discussion by][]{Selsing2016}.

Instead, the change in curvature of the reddening law most likely reflects changes in the dust grain properties in the quasars. Such changes could be due to differences in the grain composition or due to a different grain-size distribution. Particularly, a lack of larger grains (typically $0.3~\mu$m and larger) leads to steeper and more curved extinction curves \citep{Draine1984, Draine2003}. This can be caused by dust destruction or inhibited grain growth, both are viable explanations in the harsh quasar environments. However, it is then puzzling why we do not observe this type of extinction towards every quasar, as also noted by \citet{Leighly2014}. One explanation for this might be that low amounts of reddening effectively masks the curvature (i.e., the noise drowns the shape of the reddening over the full SED). This is especially the case when only a short wavelength range is available. 
However, as discussed in \citet{Goobar2008} and \citet{Leighly2014}, the extinction curve also depends critically on the assumed geometry. Since the dust in the quasar environment is hardly a uniform screen (but more likely distributed all around the central emitting region), the extinction curve can vary significantly due to multiple scattering as the photons propagate through the dusty medium.

Another explanation for the lack of observed curved reddening laws could be due to a clumpy distribution of dust in the surrounding environment where dusty clouds move in and out of the line-of-sight \citep{Leighly2015, Netzer2015}. This may explain why this type of dust is generally seen towards highly reddened quasars with large amounts of absorption and not for the regular quasar population, for which little dust is obscuring the view to central region.

The sample presented by \citet{Meusinger2016} consists of 23 quasars (out of several hundred thousand quasar spectra in SDSS and BOSS) that exhibit the strange {\it 3\,000 \AA\ break} similar to the strange targets identified in this work, see Fig.~\ref{fig:steep_extinction}. For comparison, we identify 6 such quasars in our sample of only 100 quasars (when following broadly the same classification criteria for their {\it group A}). Moreover, the authors find that the peculiar quasars have a much higher fraction of radio detections ($f_{\rm radio}=74$~\%) than the overall SDSS+BOSS sample ($f_{\rm radio}\sim8$~\% [8307/98544]; \citealt{Balokovic2012}). \citeauthor{Meusinger2016} argue that this is most probably due to a selection bias, since quasars in this part of color- and redshift-space are more frequently targeted for spectroscopic observations based on preexisting radio detections. We observe no radio detections among the six quasars from our survey; hence consistent with the overall $f_{\rm radio}$ from SDSS+BOSS.

We inspect the number of targets with radio detections in our candidate sample and find $f_{\rm radio}=10\pm1$\% (83 out of 880) in our photometric sample. In our spectroscopic sample of 100 quasars, we find a total of 7 quasars with radio detections (6 from FIRST and 1 from NVSS). The fraction of radio detections for the sample presented here is thus fully consistent with the overall $f_{\rm radio}$ for quasars in SDSS+BOSS.
When comparing this to the SDSS+BOSS spectroscopic overlap of our candidates, we observe a much higher fraction of radio detections: $25\pm3$~\% of SDSS+BOSS quasars in our candidate sample (57/232) are detected in FIRST. This demonstrates that the SDSS+BOSS sample of quasars, in this part of color-space, is biased towards radio-bright quasars. It is therefore likely that the high fraction of radio detections observed for the {\it 3\,000 \AA\ break} quasars in \citet{Meusinger2016} is due to this selection artefact and not due to some intrinsic difference in the radio properties of these peculiar quasars. This is further supported by the fact that we observe no enhanced radio detection rate for the six quasars with 3000 \AA\ breaks relative to the parent eHAQ quasars.

While the quasars in our survey are not selected to provide a complete census of the quasar population (we select based on optical properties to look for intervening dust), the spectroscopically identified quasars presented in this survey still provide important observations of the quasar population which is otherwise sparsely sampled in SDSS and BOSS.

\subsection{Implications for Quasar Selection}
With the advance in recent quasar selection algorithms \citep[e.g.,][]{Richards2015, Peters2015}, new surveys (e.g., eBOSS and LSST) will target the population of red quasars in ever greater detail. Nevertheless, the lack of spectroscopically confirmed high-redshift quasars still limits our ability to model and predict the appearance of quasars. We illustrate this by comparing our sample of quasar candidates (with a purity of around 98\%) to the photometric study by \citet{Richards2015}.
We cross-match our sample of 880 high-confidence quasars with the photometrically classified quasar catalog of \citet{Richards2015}. The matching is performed on the coordinates within a matching-radius of 1 arcsec.
Out of 880 eHAQ target candidates, a total of 504 (57\%) are recovered in the catalog of \citet{Richards2015}. However, this includes 232 quasars from BOSS which made part of the learning sample for their algorithm. Excluding those already known quasars, we obtain a total of 305 matches out of 648 candidates with no pre-existing spectra. So, Richards et al. recover 47\% of our candidates. This is also observed in the spectroscopic sample of 100 quasars identified in this work, 47 are recovered in the Richards et al. catalog.

We can compare the number of eHAQ quasars to the total number of quasars selected by Richards et al. in the same area of the sky. We do this by selecting only quasars from the catalog of Richards et al. for which UKIDSS photometry is available, since the smaller extent of the UKIDSS footprint (compared to SDSS and {\it WISE}) sets the limit for the available survey area in both studies\footnote{We remind the reader that we restrict our selection to declinations between $-4$\degr\ and $+17$\degr, see Sect.~\ref{selection}.}. In order to make a fair comparison, we only consider targets with $J_{\rm AB}<19$~mag (as this makes up the homogenous set of our sample). Moreover, since we reject low-redshift candidates ($z<1.5$) from eHAQ, we require a similar cut in the photometric redshifts derived by \citet{Richards2015}. For this purpose, we use the `{\sc zphotbestjhk}' estimate from Richards et al. and impose a conservative redshift limit of $z_{\rm phot}>1$. Using Table~2 of \citeauthor{Richards2015}, this yields a total number of 14,641 quasars. Restricting the eHAQ sample to $J_{\rm AB}<19$~mag yields 588 quasars (4\% of the total number), of which \citeauthor{Richards2015} recover about half, i.e., $\sim$2\% of quasars are not identified in the photometric catalog of \citet{Richards2015} down to a limit of $J_{\rm AB}<19$~mag. However, this fraction might be even larger due to the fact that the eHAQ sample is limited by the inclusion of the shallower bands 3 and 4 from {\it WISE}, whereas Richards et al. utilize the deeper AllWISE data for bands 1 and 2 only.

We note that the selection functions of both studies rely on optical and MIR data, however, we include additional NIR data from UKIDSS, which \citeauthor{Richards2015} only use for the estimates of photometric redshifts. Much of the difference in candidate selection will therefore be due to the inclusion of NIR data, since most of our ability to separate stars and quasars comes from the $J-K$ color cut, especially for targets with $W_1-W_2 < 1$~mag.

Other classification methods involving astrometric data \citep*{Koo1986,Heintz2015} and time variability data \citep[e.g.,][]{Koo1986,Schmidt2010,Butler2011,Graham2014} will also provide important constraints, although very large temporal baselines are needed to select high-redshift quasars through variability, due to the stretching of observed time-scales in an expanding universe.
Therefore, in order to push quasar classification to higher completeness (particularly at high redshift) deep NIR large-sky surveys will be of great importance.
Selecting quasars in these new ways will expand the population of quasars dramatically, however, spectroscopic confirmation of the candidates is still very time consuming, and exploratory surveys like our HAQ survey (Paper I and II) and others \citep{Maddox2008, Maddox2012, Glikman2006, Glikman2007, Glikman2012, Glikman2013, Banerji2013, Hainline2014} provide important sampling of the parameter space for the advanced classification algorithms.

\subsection{Dusty Absorption Systems}
Only two absorbers are detected in Ly$\alpha$ both of which are consistent with having $\log N_{\rm H \textsc{i}}>20.3$ but due to the poor spectral quality it is not possible to determine their column densities precisely. The amount of reddening inferred from the spectral fit for these two Ly$\alpha$ absorbers (eHAQ0930+0148 and eHAQ2359+1354) is ${\rm A(V)} = 0.24$ and ${\rm A(V)}=0.09$, respectively. Nonetheless, we find no strong evidence for the dust reddening being caused by the absorbers. Thus, we conclude that no dusty DLAs are observed in this survey. However, due to the large fraction of BAL quasars, only 50 (out of a total of 100) quasars are not heavily absorbed by BAL features, which hampers the detection of intervening absorption lines. Of these 50, only 28 are at sufficiently high redshift to detect Ly$\alpha$ ($z_{\rm abs}>1.8$). Thus the sample size is too small to draw firm conclusions about the incidence rate of dusty DLAs in this part of color-space.

We do however detect two absorption systems with evidence for the 2175\AA\ dust feature. One of these is at too low redshift for us to detect Ly$\alpha$, and for the other absorber, the Ly$\alpha$ line falls right at the edge of the spectrum where the noise is very high. We are therefore not able to securely say anything about the column density of \ion{H}{1} for these absorption systems.

For the remaining absorption systems, we do not see any clear evidence for the dust being in the absorber, i.e., we do not see any significant evidence for the 2175~\AA\ bump in the rest-frame of the absorption systems; Nevertheless, this cannot be completely ruled out with the current data, since SMC-type extinction in the absorber would be impossible to distinguish from SMC-type dust in the quasar with the limited data available here \citep[e.g.,][]{Krogager2016}.

The non-detection of dusty DLAs in our sample shows that these are rare and do not make up a significant part of the population in terms of the incidence rate of neutral hydrogen. However, their importance for studies of cold gas and molecules at high redshift \citep[e.g.,][]{Noterdaeme07, Srianand2008a, Ledoux2015, Krogager2016} still needs to be quantified with larger samples in order to be conclusive. These cold absorbers hold crucial information about the \ion{H}{1}-to-H$_2$ transition and the interplay between dust and gas-phase chemistry \citep{Noterdaeme2015b, Noterdaeme2016_prep}. With more and more complex quasar selection methods, using various wavelength baselines correlated with astrometric and temporal-variability data, the discovery of dusty and reddened intervening absorbers will be more frequent due to a more complete parent quasar sample.\\

\acknowledgements

We thank the anonymous referee for the constructive comments that helped to improve the quality of this work.
The authors thank the many students who dedicated their time
to observing quasars during the summer school at the Nordic Optical Telescope in 2015. JK acknowledges financial support from the Danish Council for Independent Research (EU-FP7 under the Marie-Curie grant agreement no. 600207) with reference DFF-MOBILEX--5051-00115. 
JPUF acknowledges support from the ERC-StG grant EGGS-278202.
PN acknowledges support from the Agence Nationale de la Recherche under grant ANR-12-BS05-0015.
MV gratefully acknowledges financial support from the Danish Council for Independent Research via grant no. DFF 4002-00275.
Based on observations made with the Nordic Optical Telescope, operated on the island of La Palma jointly by Denmark, Finland, Iceland, Norway, and Sweden, in the Spanish Observatorio del Roque de los Muchachos of the Instituto de Astrof\'isica de Canarias. 
Funding for SDSS-III has been provided by the Alfred P. Sloan Foundation, the Participating Institutions, the National Science Foundation, and the U.S. Department of Energy Office of Science. The SDSS-III web site is \url{http://www.sdss3.org/}.
SDSS-III is managed by the Astrophysical Research Consortium for the Participating Institutions of the SDSS-III Collaboration including the University of Arizona, the Brazilian Participation Group, Brookhaven National Laboratory, Carnegie Mellon University, University of Florida, the French Participation Group, the German Participation Group, Harvard University, the Instituto de Astrof\'isica de Canarias, the Michigan State/Notre Dame/JINA Participation Group, Johns Hopkins University, Lawrence Berkeley National Laboratory, Max Planck Institute for Astrophysics, Max Planck Institute for Extraterrestrial Physics, New Mexico State University, New York University, Ohio State University, Pennsylvania State University, University of Portsmouth, Princeton University, the Spanish Participation Group, University of Tokyo, University of Utah, Vanderbilt University, University of Virginia, University of Washington, and Yale University. 
This publication makes use of data products from the {\it Wide-field Infrared Survey Explorer}, which is a joint project of the University of California, Los Angeles, and the Jet Propulsion Laboratory/California Institute of Technology, funded by the National Aeronautics and Space Administration.

\bibliographystyle{apj}

\appendix

\section{Tables}
\label{appendix:tables}
\input{./sample_overview}
\newpage
\input{./sample_data}

\newpage

\section{Stellar Contaminants}
\label{appendix:stars}

The individual spectral classifications of stellar contaminants are given in Table~\ref{tab:stars}. Their spectra and broad-band photometry are shown in Fig.~\ref{fig:stars}.

\begin{deluxetable}{ll}
\tabletypesize{\small}
\tablecaption{Stellar Classification
\label{tab:stars}}
\tablehead{ \colhead{Target} & \colhead{Spectral Type} }
\startdata
eHAQ0835+0830	&   K4\, {\sc v} \\
eHAQ0915+0115	&   K5\, {\sc v} \\
eHAQ1002+0406	&   M2\, {\sc v} \\
eHAQ1109+0135	&   M2\, {\sc v} \\
eHAQ1111+0151	&   K7\, {\sc v} \\
eHAQ1120+0812	&   K5\, {\sc iii} \\
eHAQ1132+1243	&   K4\, {\sc iii} {\small (metal-rich)}; tentative \\
eHAQ1331+1304	&   M1\, {\sc v}	 \\
\enddata
\tablecomments{The spectral types have been determined from a template matching algorithm.}
\end{deluxetable}

\begin{figure}[h]
\figurenum{B1}
\gridline{\fig{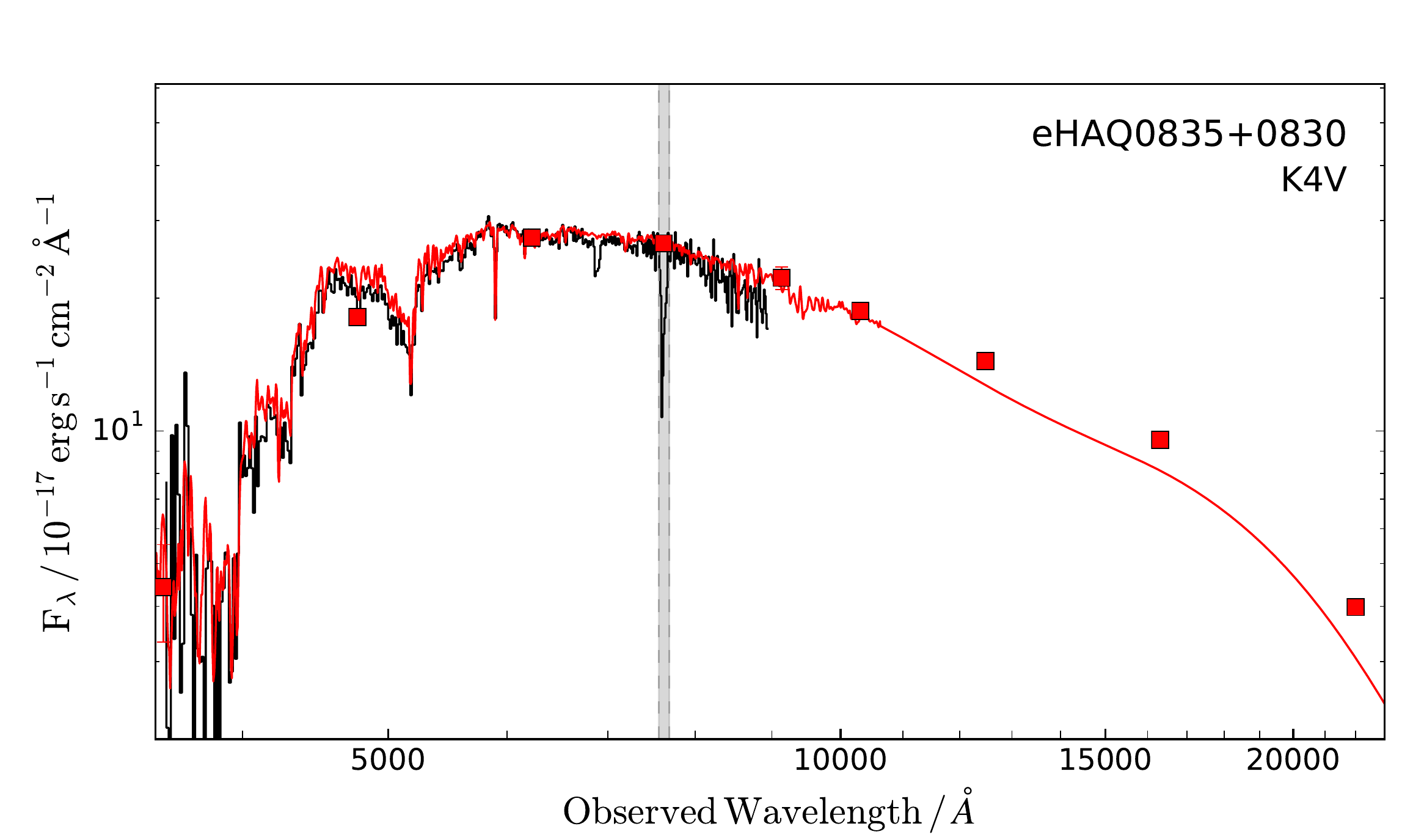}{0.45\textwidth}{(a)}
          \fig{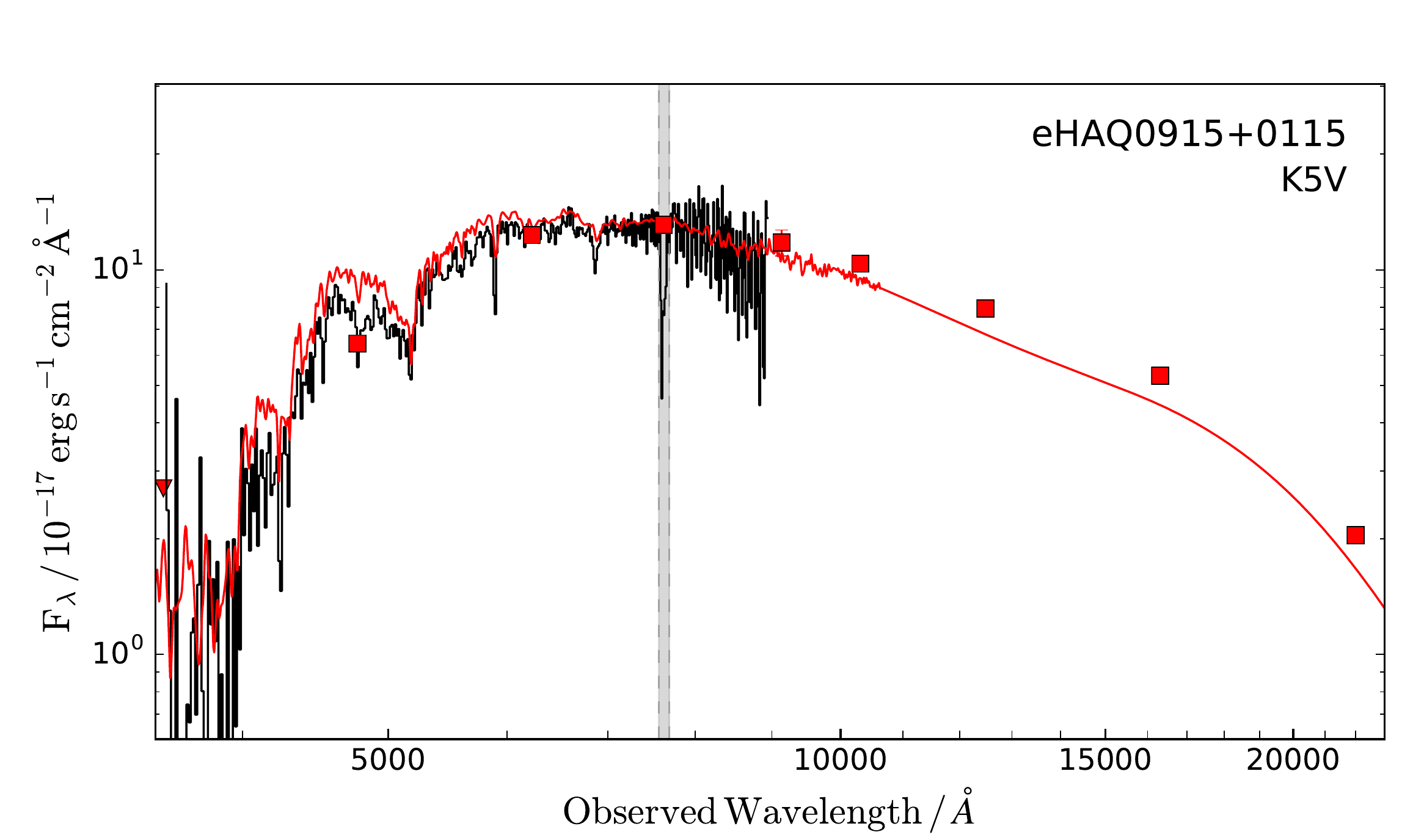}{0.45\textwidth}{(b)}
          }
\gridline{\fig{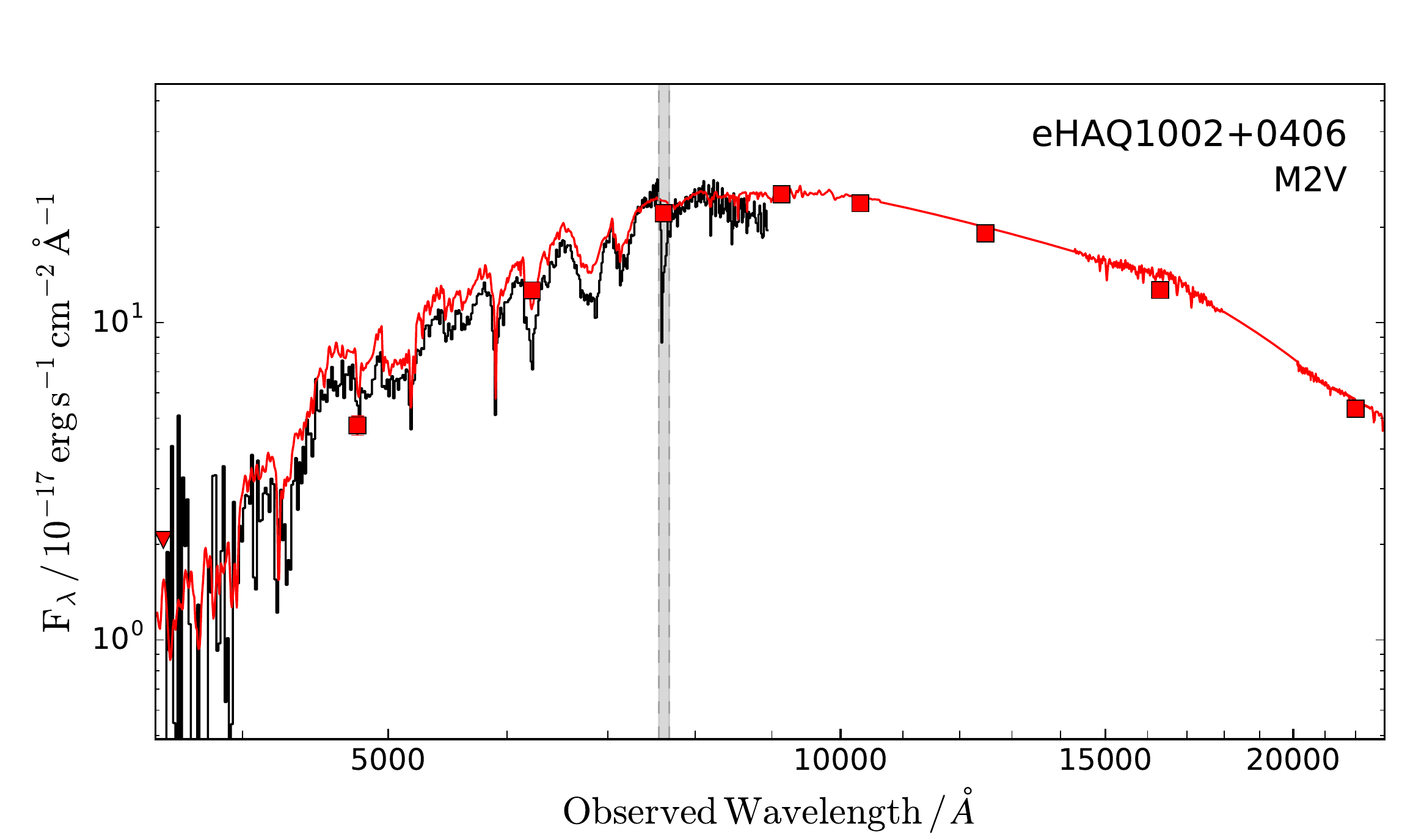}{0.45\textwidth}{(c)}
          \fig{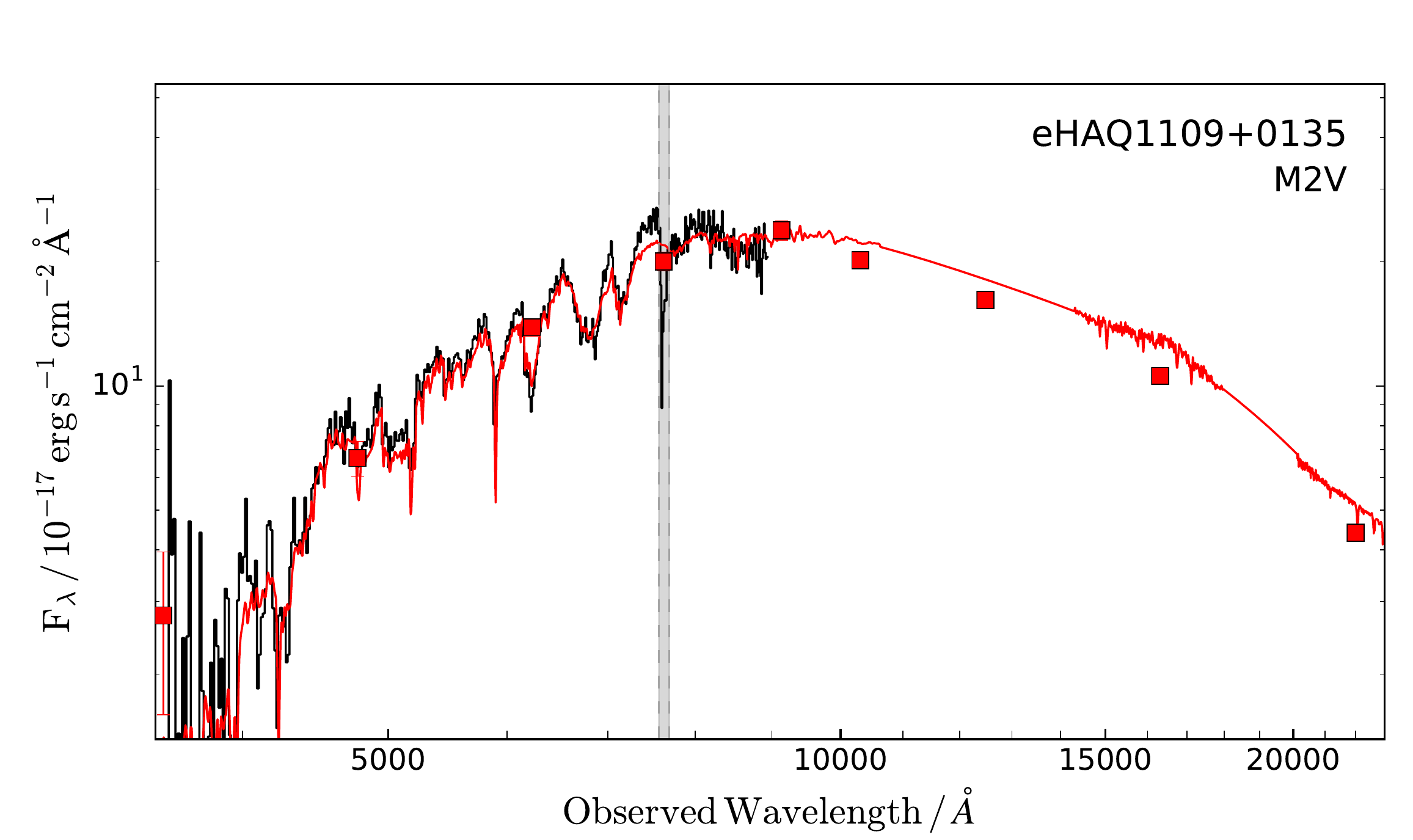}{0.45\textwidth}{(d)}
          }
\caption{Spectra of the objects classified as stars. The red line shows the best-fit stellar template. The spectral type is indicated in the upper right corner of each panel.
\label{fig:stars}}
\end{figure}

\newpage

\begin{figure}[h]
\figurenum{B1}
\gridline{\fig{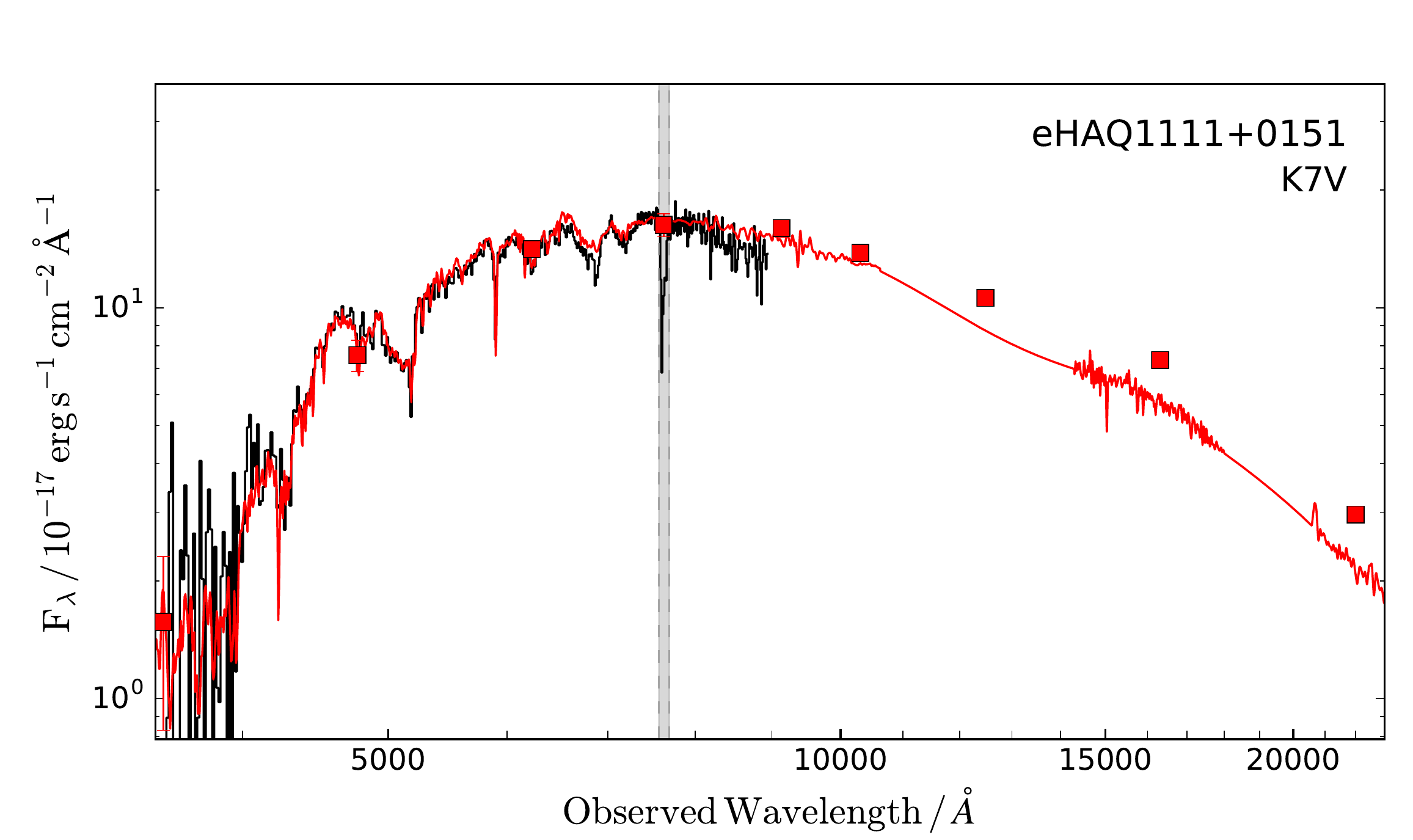}{0.45\textwidth}{(e)}
          \fig{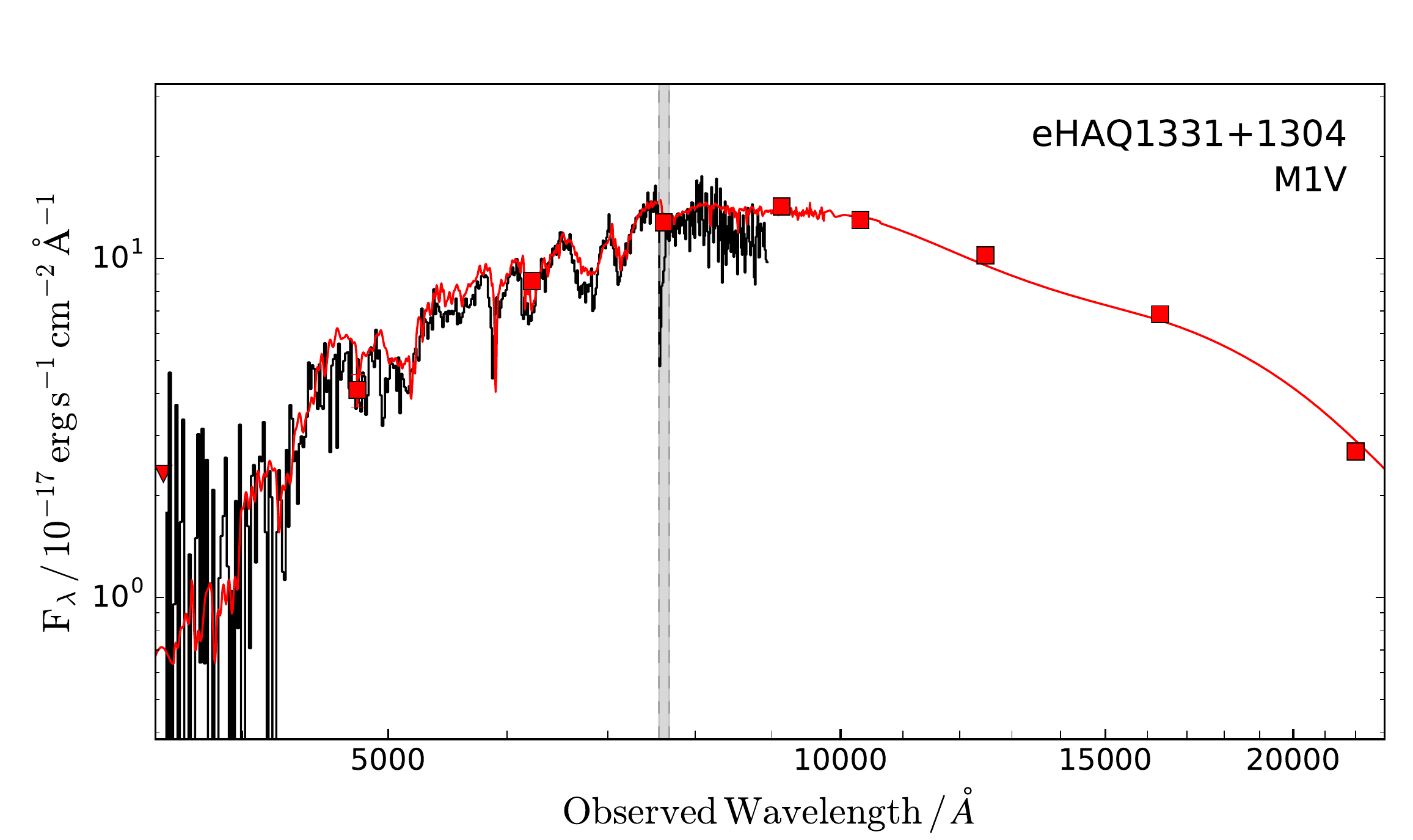}{0.45\textwidth}{(f)}
          }
\gridline{\fig{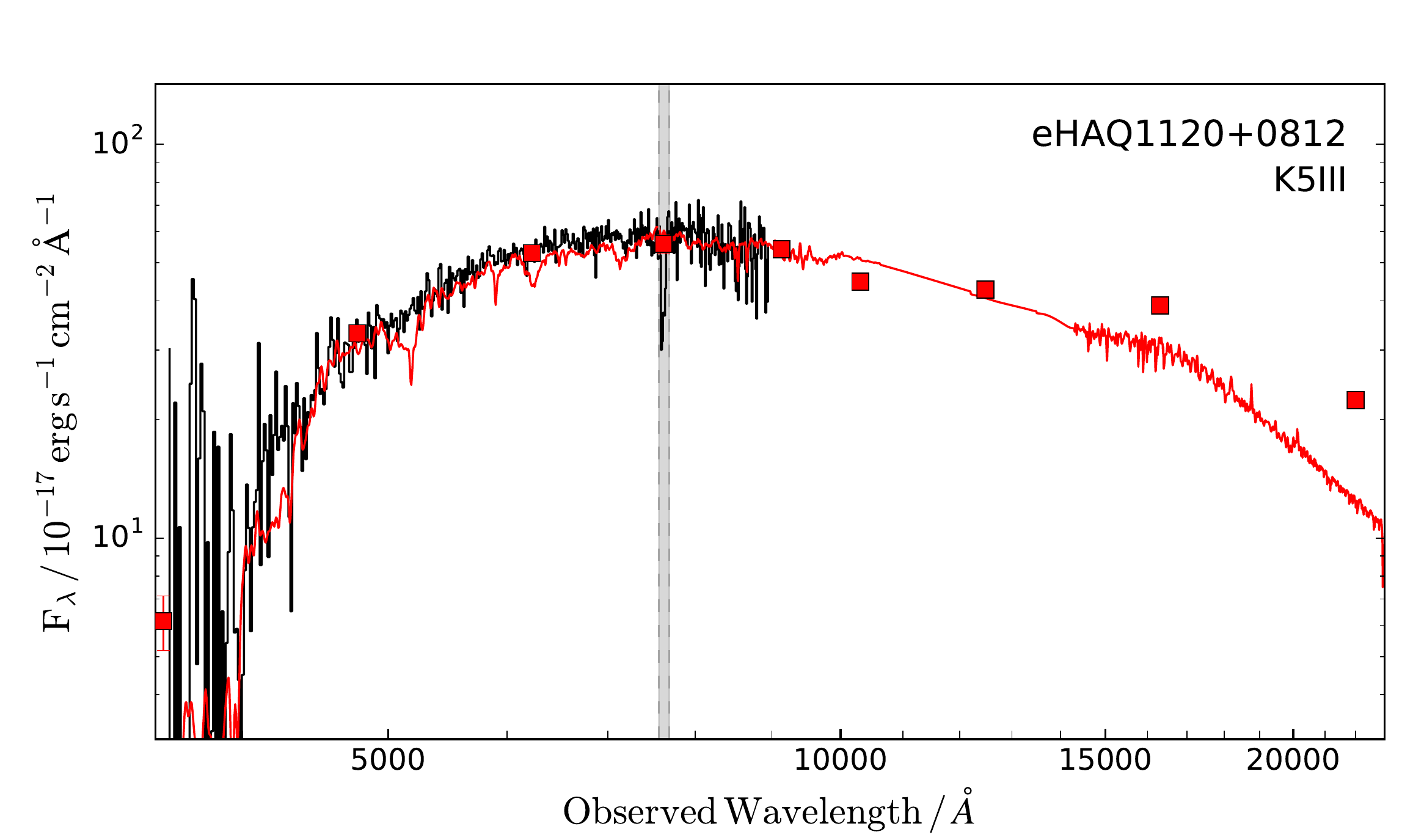}{0.45\textwidth}{(g)}
          \fig{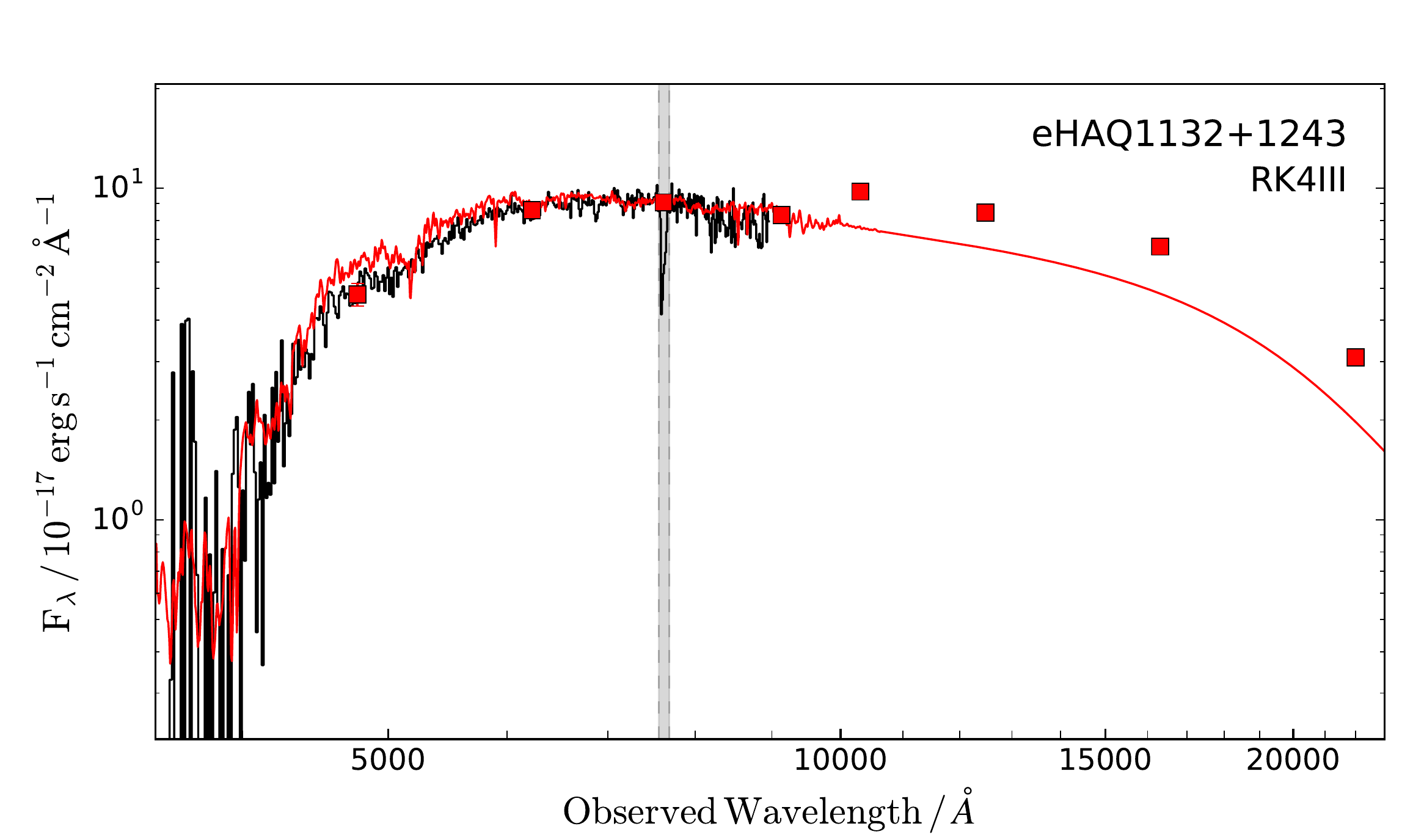}{0.45\textwidth}{(h)}
          }
\caption{(Continued.) The bottom row (panels {\it g} and {\it h}) shows targets with $J-K>-0.05$~mag.}
\end{figure}

\newpage

\section{Individual Fits for Dust in Intervening Absorbers}
\label{appendix:absorber-dust}
\begin{figure}[h]
\gridline{\fig{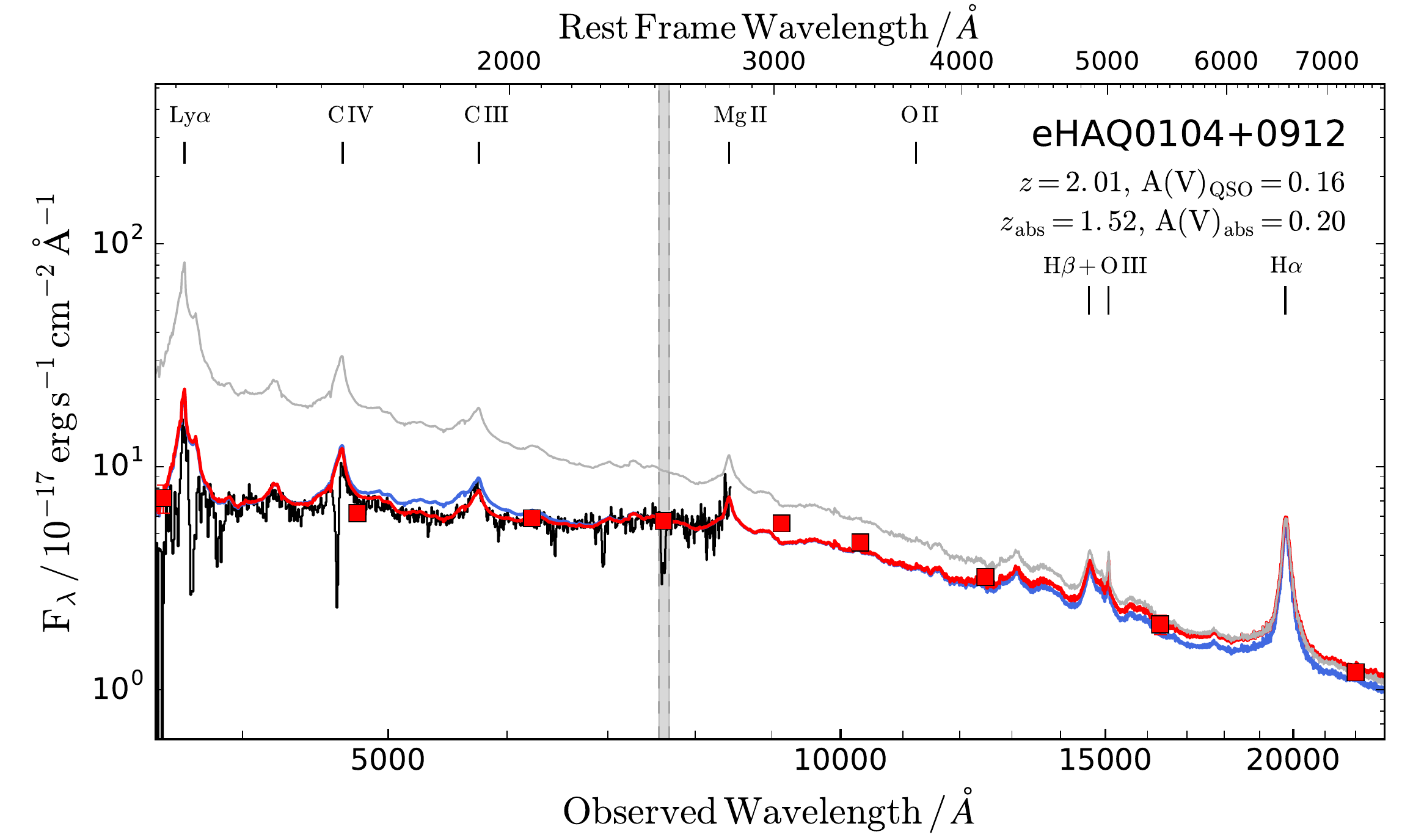}{0.5\textwidth}{(a)}
          \fig{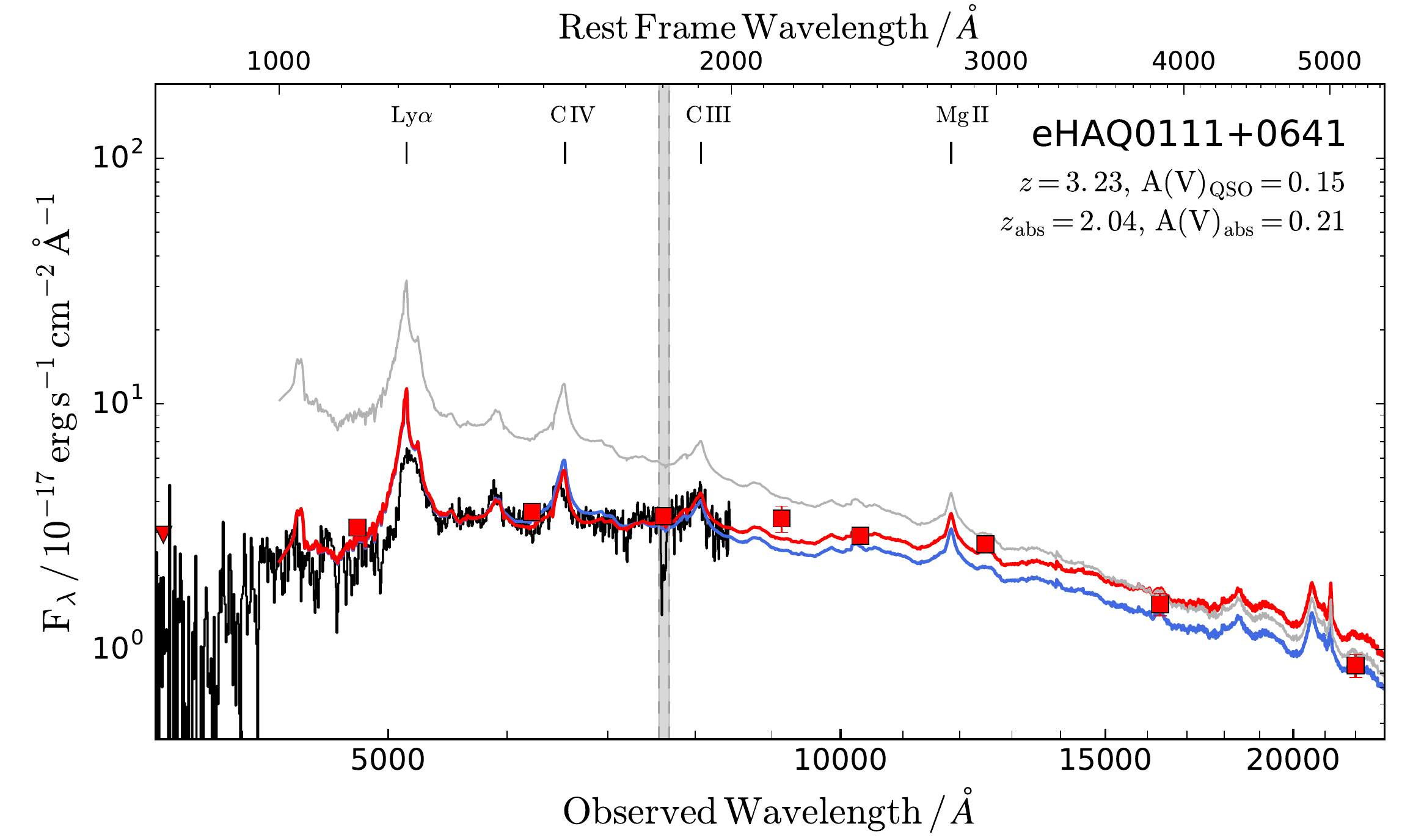}{0.5\textwidth}{(b)}
          }
\gridline{\fig{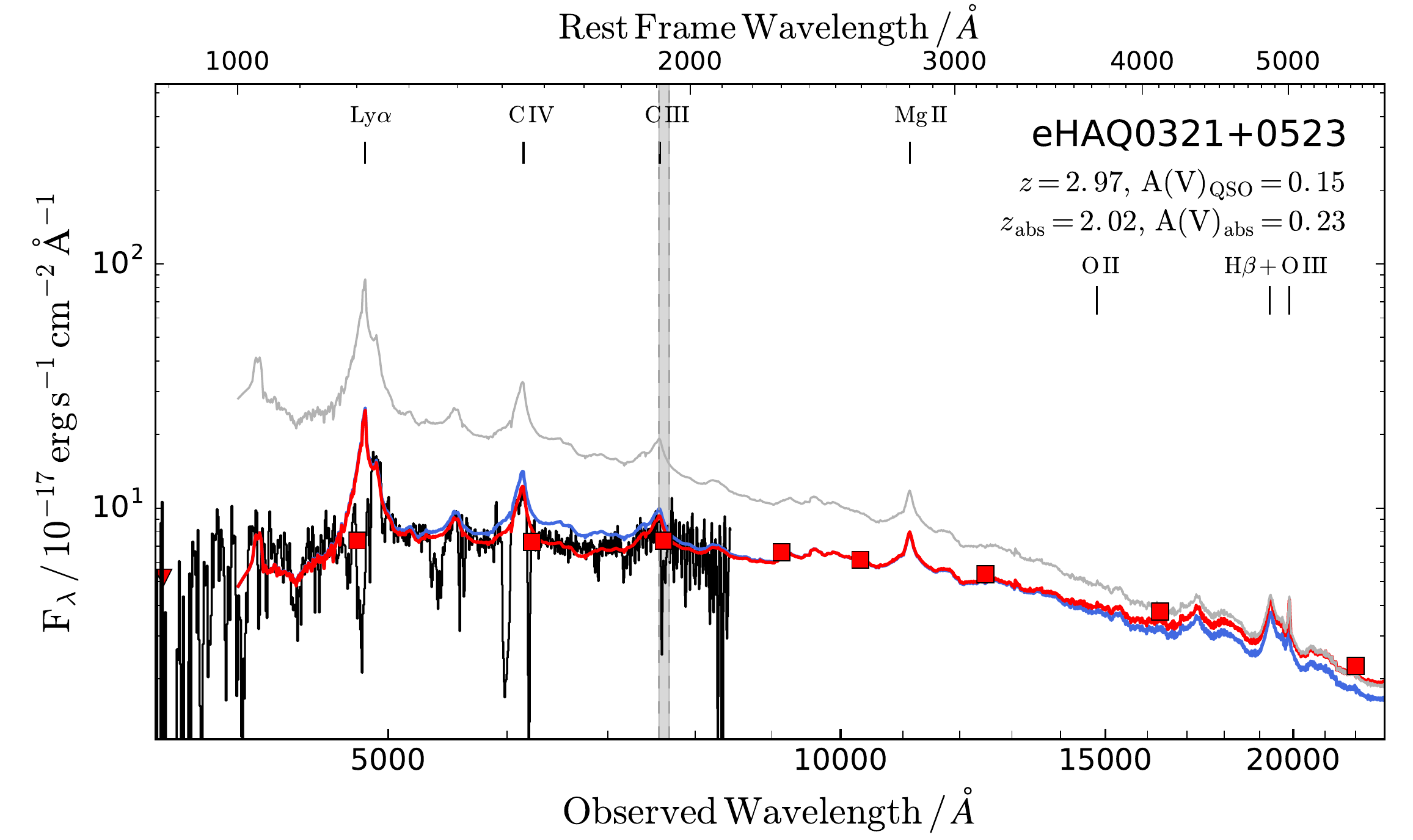}{0.5\textwidth}{(c)}
          \fig{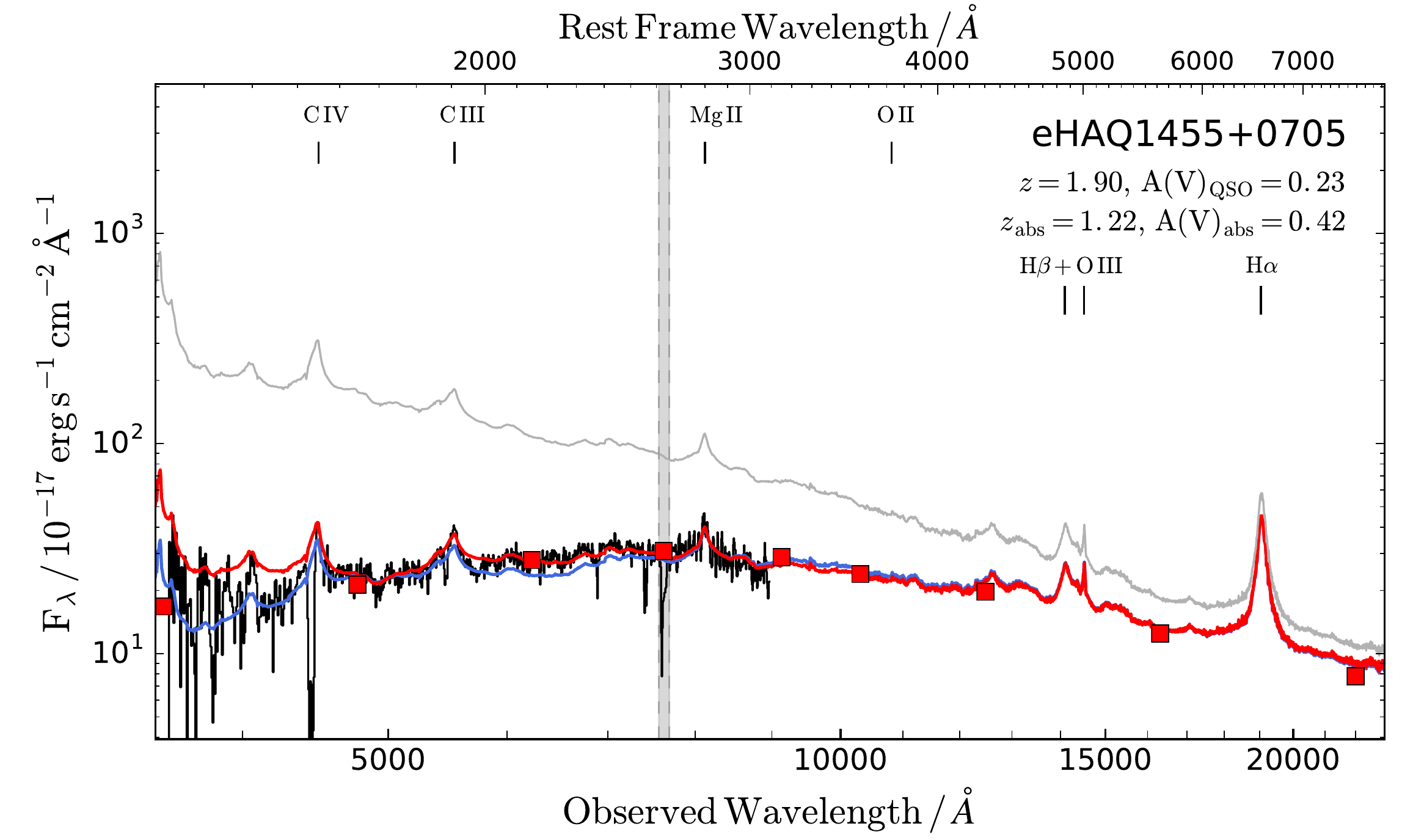}{0.5\textwidth}{(d)}
          }
\caption{Statistically identified intervening absorption systems
	with 2175~\AA\ dust bump. The upper row [panels (a) and (b)] shows the cases where an absorption line system has been identified with a spectroscopic redshift consistent with the best-fit $z_{\rm abs}$. No absorption line systems with a corresponding spectroscopic redshift to $z_{\rm abs}$ have been identified for the cases in the lower row [panels (c) and (d)].
	The observed spectrum is shown as a solid black curve.
	The filled squares indicate the SDSS and UKIDSS photometric data points.
	In the upper right corner we provide the estimated emission redshift,
	$z$, the amount of (SMC-type) extinction in the quasar's rest-frame, A(V)$_{\rm QSO}$, the best-fit absorption redshift, $z_{\rm abs}$, and amount of (LMC-type) extinction in the absorber's rest-frame, A(V)$_{\rm abs}$.
	The grey curve shows the quasar template at the estimated redshift of
	the quasar. The red template shows the best-fit dust model with an intervening dusty absorber at $z_{\rm abs}$. For comparison, the blue template indicates the best-fit dust model assuming that all dust is in the quasar.
	Note that the spectra have not been corrected for telluric absorption (marked with a grey band at $\sim7600$~\AA).
	\label{fig:absorber-dust}}
\end{figure}

In Fig.~\ref{fig:absorber-dust}, the spectra and broad-band photometry are shown together with the best-fit dust models for the four cases where an intervening system has been identified statistically. The best-fit parameters are summarized in Table~\ref{tab:intervening_dust}.

\newpage

\begin{deluxetable}{lcccccccr}
\tabletypesize{\small}
\tablecaption{Fit Parameters for Intervening Dust Models
\label{tab:intervening_dust}}
\tablehead{
\colhead{Target} &
\colhead{$z_{\rm QSO}$} &
\colhead{${\rm A(V)_{QSO}}$} &
\colhead{$z_{\rm abs}$} &
\colhead{${\rm A(V)_{abs}}$} &
\colhead{$L$\tablenotemark{(a)}} &
\colhead{$P_{\rm KS}$\tablenotemark{(b)}} &
\colhead{$p$\tablenotemark{(c)}}
}
\startdata
eHAQ0104+09 & 2.01 & $0.16\pm0.01$ & $1.52\pm0.03$ & $0.20\pm0.02$ & 92.0 & 0.20 & $1.1 \times 10^{-20}$ \\
eHAQ0111+06 & 3.23 & $0.15\pm0.02$ & $2.04\pm0.09$ & $0.21\pm0.04$ & 91.1 & 0.12 & $1.7 \times 10^{-20}$ \\
eHAQ0321+05 & 2.97 & $0.16\pm0.01$ & $2.02\pm0.02$ & $0.23\pm0.02$ & 89.3 & 0.24 & $4.1 \times 10^{-20}$ \\
eHAQ1455+07 & 1.90 & $0.23\pm0.02$ & $1.22\pm0.02$ & $0.42\pm0.03$ & 152.2 & 0.17 & $8.9 \times 10^{-34}$ \\
\enddata
\tablenotetext{$a$}{ The logarithm of the likelihood ratio of the maximum likelihood of the null model, $\Lambda_0$, and of the general model, $\Lambda_G$; $L = -2 \ln (\Lambda_0/\Lambda_G$).}
\tablenotetext{$b$}{ $P$-value from a Kolmogorov--Smirnov test of the normalized residuals from the best-fit general model.}
\tablenotetext{$c$}{ The chance probability of the observed increase in likelihood given the two extra free parameters in the general model.}
\tablecomments{The `null model' refers to the model with only dust in the quasar (i.e., $z_{\rm abs}=0$ and ${\rm A(V)_{abs}}=0$). The `general model' refers to the model where both dust in the quasar and in the absorber are fitted simultaneously (i.e., with variable ${\rm A(V)_{QSO}}$, ${\rm z_{abs}}$, and ${\rm A(V)_{abs}}$). For details, see Sect.~4.1.1 by \citet{Krogager2015}. The general model is accepted when the following conditions are met: $L>28.75$ (5$\sigma$ threshold) and $P_{\rm KS}>0.10$.}
\end{deluxetable}

\section{Notes on Individual Objects}
\label{appendix:notes}

\input{individual_notes}

\newpage

\section{Figures}
\label{appendix:figureset}
Figure~E2 shows the full figure set previewed in Fig.~\ref{fig:spectra} of Sect.~\ref{results}.

\begin{figure}[h]
\gridline{\fig{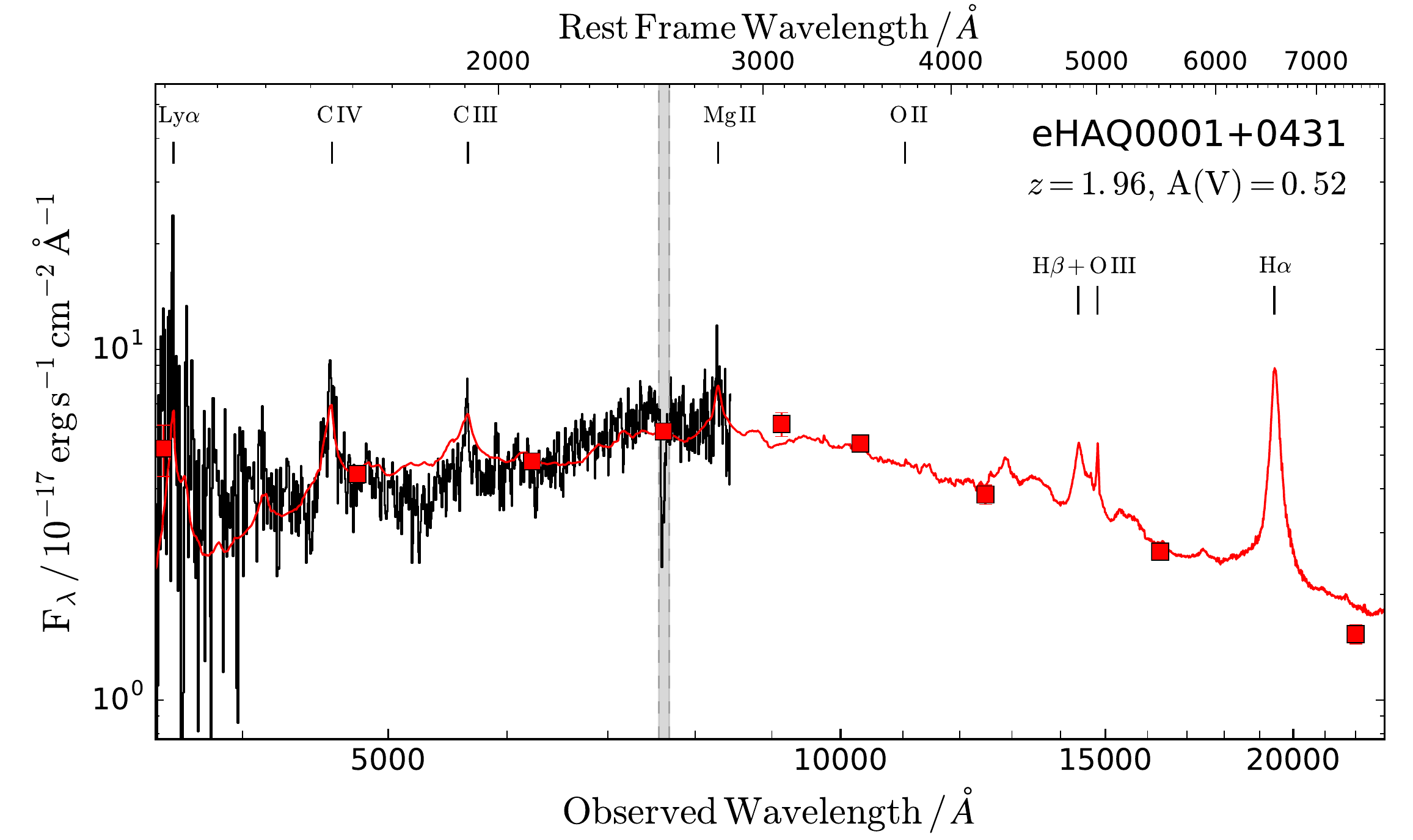}{0.5\textwidth}{}
          \fig{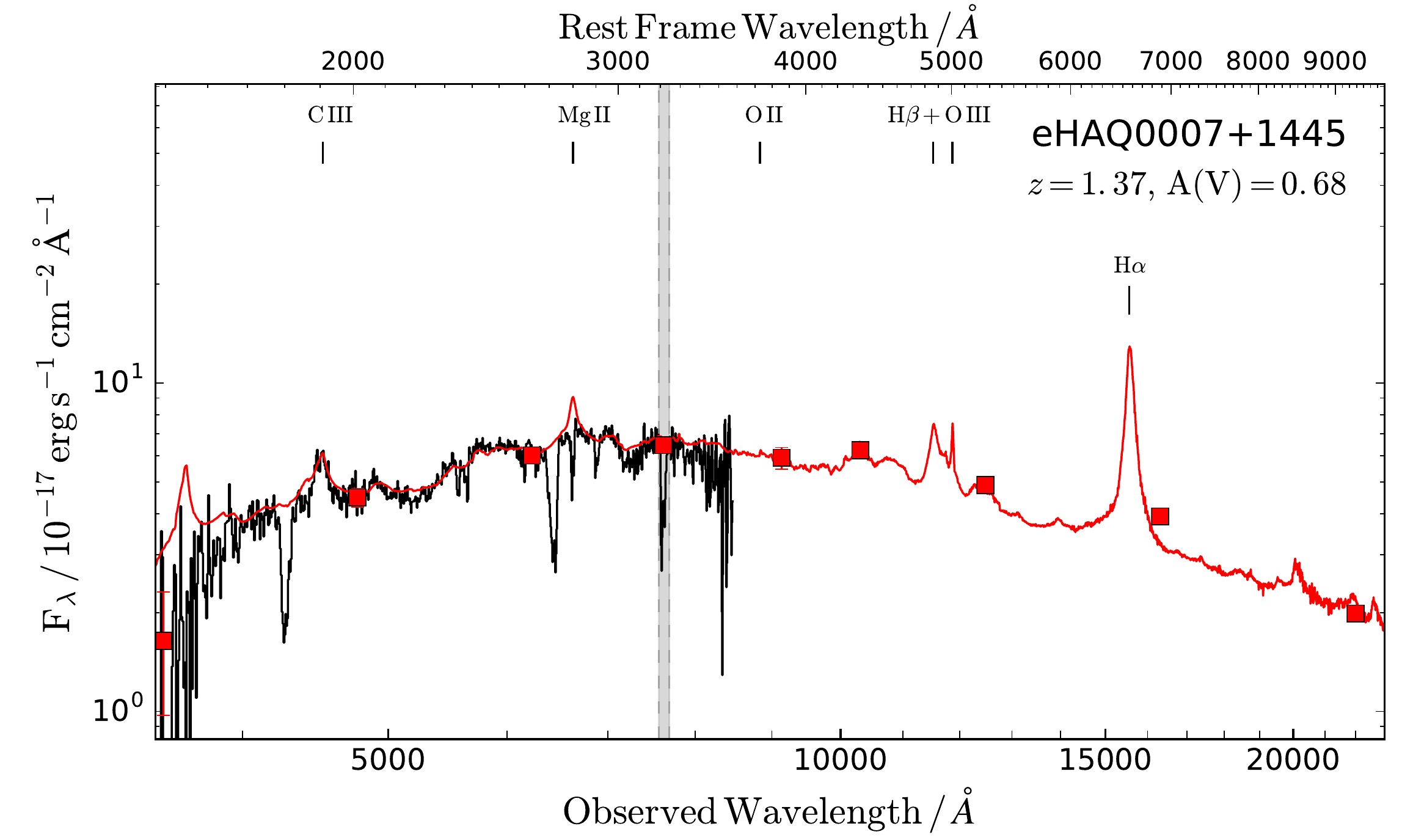}{0.5\textwidth}{}}
\gridline{\fig{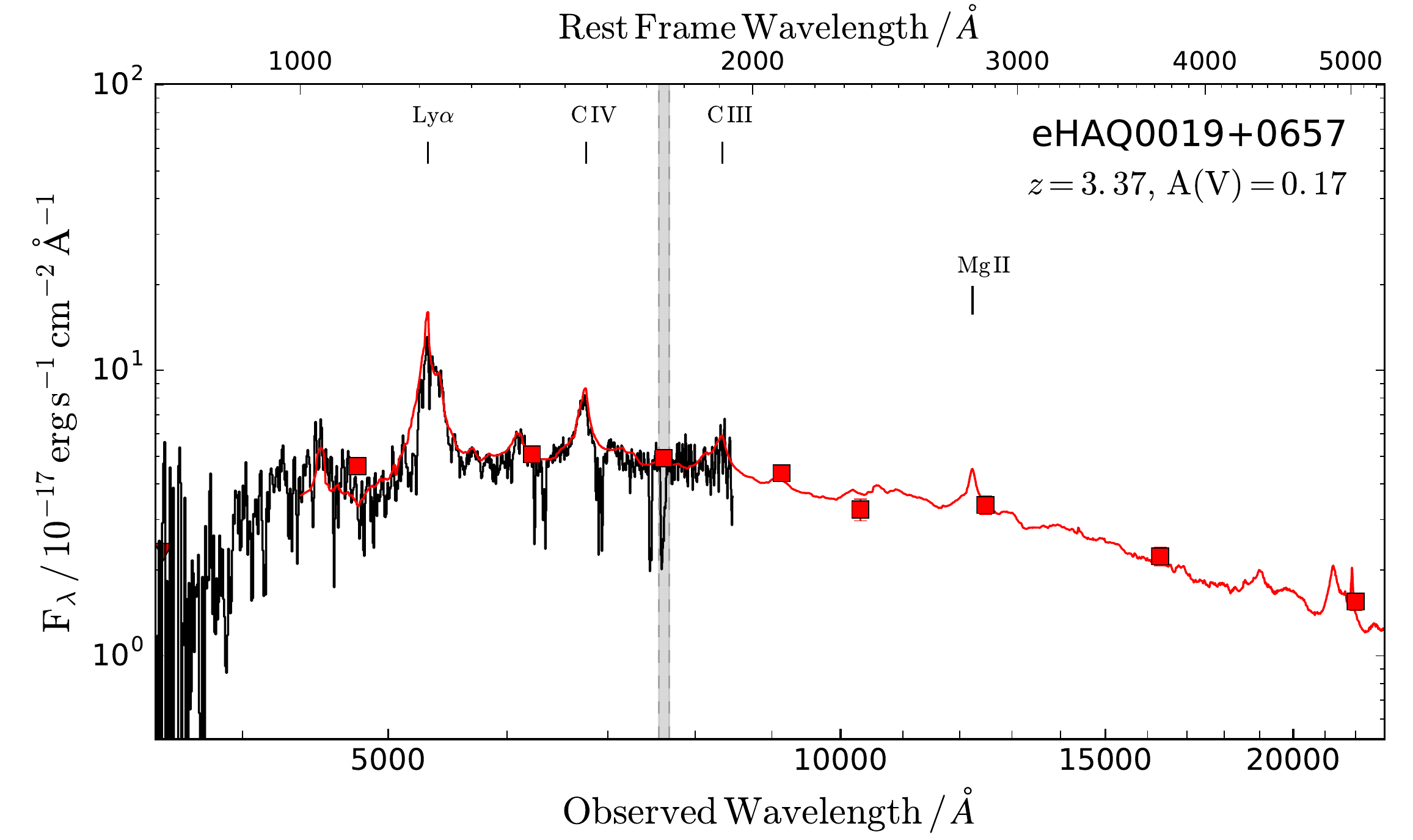}{0.5\textwidth}{}
          \fig{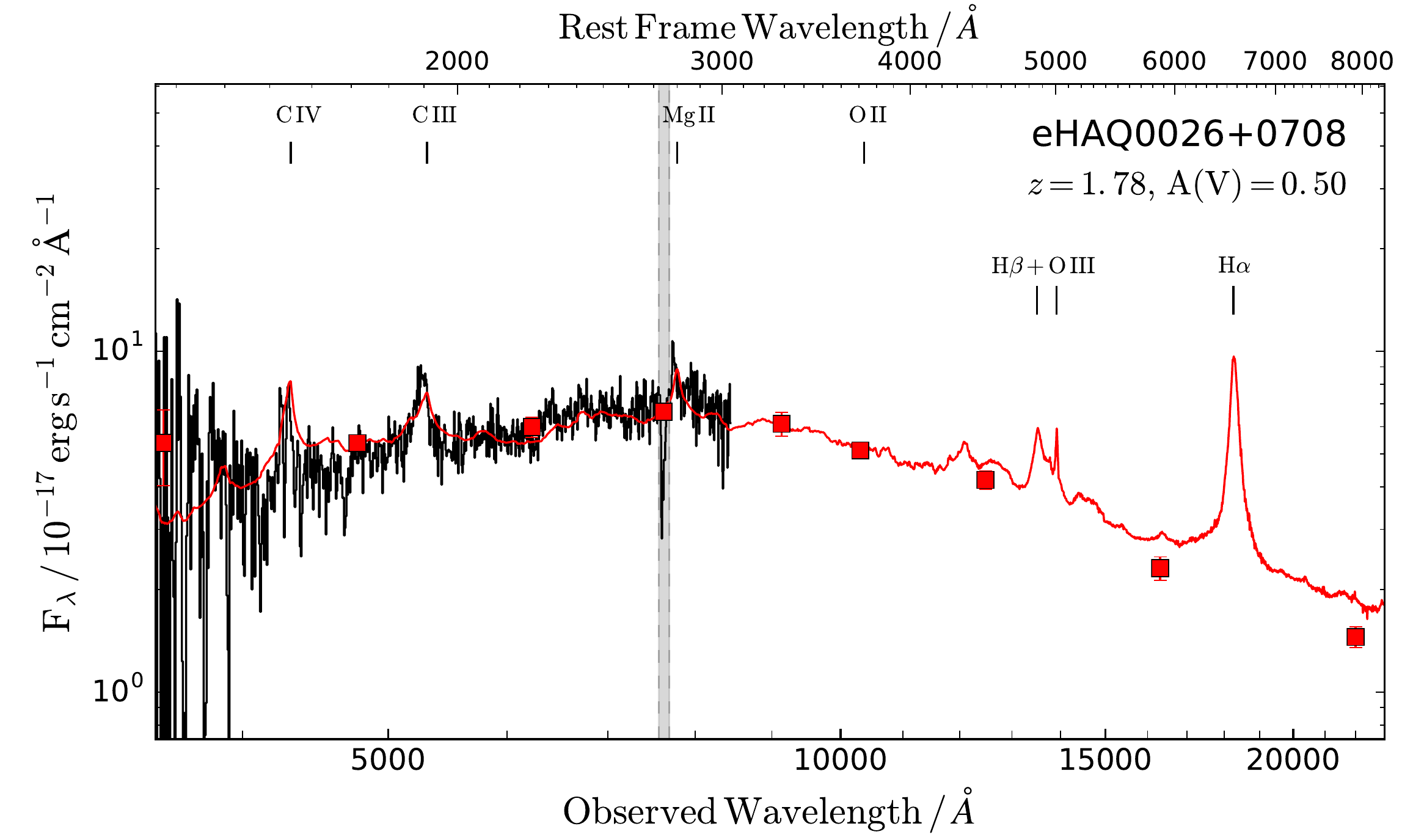}{0.5\textwidth}{}}
\gridline{\fig{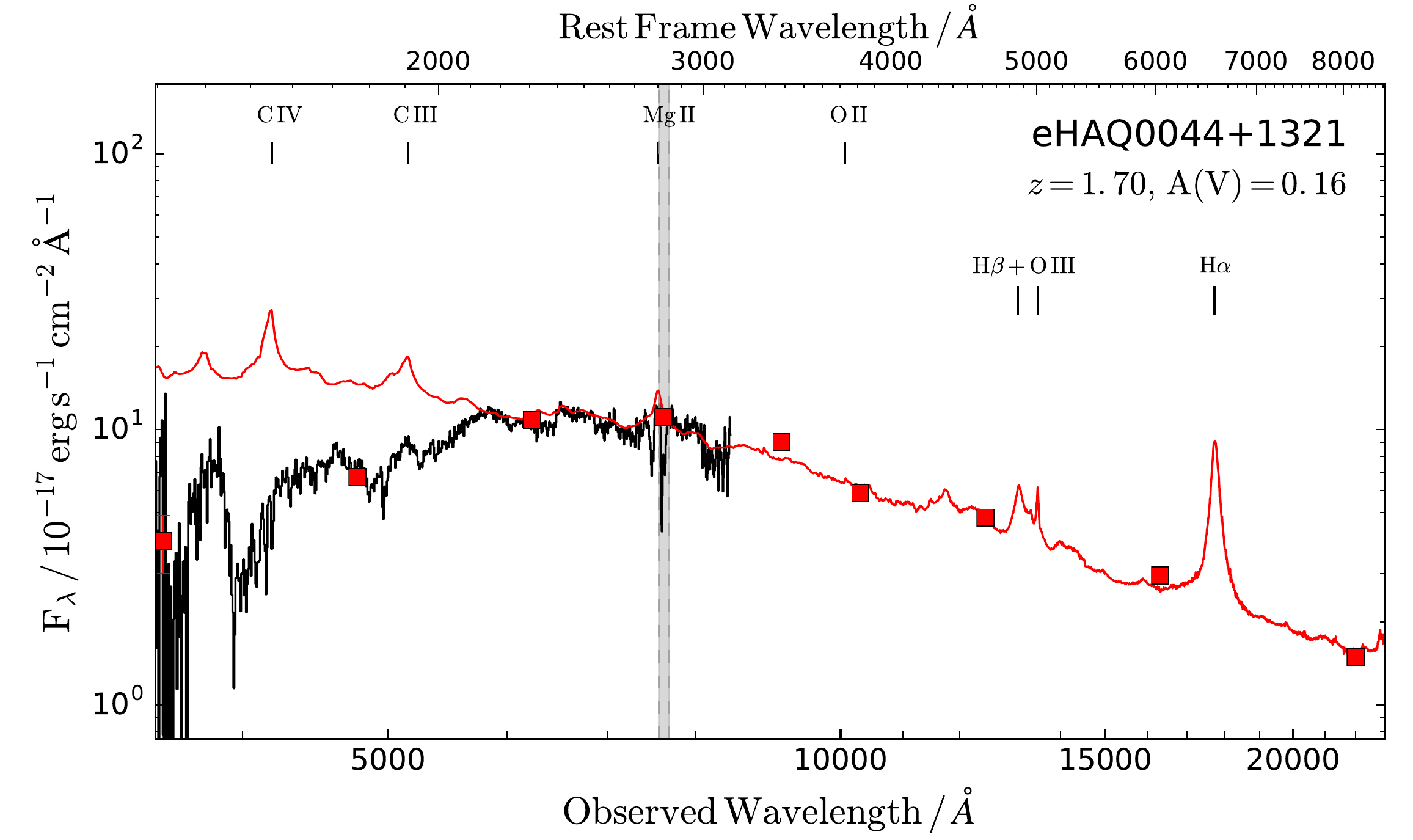}{0.5\textwidth}{}
          \fig{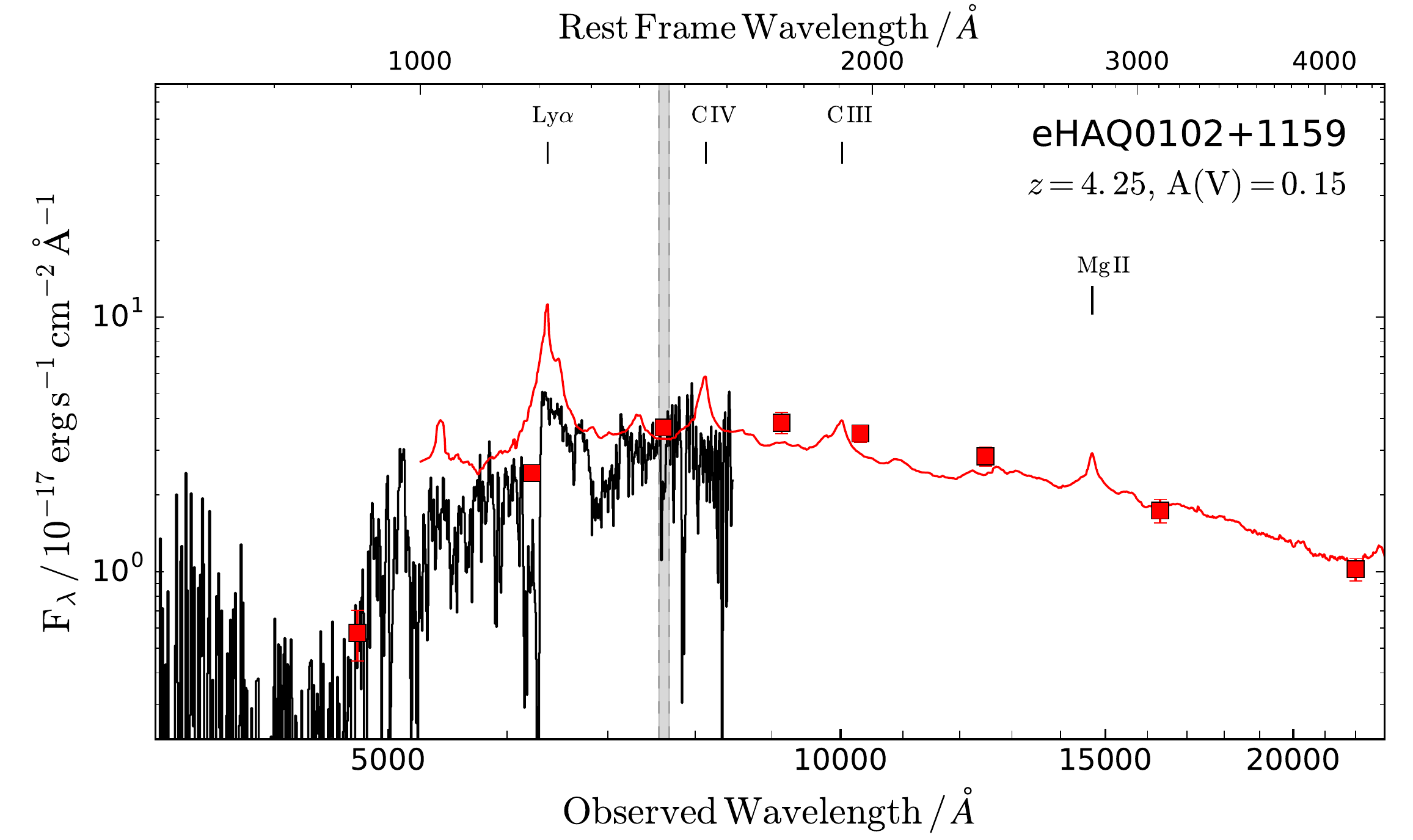}{0.5\textwidth}{}}
\caption{(Figure 3 continued.) The observed spectrum is shown as a solid black curve.
	For spectra observed with the INT, the spectra are binned by a factor of 8
	for visual clarity. These cases are marked by the label `{\sc int}' in the lower
	left corner of the figure.
	The filled, red squares indicate the SDSS and UKIDSS photometric
	data points.
	In the upper right corner we provide the estimated emission redshift
	and rest-frame $V$-band extinction. The red curve shows the redshifted
	composite quasar template of \citet{Selsing2016} reddened by the best-fit
	dust model (indicated in the upper right corner).
	The dust reddening applied in the rest-frame of
	the quasar, A(V), and absorber, A(V)$_{\rm abs}$, assumes the SMC and
	LMC extinction laws, respectively.
	In cases where the template is also shown in blue, we were not able to
	fit both optical and infrared data simultaneously; The blue template
	then indicates the best fit to the infrared data only, and the
	corresponding A(V) is given as ${\rm A(V)_{blue}}$. Note that the spectra
	have not been corrected for telluric absorption (marked with a grey band
	at $\sim7600$~\AA).
	\label{fig:figureset}}
\end{figure}

\begin{figure}
\figurenum{E2}
    \includegraphics[width=0.49\textwidth]{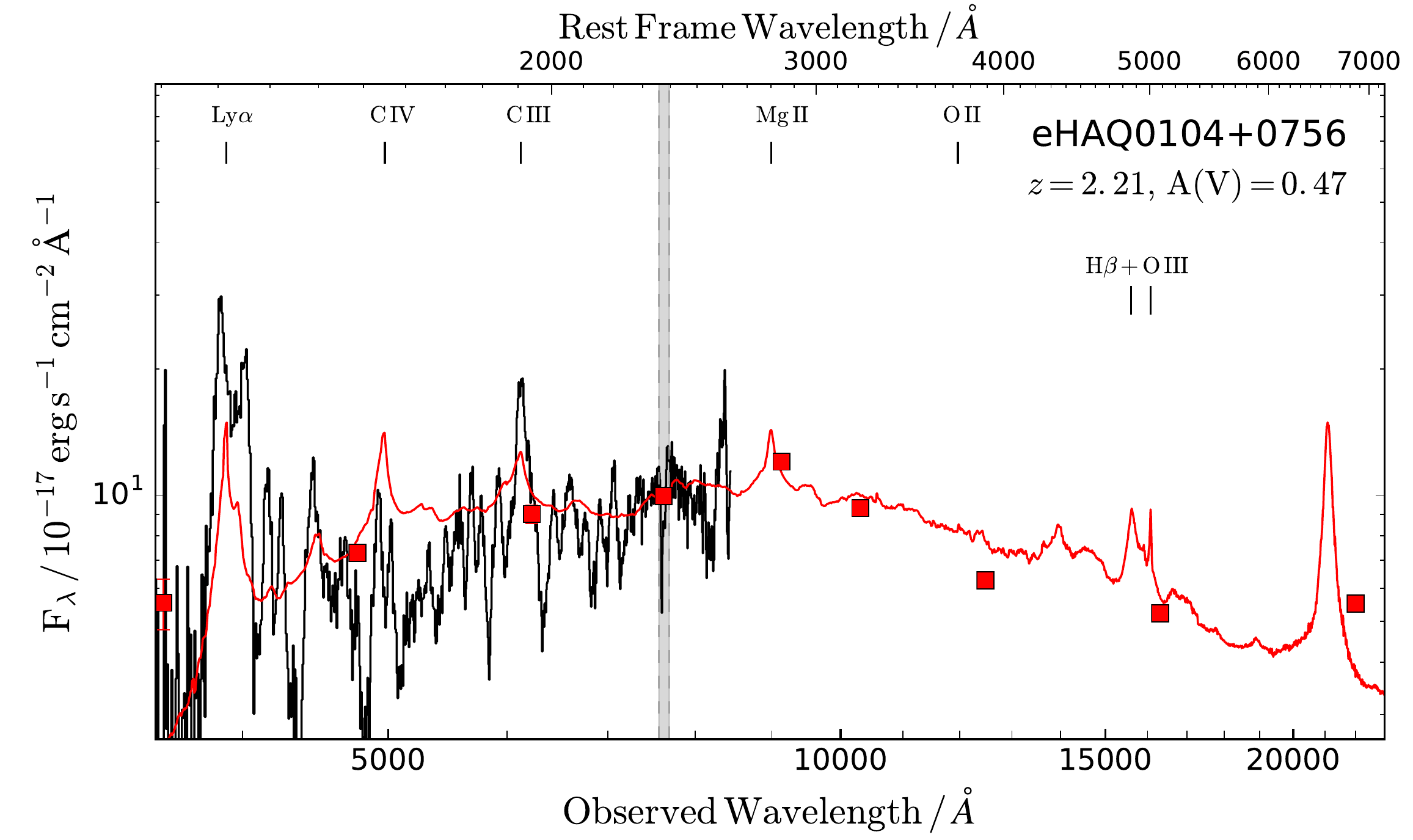}
    \includegraphics[width=0.49\textwidth]{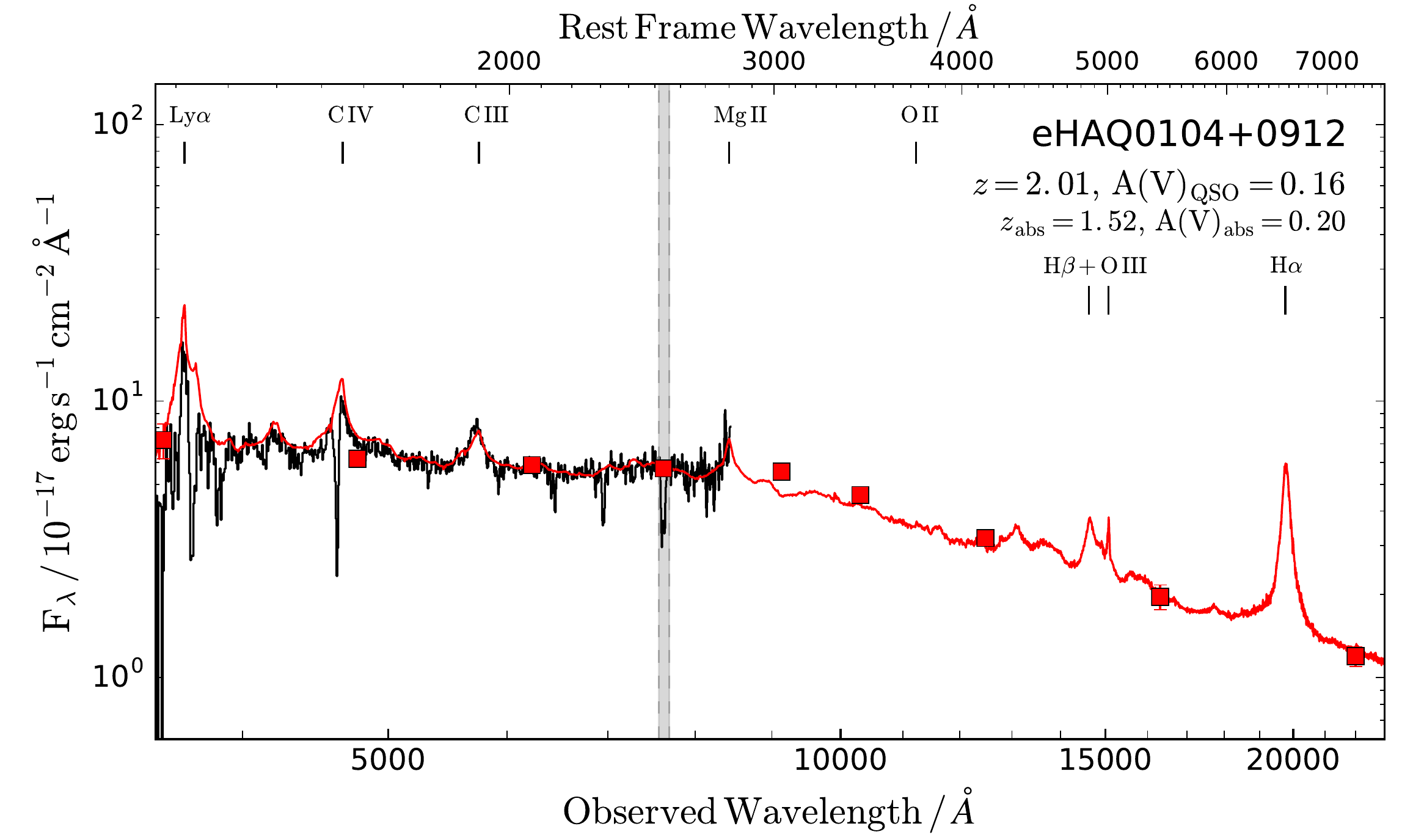}
    \vspace{4mm}

    \includegraphics[width=0.49\textwidth]{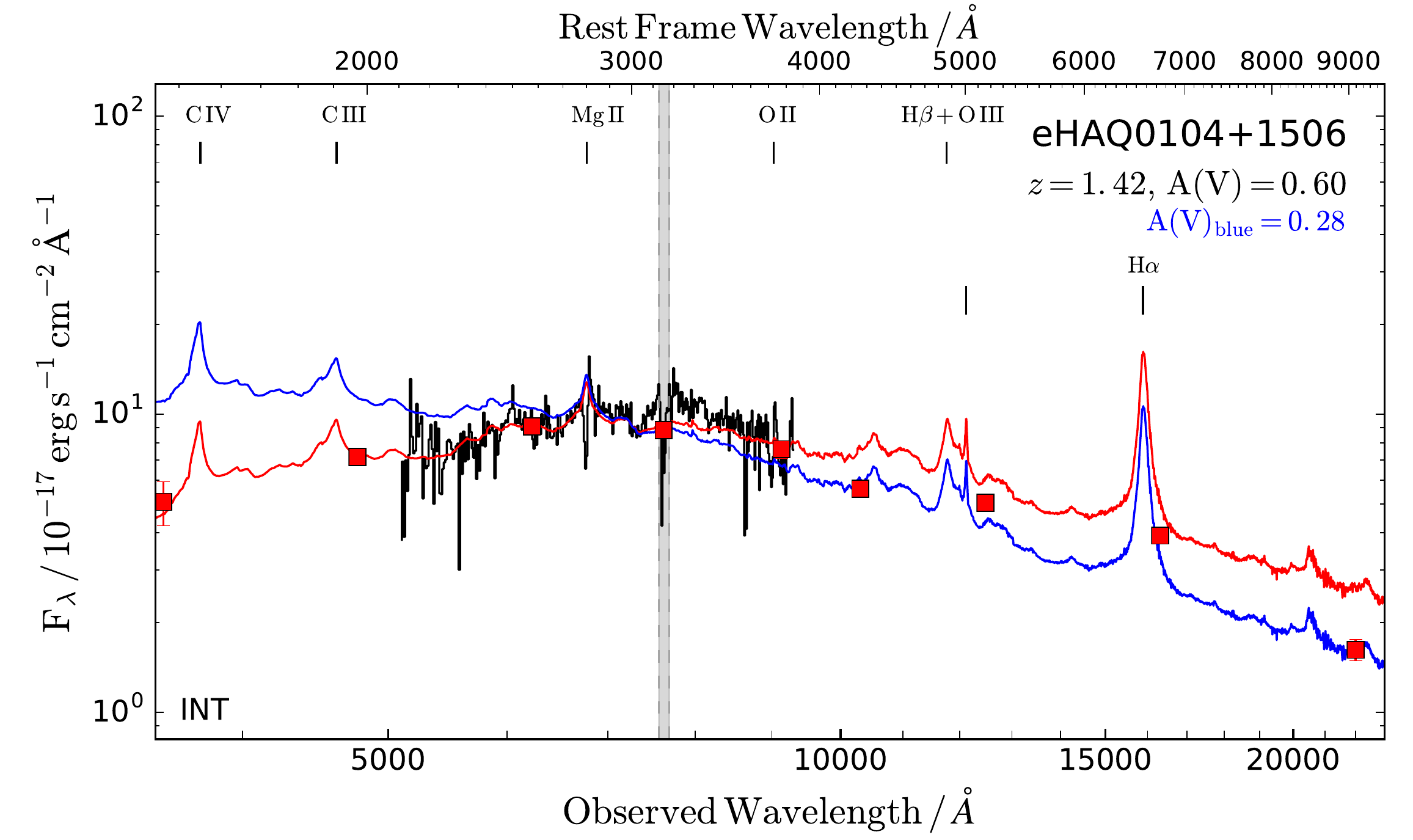}
    \includegraphics[width=0.49\textwidth]{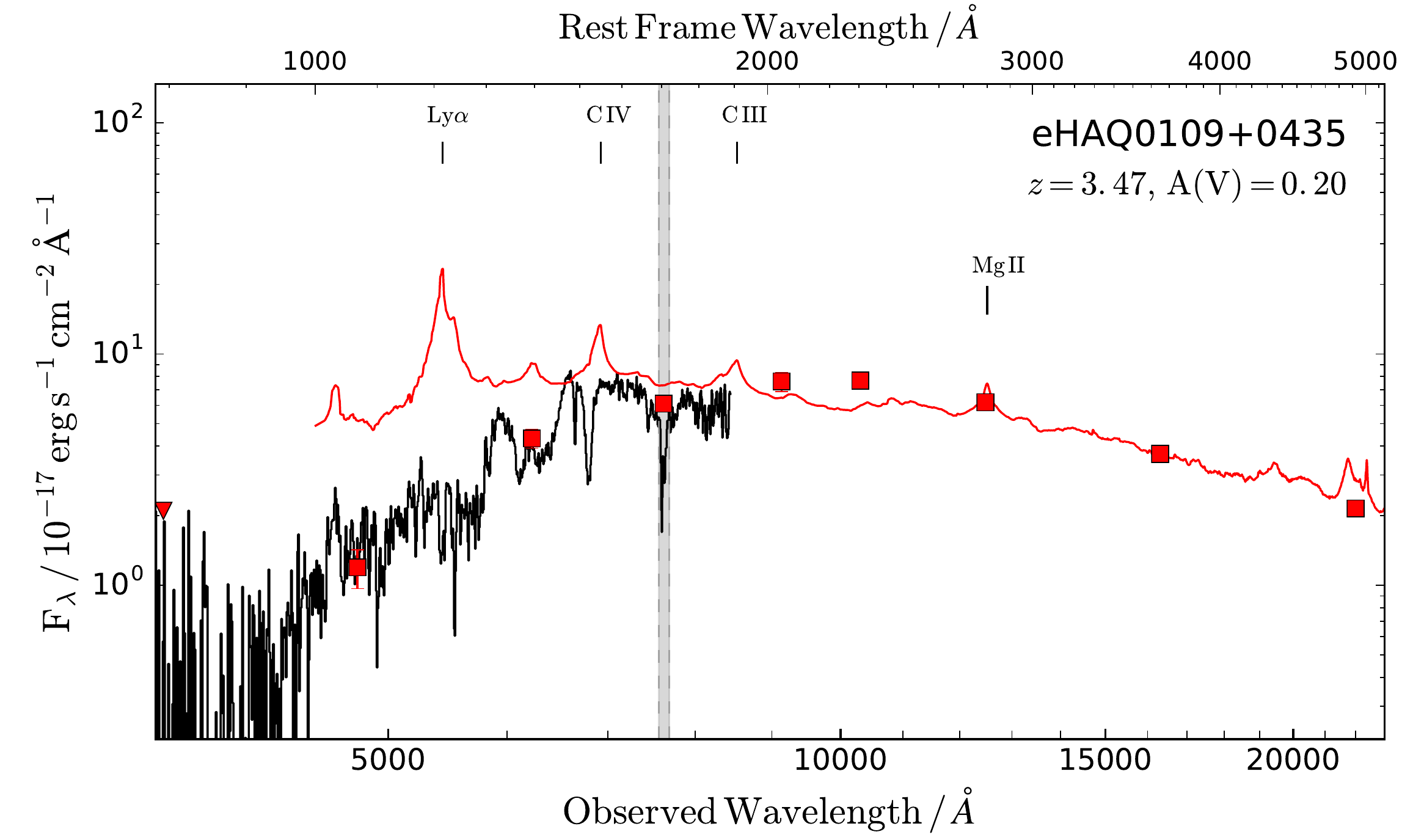}
    \vspace{4mm}

    \includegraphics[width=0.49\textwidth]{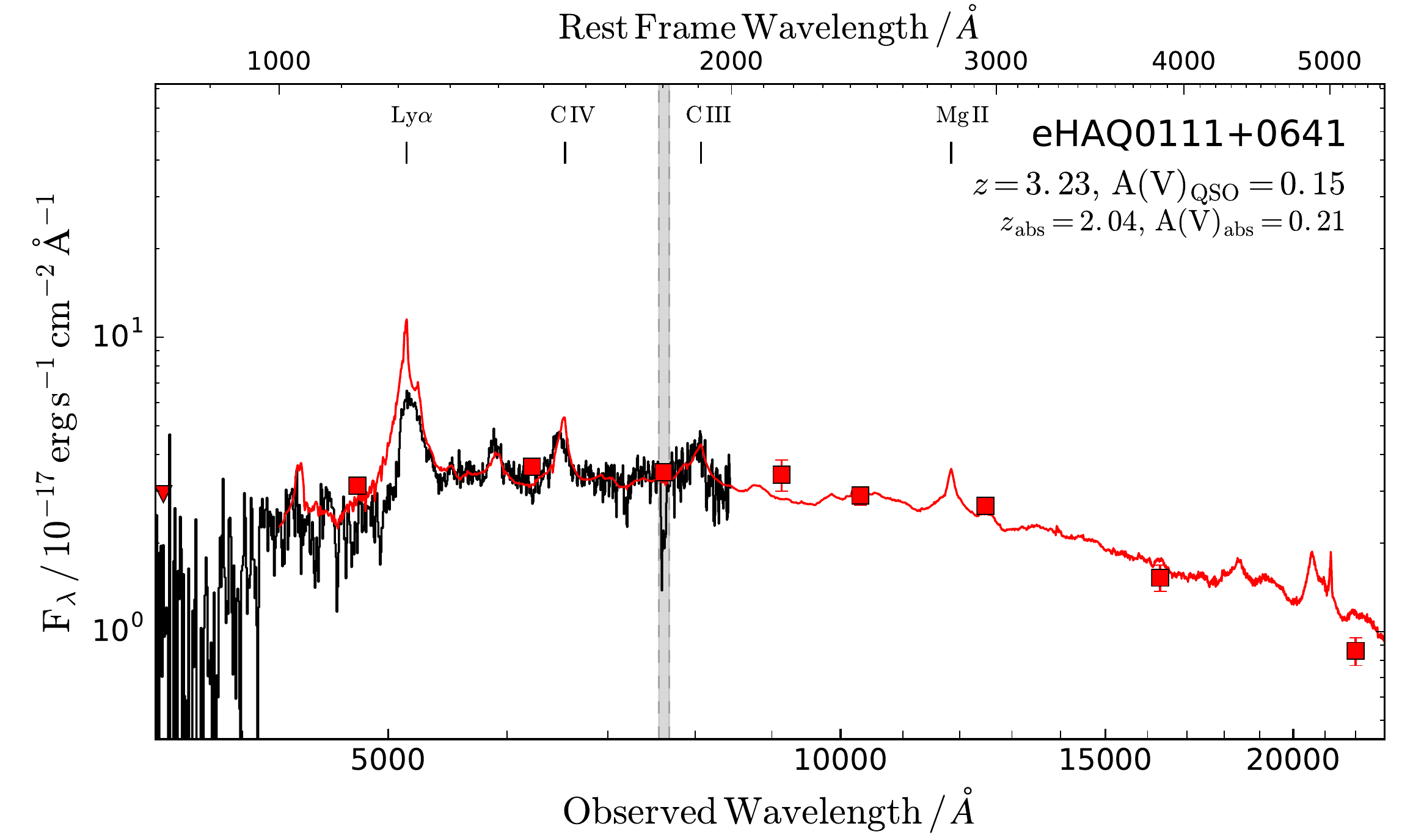}
    \includegraphics[width=0.49\textwidth]{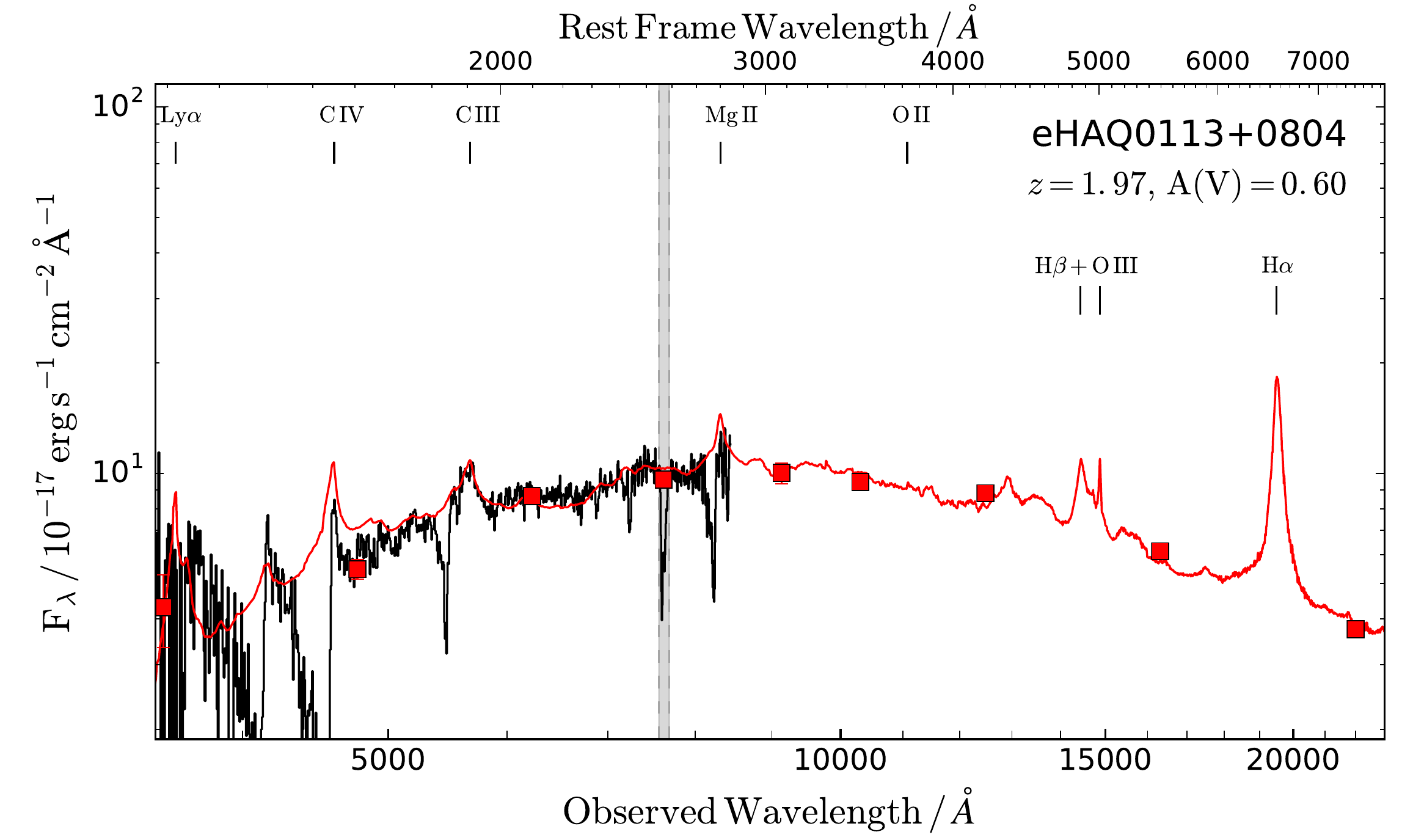}
    \vspace{4mm}

    \includegraphics[width=0.49\textwidth]{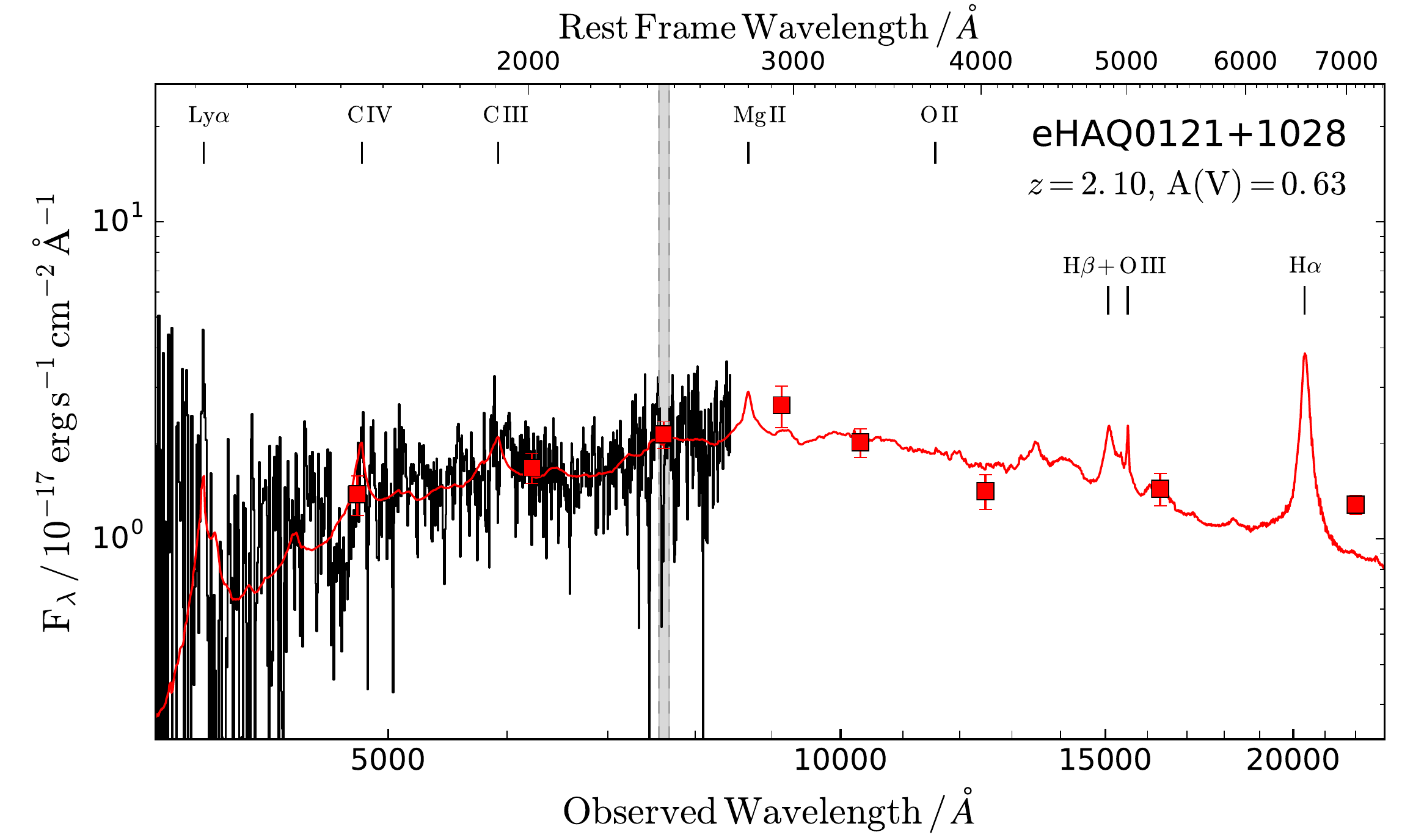}
    \includegraphics[width=0.49\textwidth]{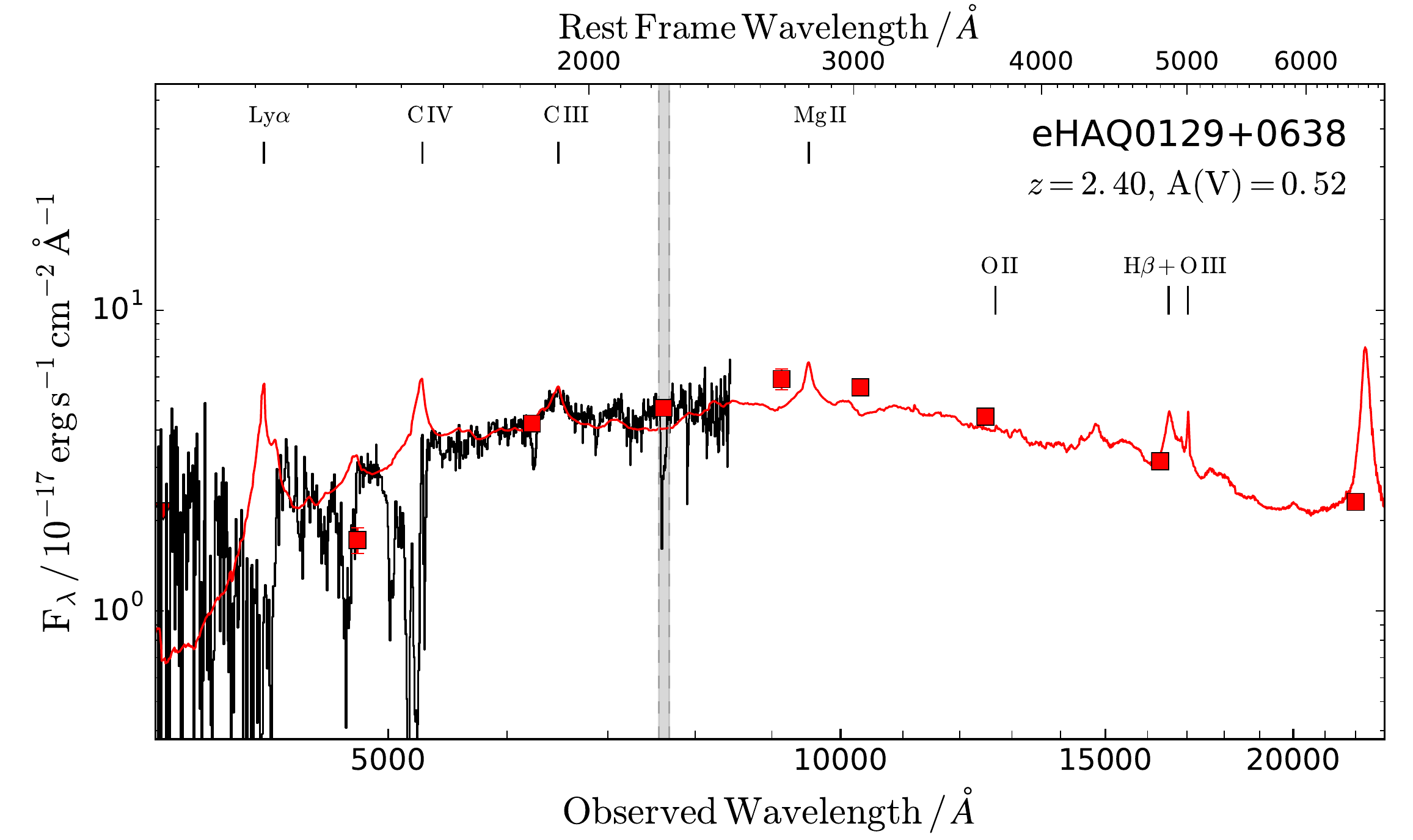}
\caption{(Continued.)}
\end{figure}

\begin{figure}
\figurenum{E2}
    \includegraphics[width=0.49\textwidth]{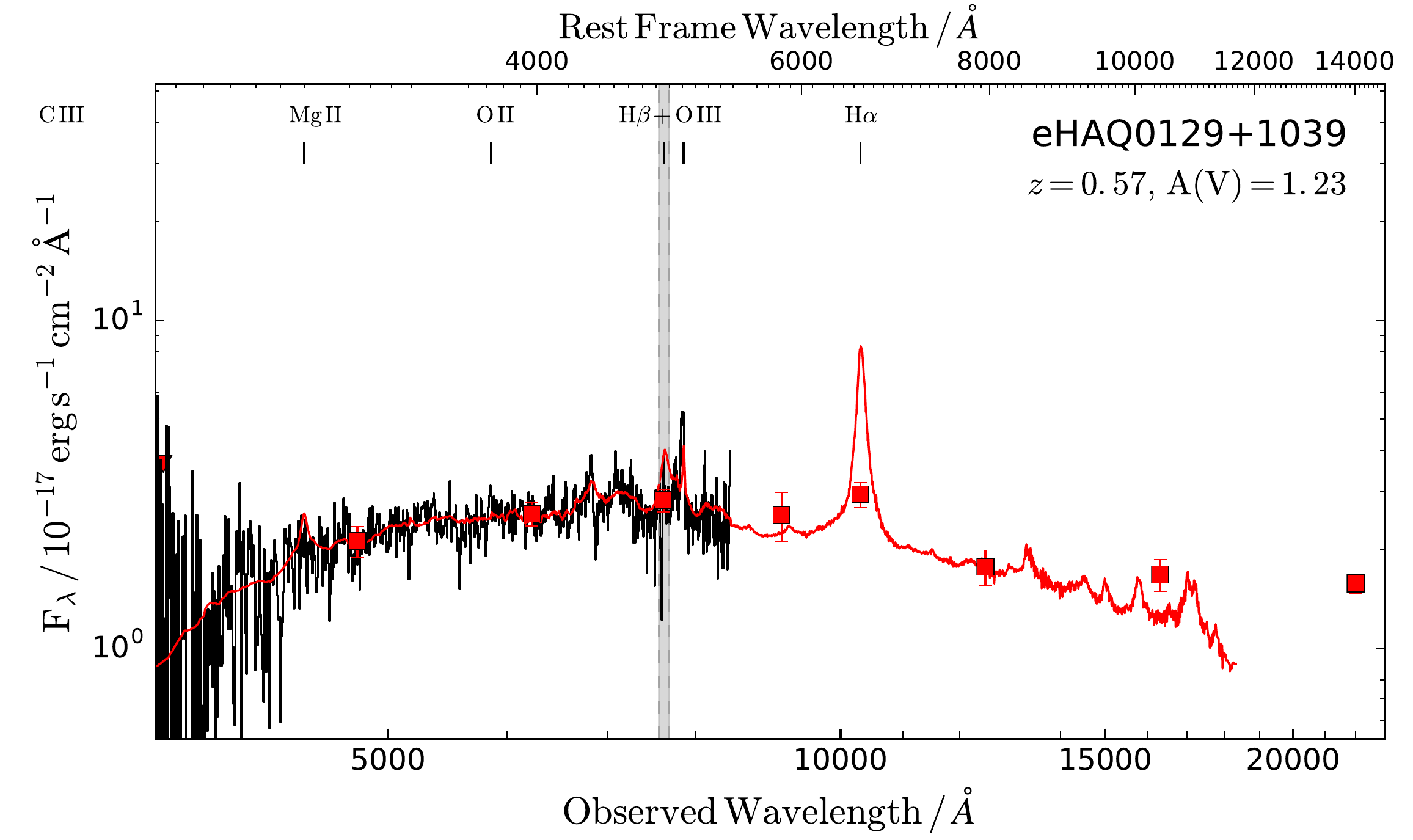}
    \includegraphics[width=0.49\textwidth]{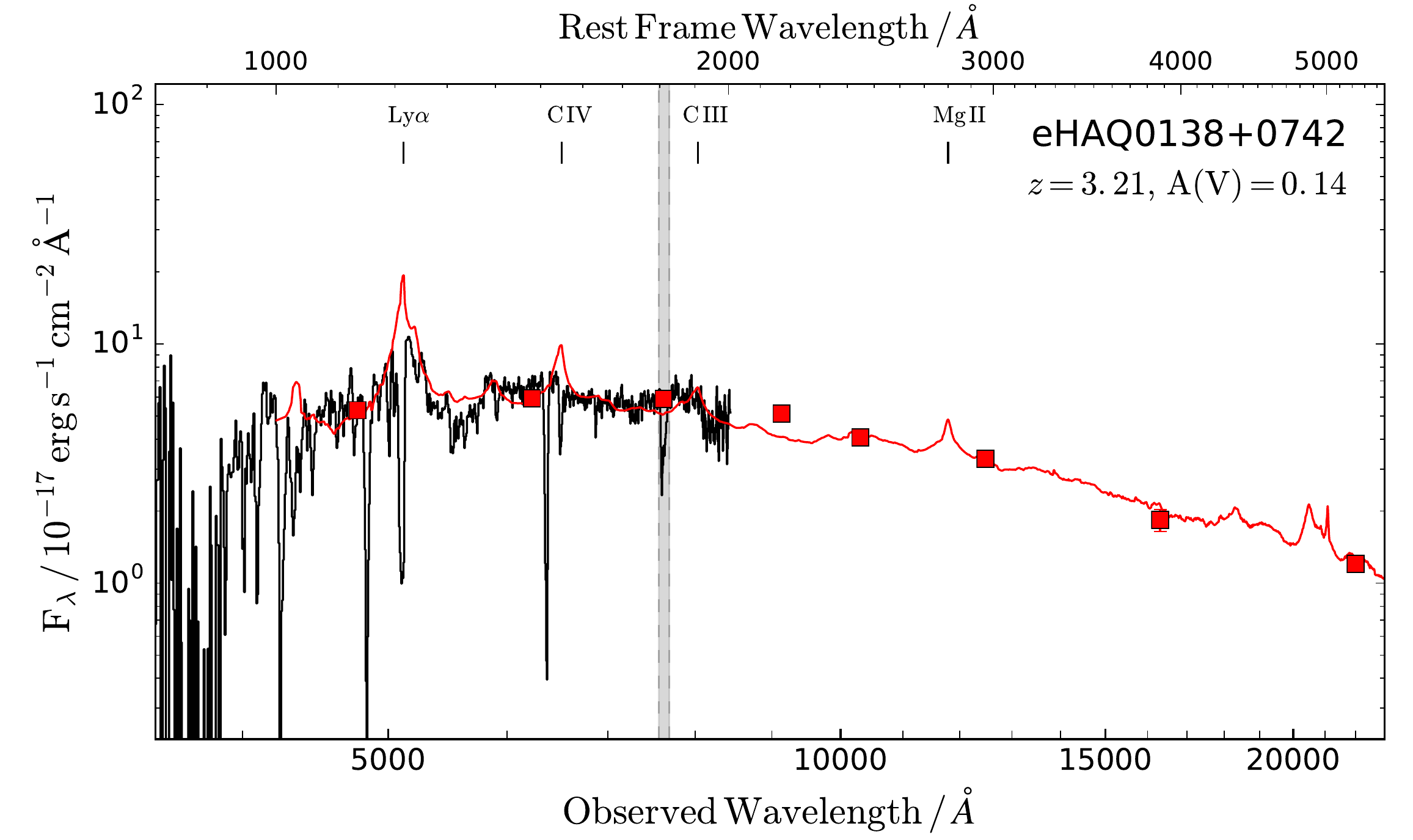}
    \vspace{4mm}

    \includegraphics[width=0.49\textwidth]{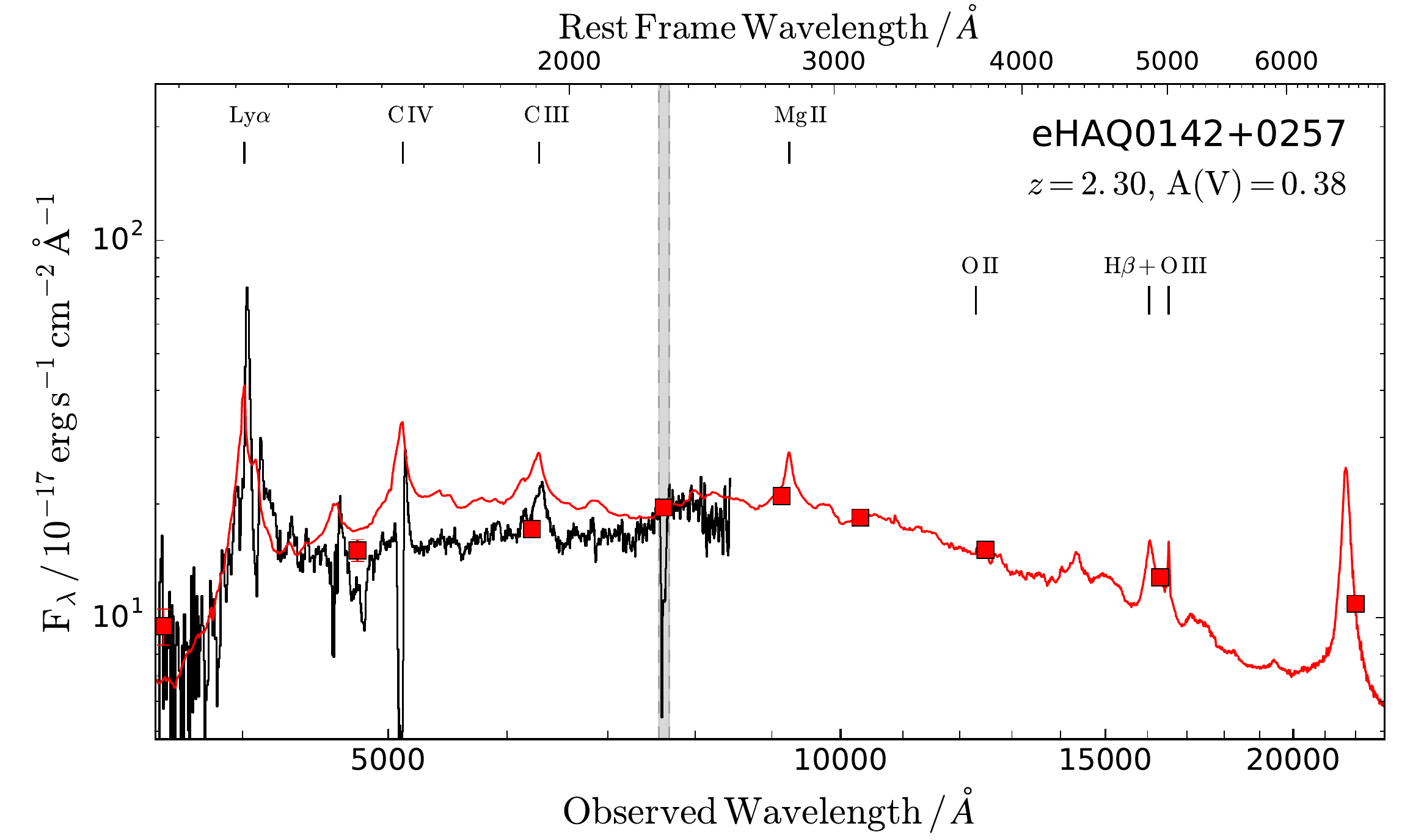}
    \includegraphics[width=0.49\textwidth]{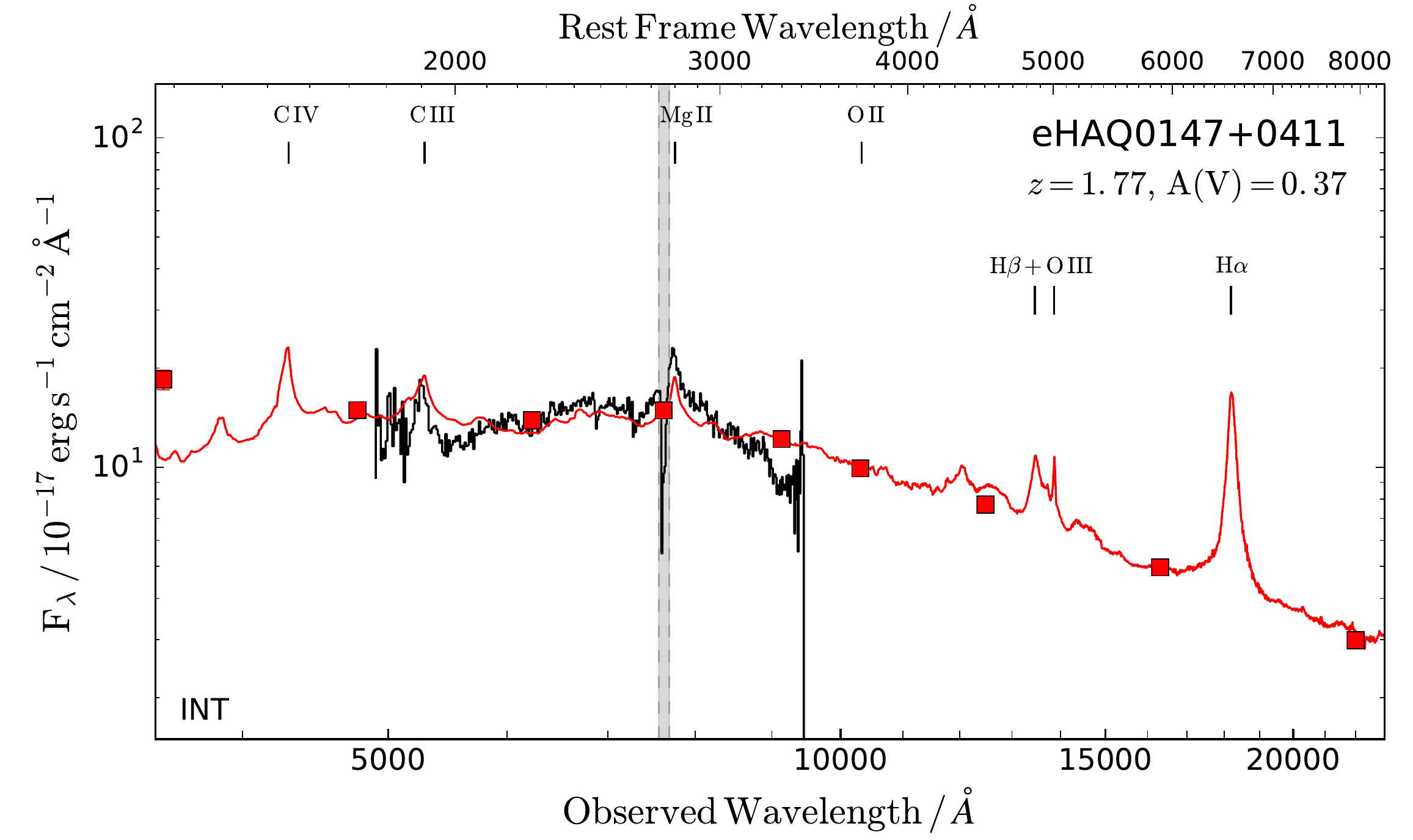}
    \vspace{4mm}

    \includegraphics[width=0.49\textwidth]{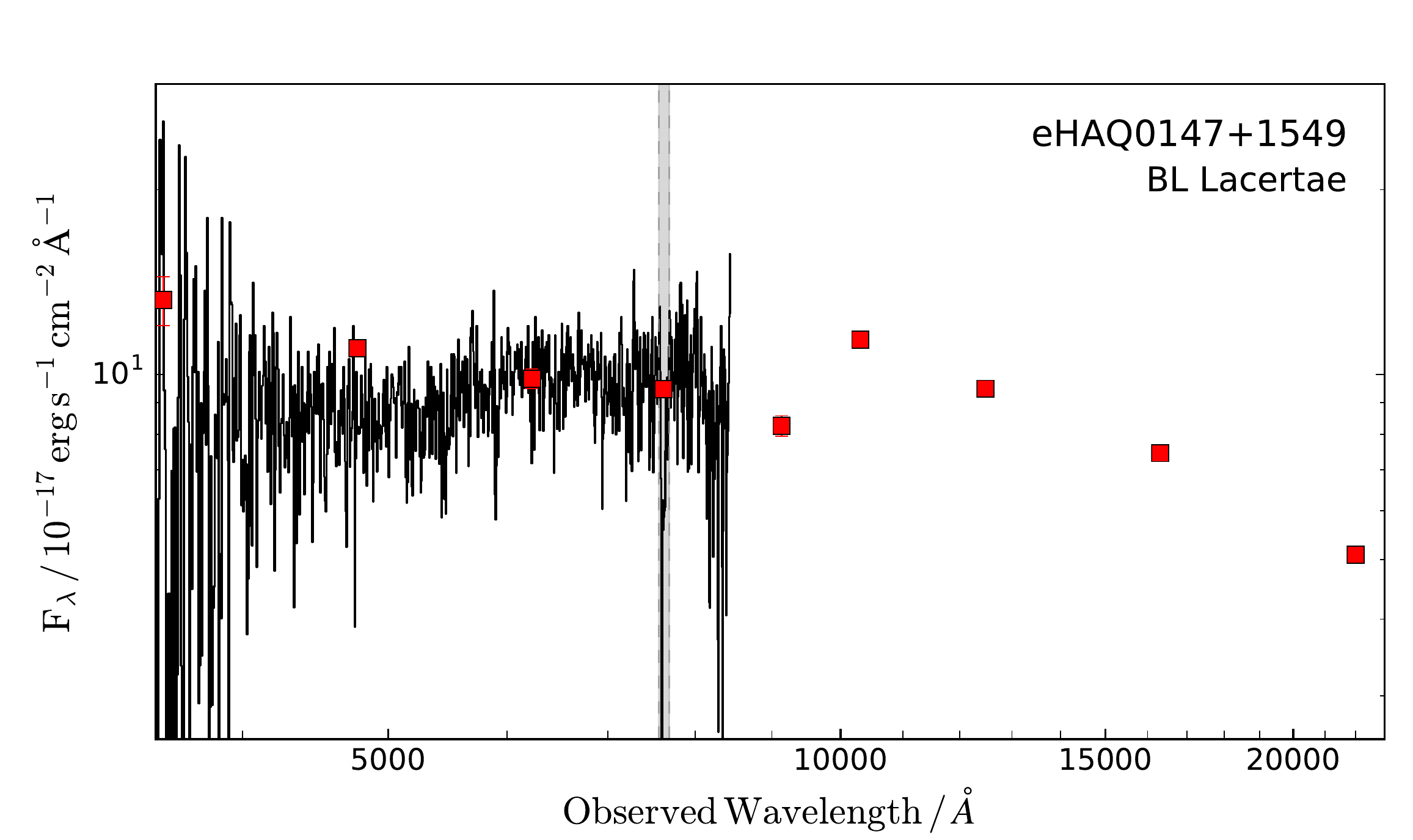}
    \includegraphics[width=0.49\textwidth]{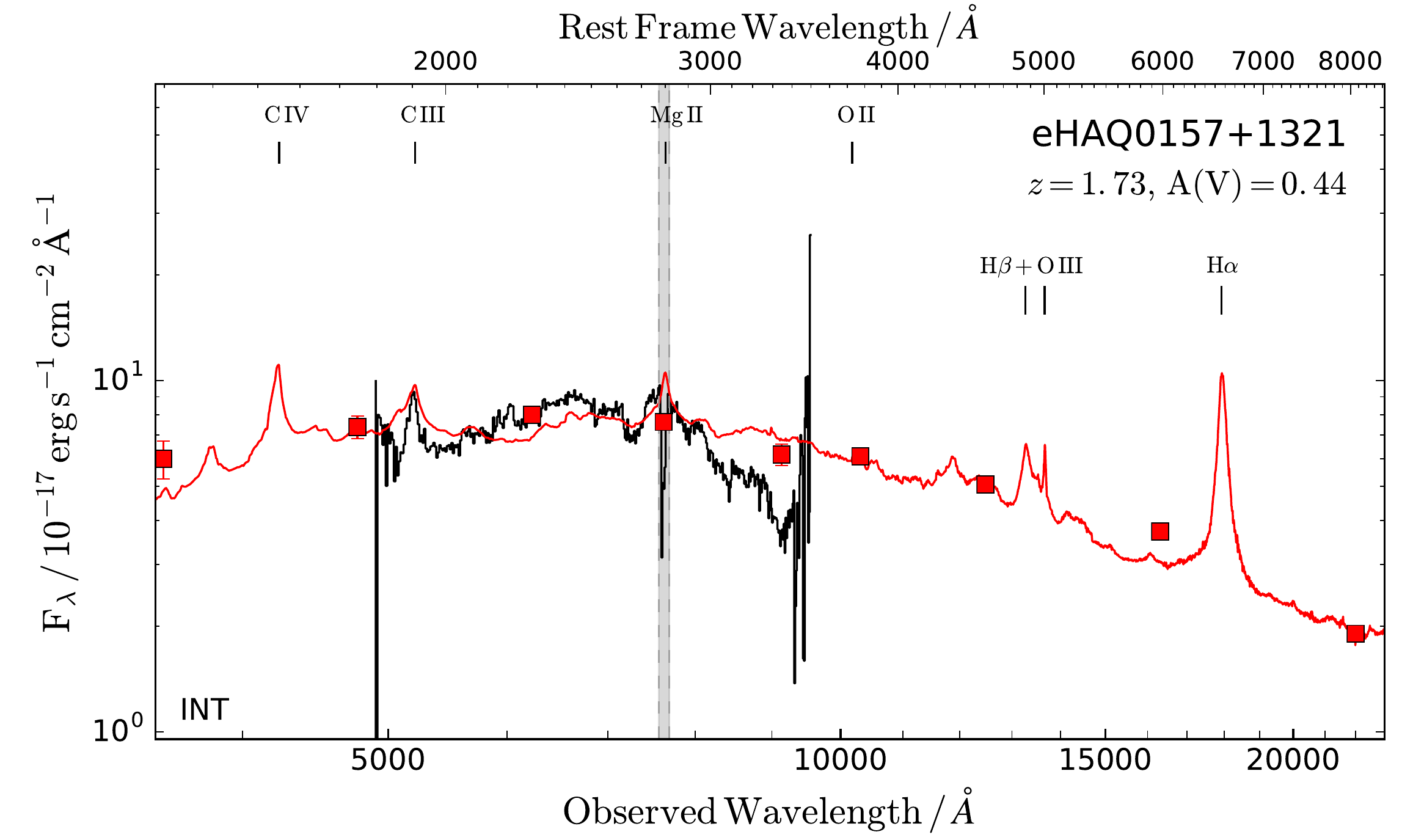}
    \vspace{4mm}

    \includegraphics[width=0.49\textwidth]{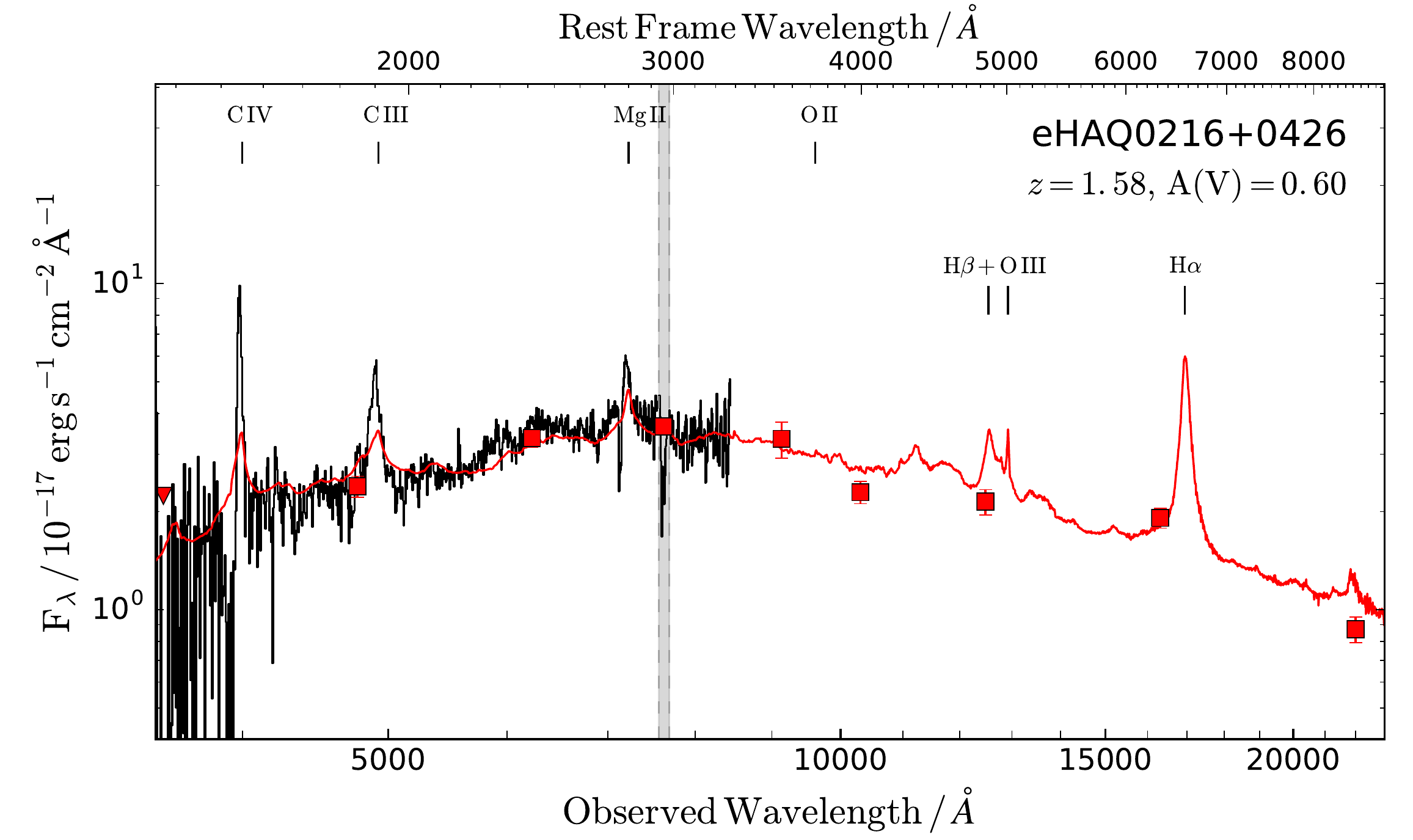}
    \includegraphics[width=0.49\textwidth]{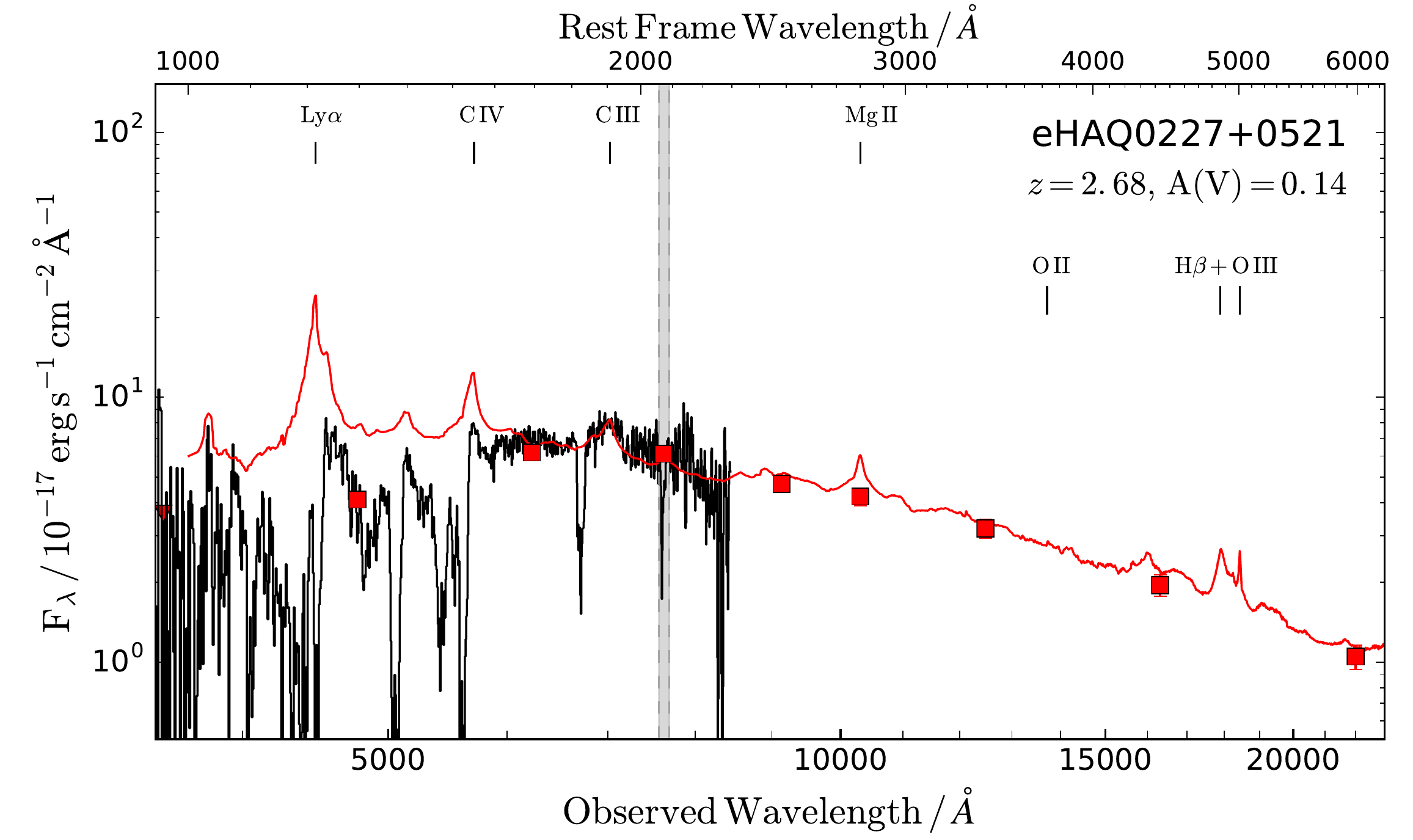}
\caption{(Continued.)}
\end{figure}

\begin{figure}
\figurenum{E2}
    \includegraphics[width=0.49\textwidth]{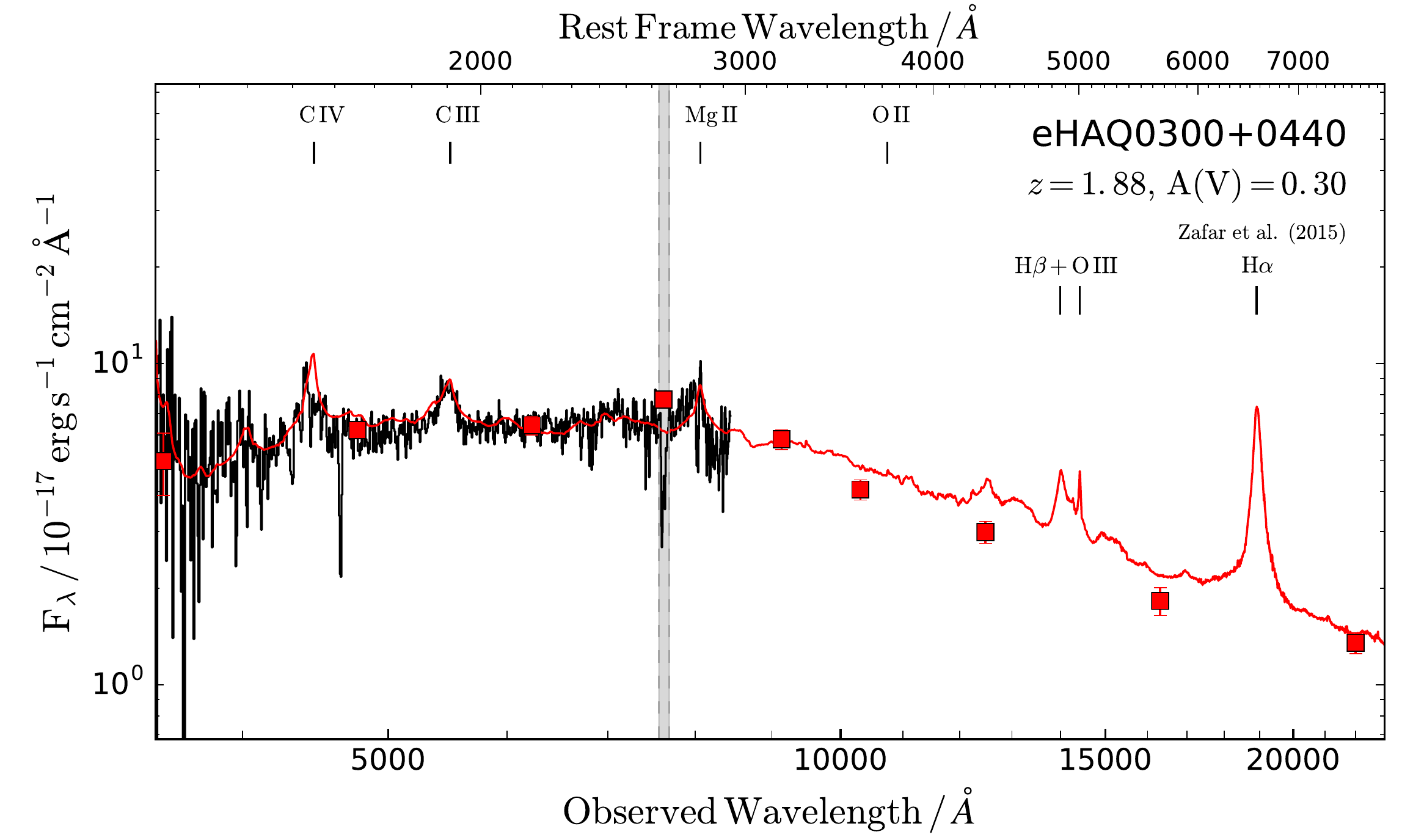}
    \includegraphics[width=0.49\textwidth]{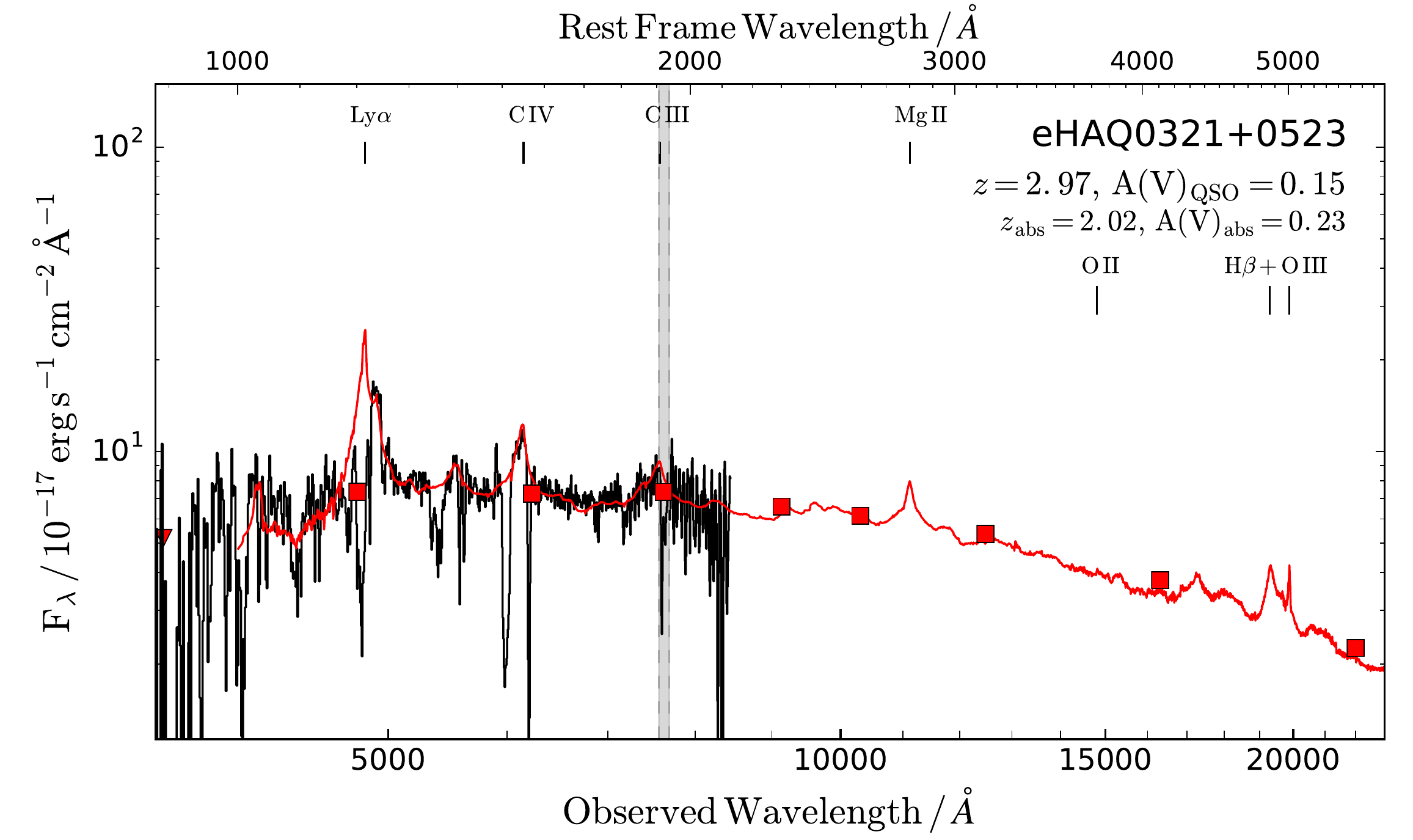}
    \vspace{4mm}

    \includegraphics[width=0.49\textwidth]{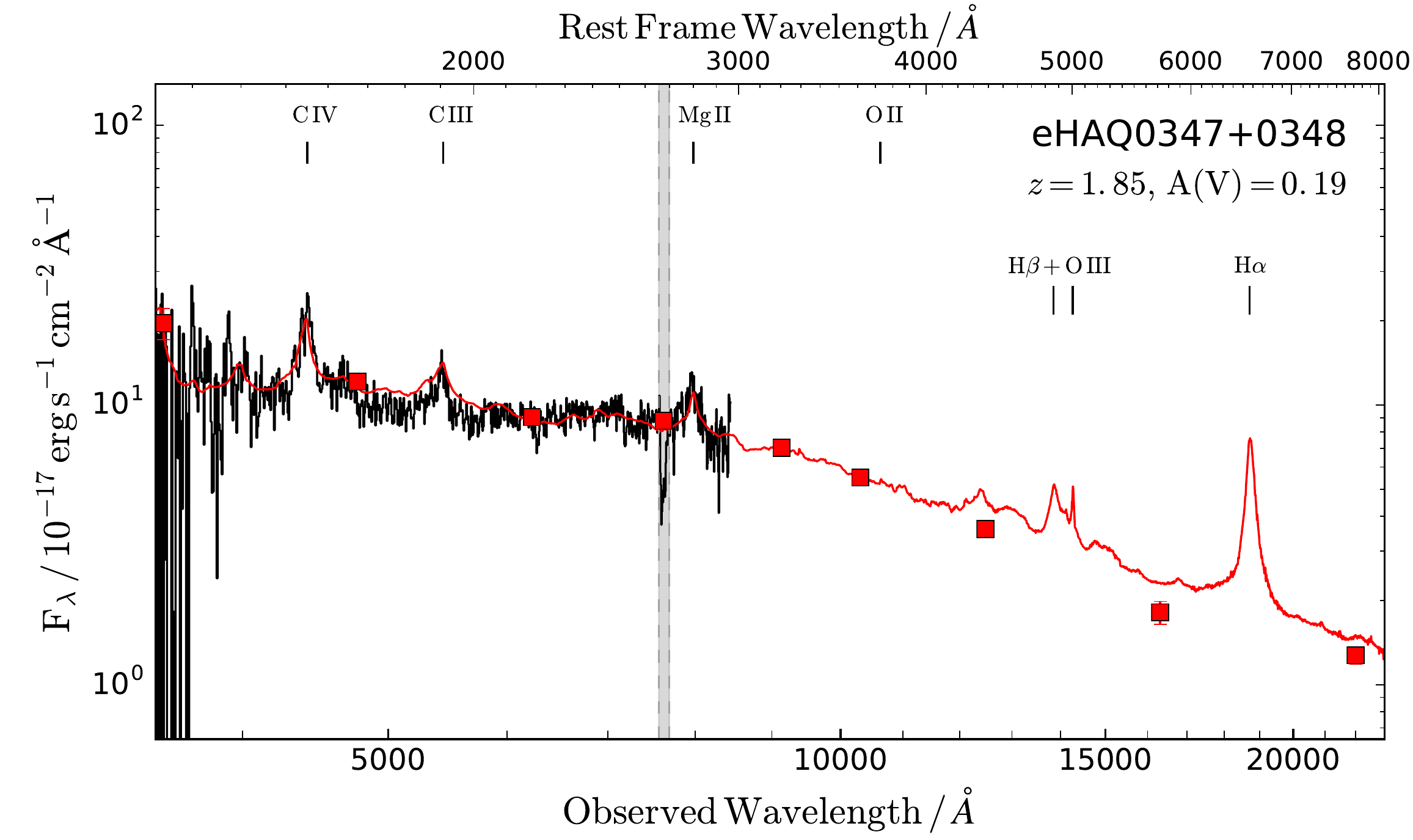}
    \includegraphics[width=0.49\textwidth]{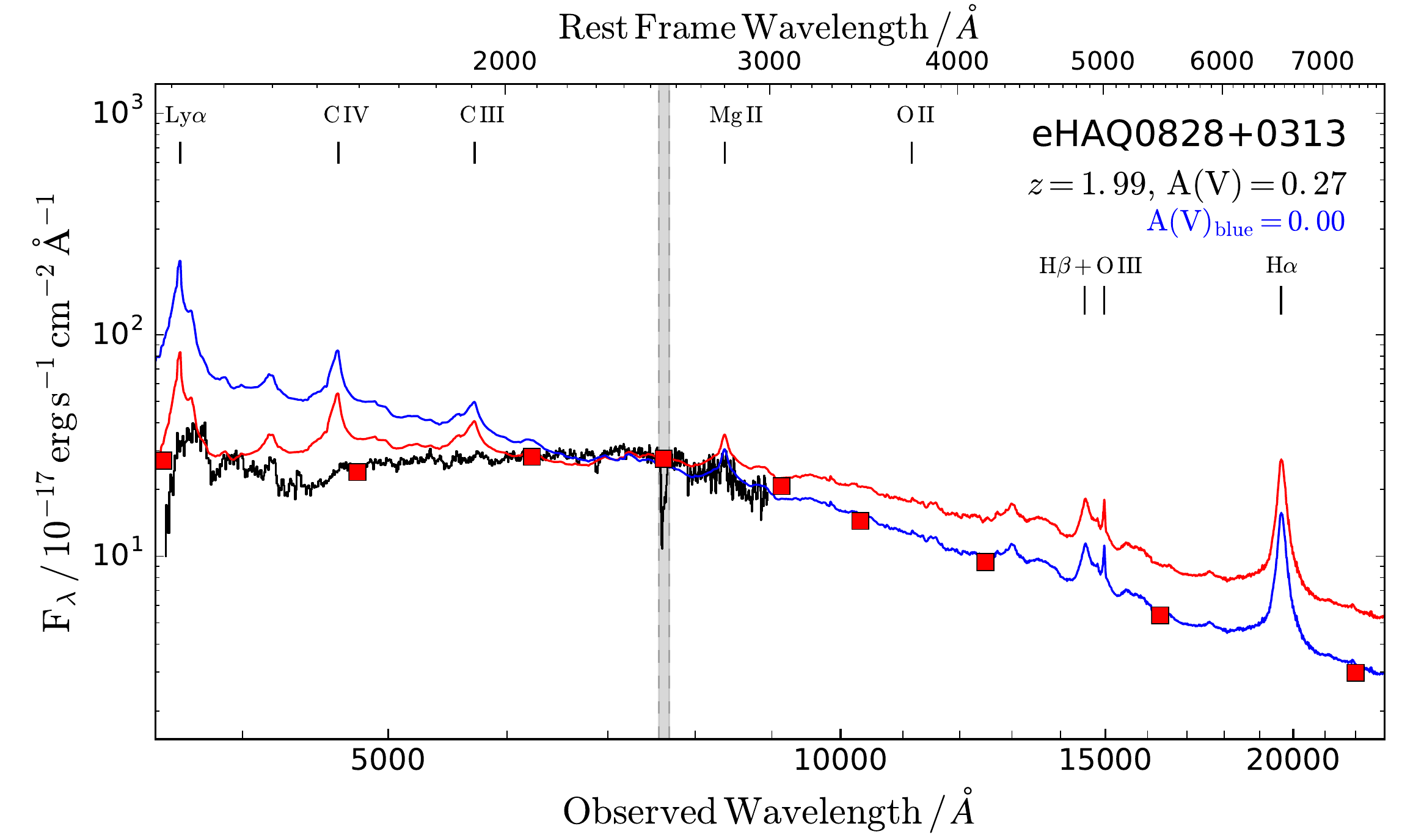}
    \vspace{4mm}

    \includegraphics[width=0.49\textwidth]{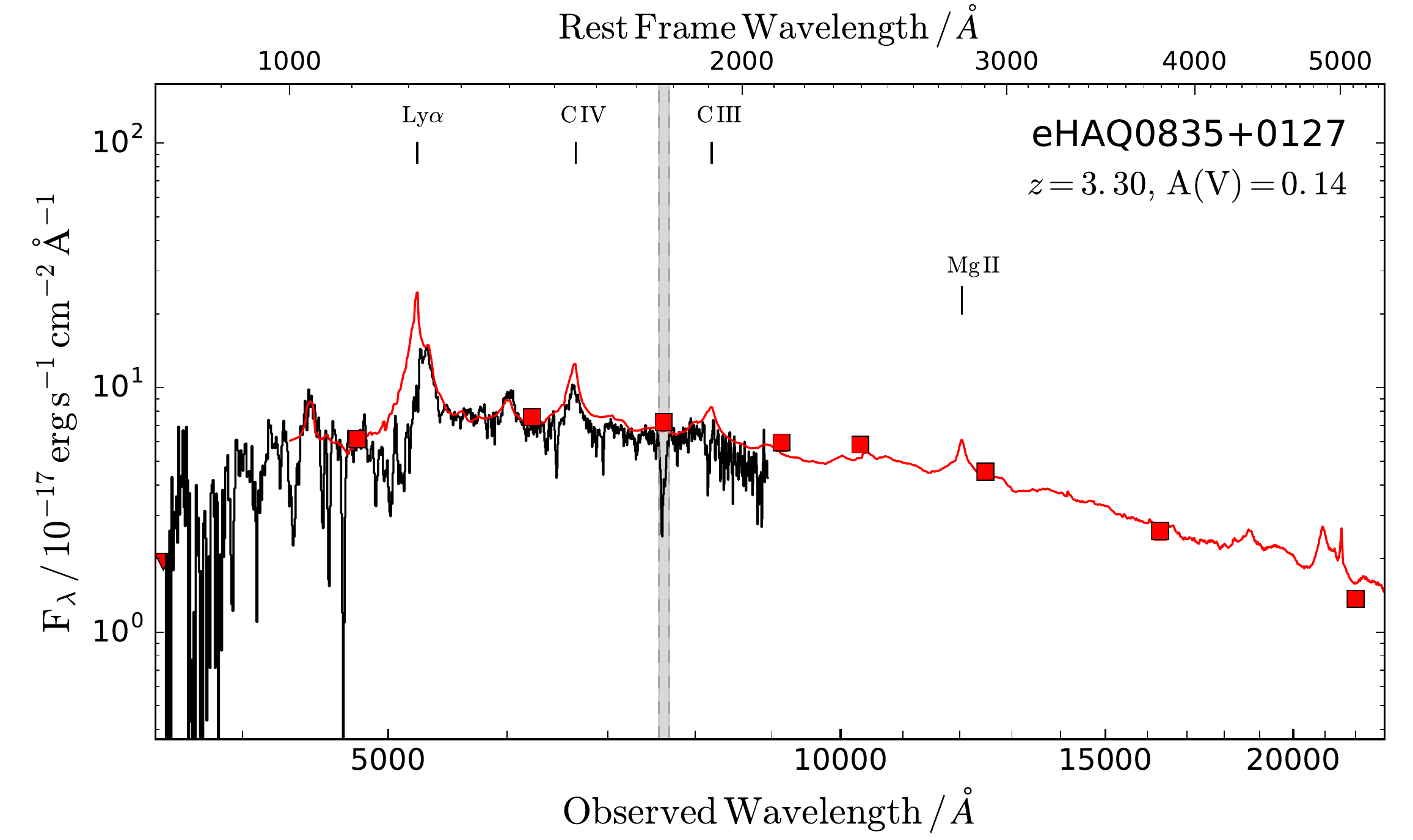}
    \includegraphics[width=0.49\textwidth]{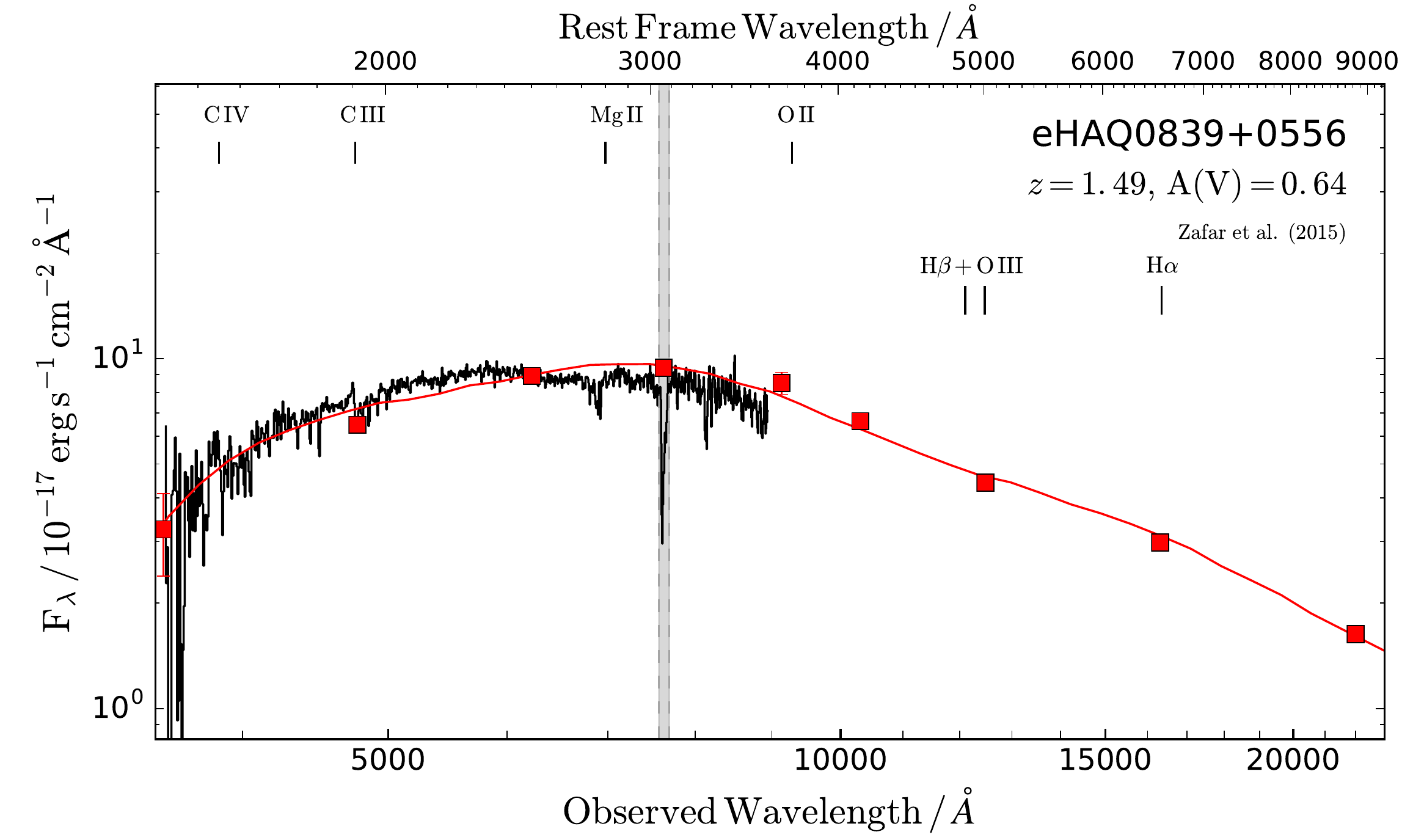}
    \vspace{4mm}

    \includegraphics[width=0.49\textwidth]{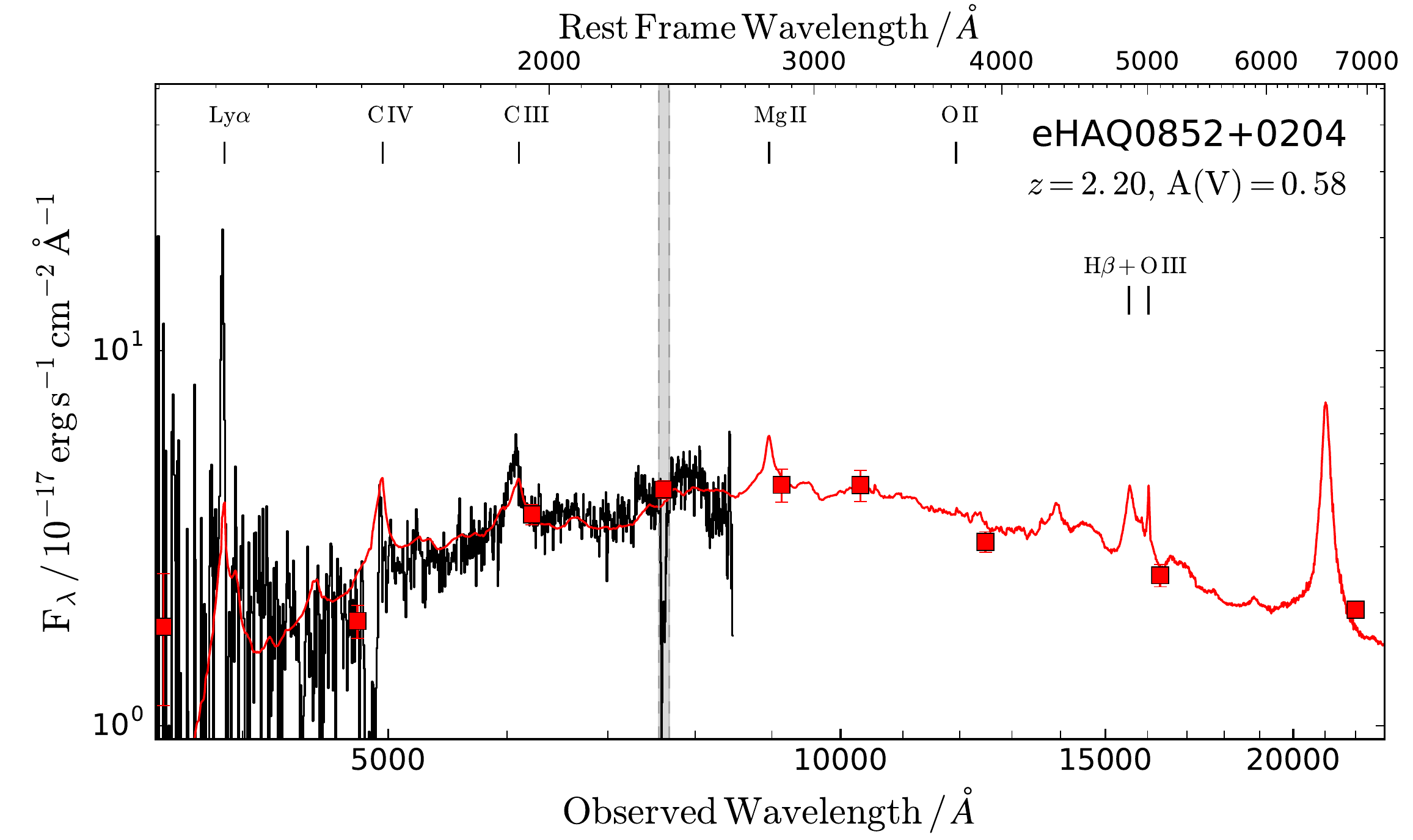}
    \includegraphics[width=0.49\textwidth]{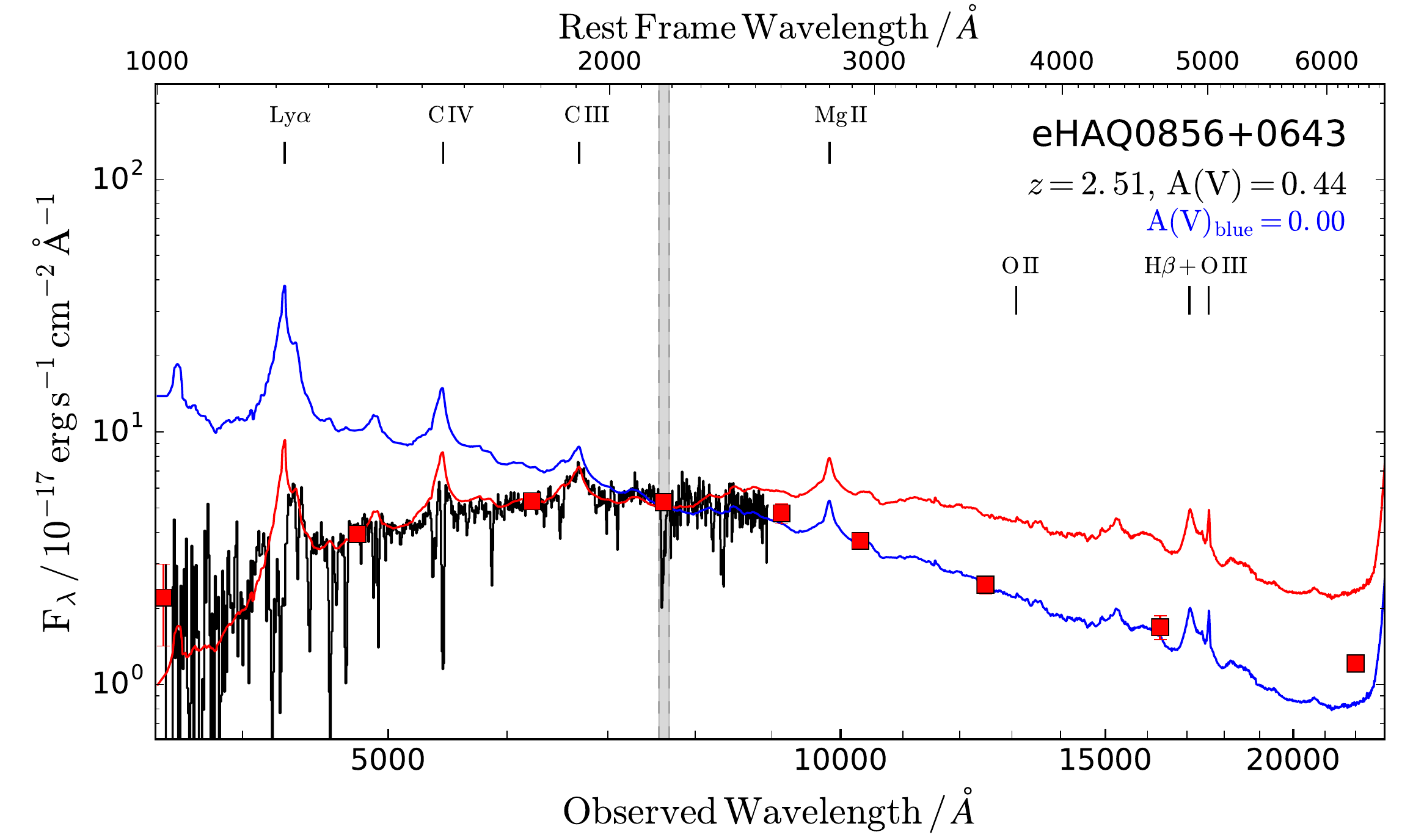}
\caption{(Continued.)}
\end{figure}

\begin{figure}
\figurenum{E2}
    \includegraphics[width=0.49\textwidth]{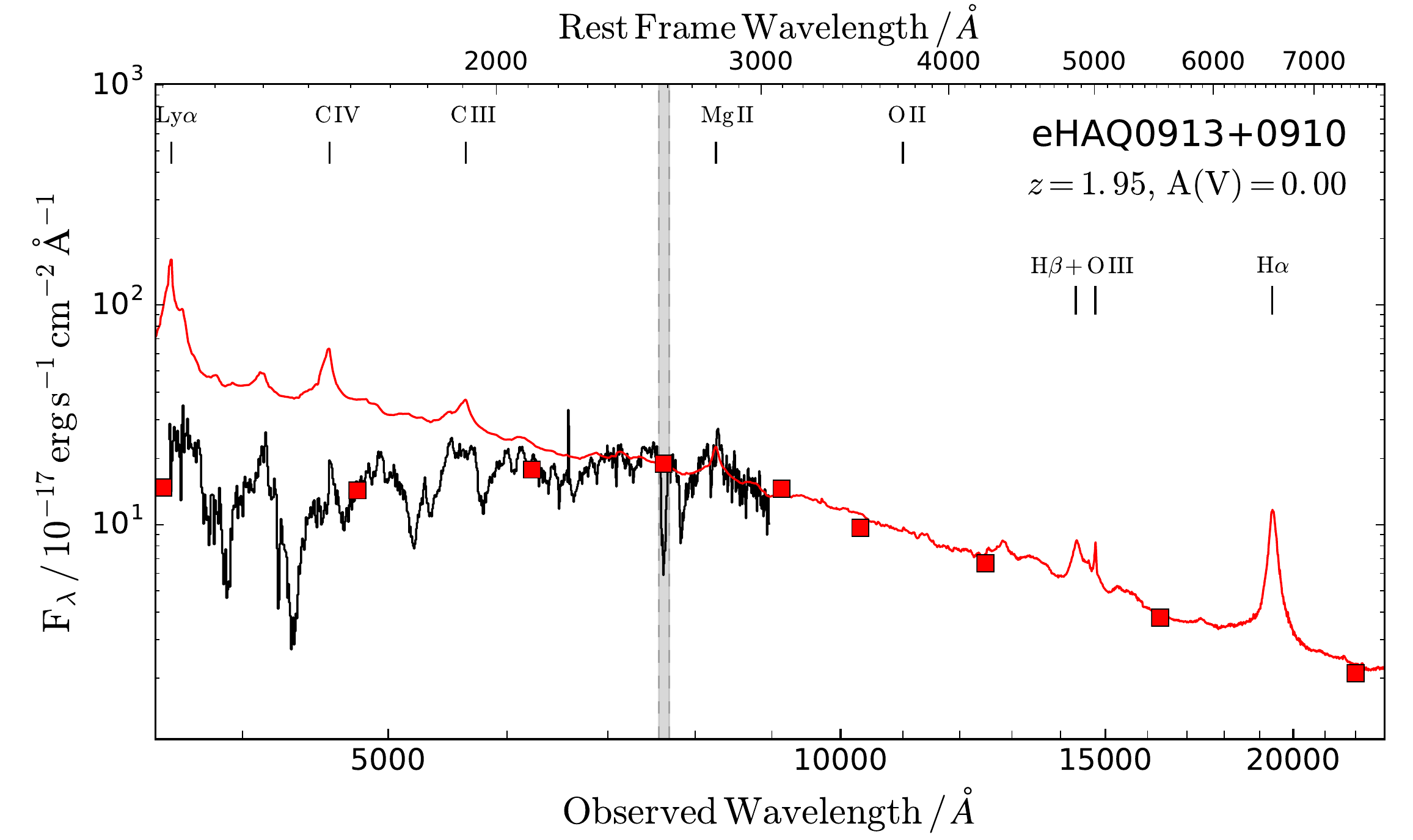}
    \includegraphics[width=0.49\textwidth]{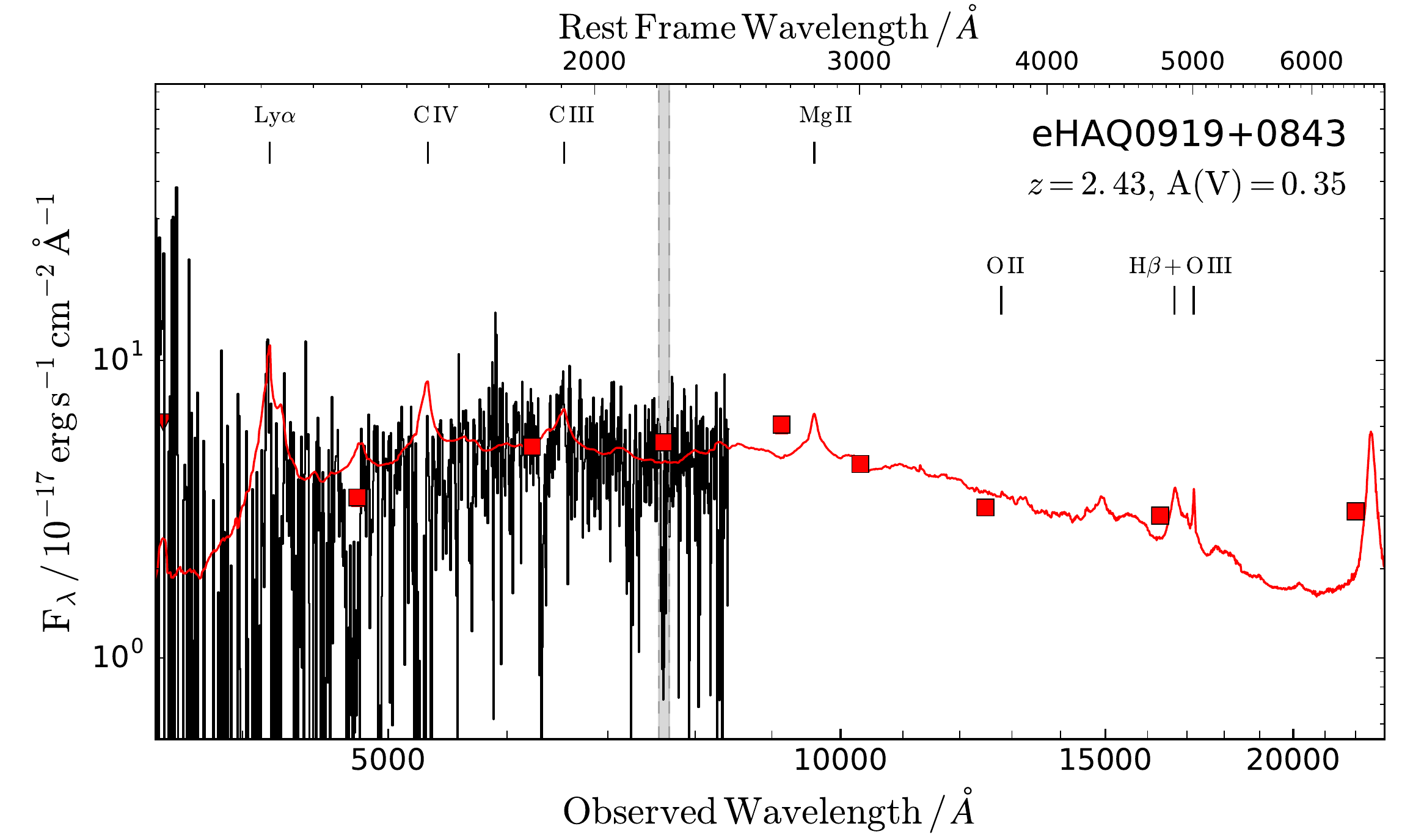}
    \vspace{4mm}

    \includegraphics[width=0.49\textwidth]{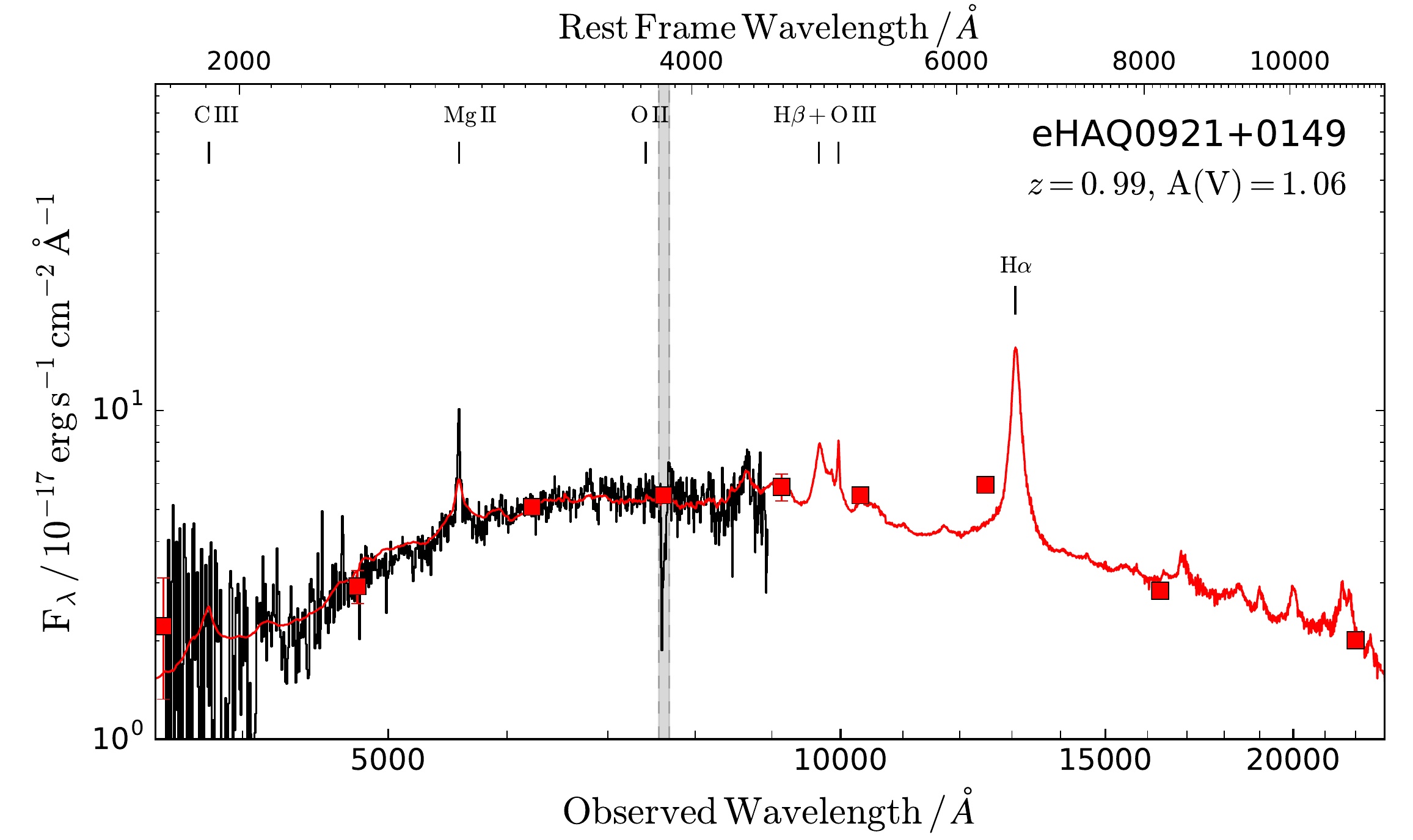}
    \includegraphics[width=0.49\textwidth]{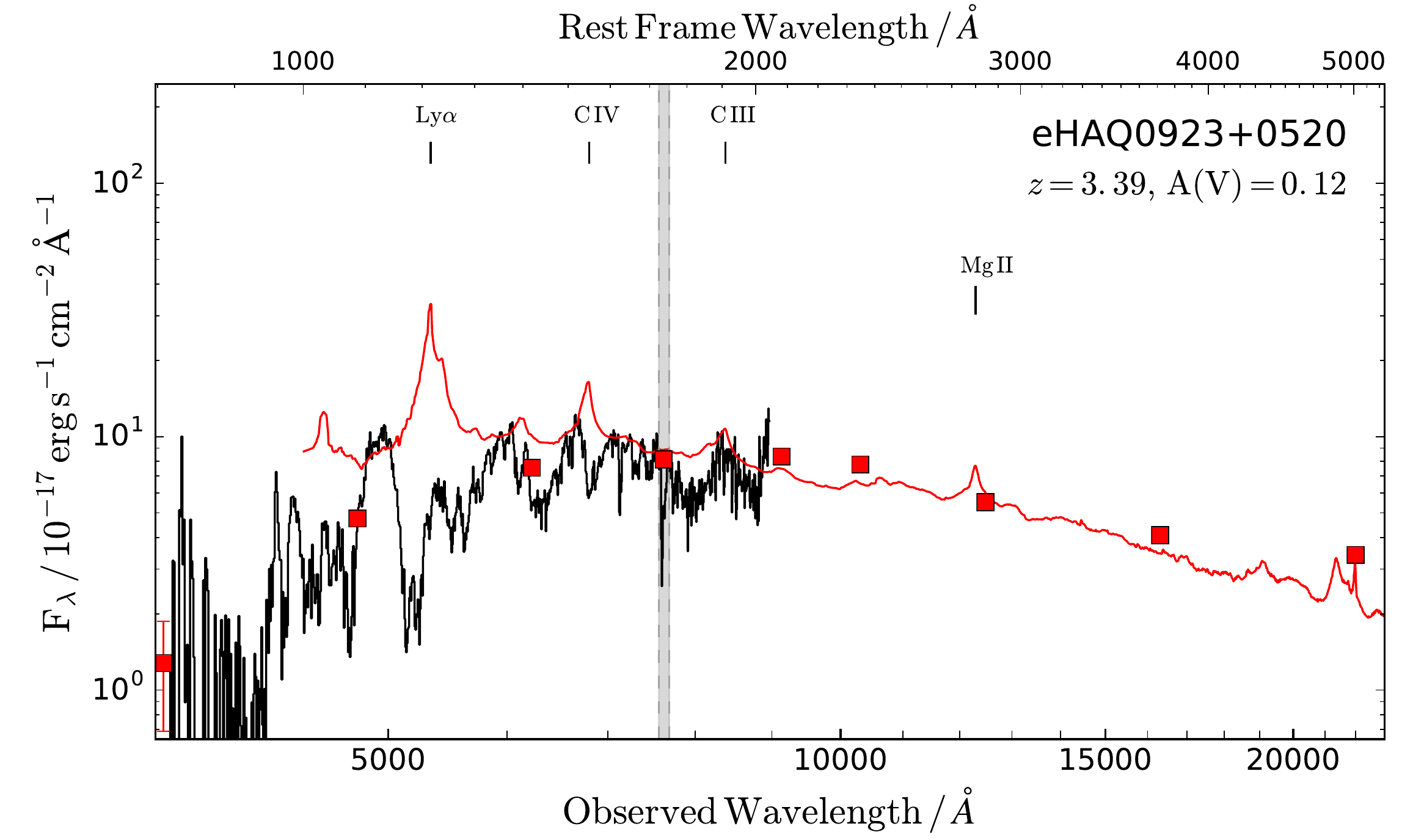}
    \vspace{4mm}

    \includegraphics[width=0.49\textwidth]{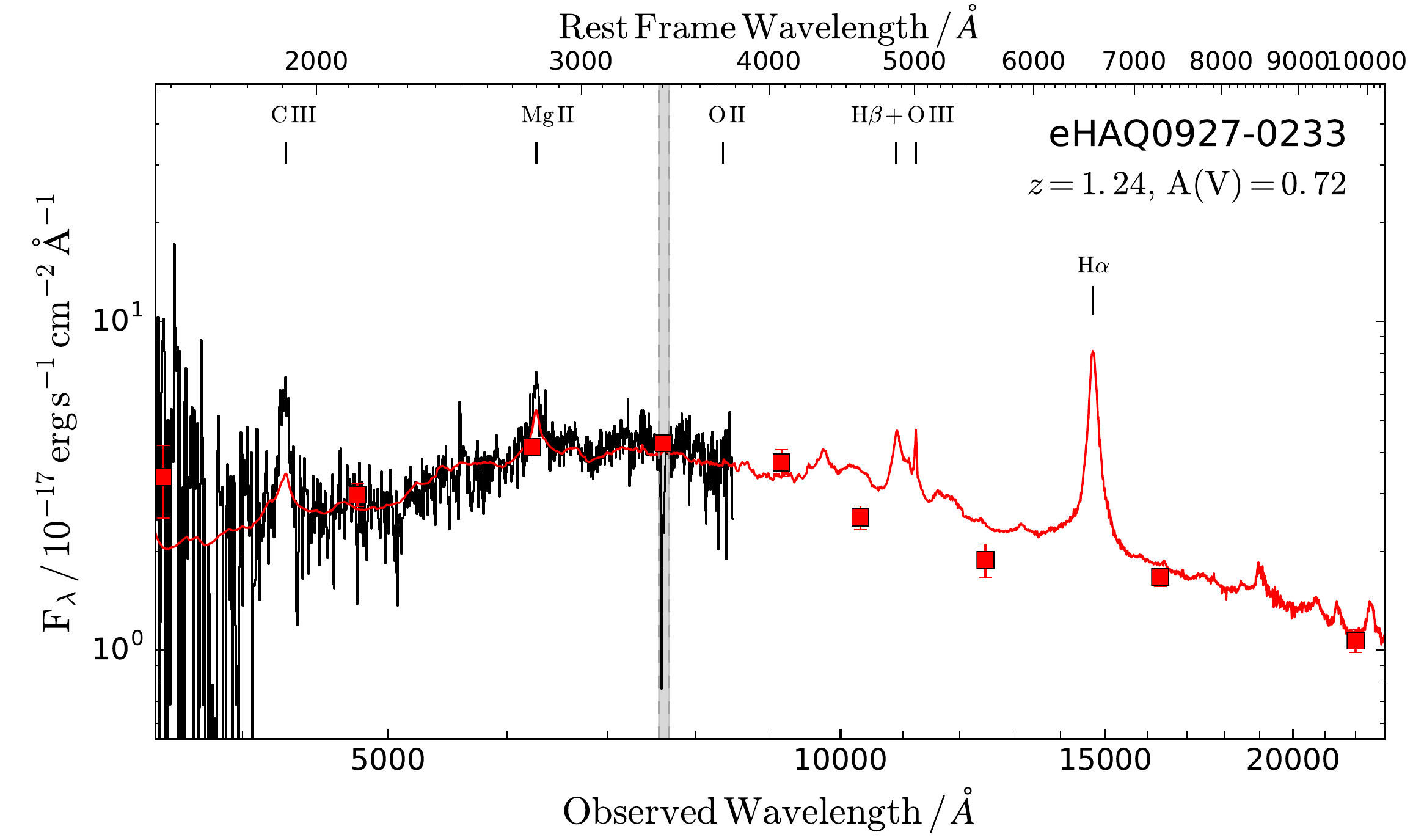}
    \includegraphics[width=0.49\textwidth]{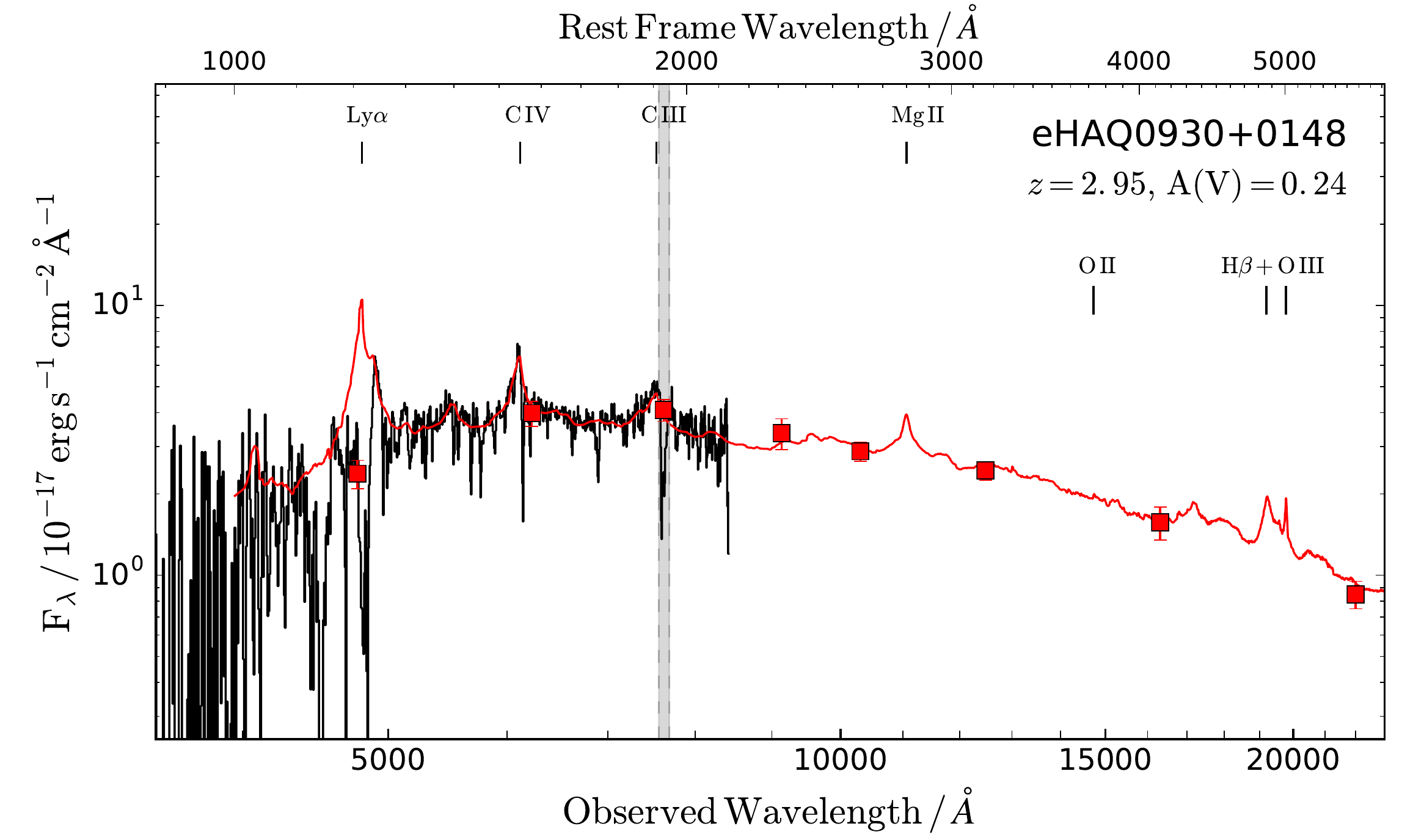}
    \vspace{4mm}

    \includegraphics[width=0.49\textwidth]{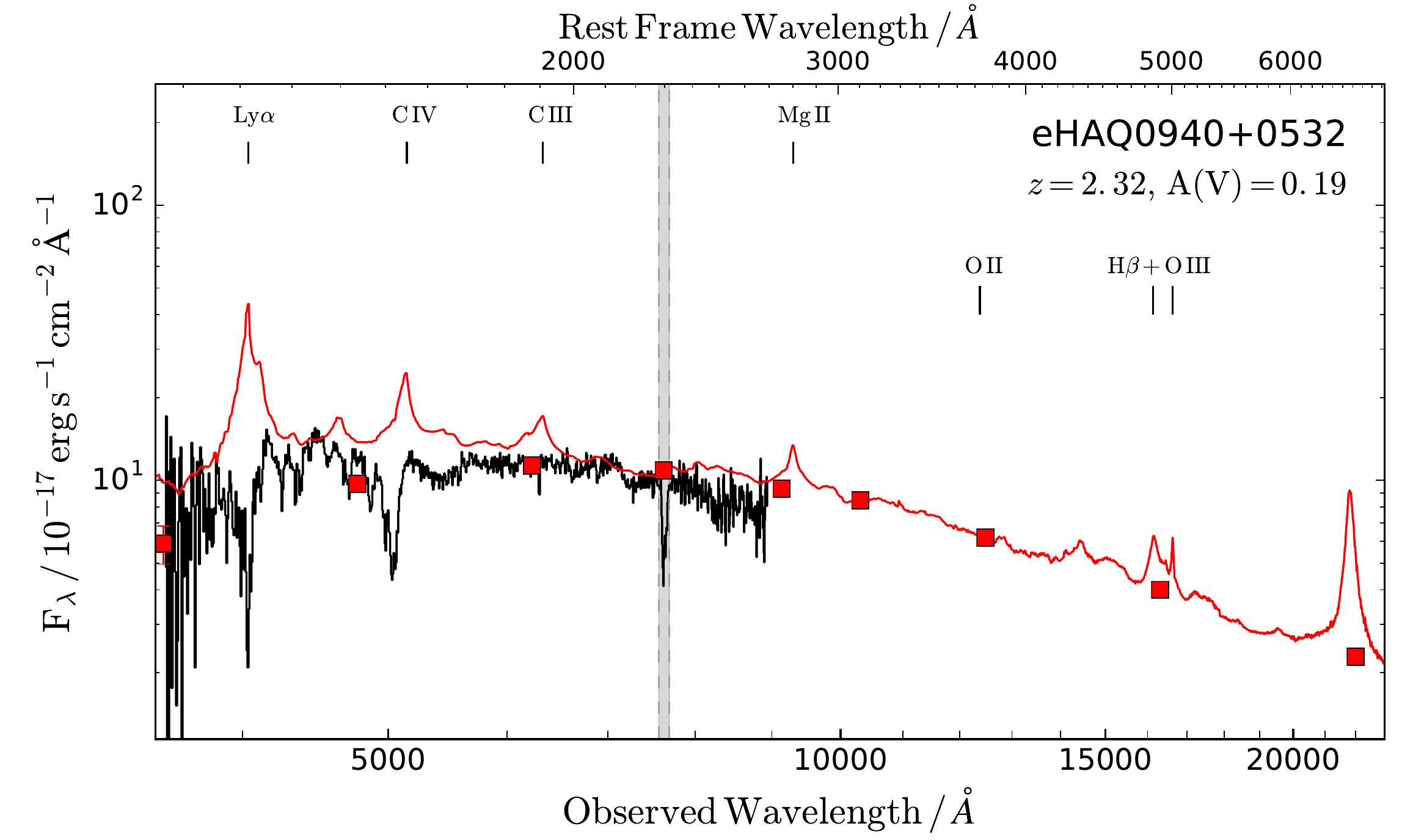}
    \includegraphics[width=0.49\textwidth]{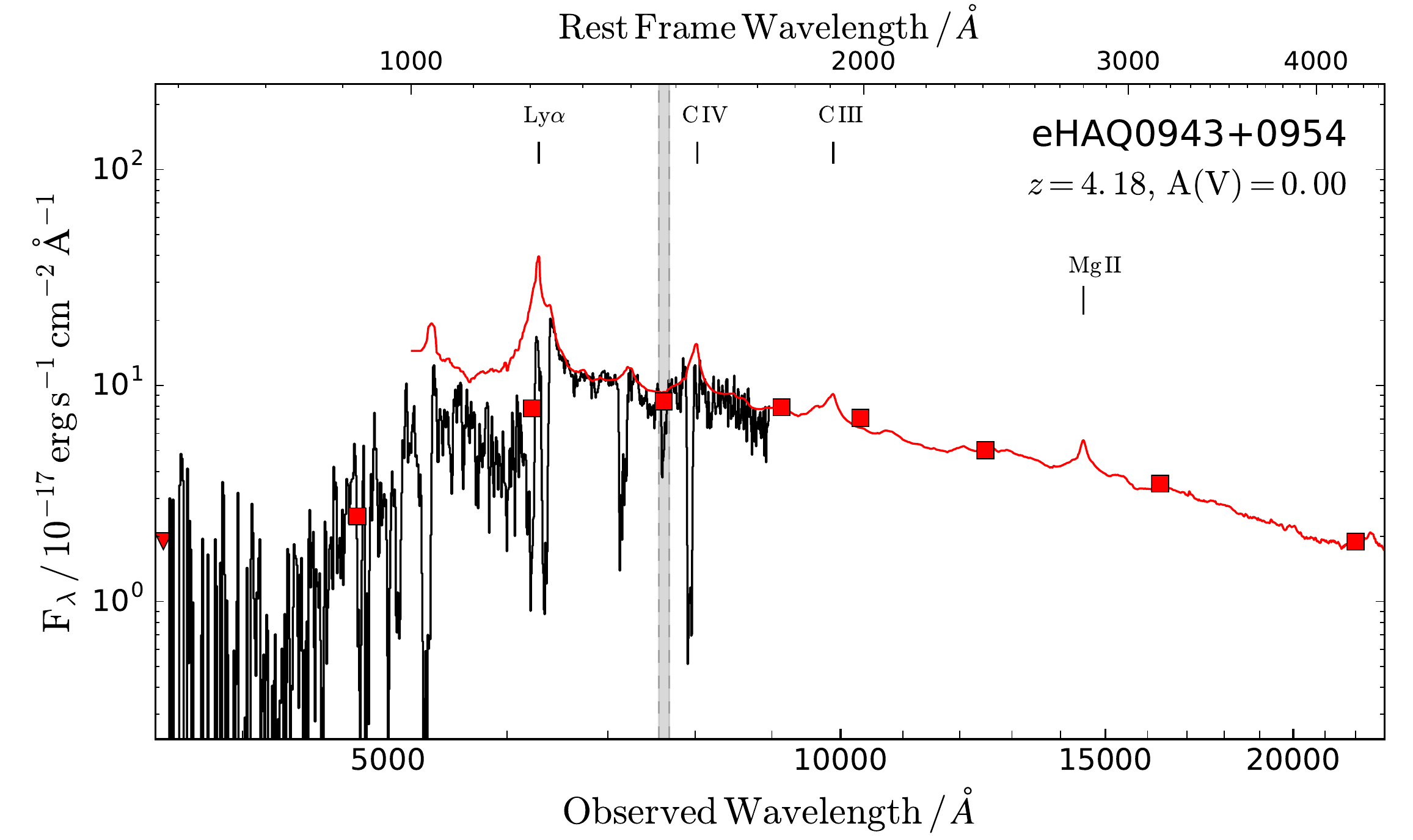}
\caption{(Continued.)}
\end{figure}

\begin{figure}
\figurenum{E2}
    \includegraphics[width=0.49\textwidth]{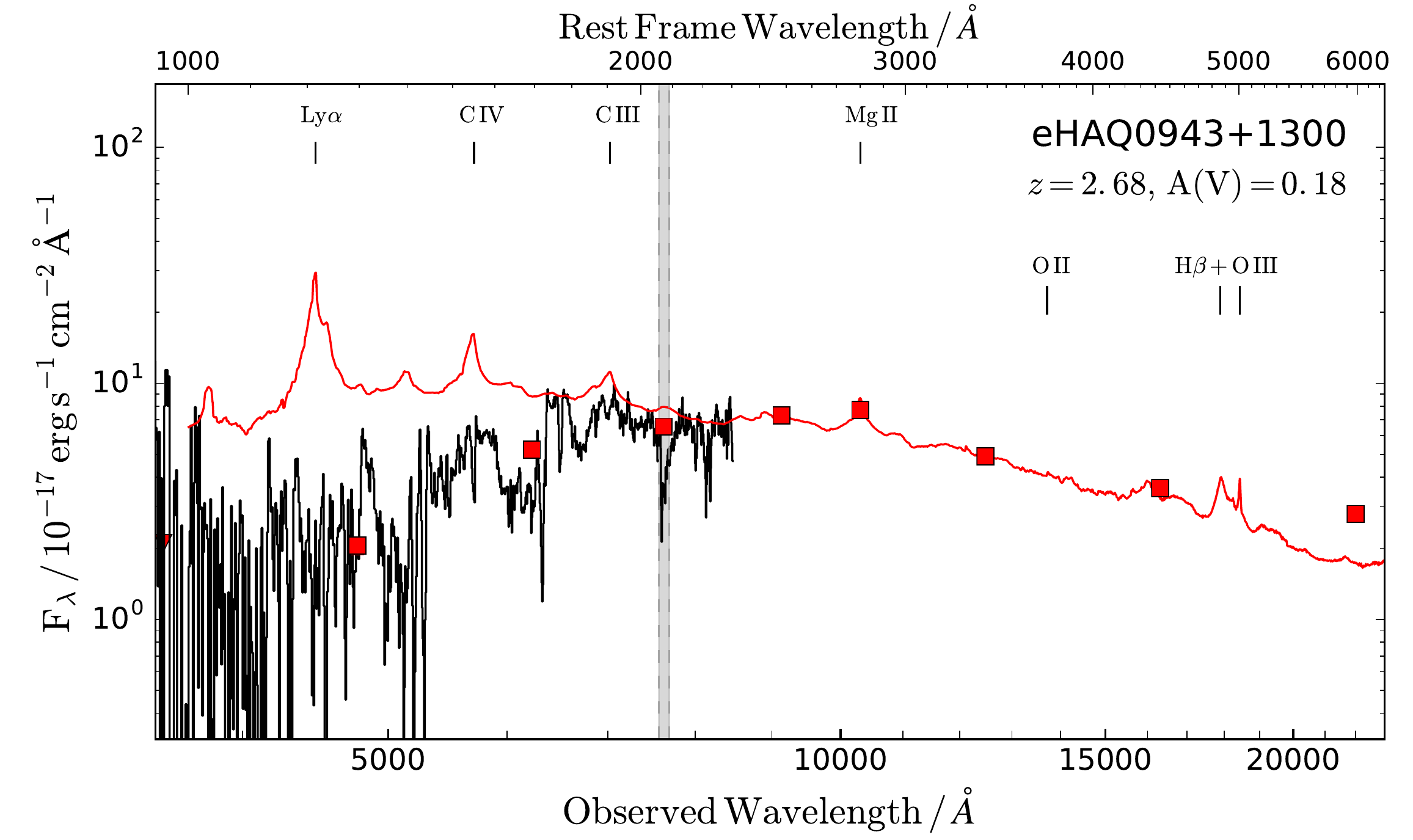}
    \includegraphics[width=0.49\textwidth]{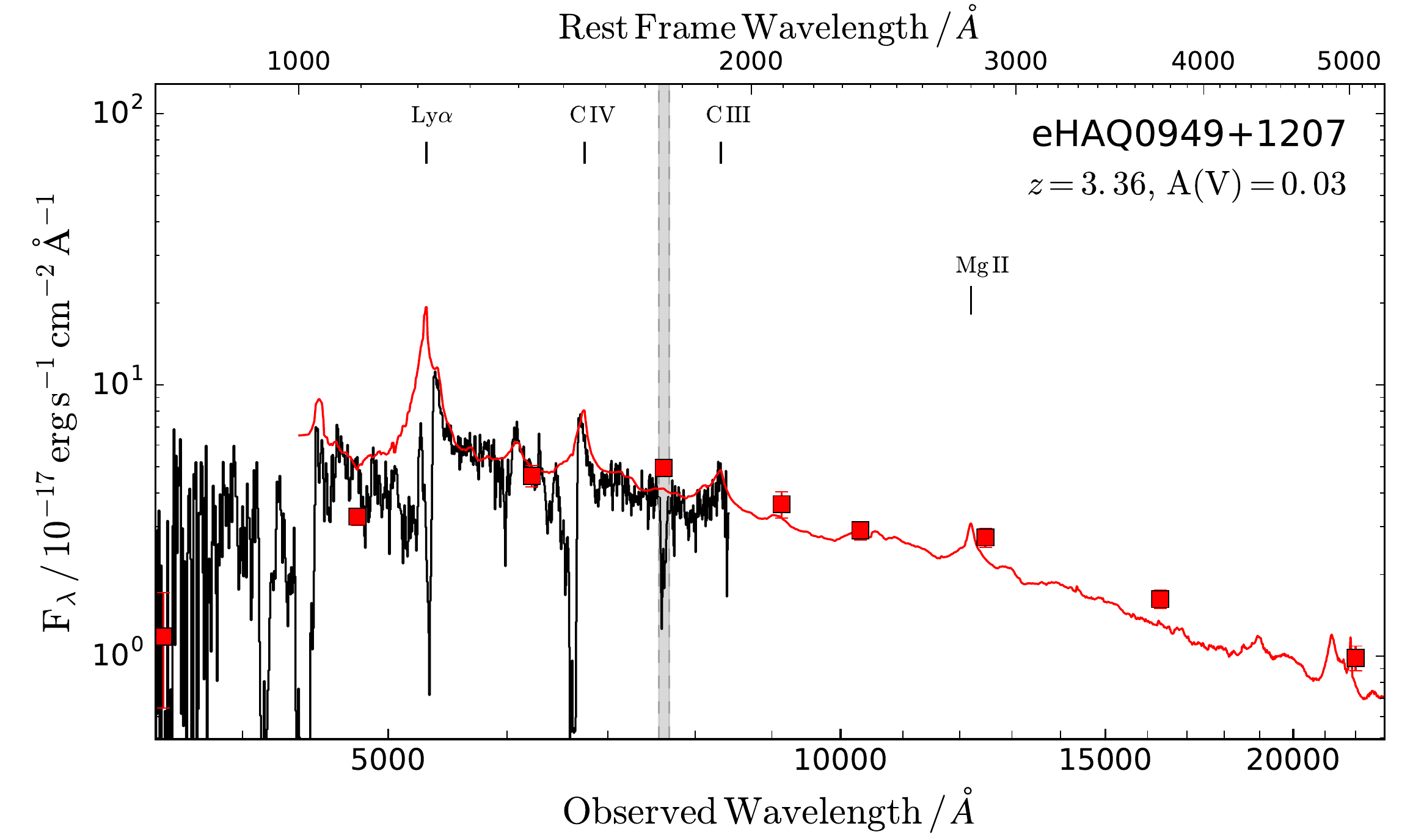}
    \vspace{4mm}

    \includegraphics[width=0.49\textwidth]{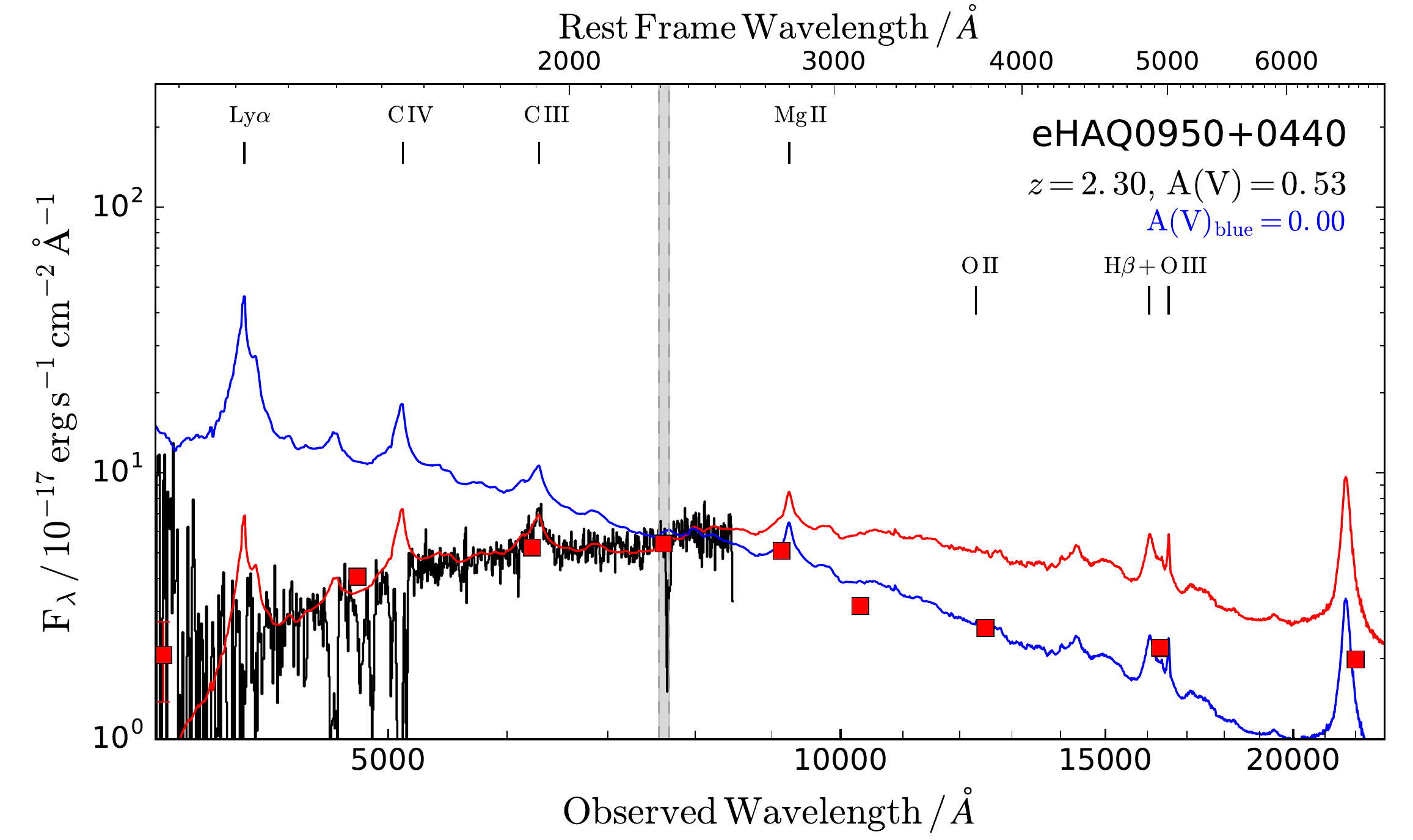}
    \includegraphics[width=0.49\textwidth]{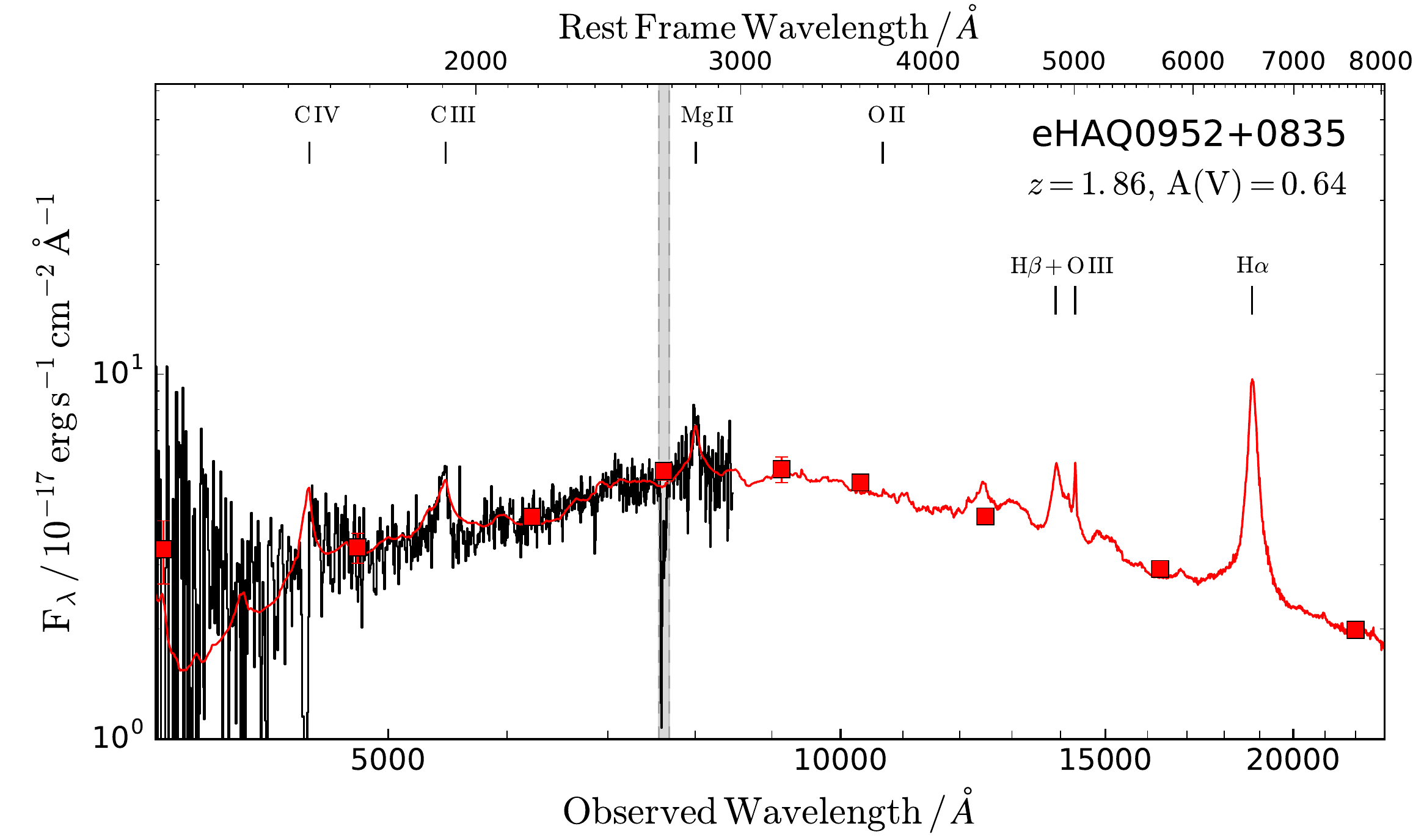}
    \vspace{4mm}

    \includegraphics[width=0.49\textwidth]{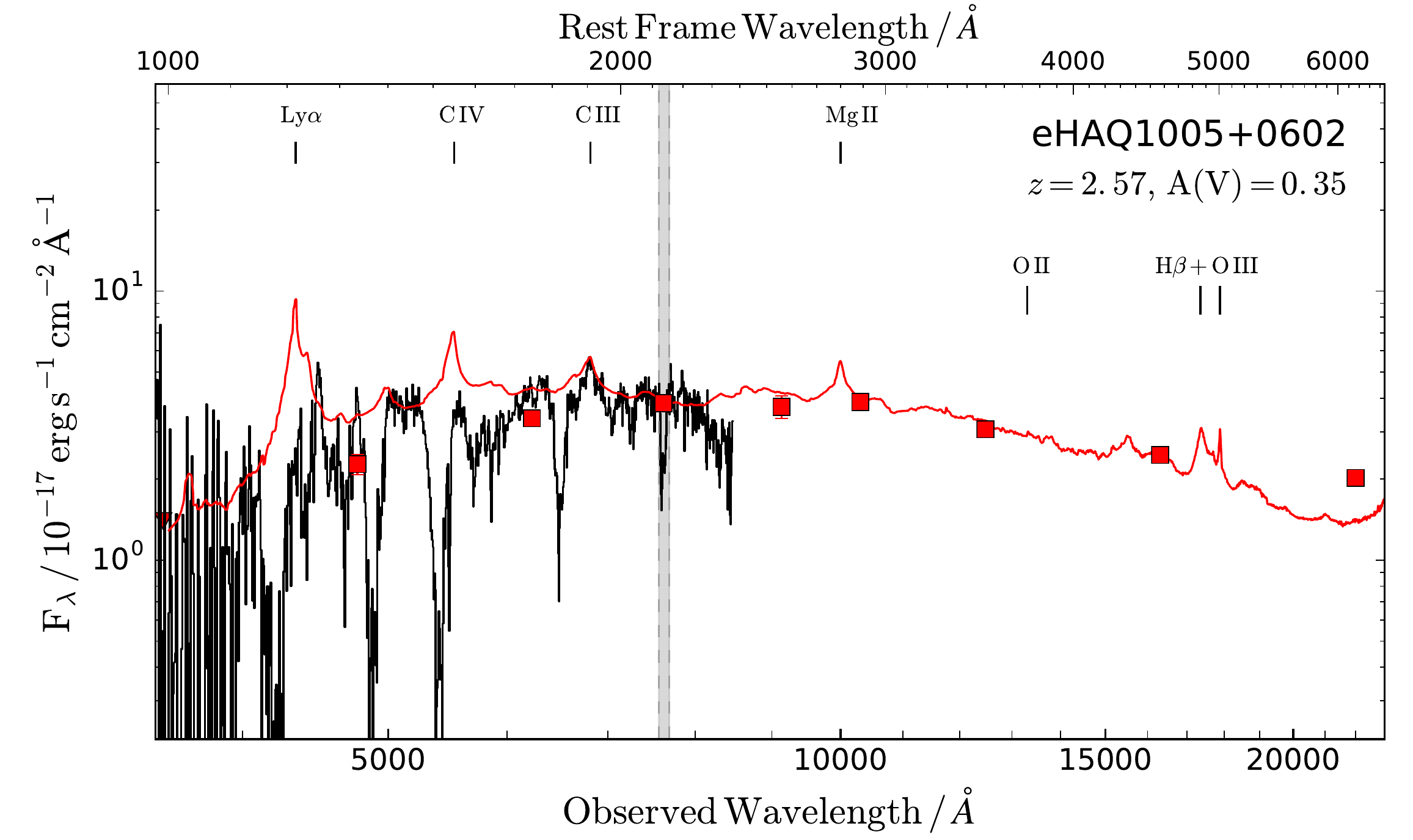}
    \includegraphics[width=0.49\textwidth]{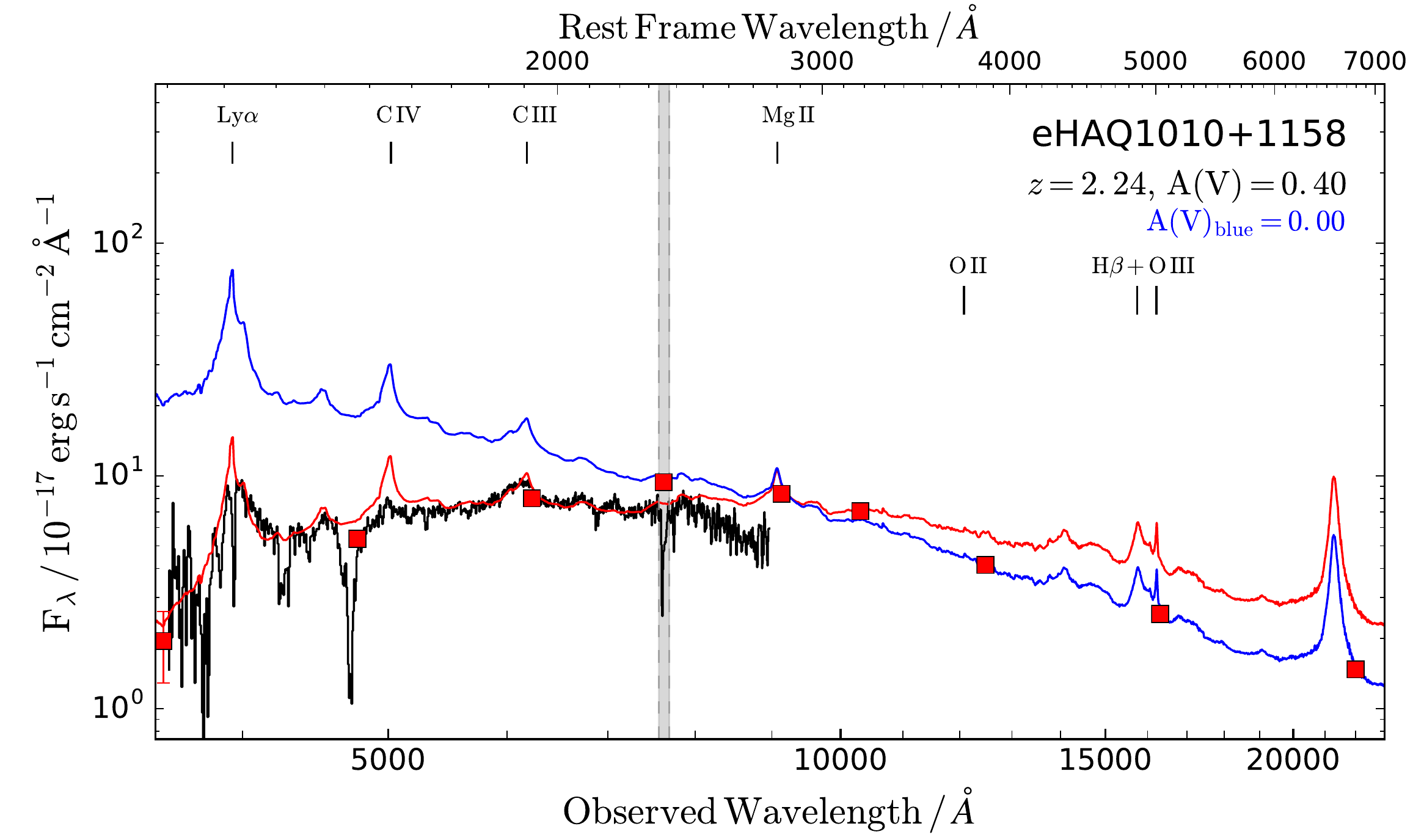}
    \vspace{4mm}

    \includegraphics[width=0.49\textwidth]{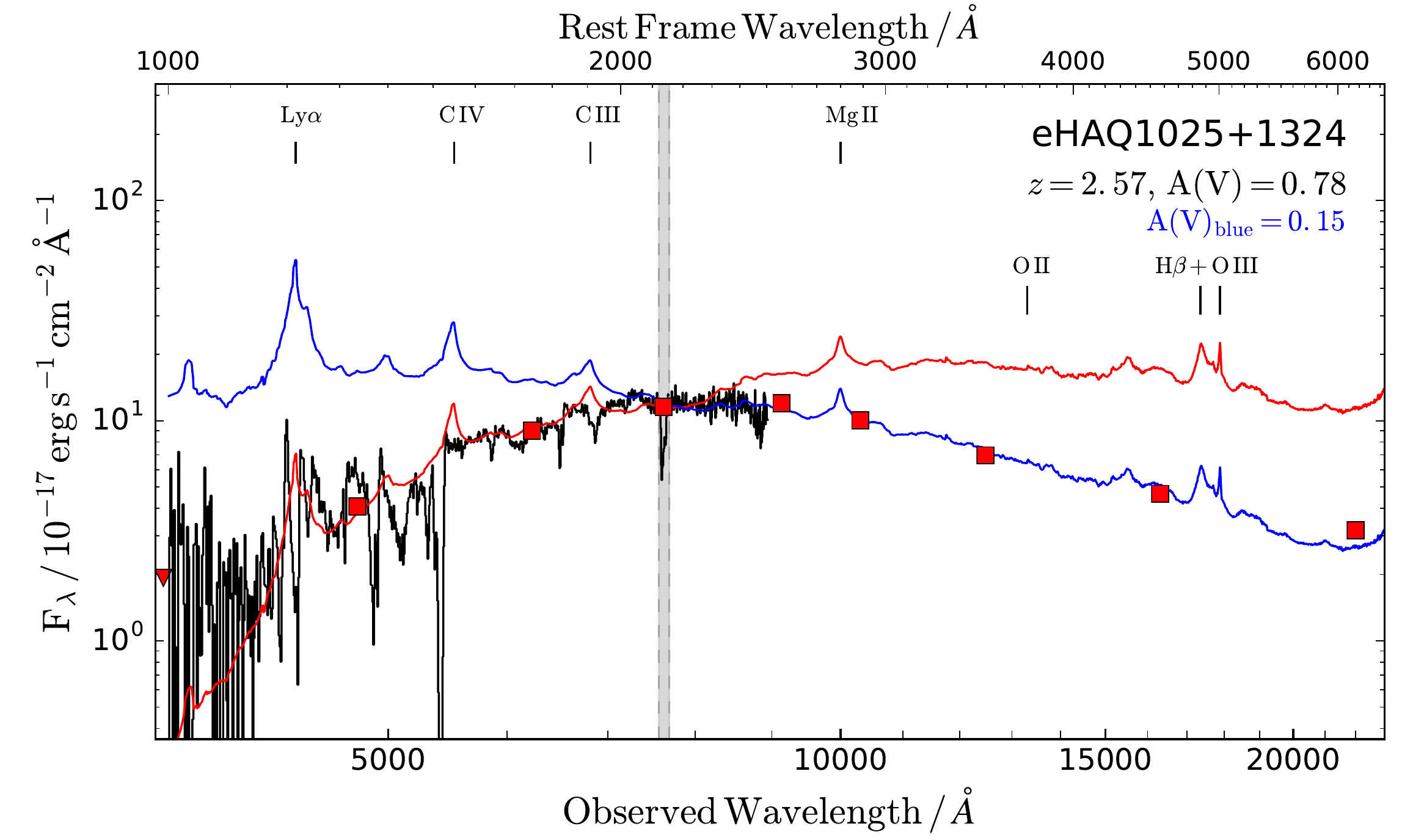}
    \includegraphics[width=0.49\textwidth]{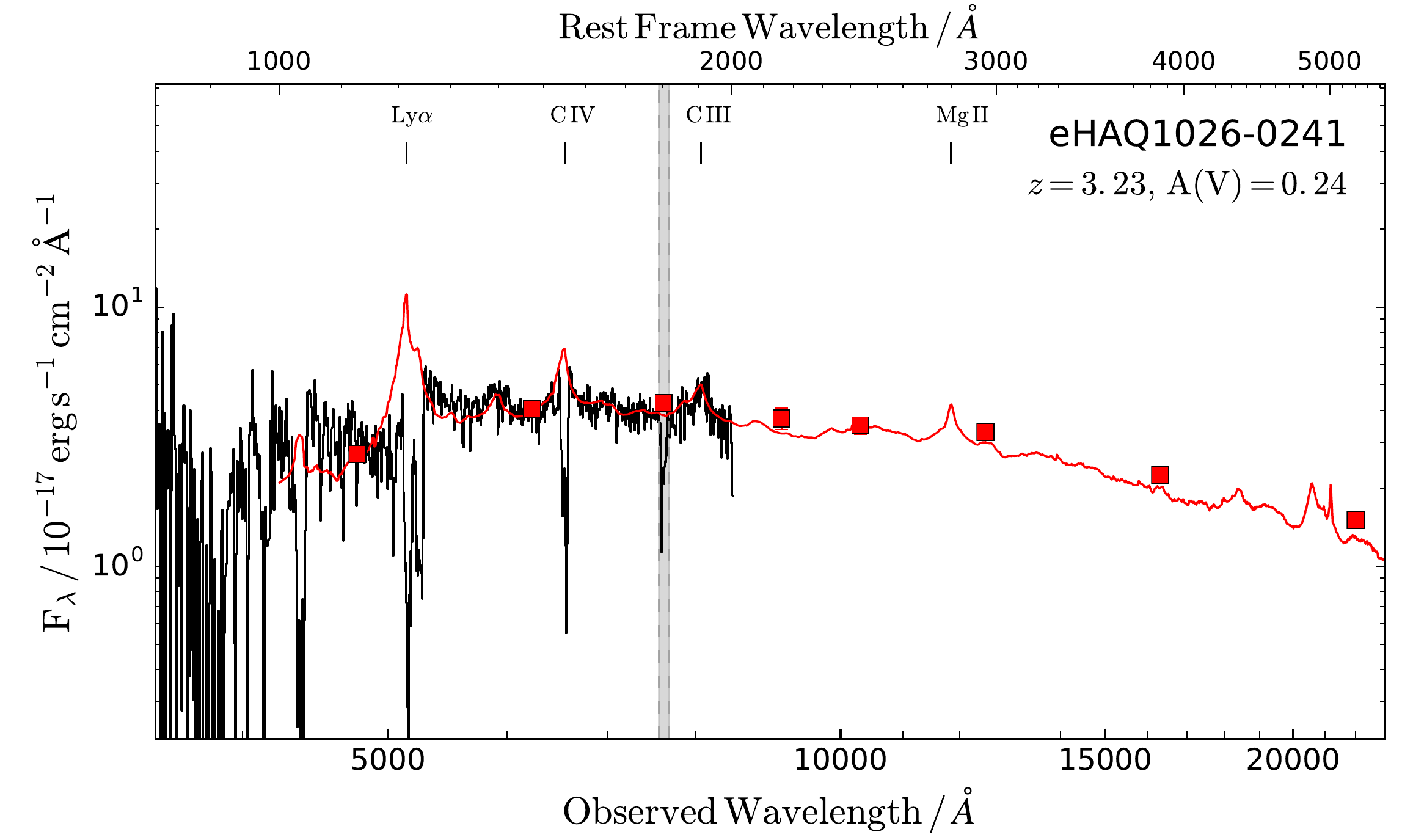}
\caption{(Continued.)}
\end{figure}

\begin{figure}
\figurenum{E2}
    \includegraphics[width=0.49\textwidth]{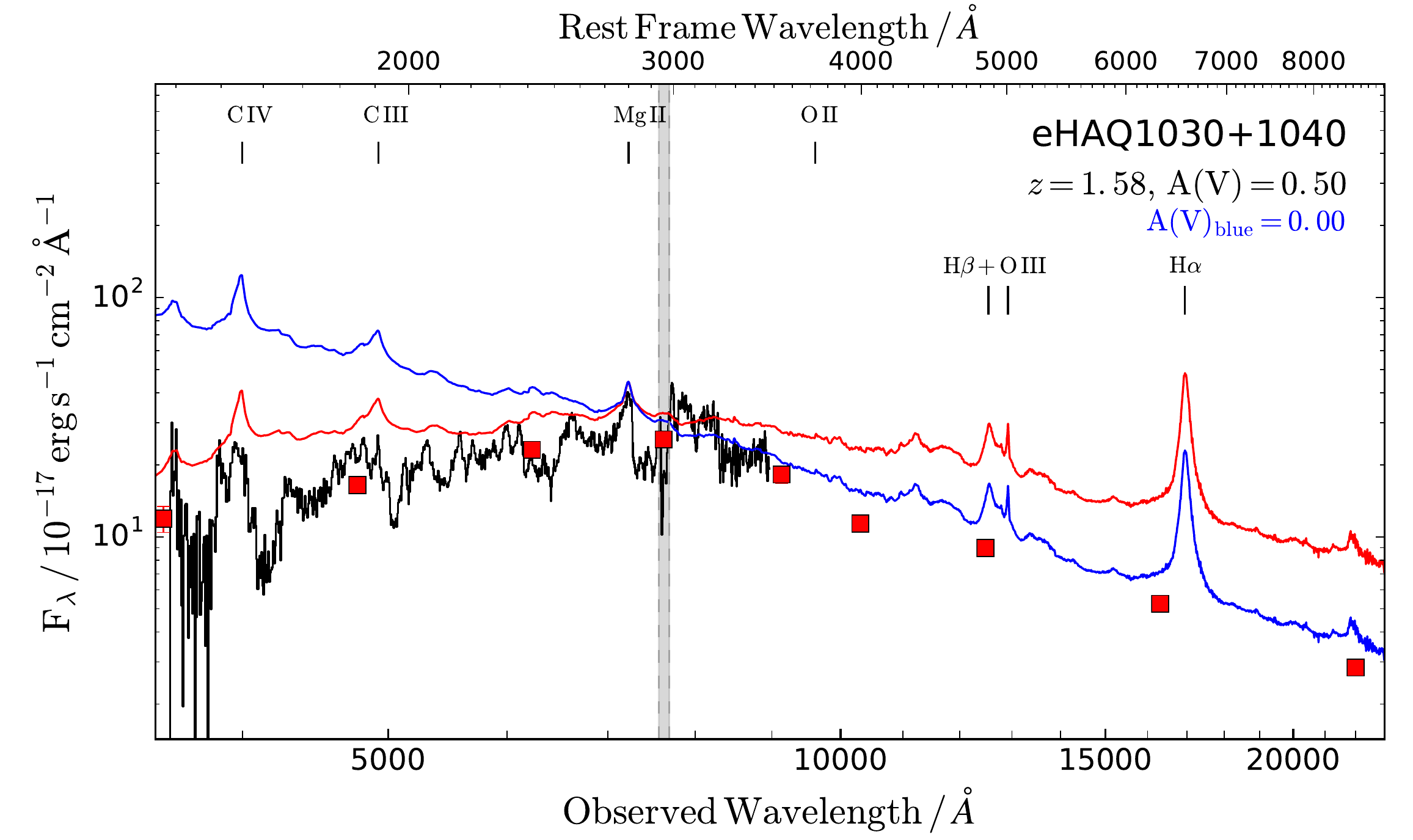}
    \includegraphics[width=0.49\textwidth]{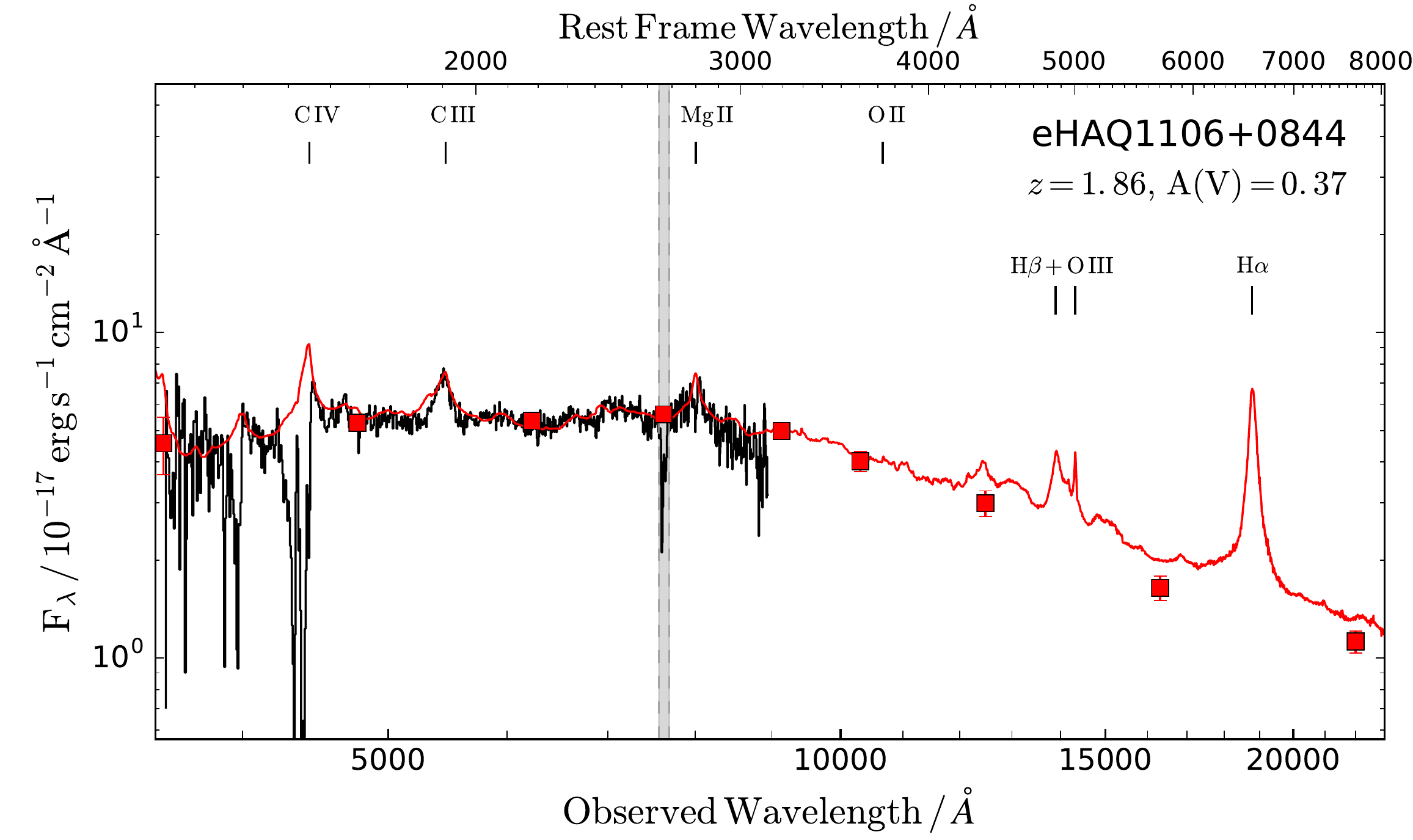}
    \vspace{4mm}

    \includegraphics[width=0.49\textwidth]{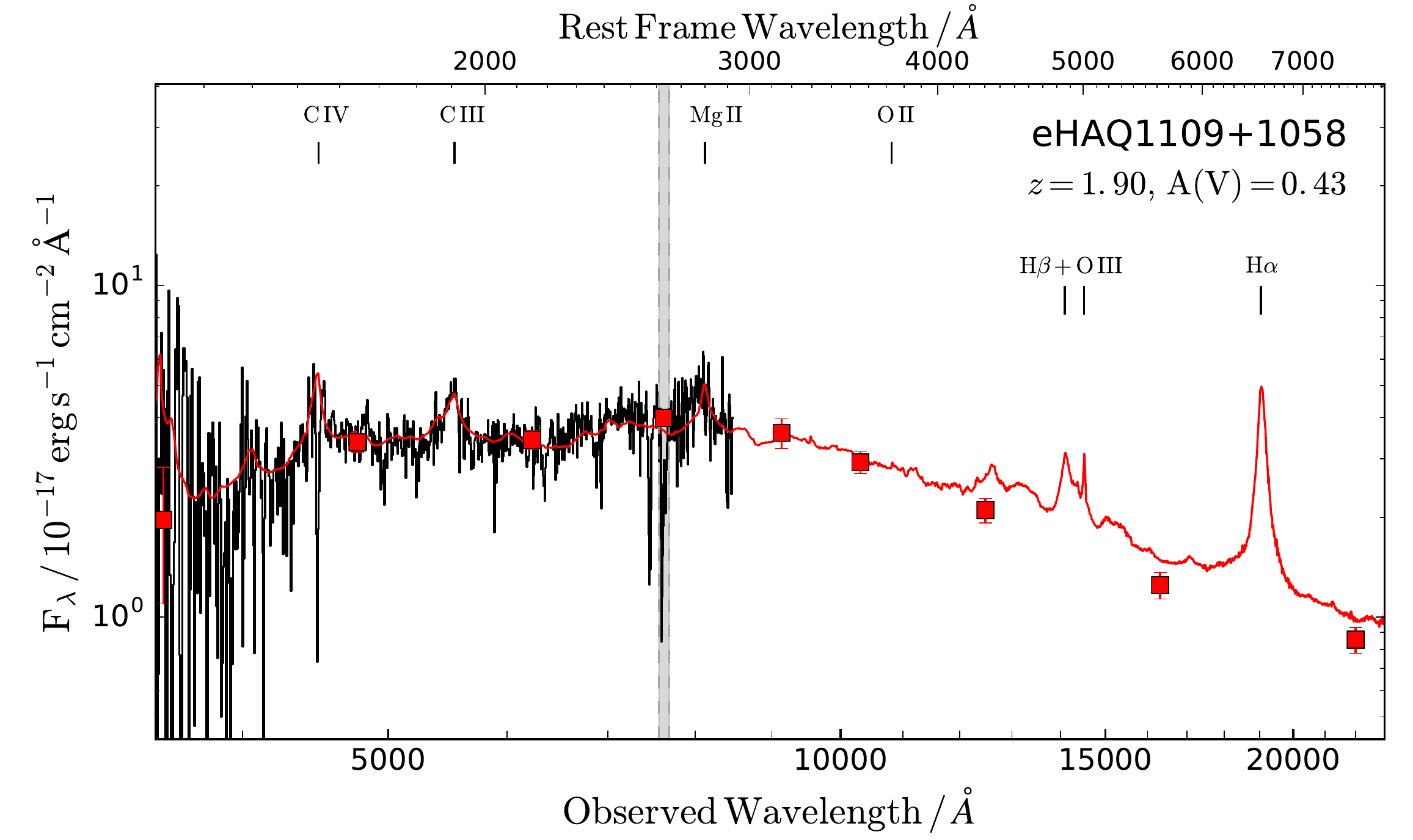}
    \includegraphics[width=0.49\textwidth]{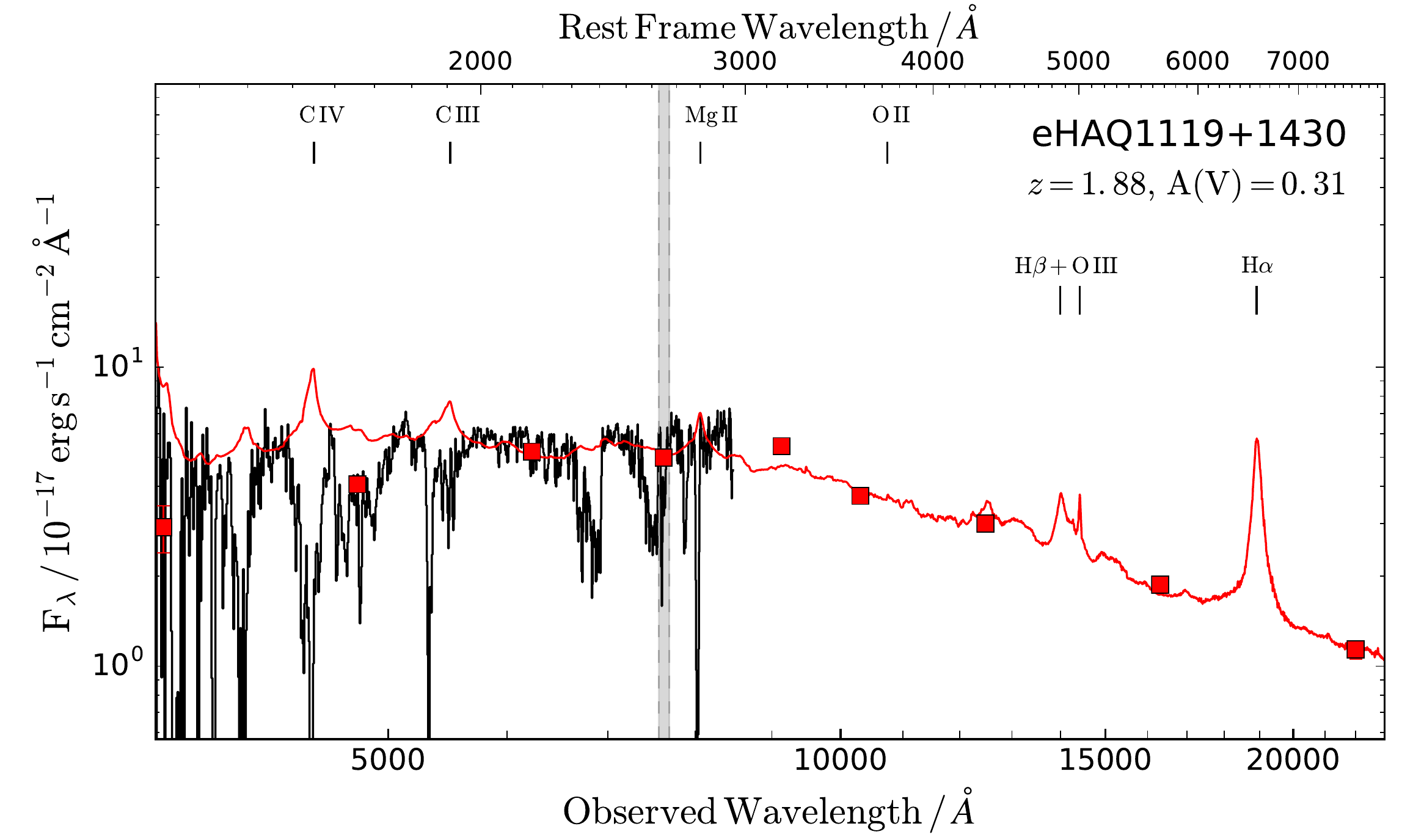}
    \vspace{4mm}

    \includegraphics[width=0.49\textwidth]{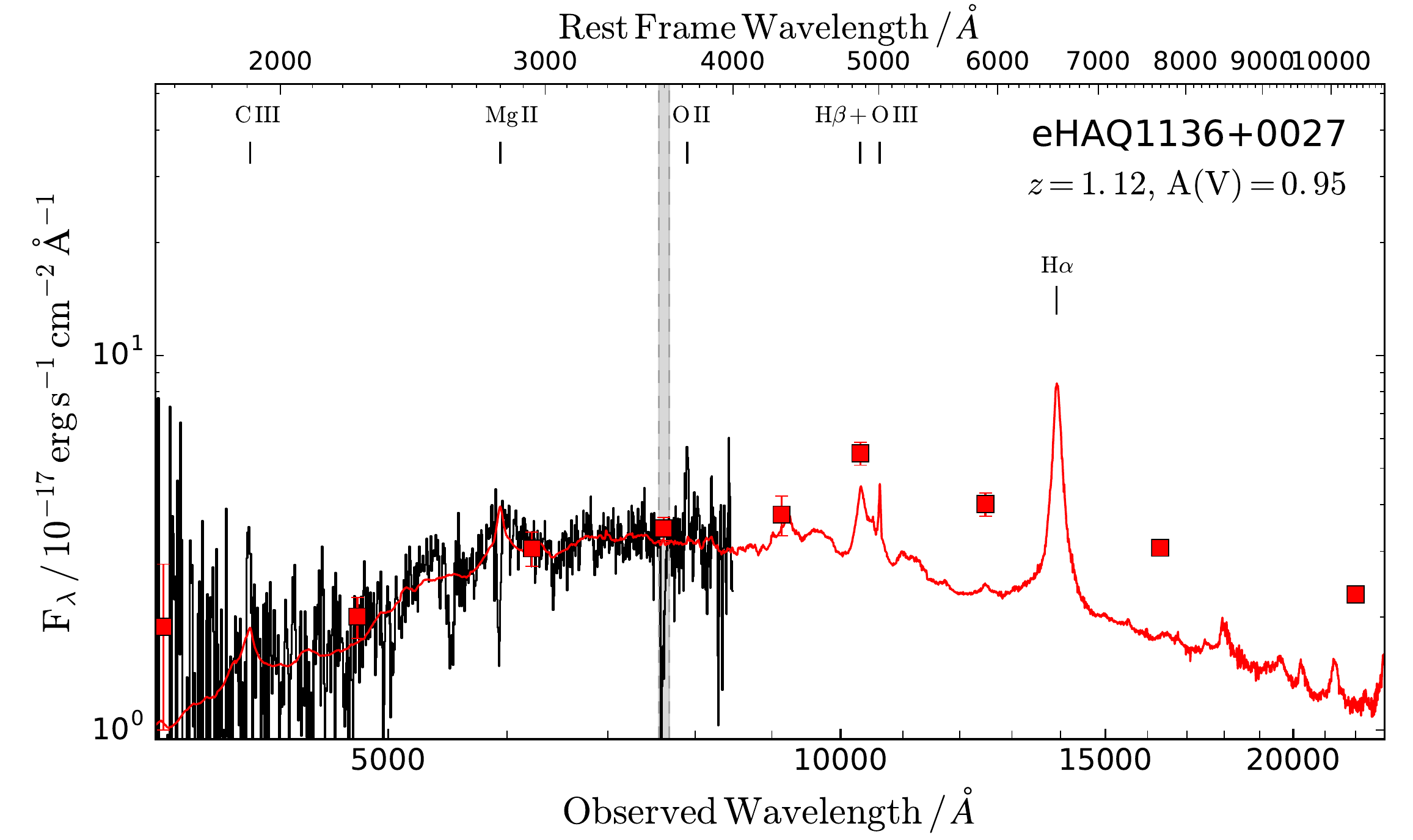}
    \includegraphics[width=0.49\textwidth]{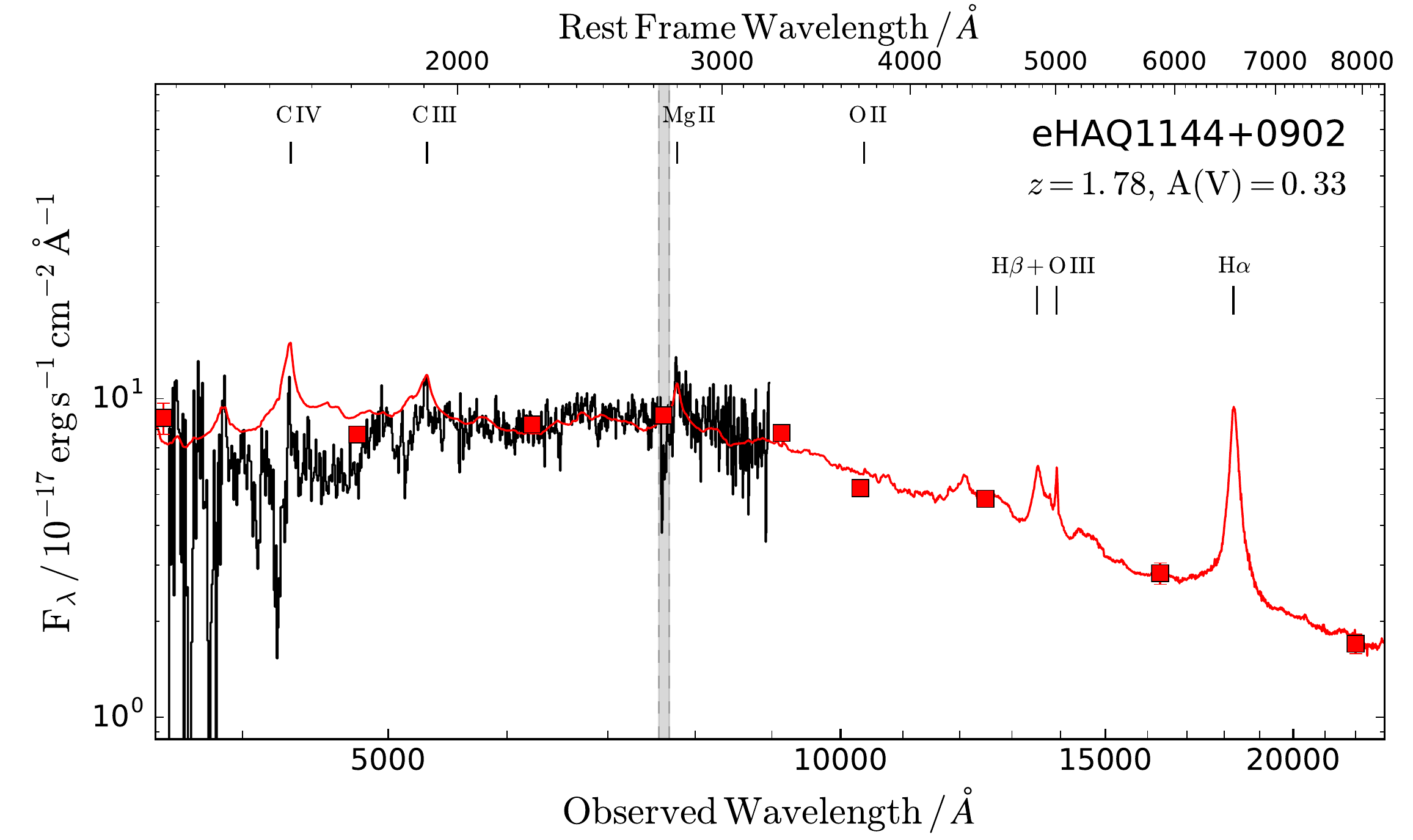}
    \vspace{4mm}

    \includegraphics[width=0.49\textwidth]{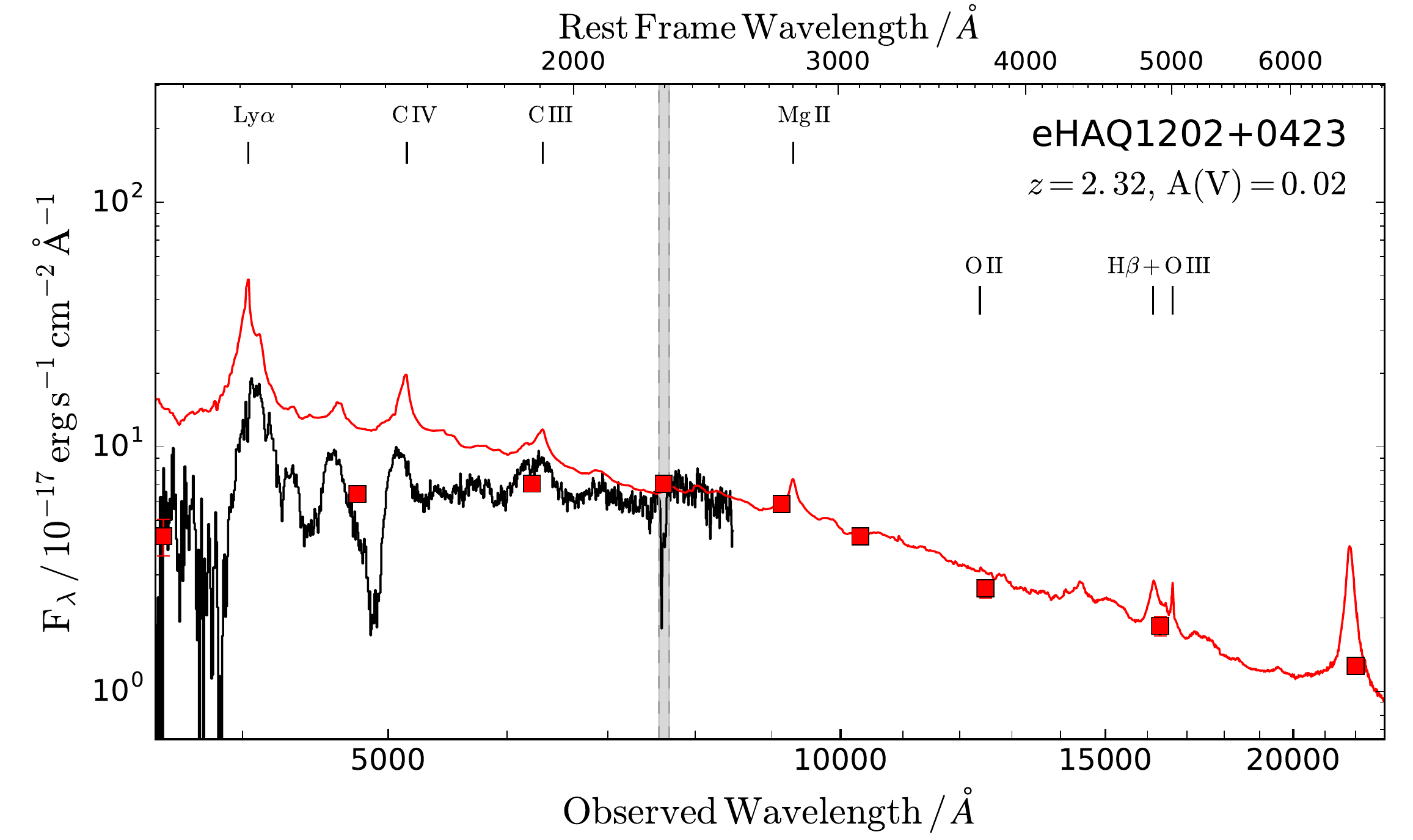}
    \includegraphics[width=0.49\textwidth]{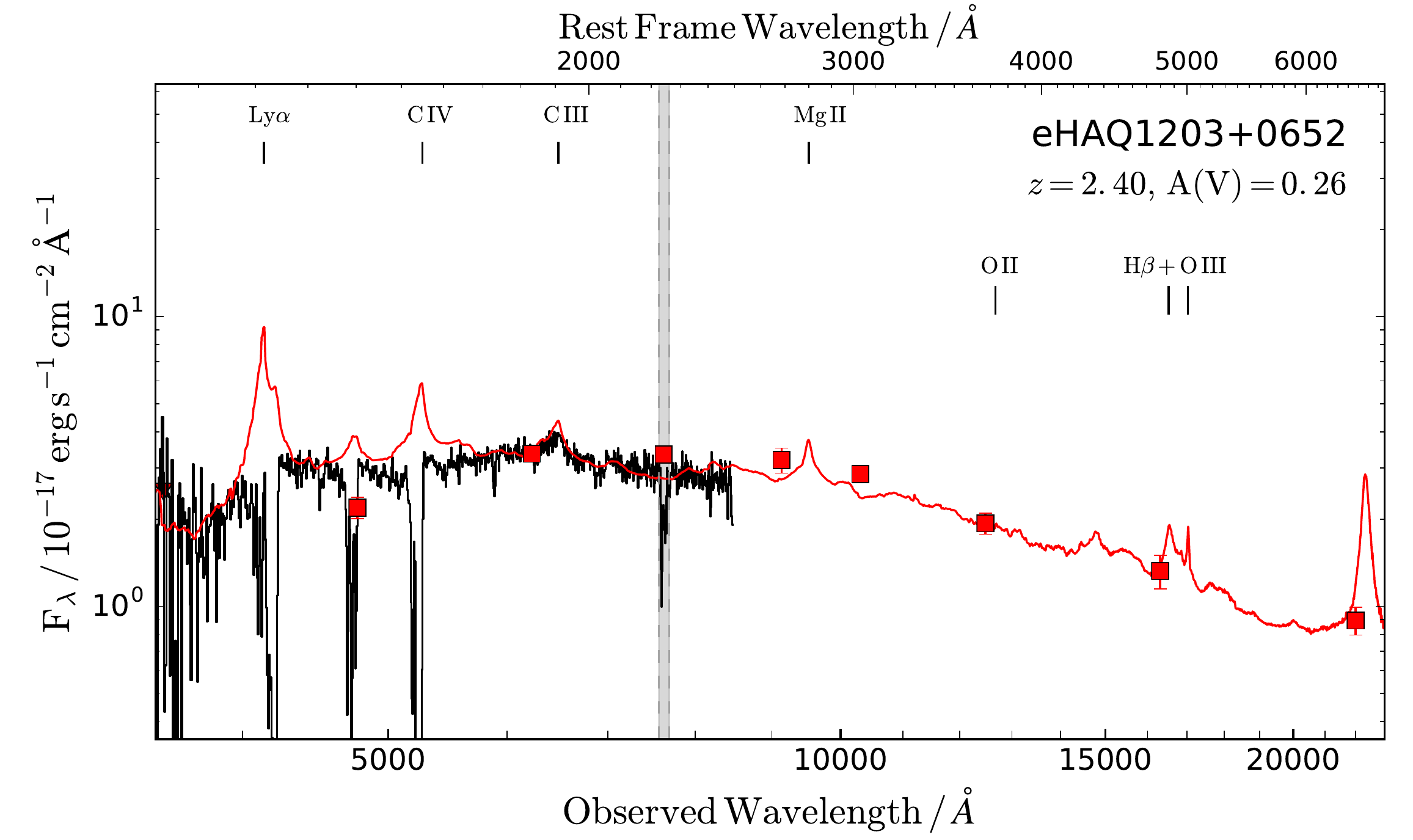}
\caption{(Continued.)}
\end{figure}

\begin{figure}
\figurenum{E2}
    \includegraphics[width=0.49\textwidth]{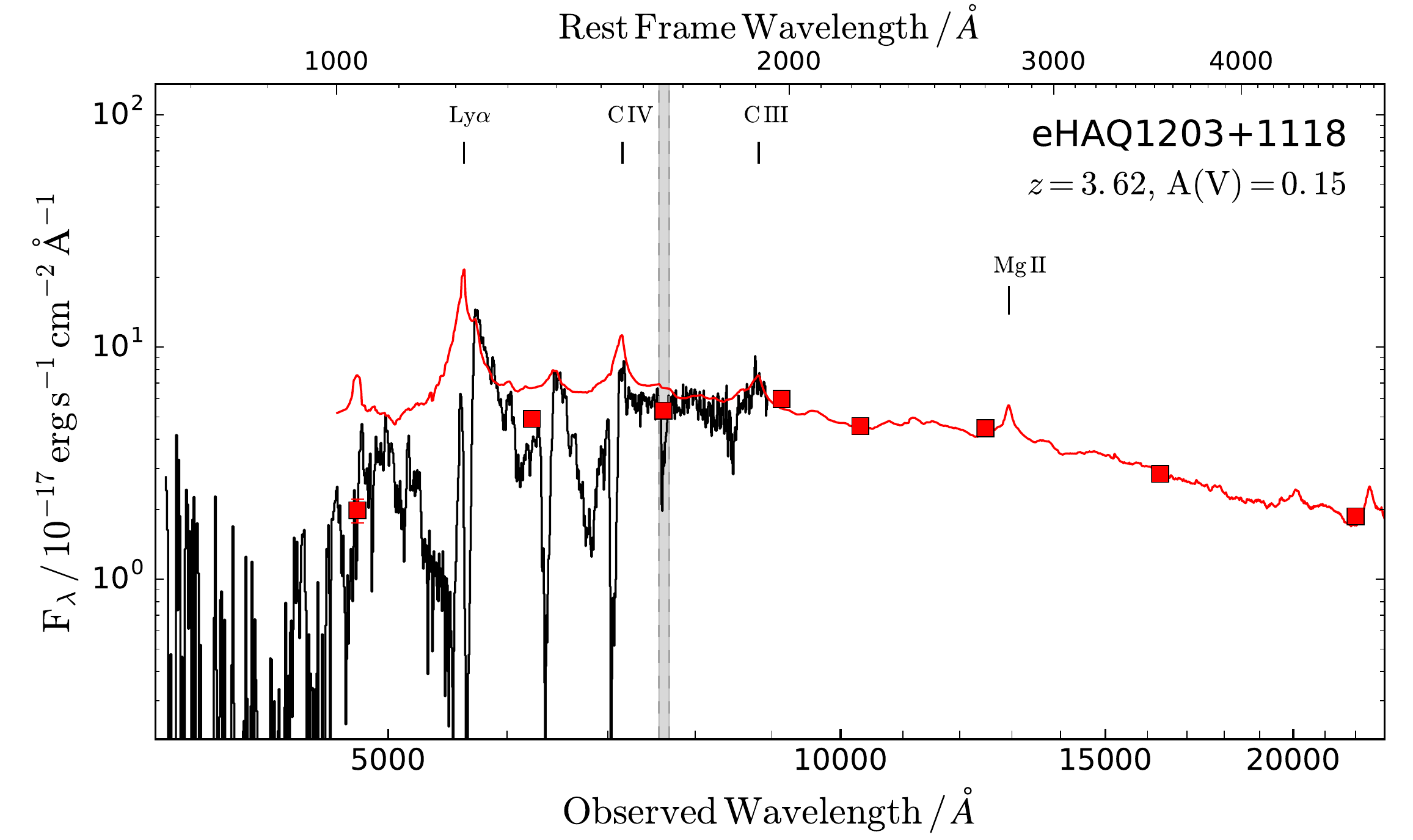}
    \includegraphics[width=0.49\textwidth]{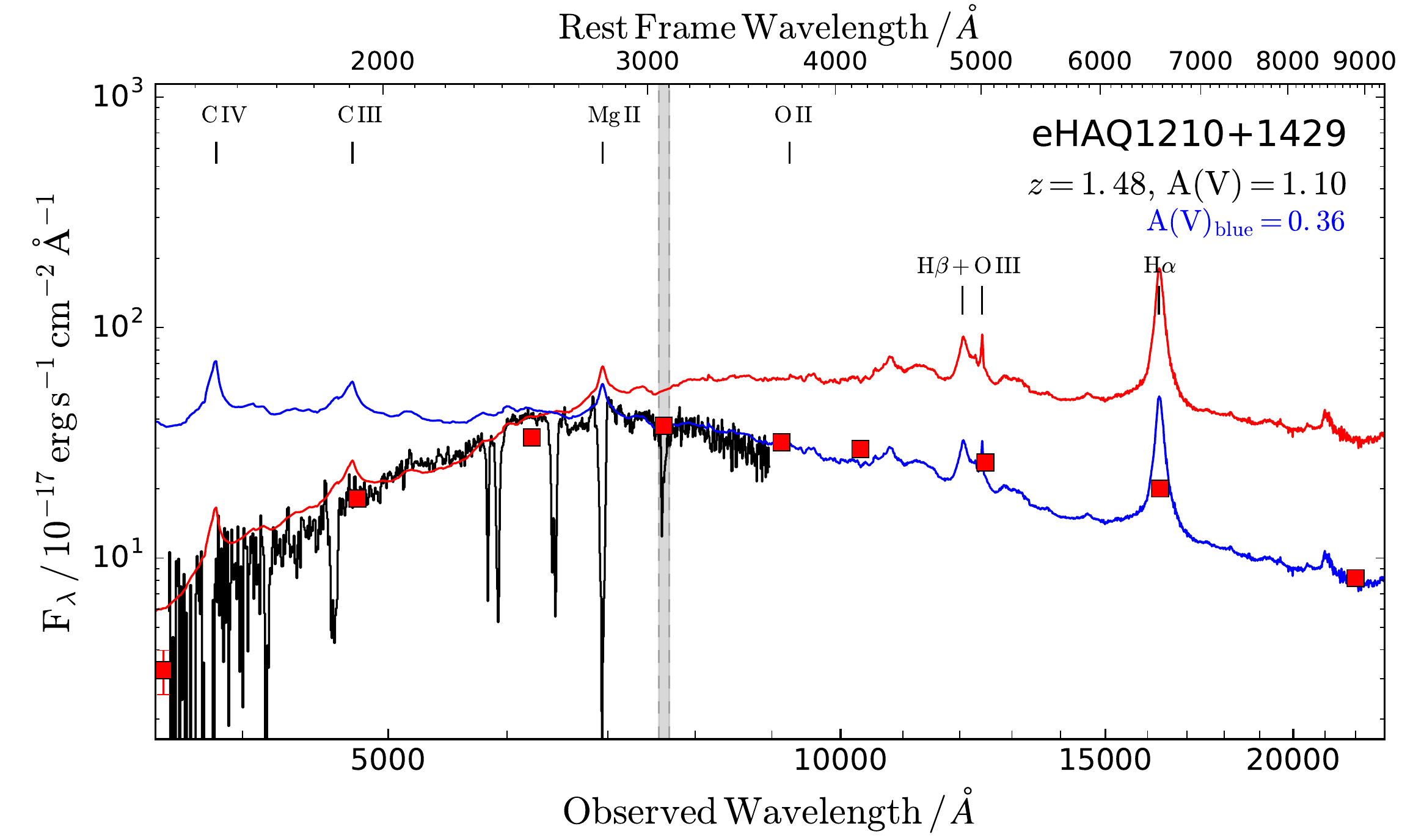}
    \vspace{4mm}

    \includegraphics[width=0.49\textwidth]{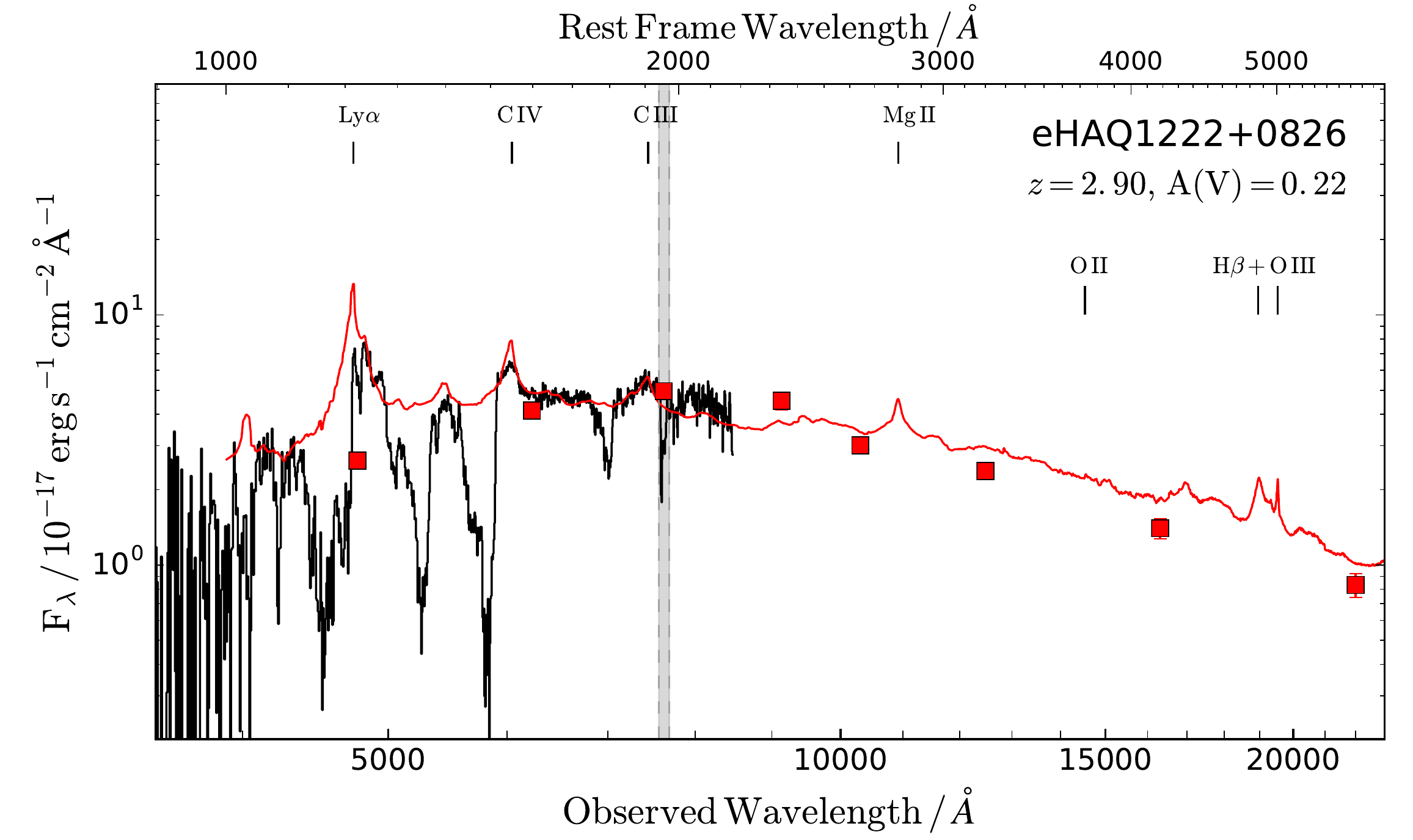}
    \includegraphics[width=0.49\textwidth]{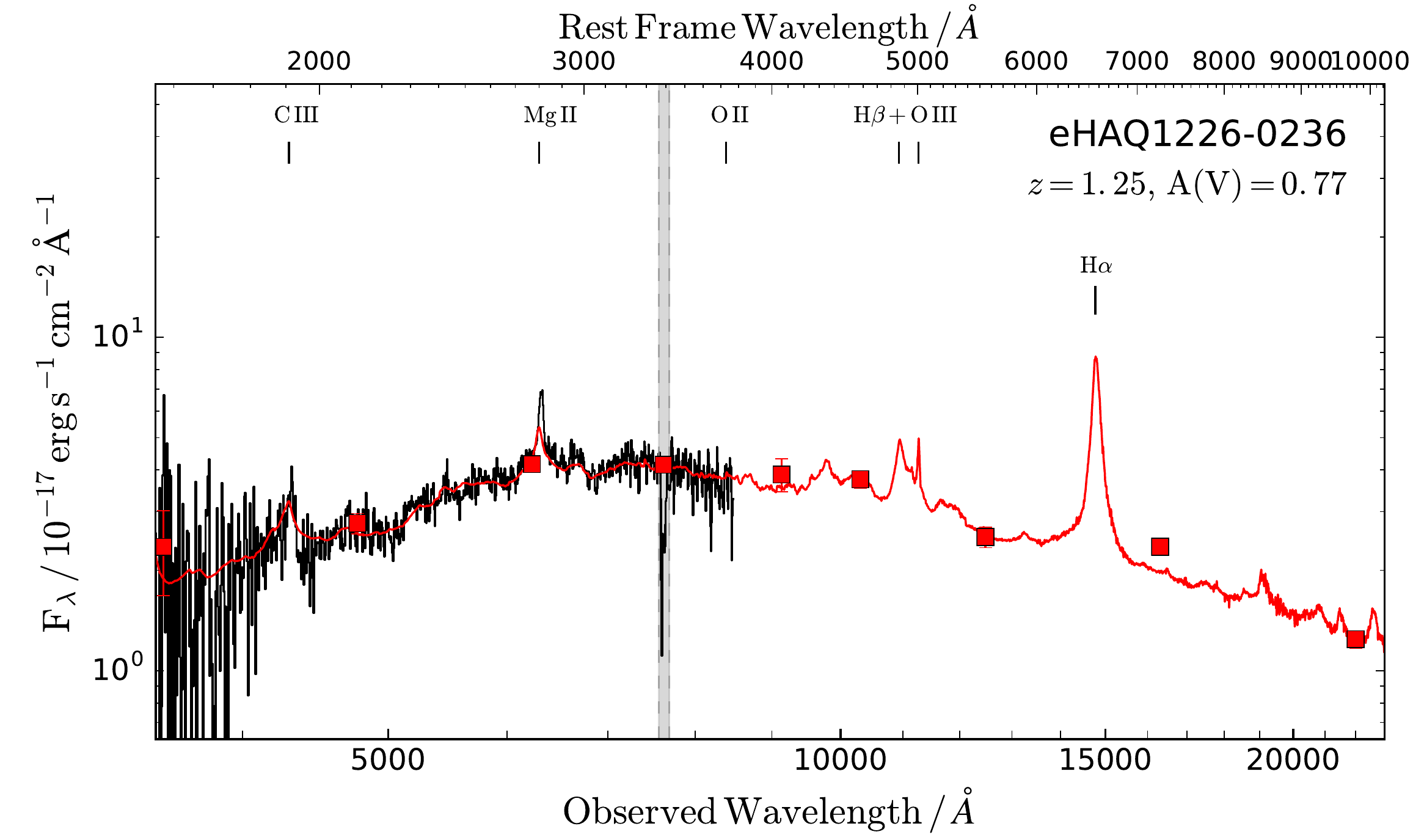}
    \vspace{4mm}

    \includegraphics[width=0.49\textwidth]{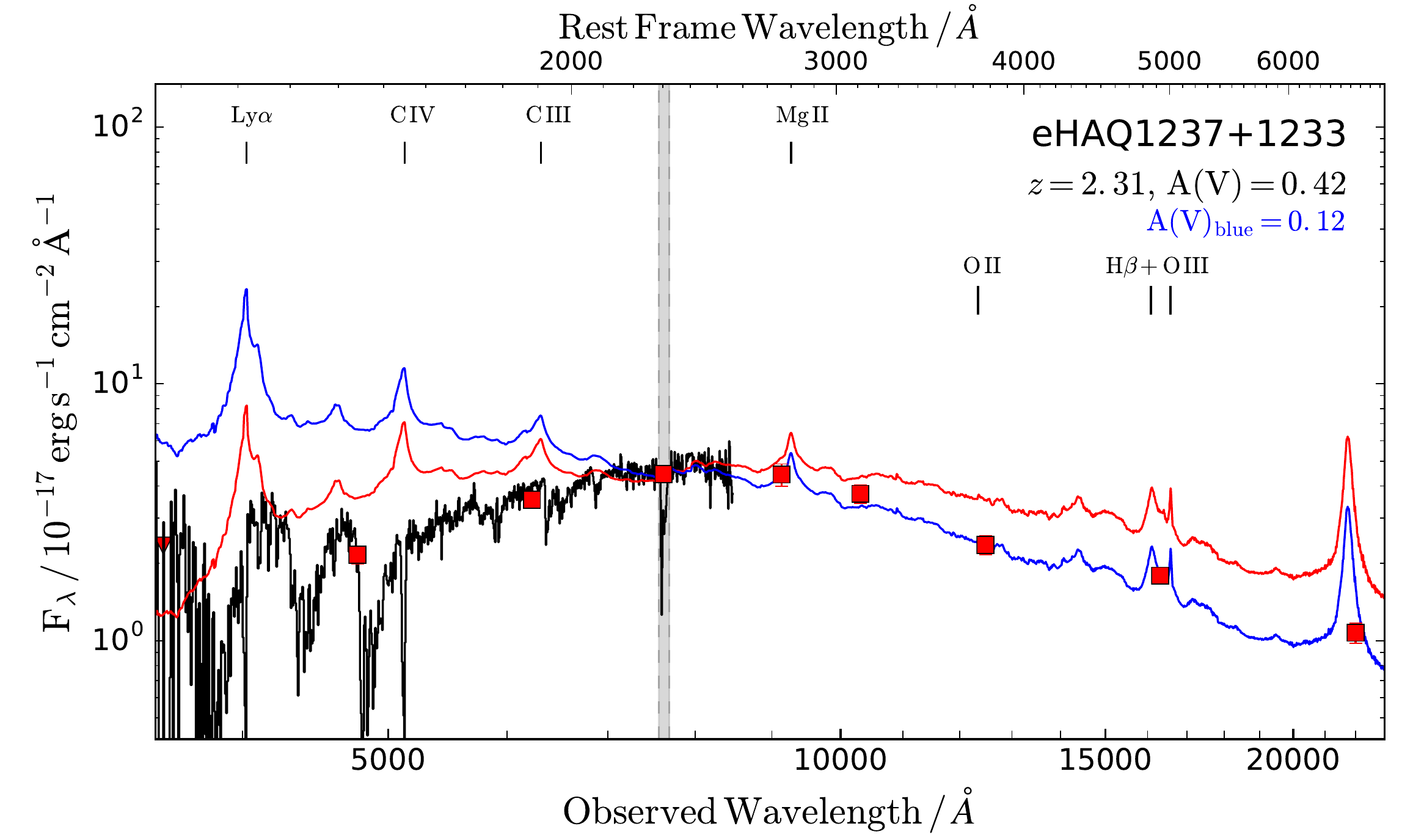}
    \includegraphics[width=0.49\textwidth]{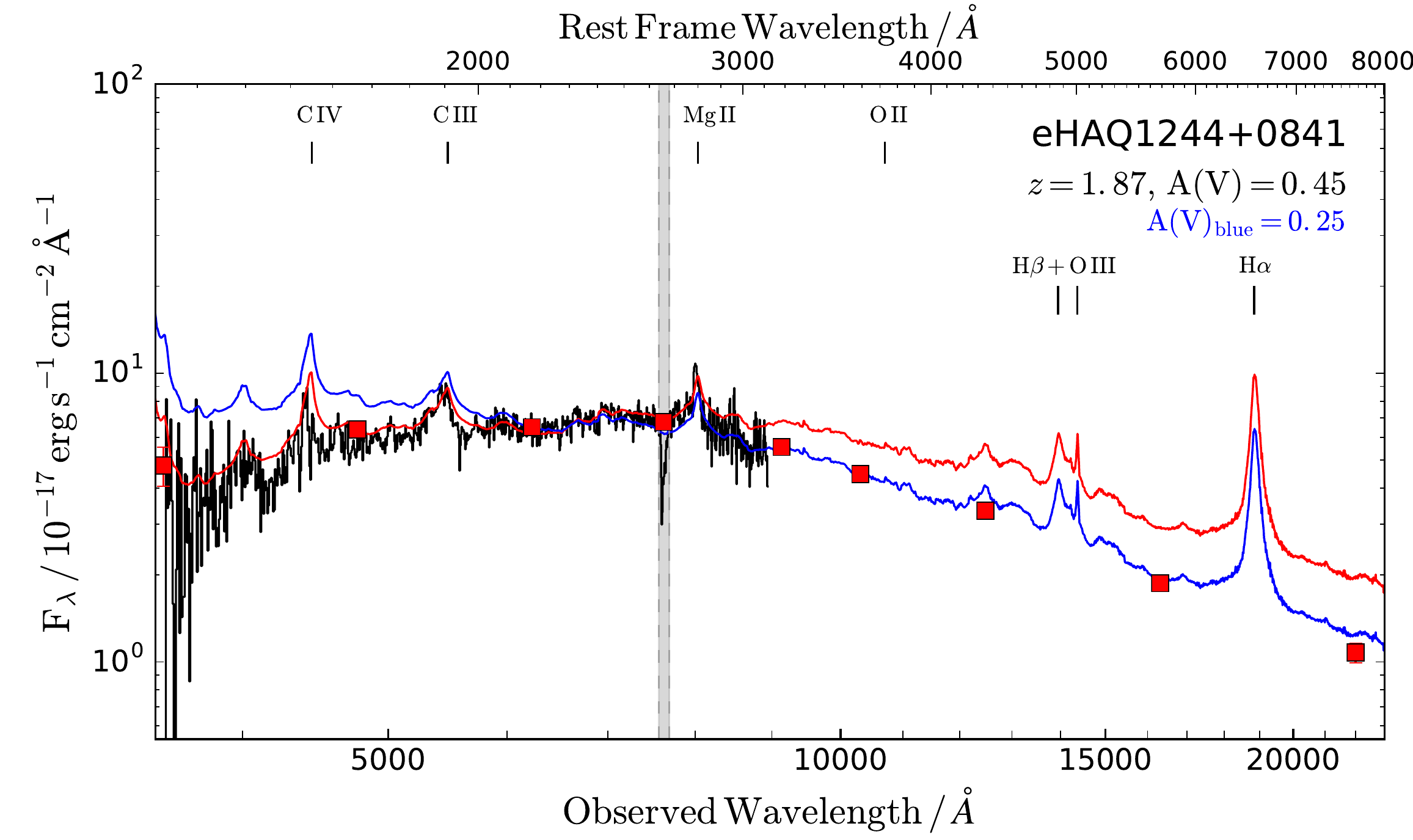}
    \vspace{4mm}

    \includegraphics[width=0.49\textwidth]{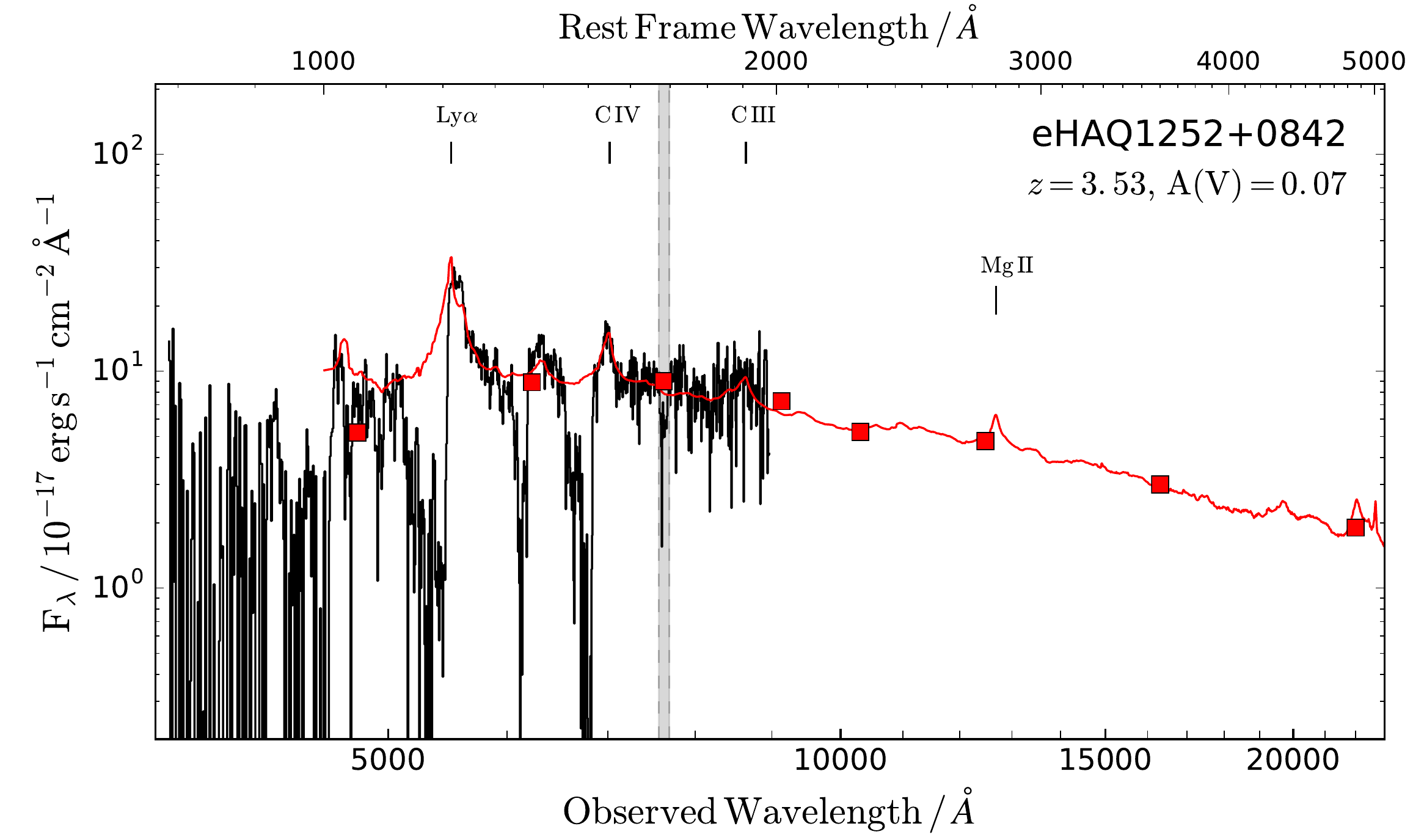}
    \includegraphics[width=0.49\textwidth]{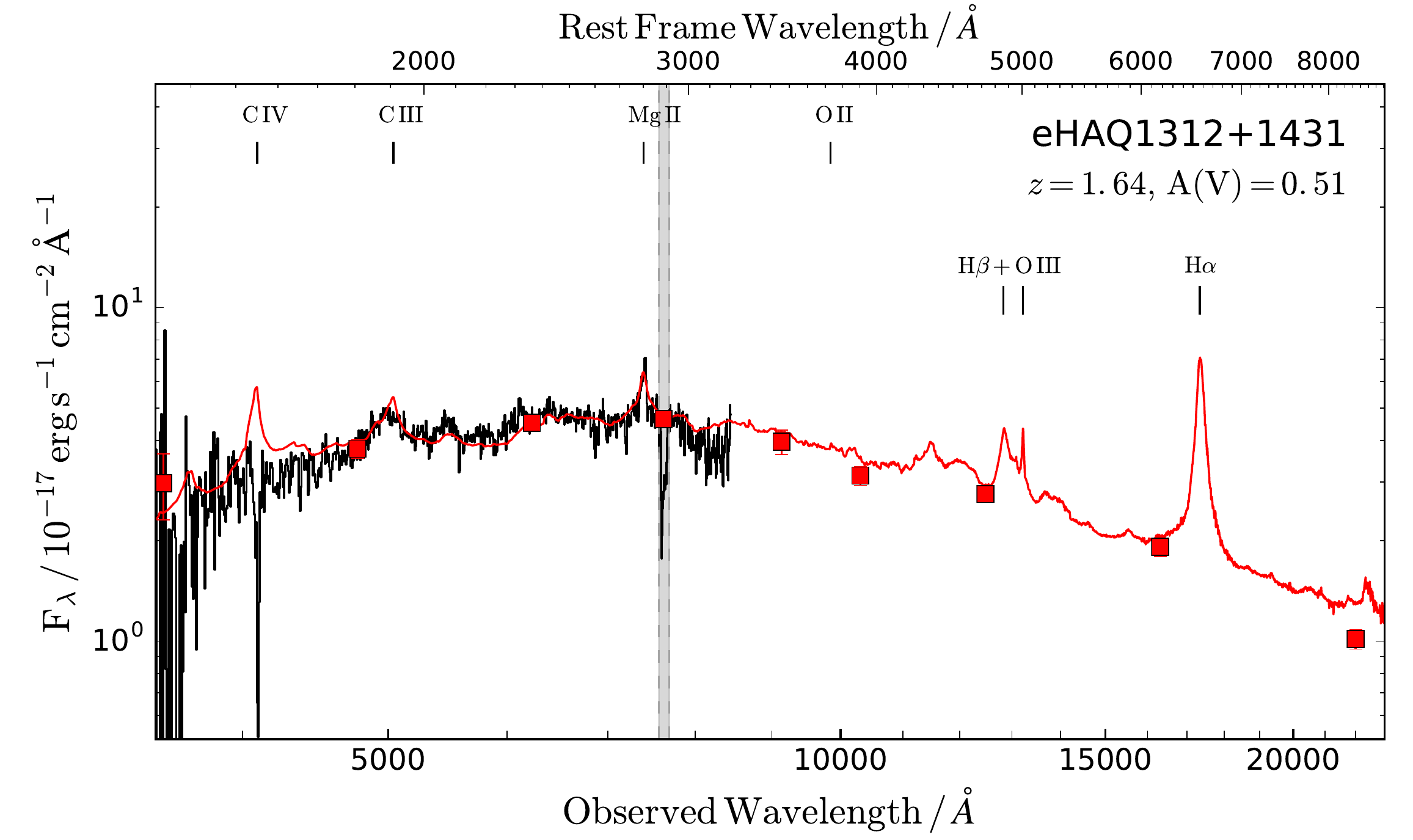}
\caption{(Continued.)}
\end{figure}

\begin{figure}
\figurenum{E2}
    \includegraphics[width=0.49\textwidth]{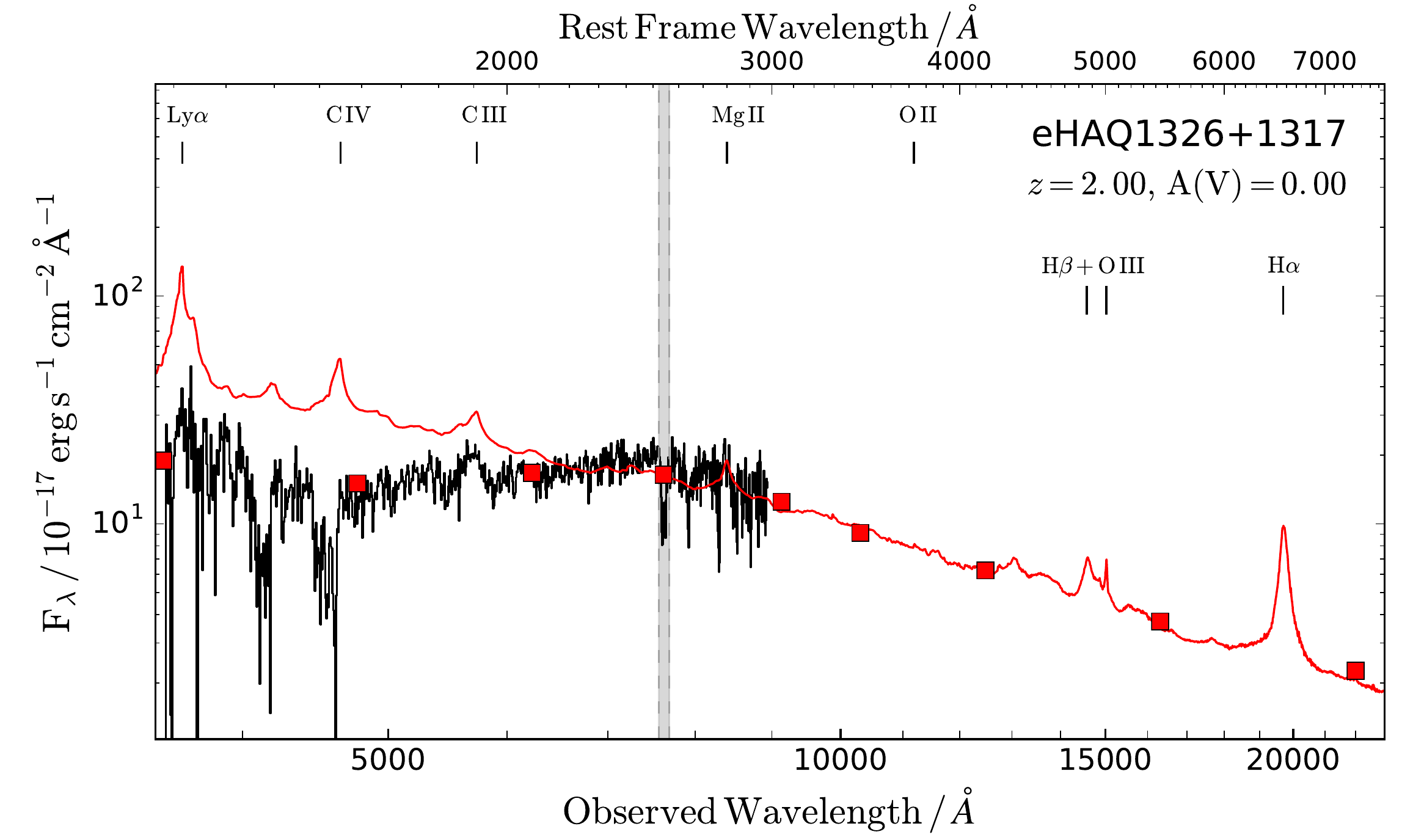}
    \includegraphics[width=0.49\textwidth]{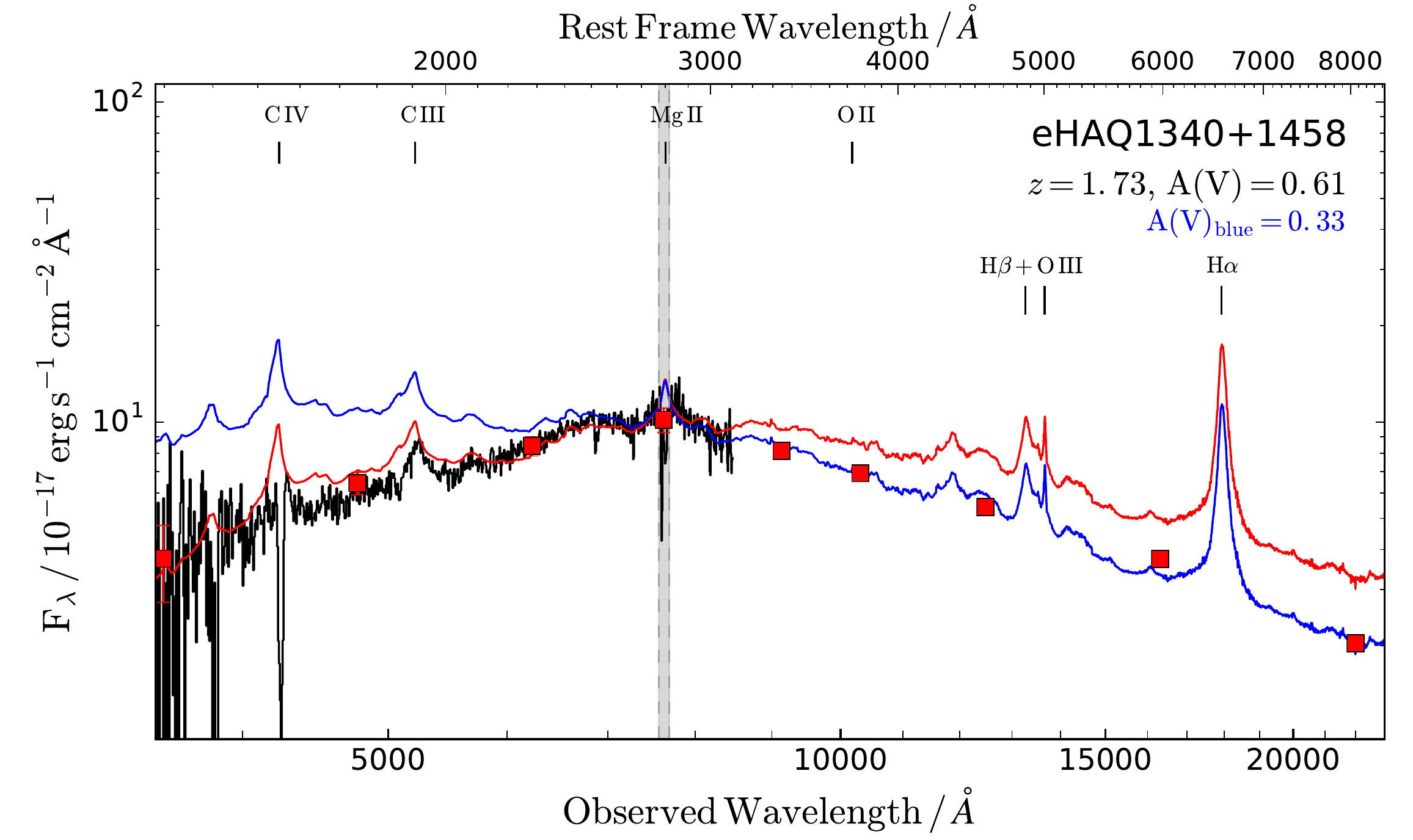}
    \vspace{4mm}

    \includegraphics[width=0.49\textwidth]{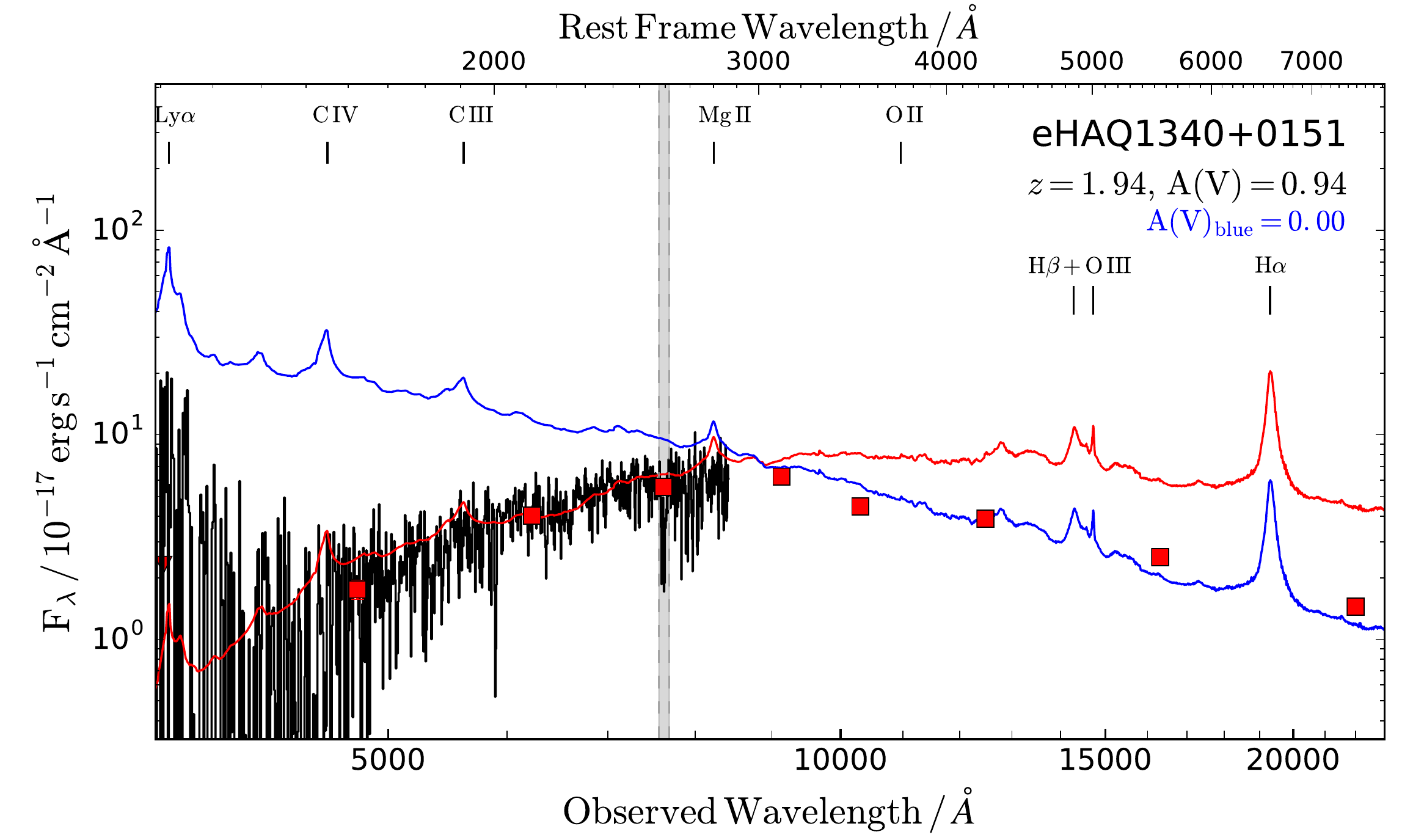}
    \includegraphics[width=0.49\textwidth]{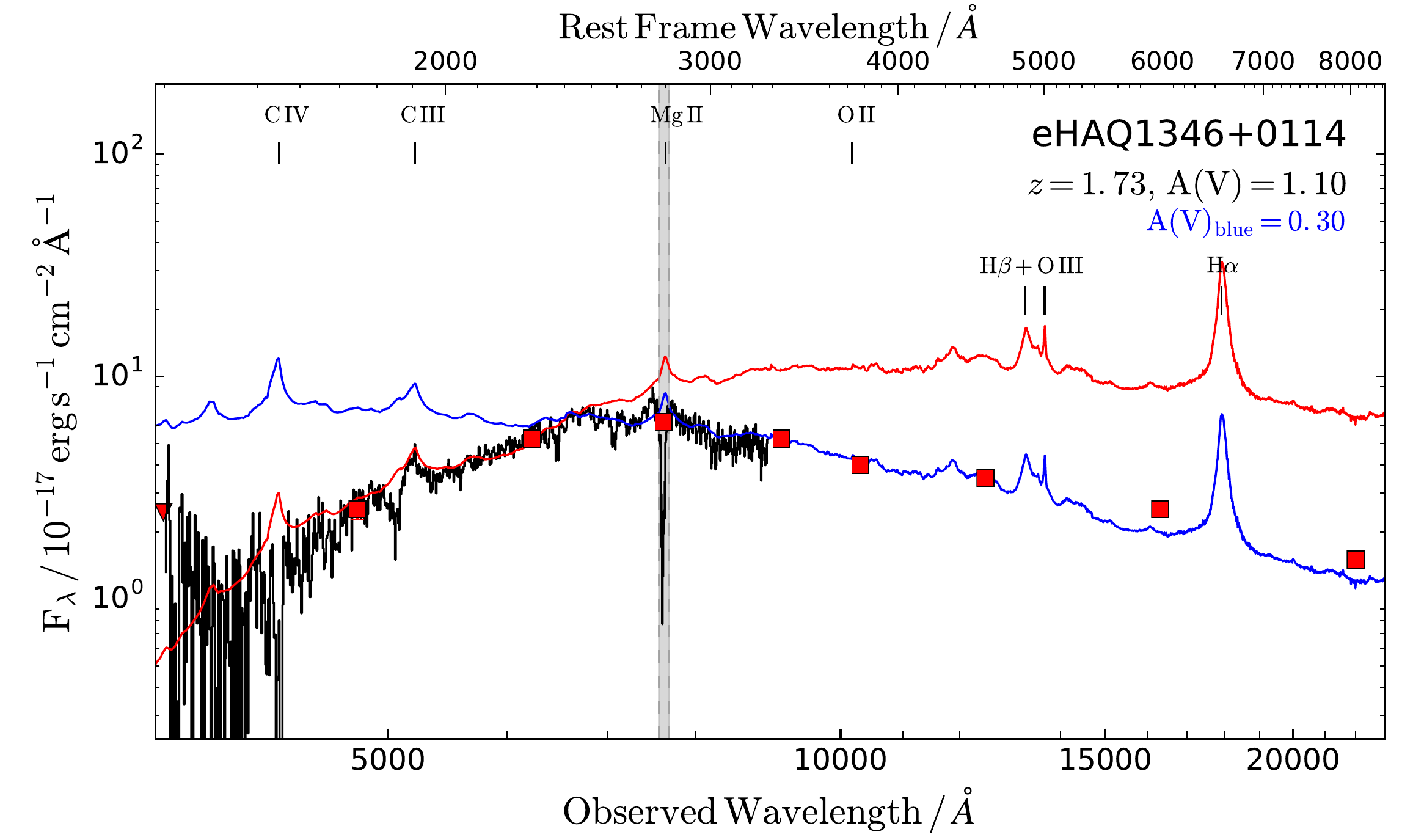}
    \vspace{4mm}

    \includegraphics[width=0.49\textwidth]{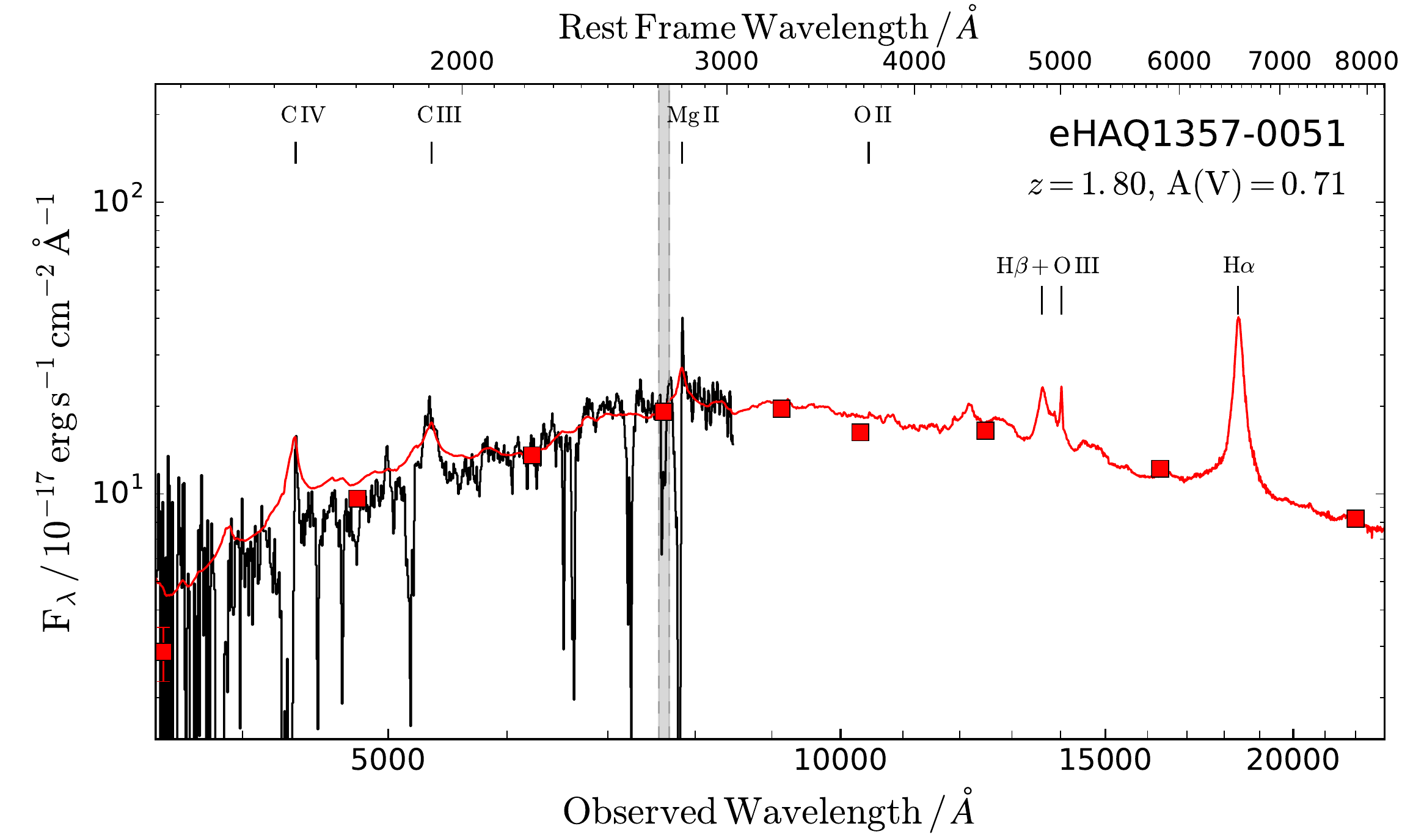}
    \includegraphics[width=0.49\textwidth]{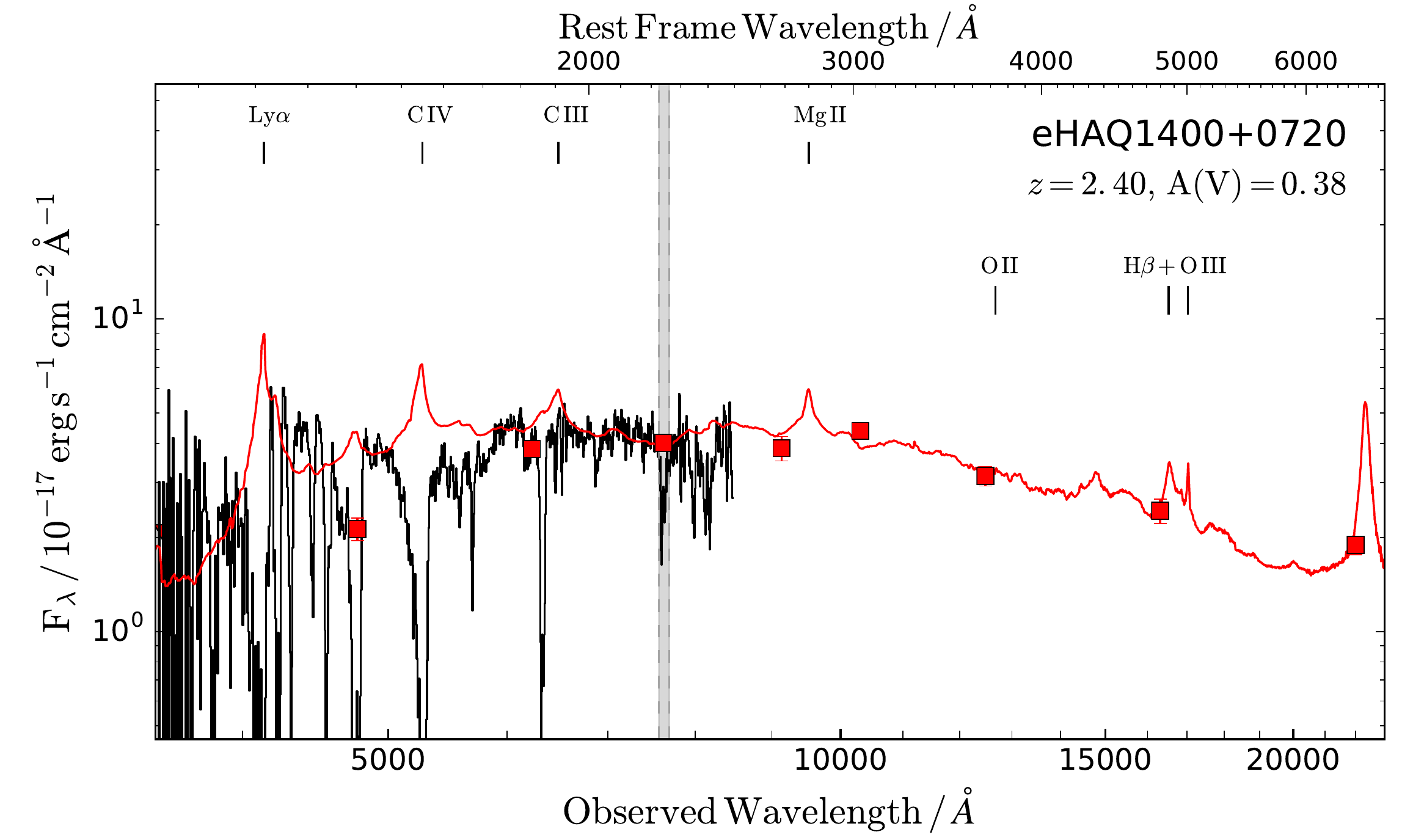}
    \vspace{4mm}

    \includegraphics[width=0.49\textwidth]{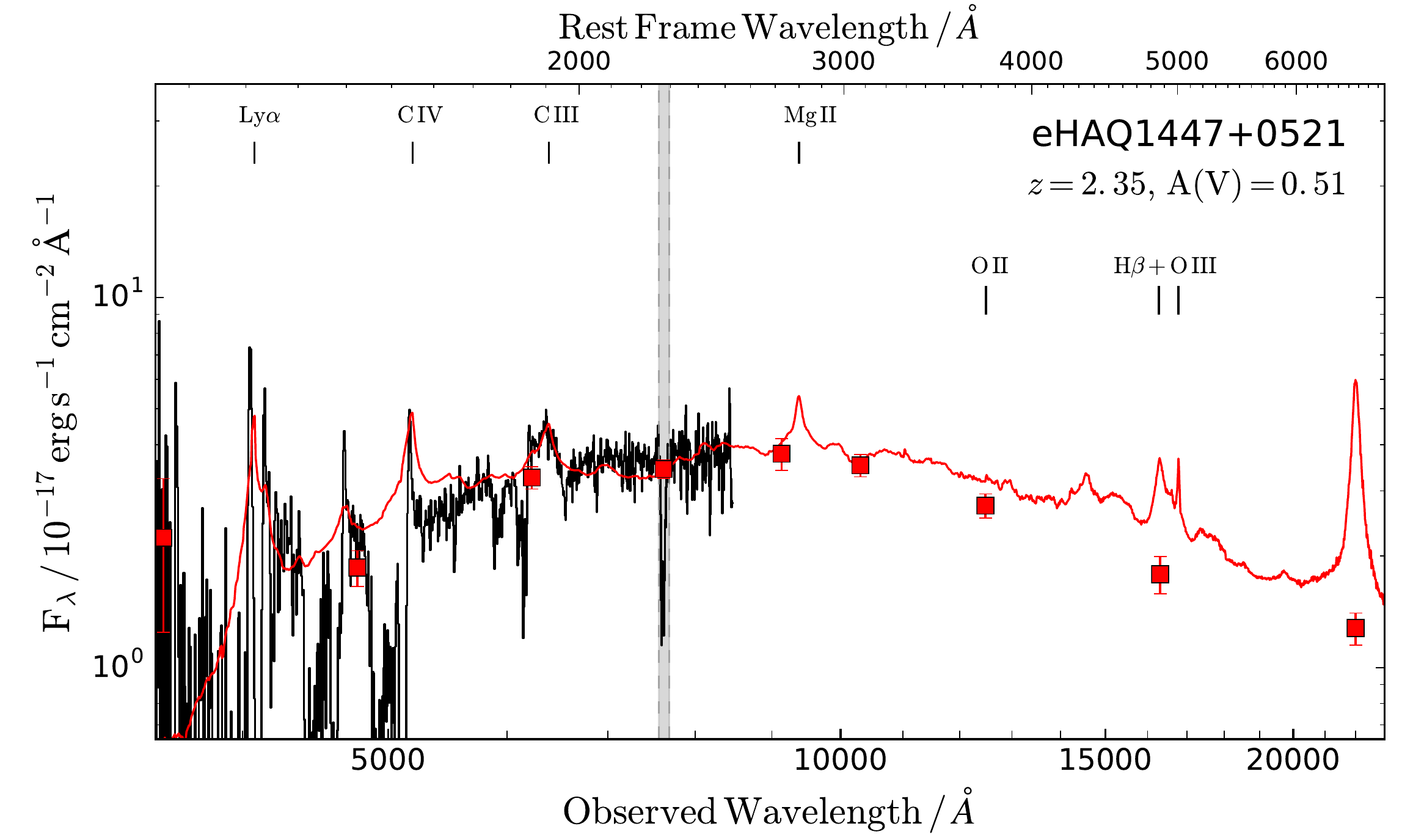}
    \includegraphics[width=0.49\textwidth]{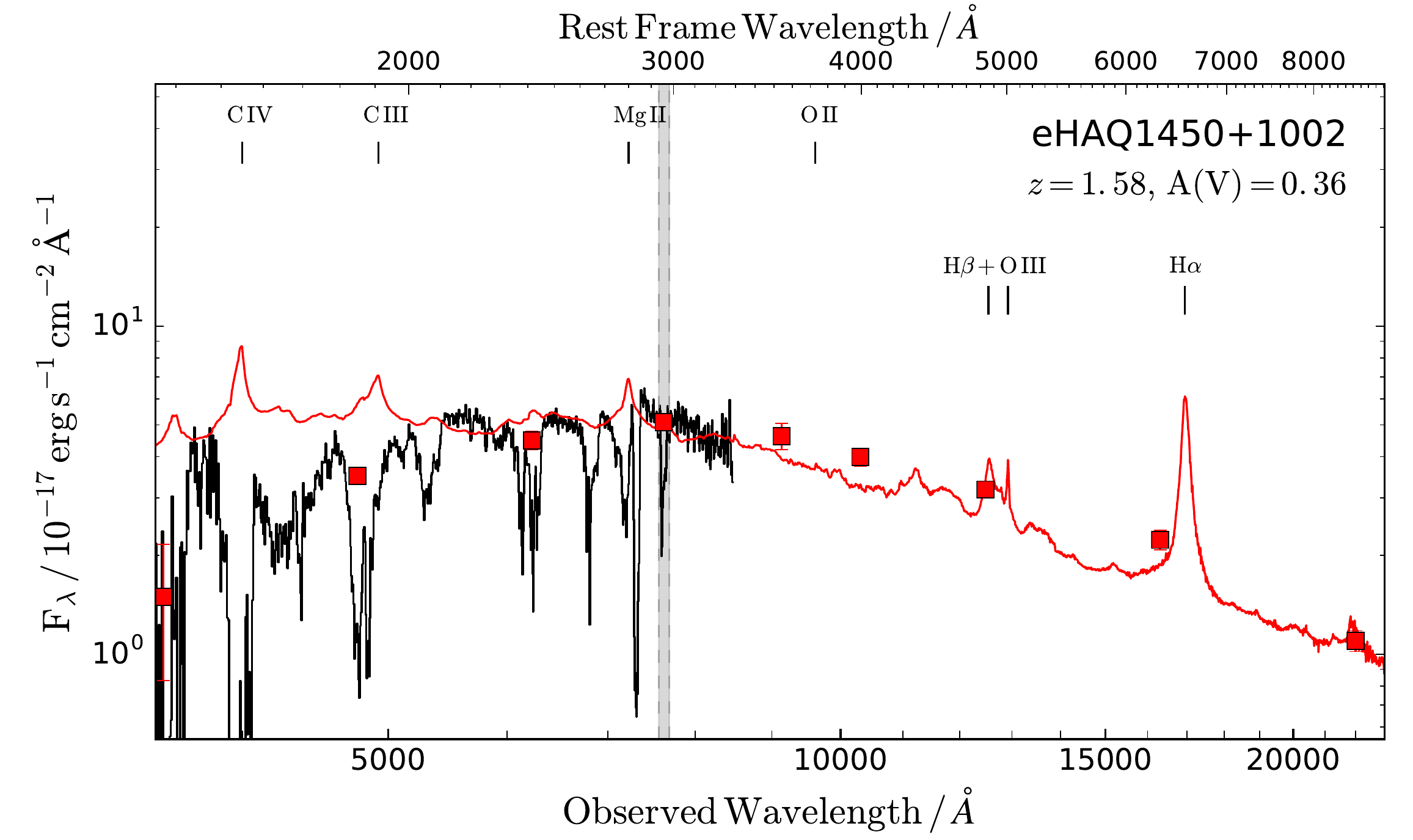}
\caption{(Continued.)}
\end{figure}

\begin{figure}
\figurenum{E2}
    \includegraphics[width=0.49\textwidth]{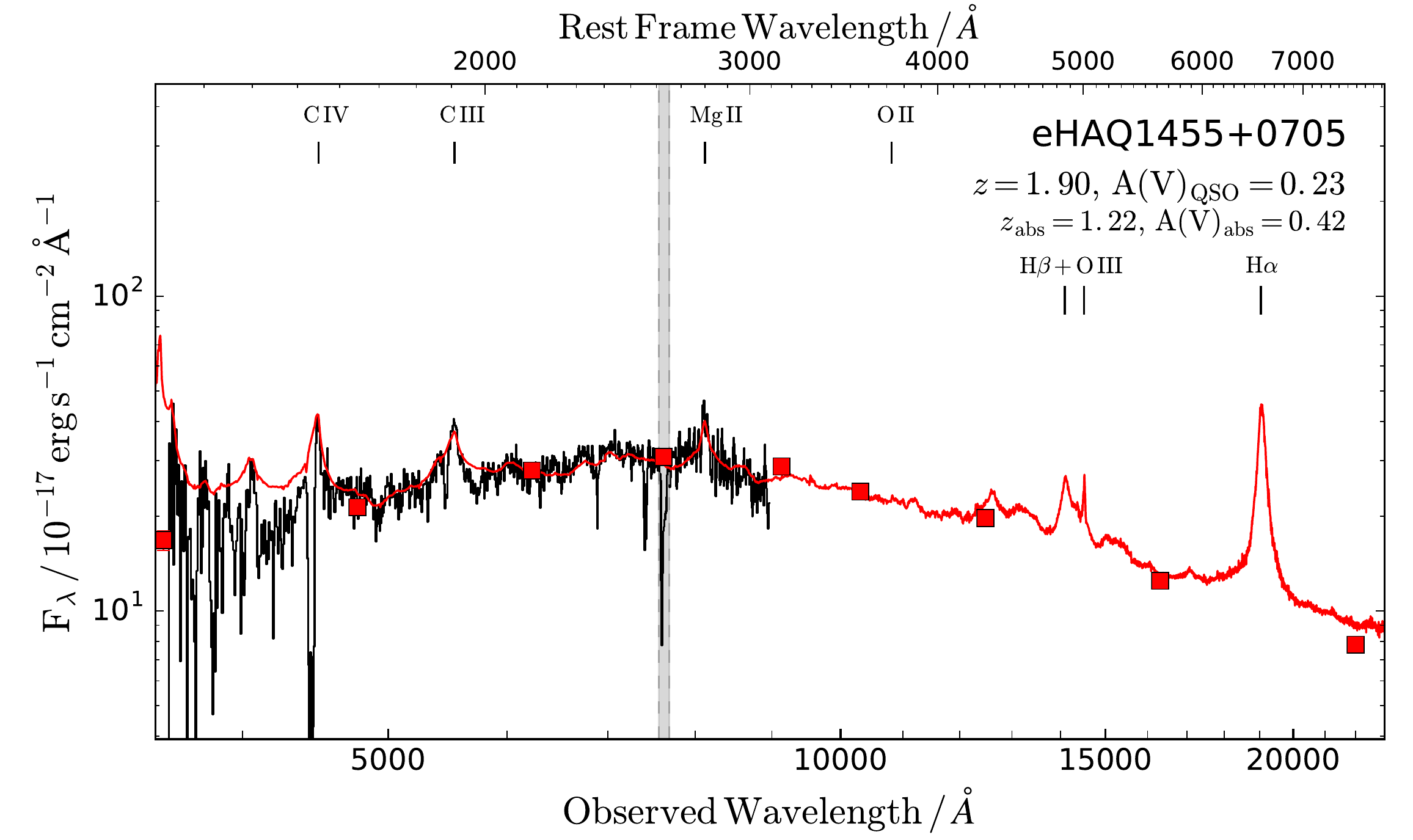}
    \includegraphics[width=0.49\textwidth]{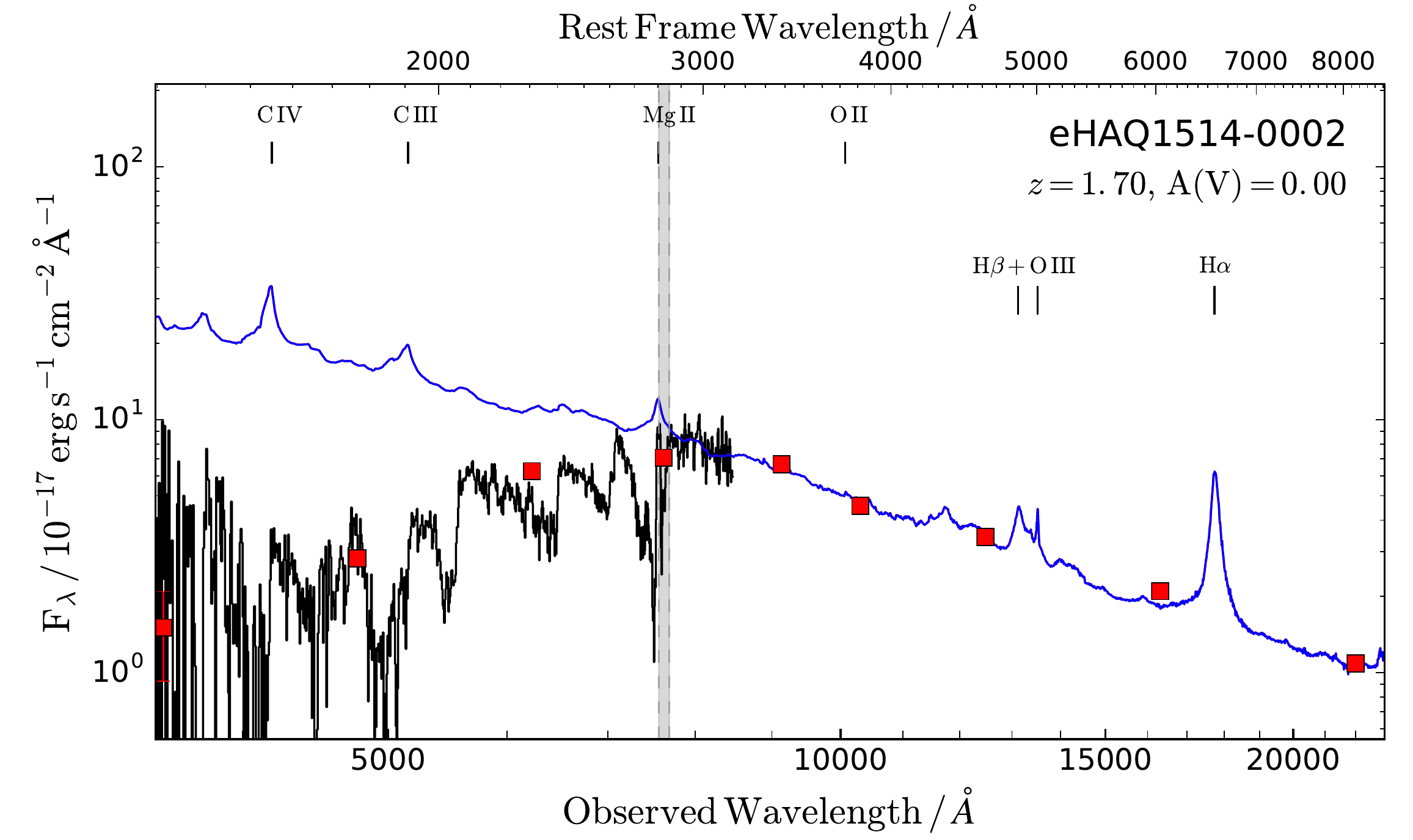}
    \vspace{4mm}

    \includegraphics[width=0.49\textwidth]{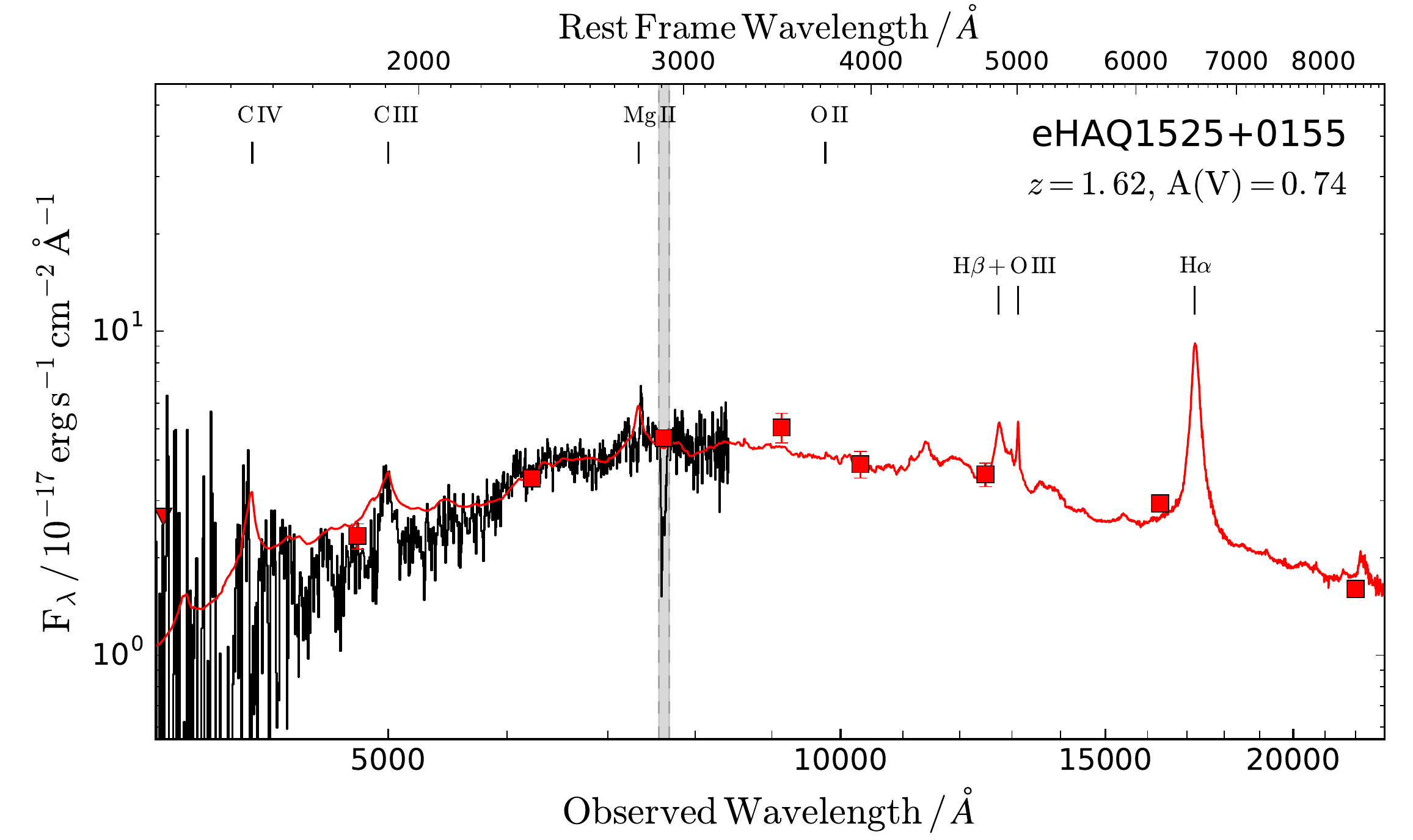}
    \includegraphics[width=0.49\textwidth]{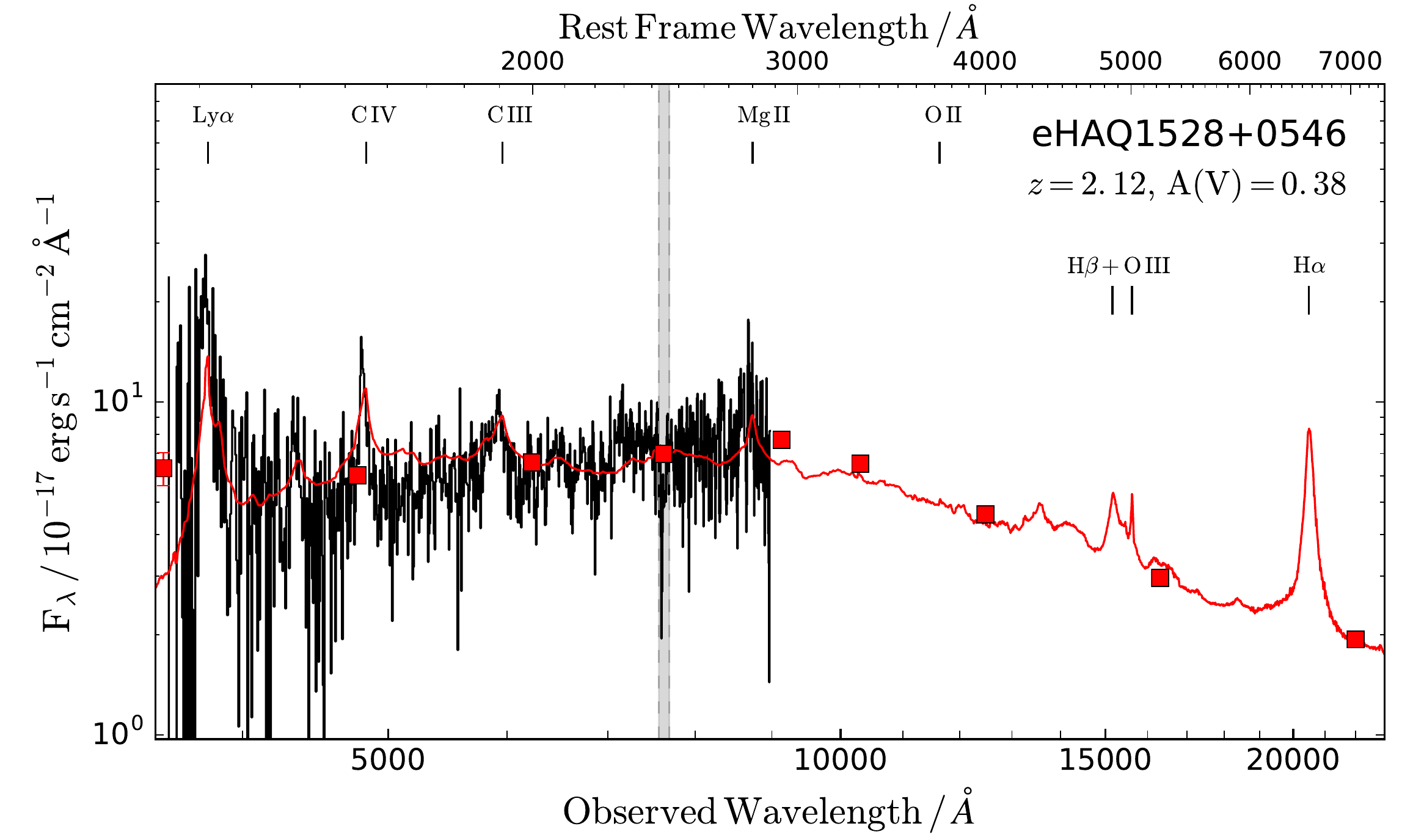}
    \vspace{4mm}

    \includegraphics[width=0.49\textwidth]{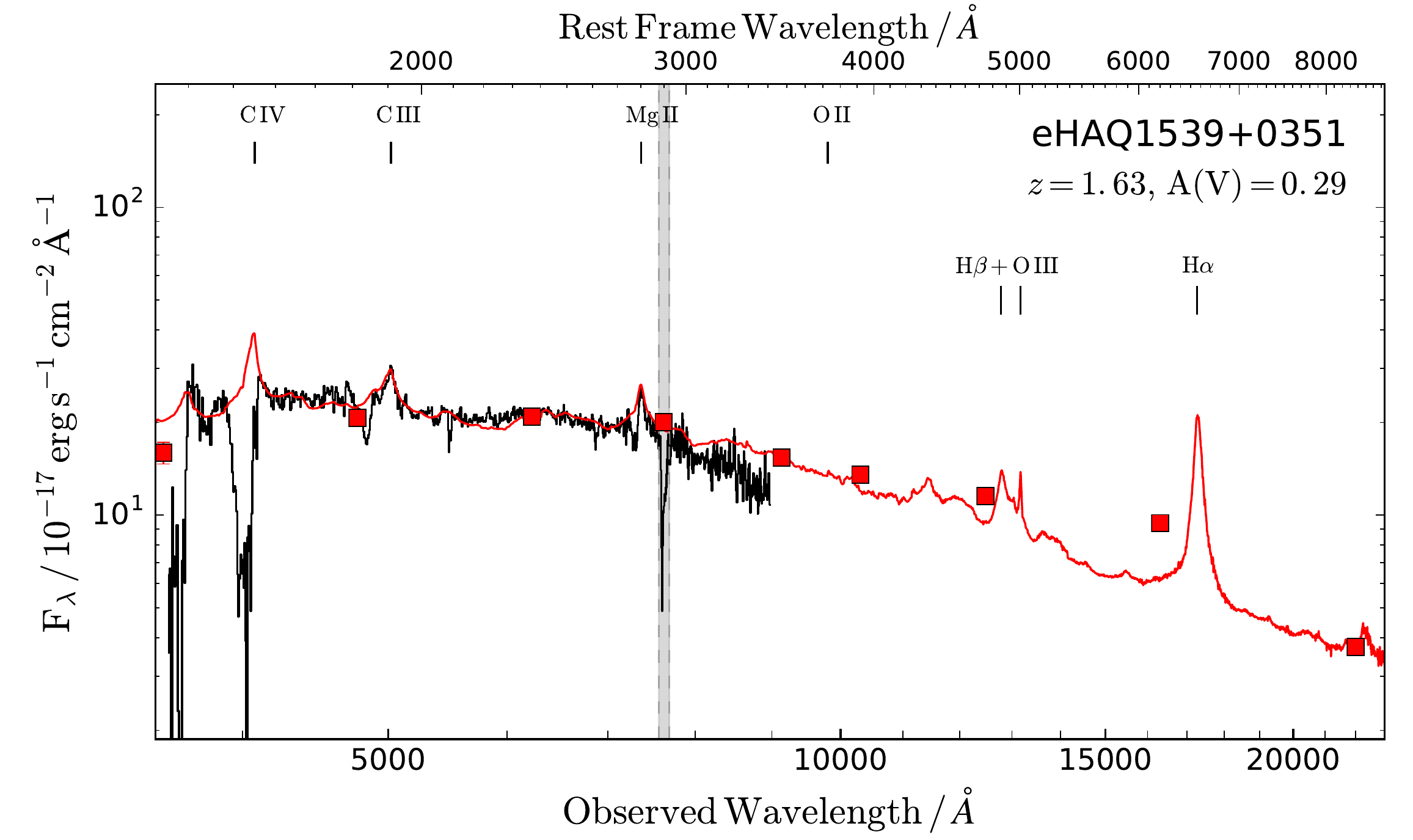}
    \includegraphics[width=0.49\textwidth]{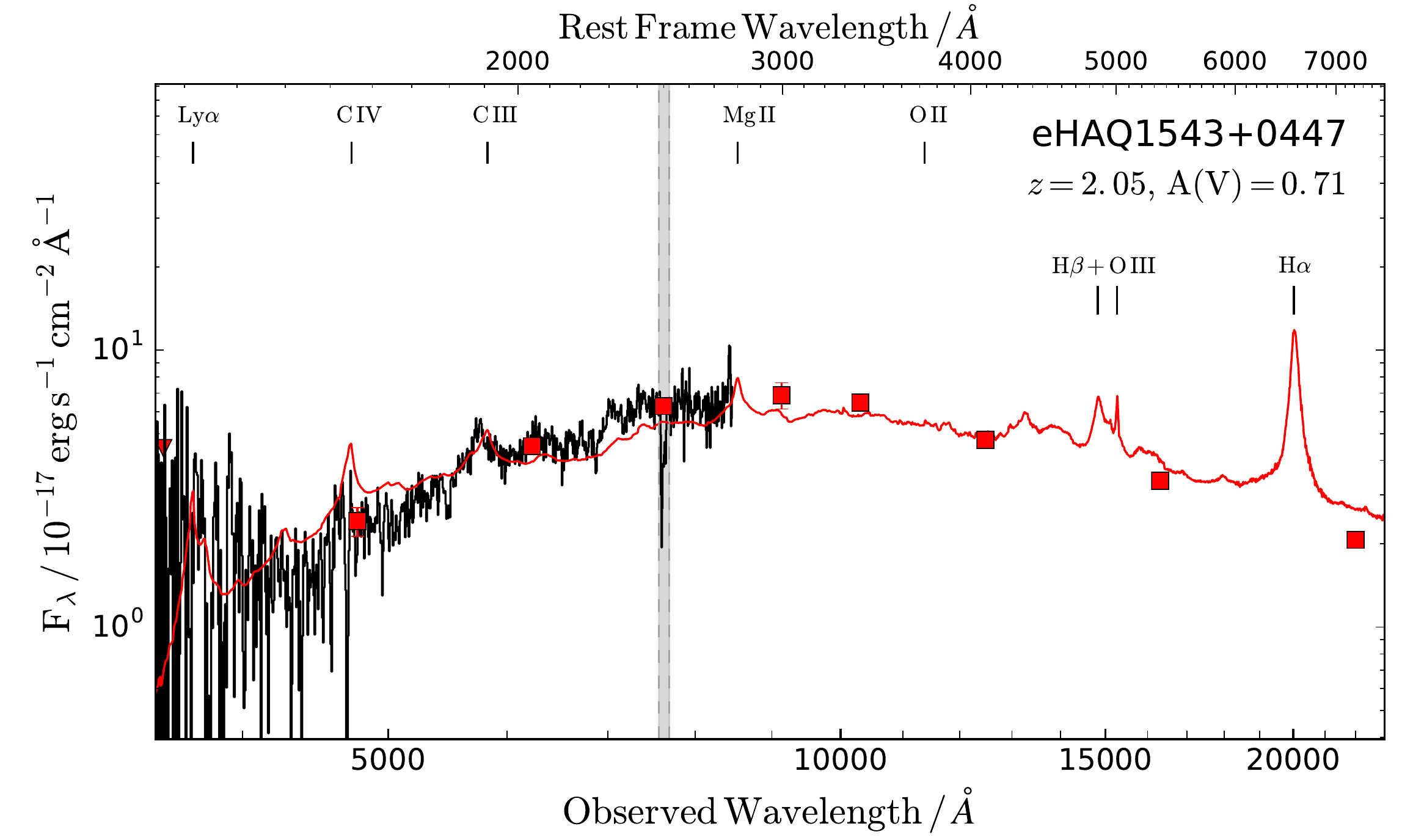}
    \vspace{4mm}

    \includegraphics[width=0.49\textwidth]{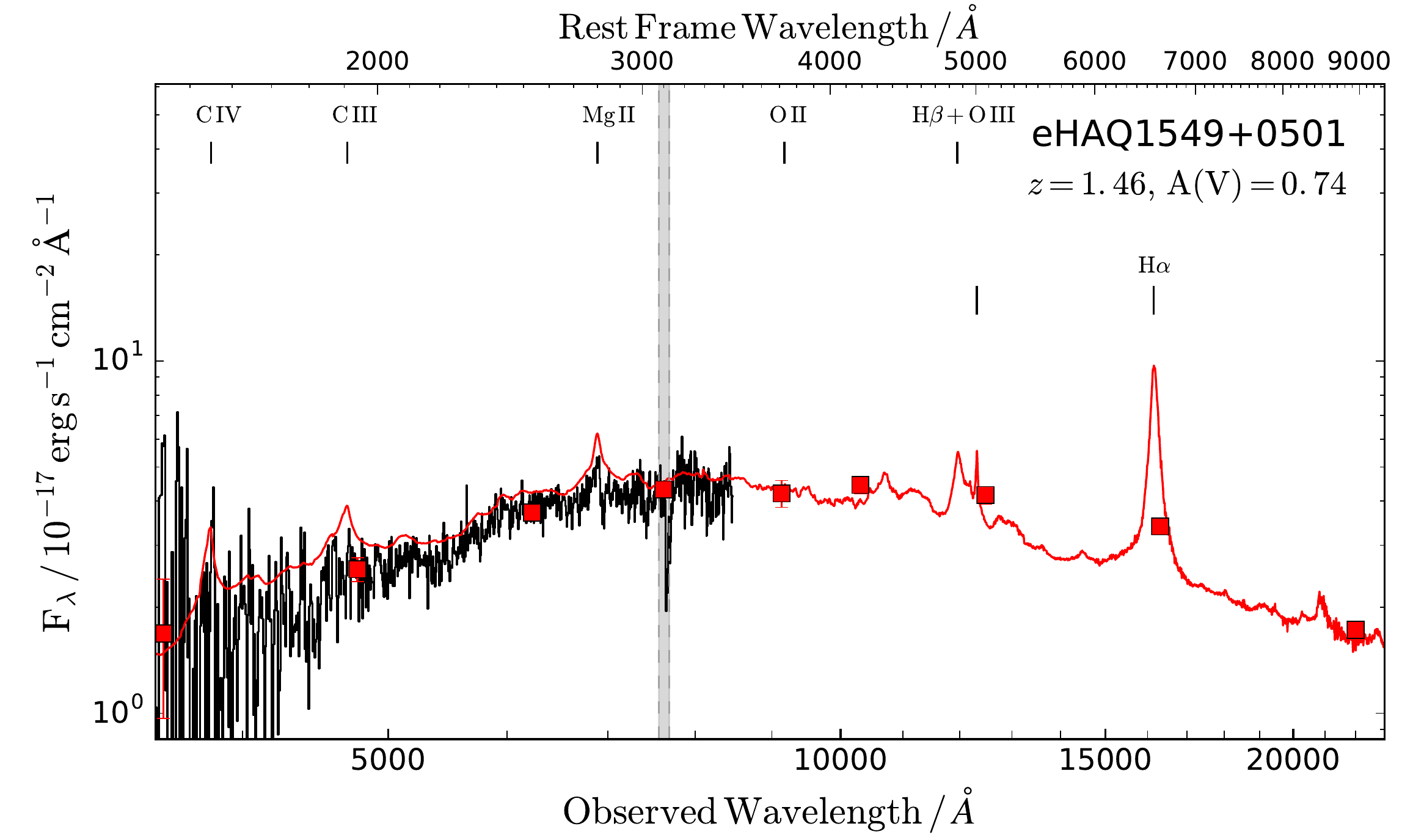}
    \includegraphics[width=0.49\textwidth]{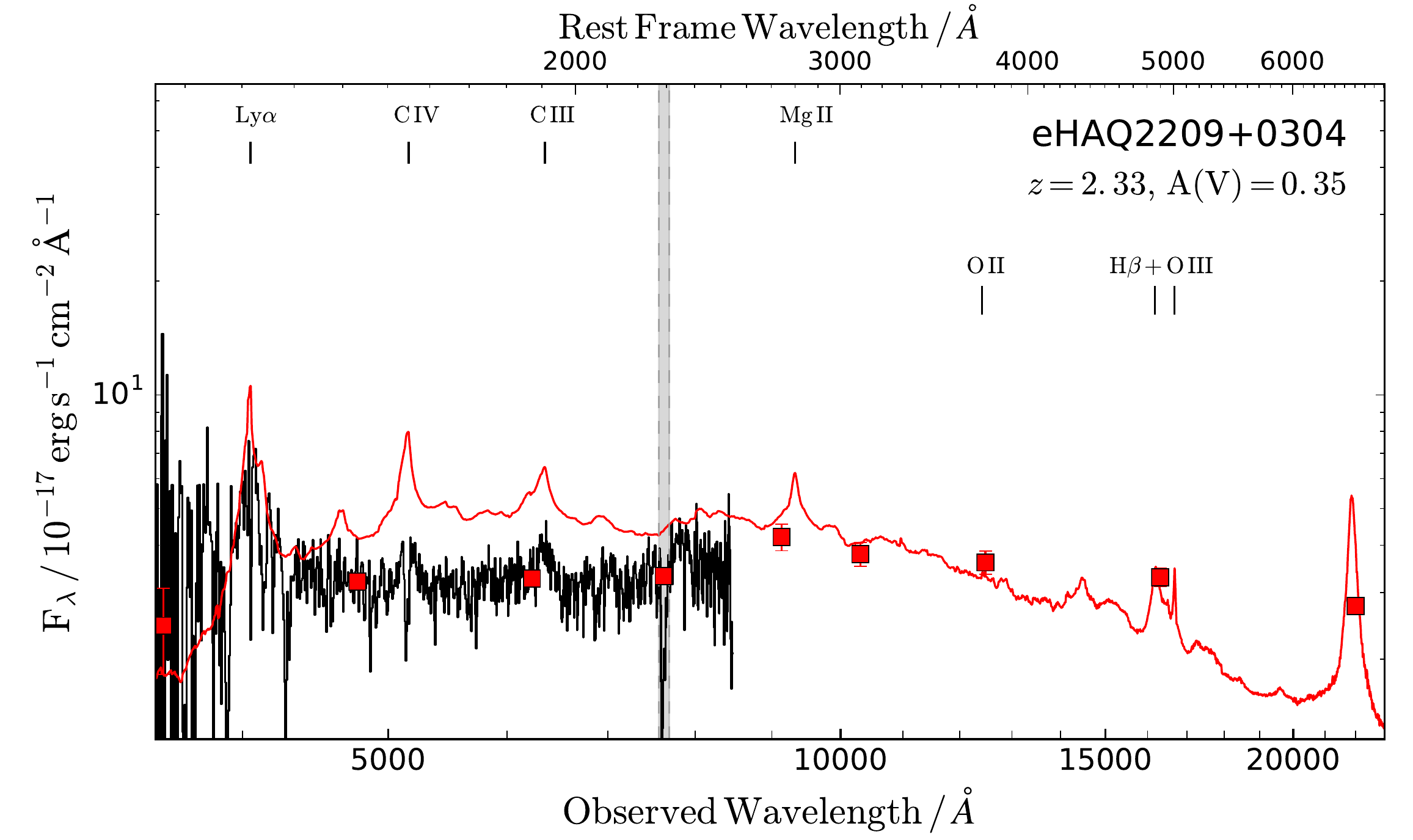}
\caption{(Continued.)}
\end{figure}

\begin{figure}
\figurenum{E2}
    \includegraphics[width=0.49\textwidth]{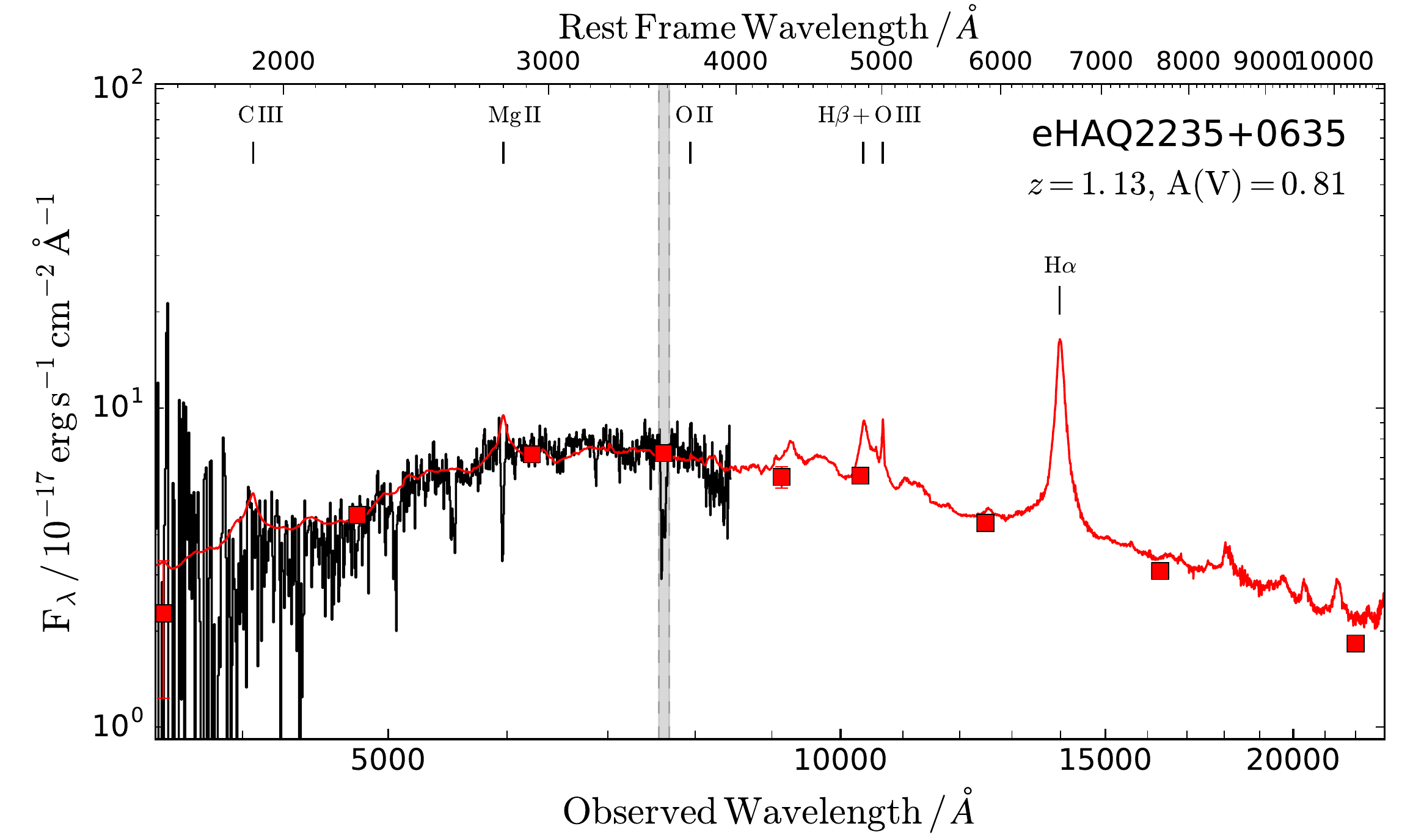}
    \includegraphics[width=0.49\textwidth]{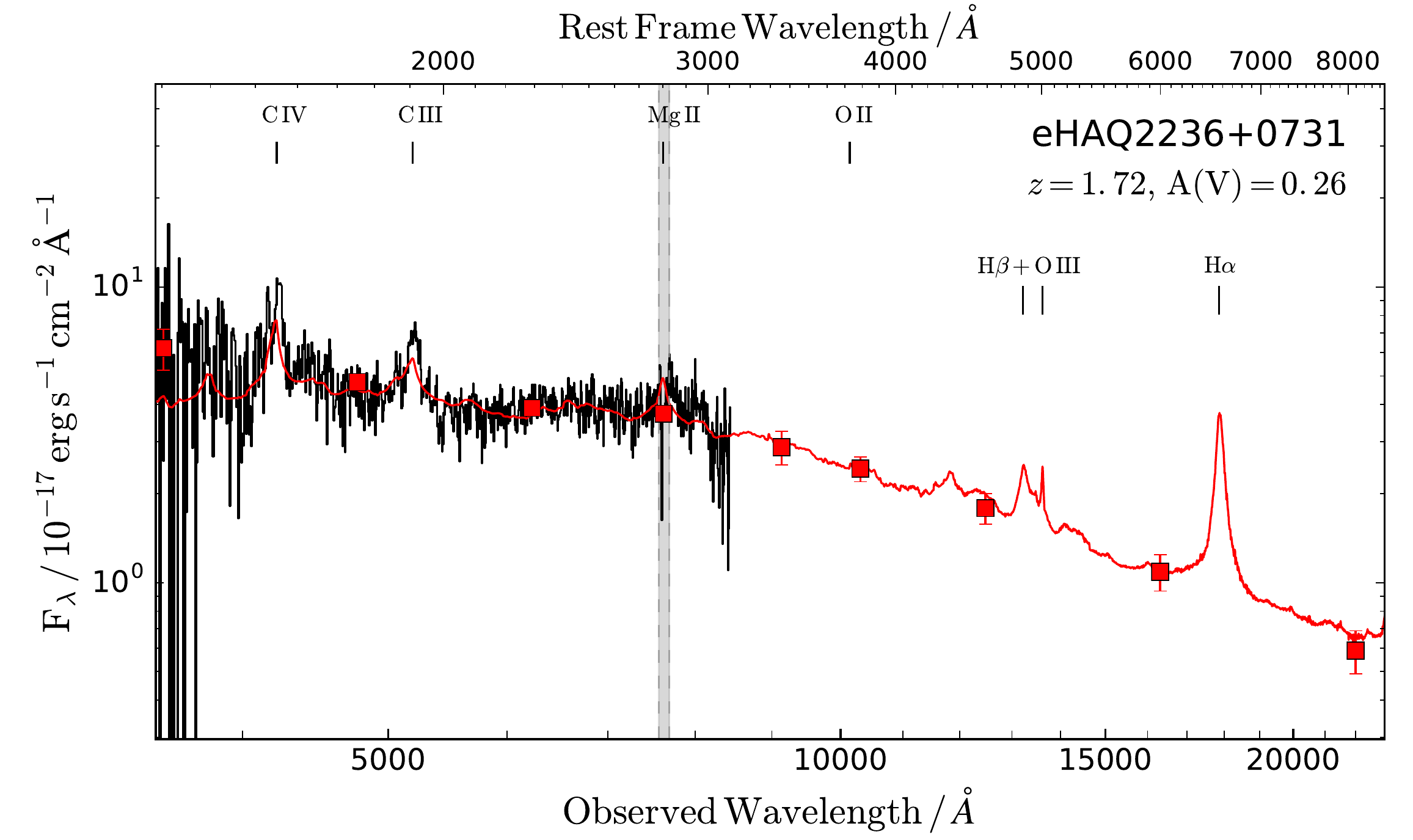}
    \vspace{4mm}

    \includegraphics[width=0.49\textwidth]{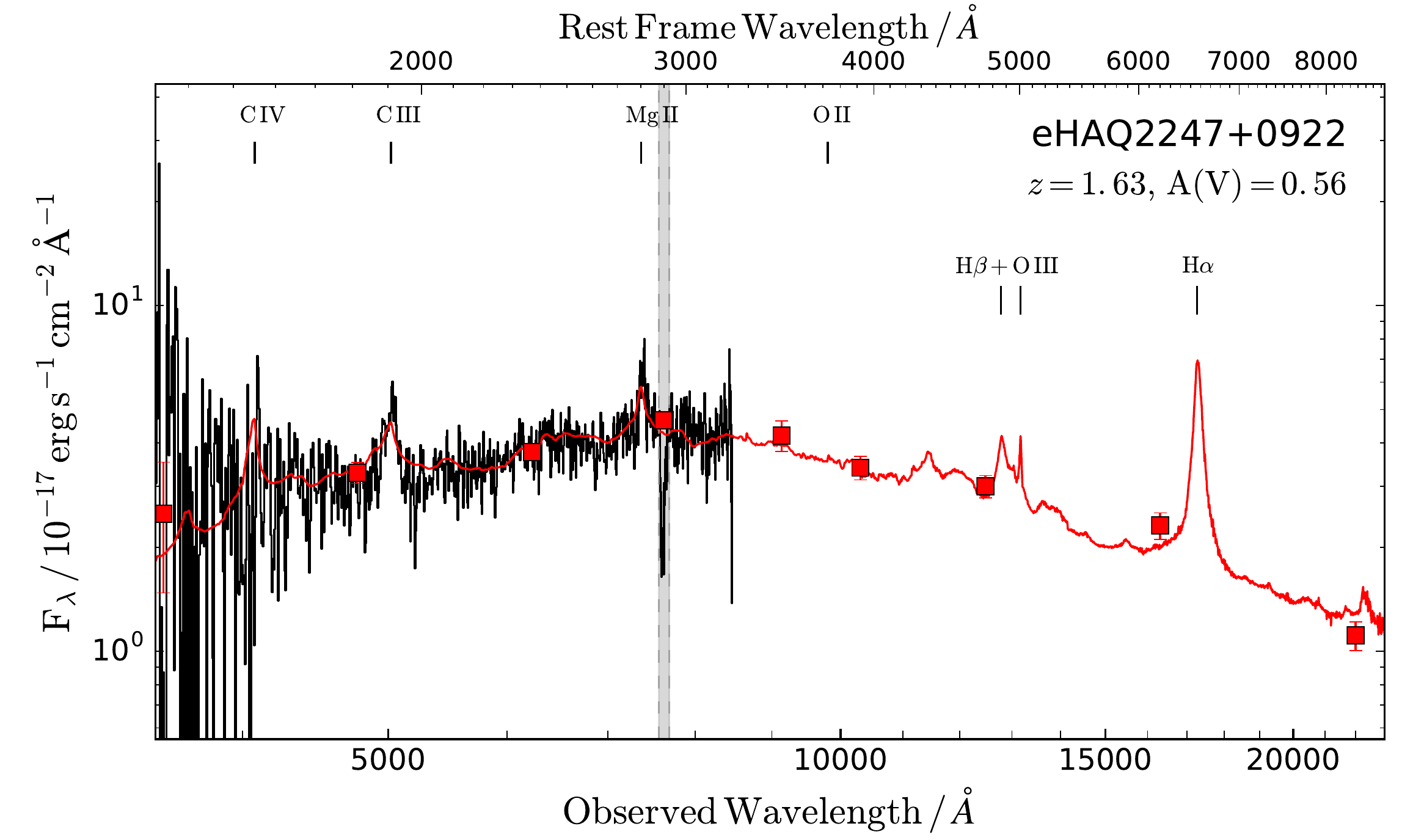}
    \includegraphics[width=0.49\textwidth]{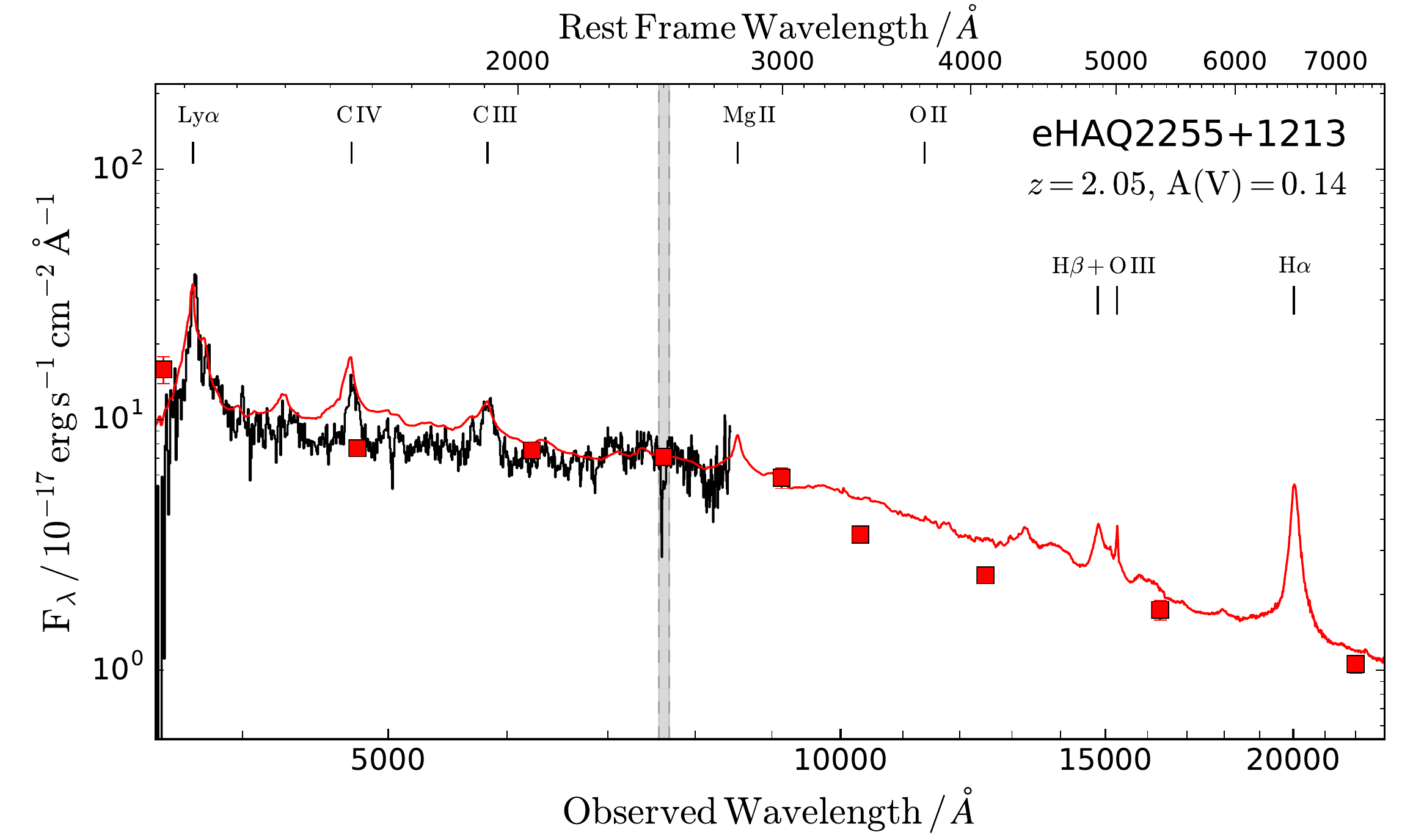}
    \vspace{4mm}

    \includegraphics[width=0.49\textwidth]{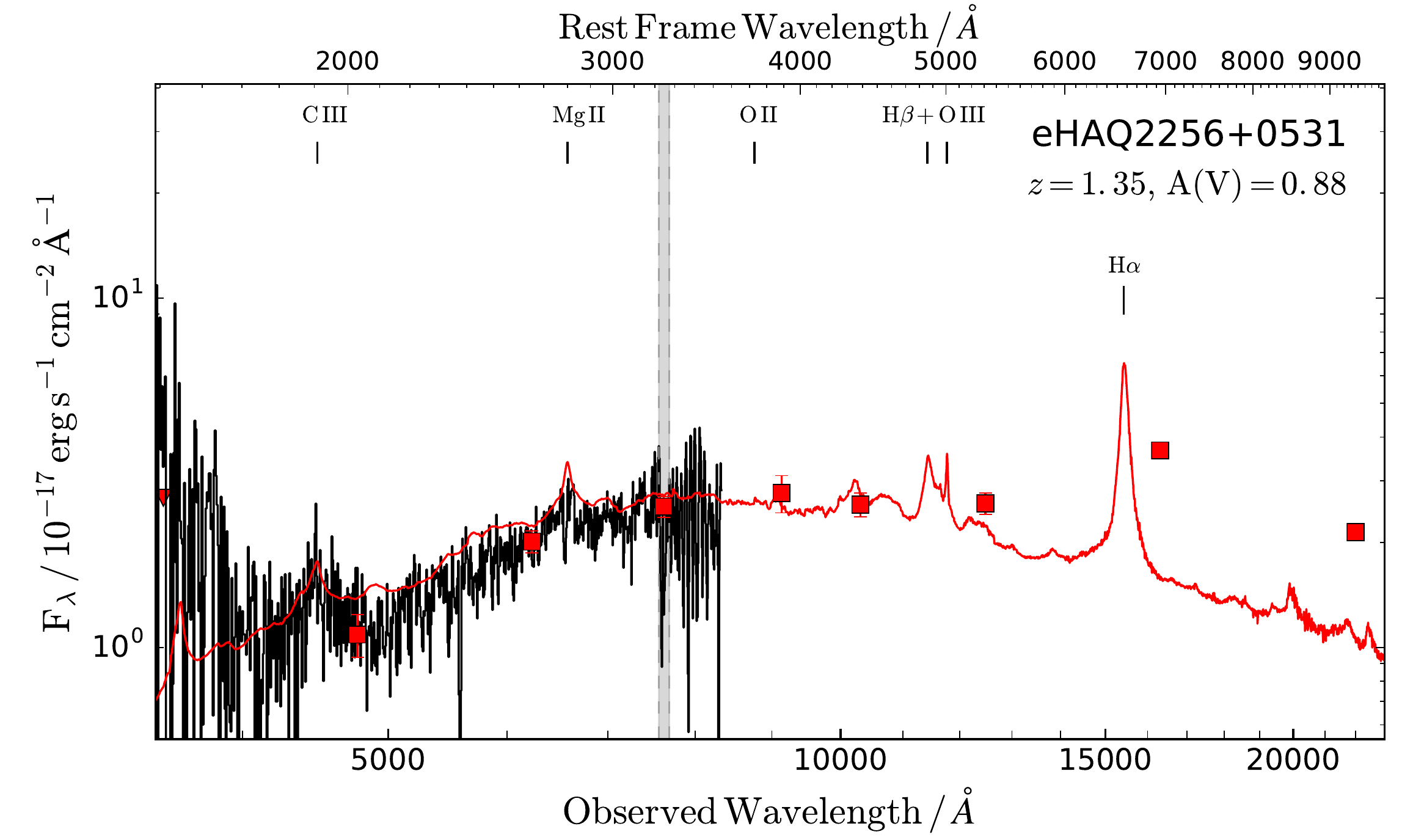}
    \includegraphics[width=0.49\textwidth]{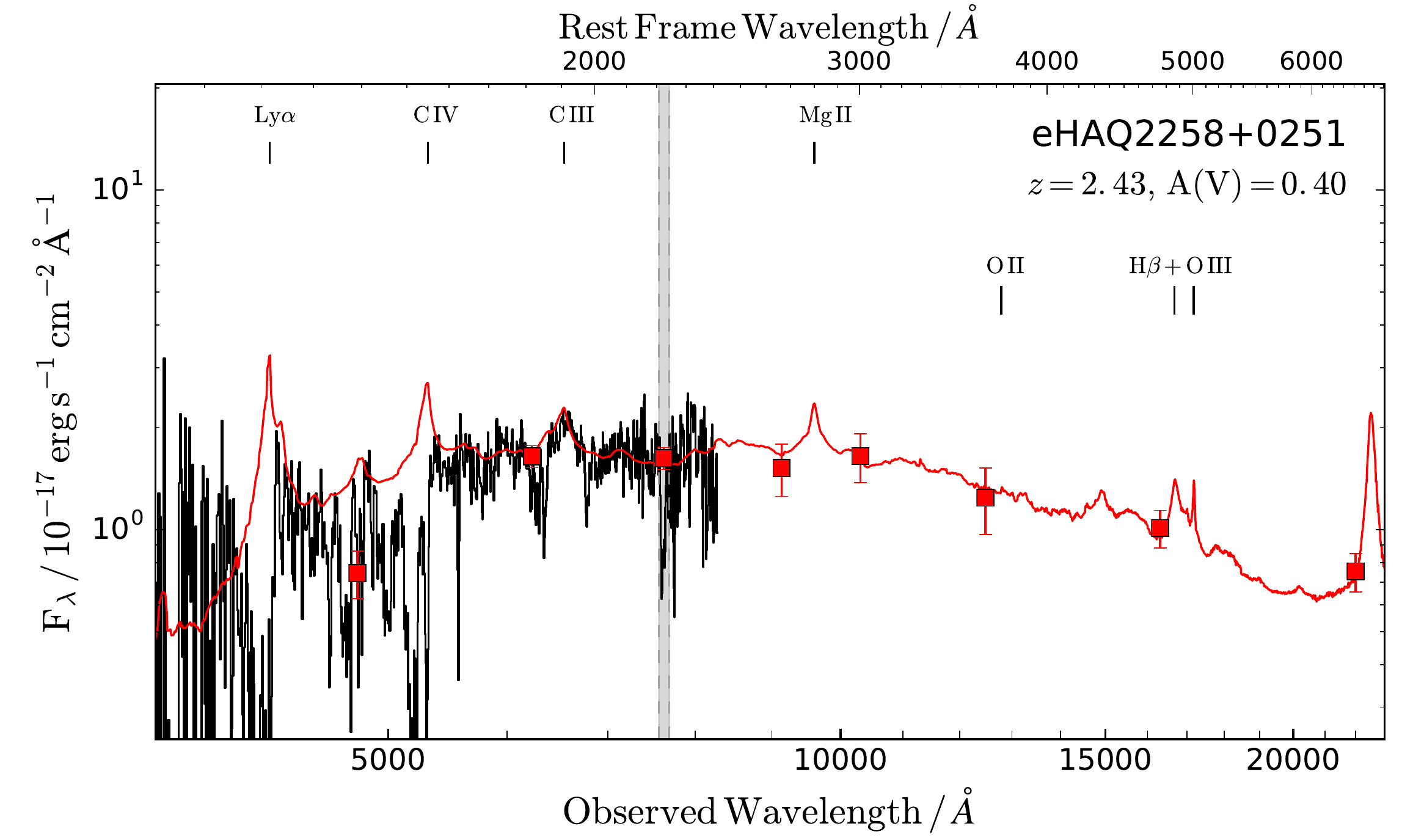}
    \vspace{4mm}

    \includegraphics[width=0.49\textwidth]{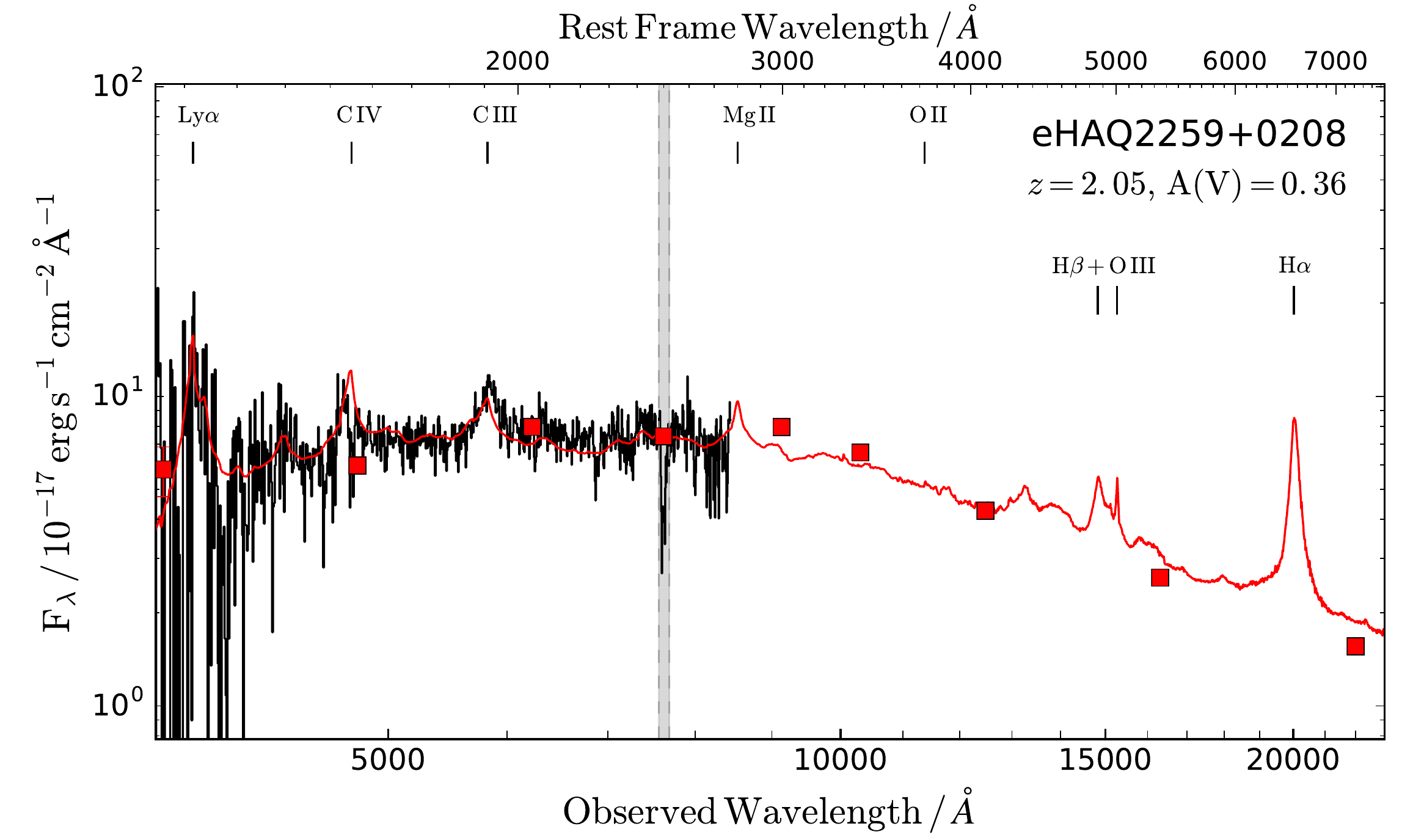}
    \includegraphics[width=0.49\textwidth]{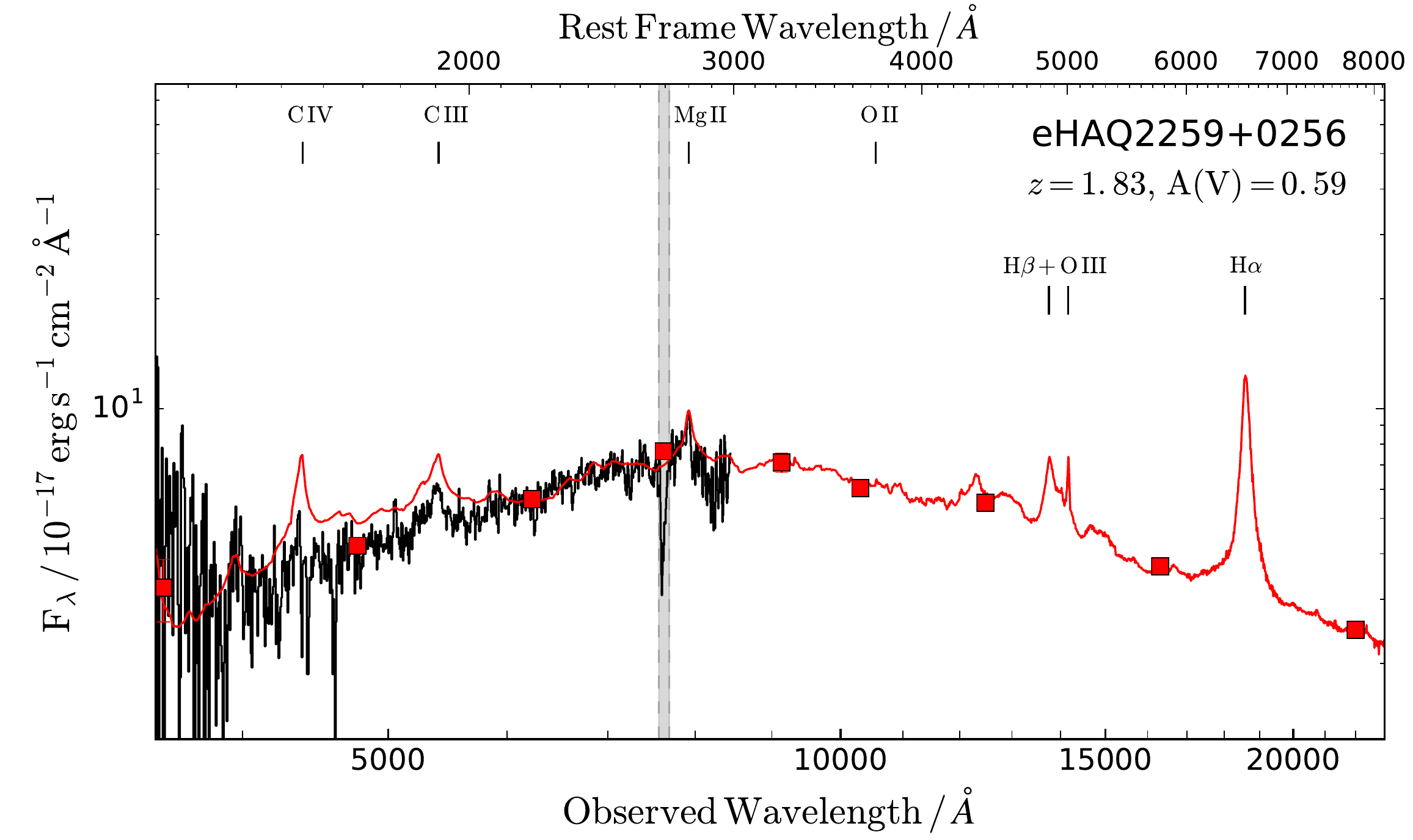}
\caption{(Continued.)}
\end{figure}

\begin{figure}
\figurenum{E2}
    \includegraphics[width=0.49\textwidth]{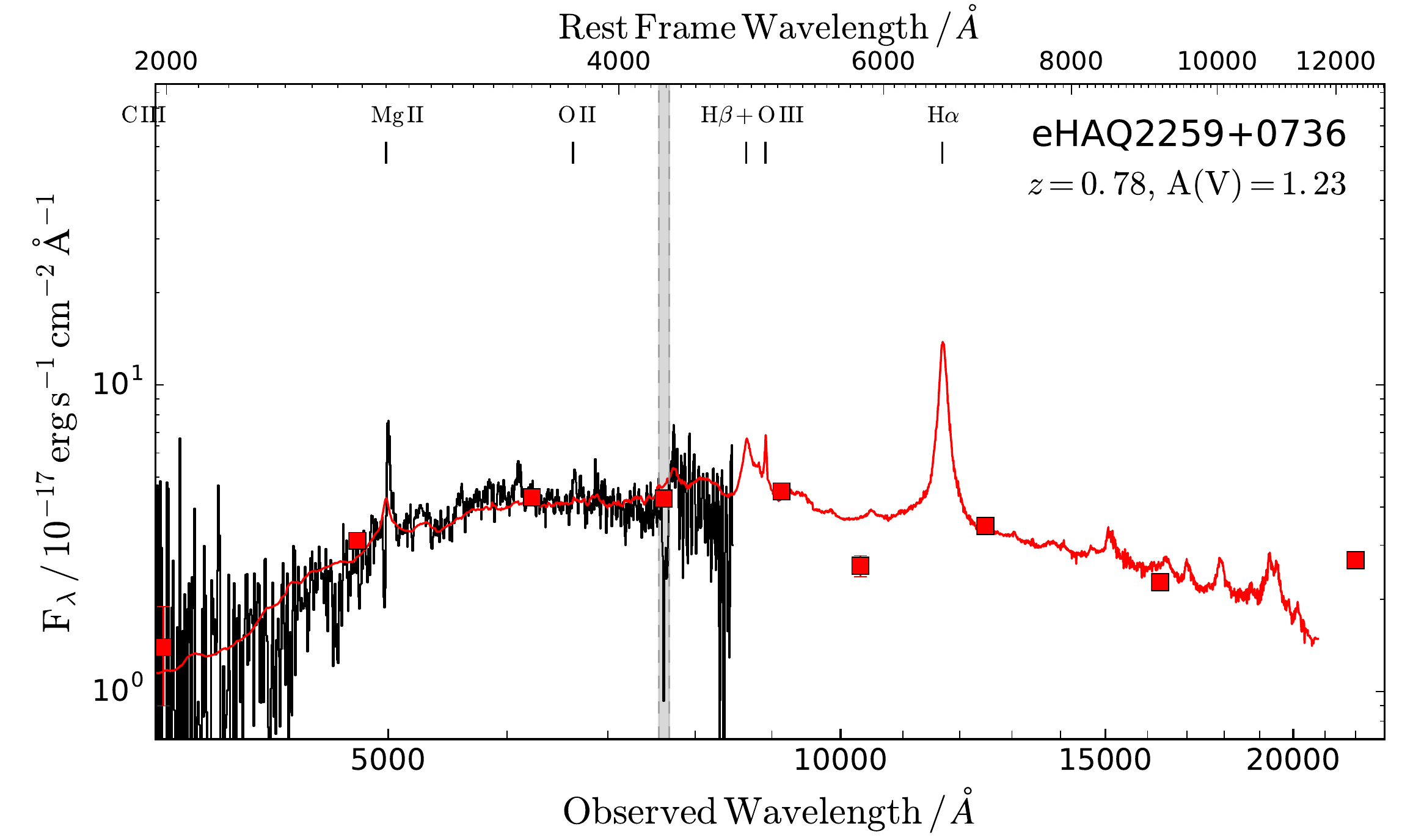}
    \includegraphics[width=0.49\textwidth]{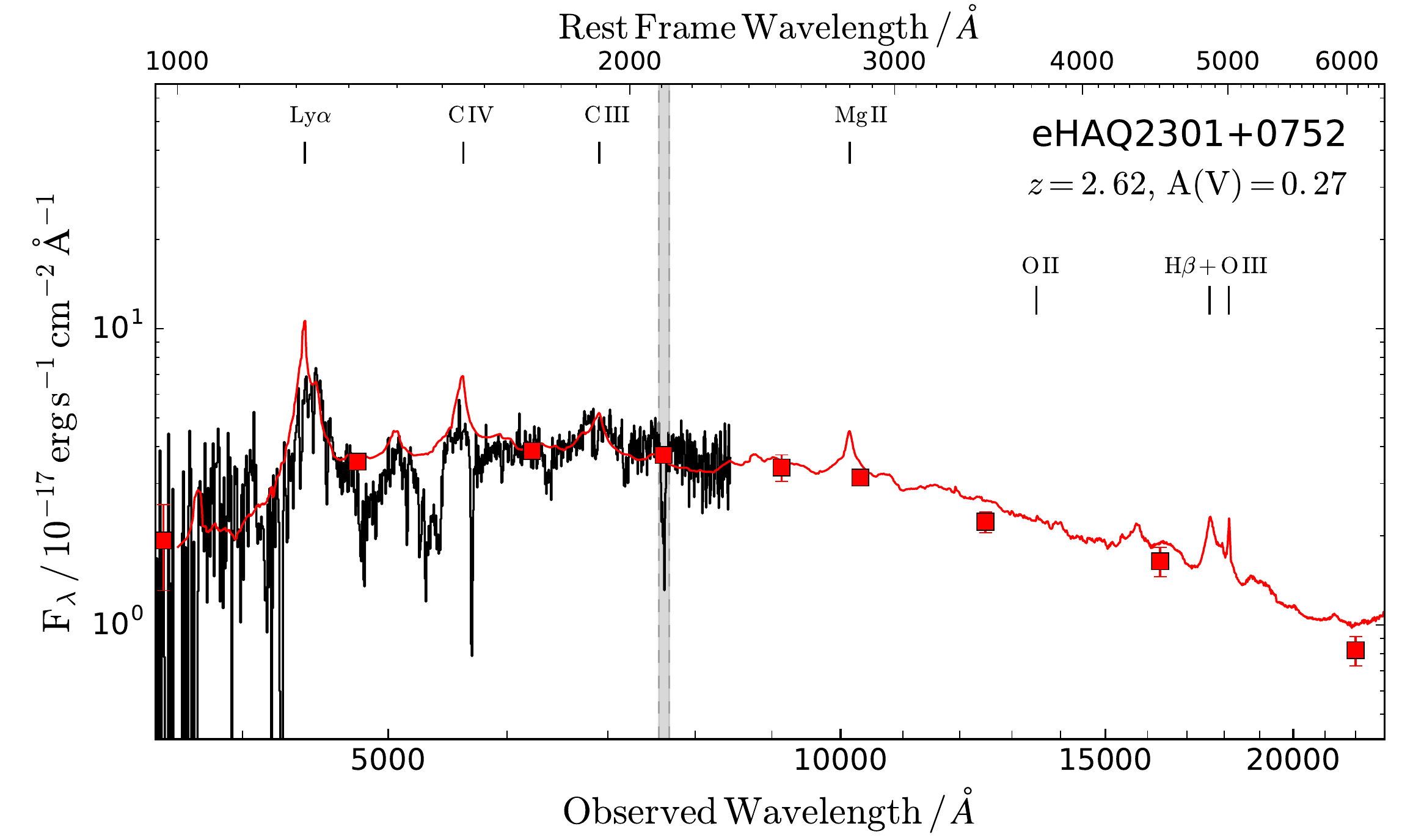}
    \vspace{4mm}

    \includegraphics[width=0.49\textwidth]{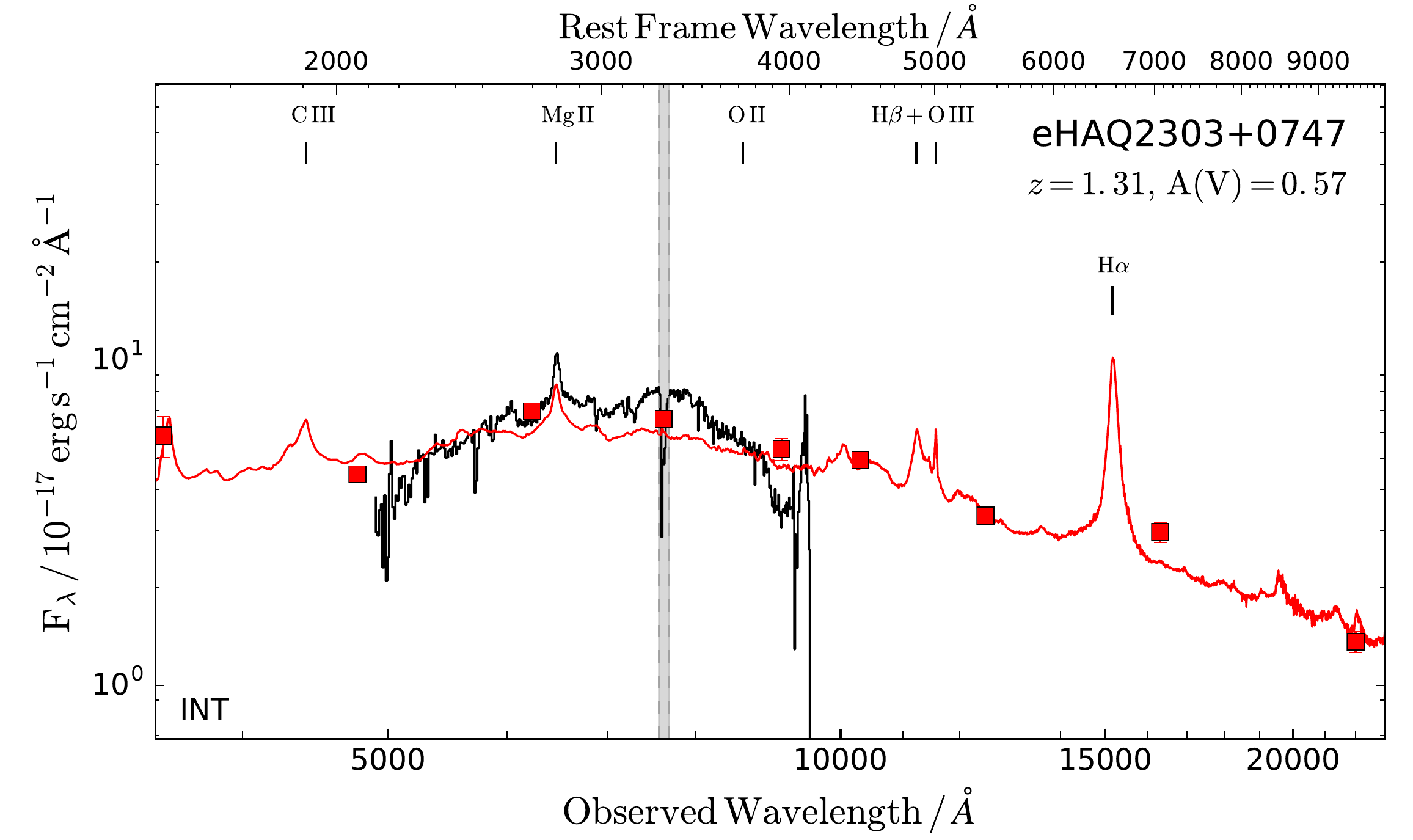}
    \includegraphics[width=0.49\textwidth]{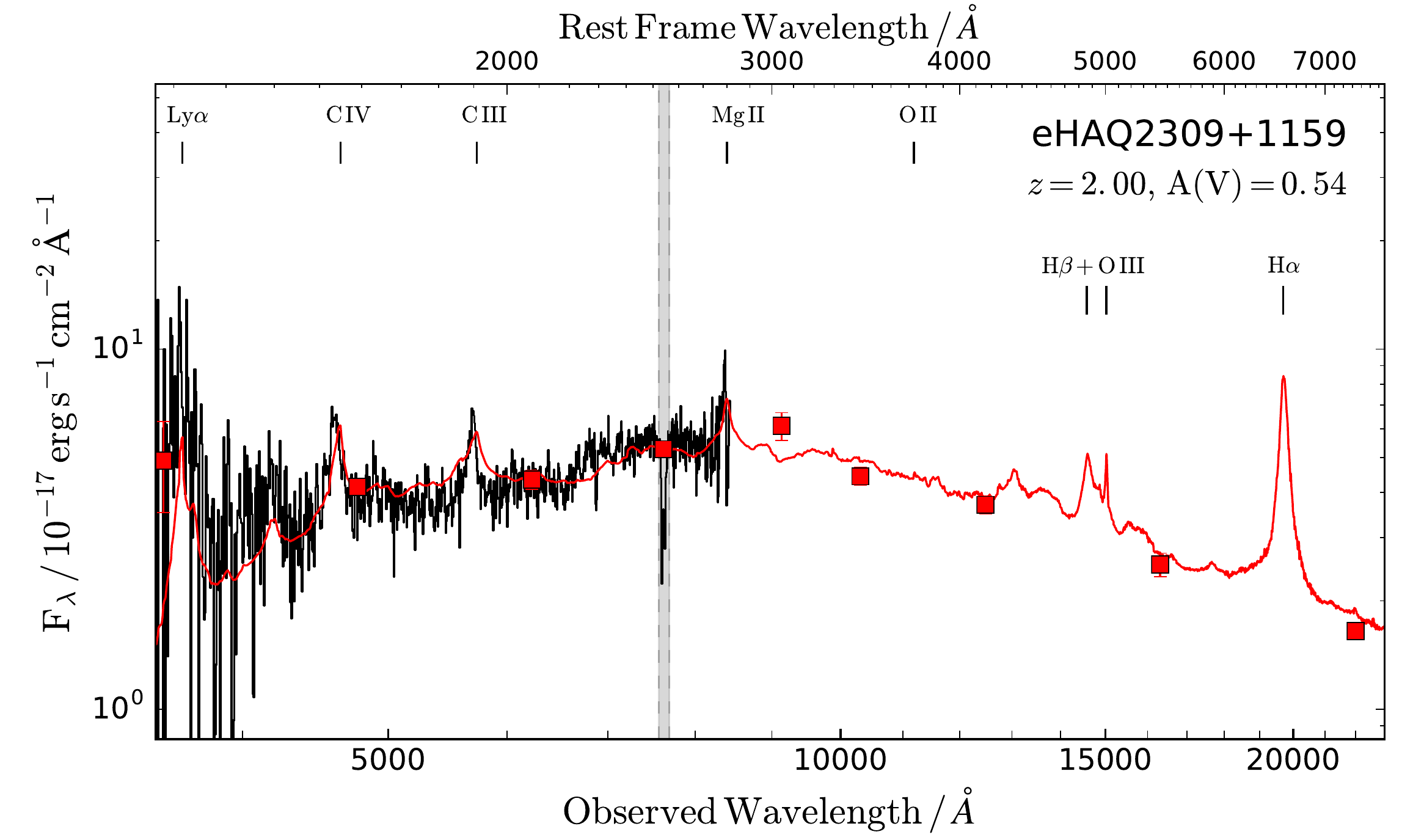}
    \vspace{4mm}

    \includegraphics[width=0.49\textwidth]{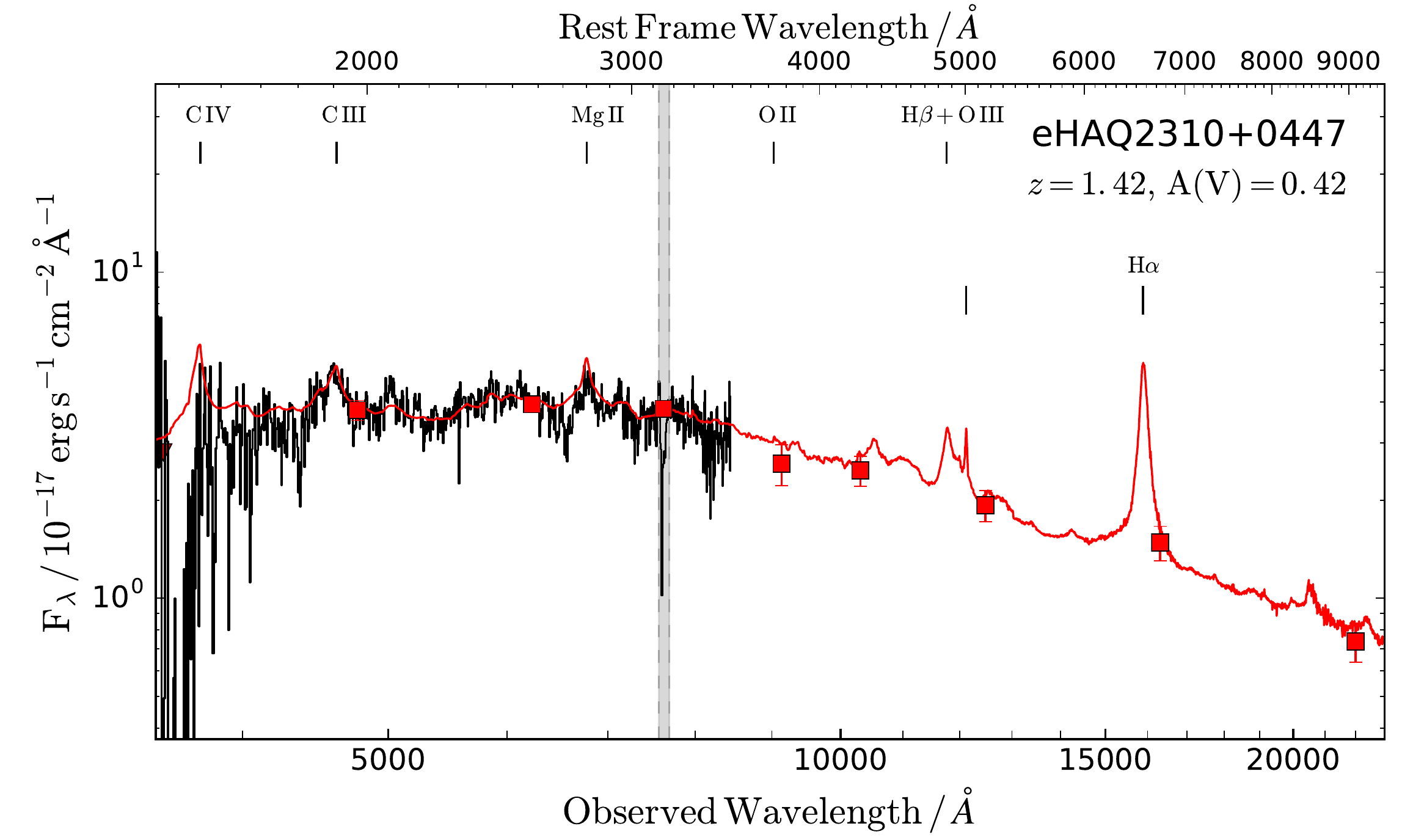}
    \includegraphics[width=0.49\textwidth]{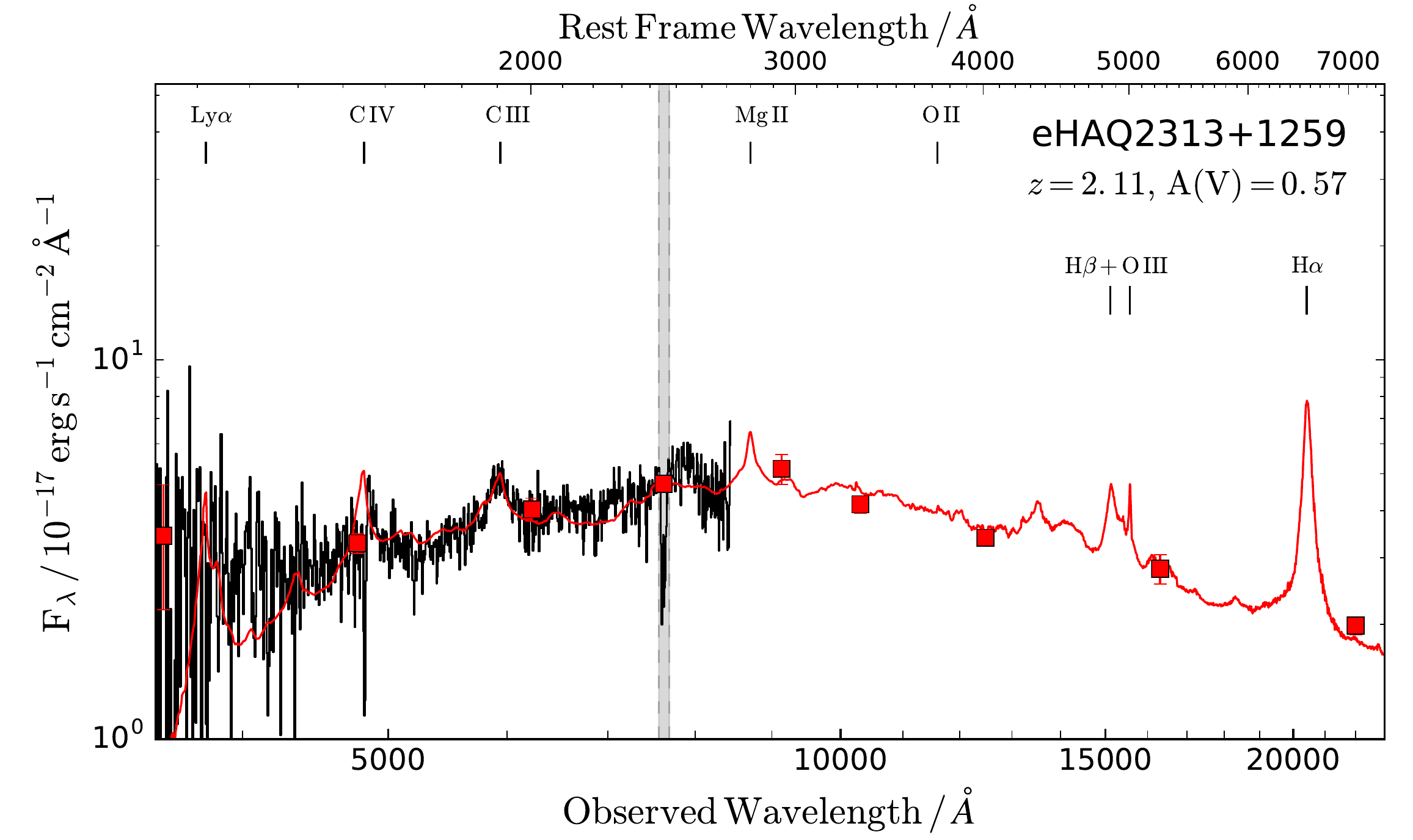}
    \vspace{4mm}

    \includegraphics[width=0.49\textwidth]{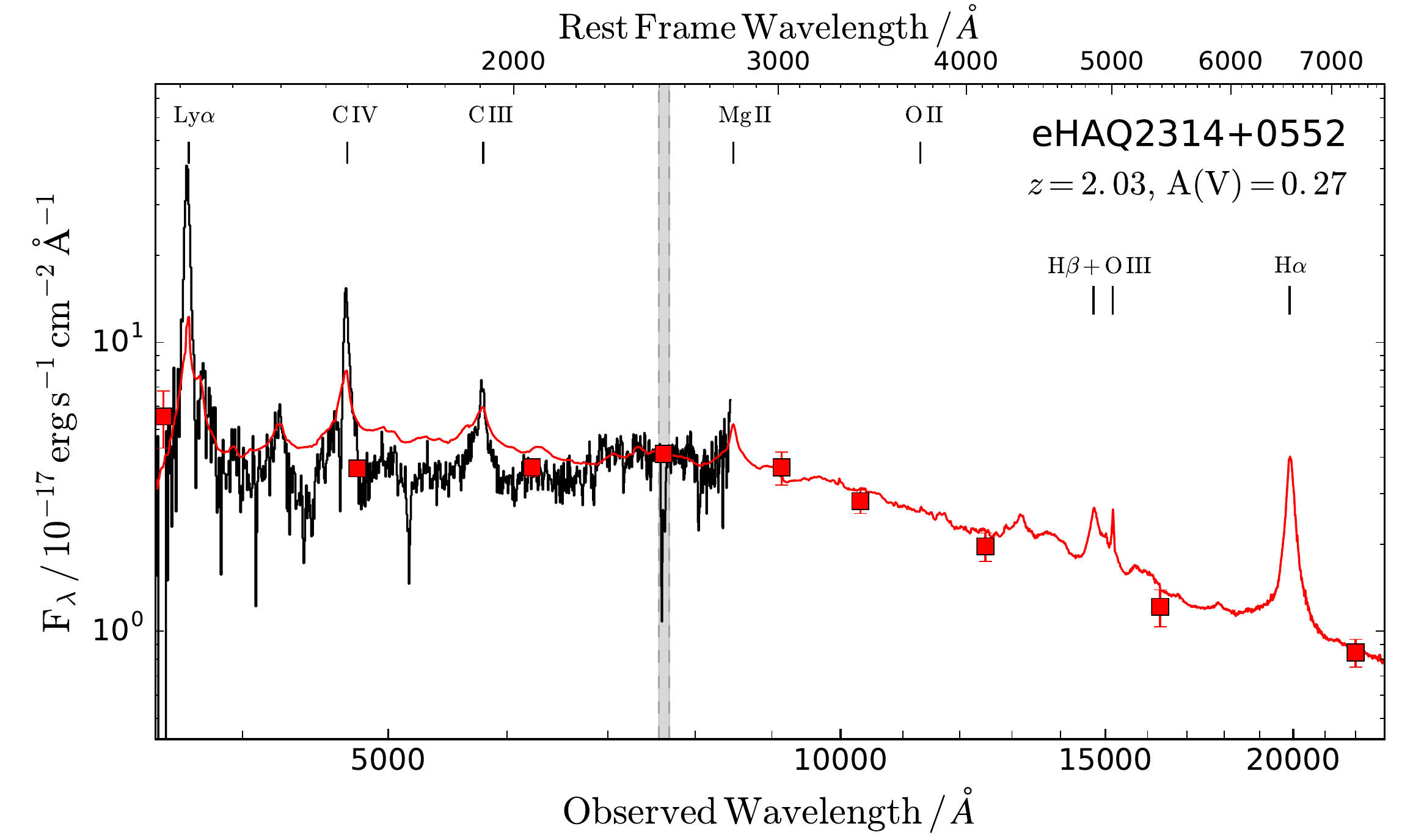}
    \includegraphics[width=0.49\textwidth]{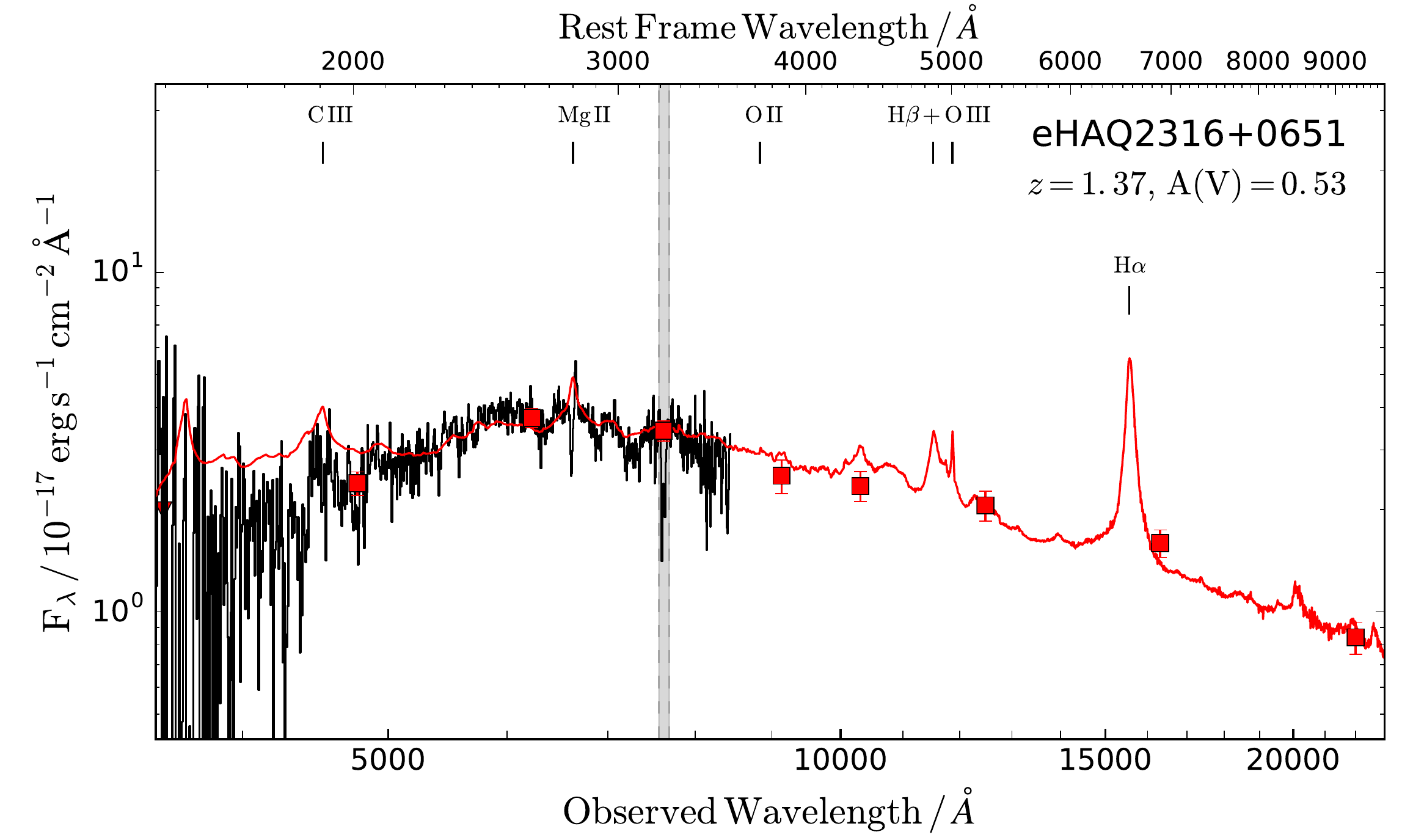}
\caption{(Continued.)}
\end{figure}

\begin{figure}
\figurenum{E2}
    \includegraphics[width=0.49\textwidth]{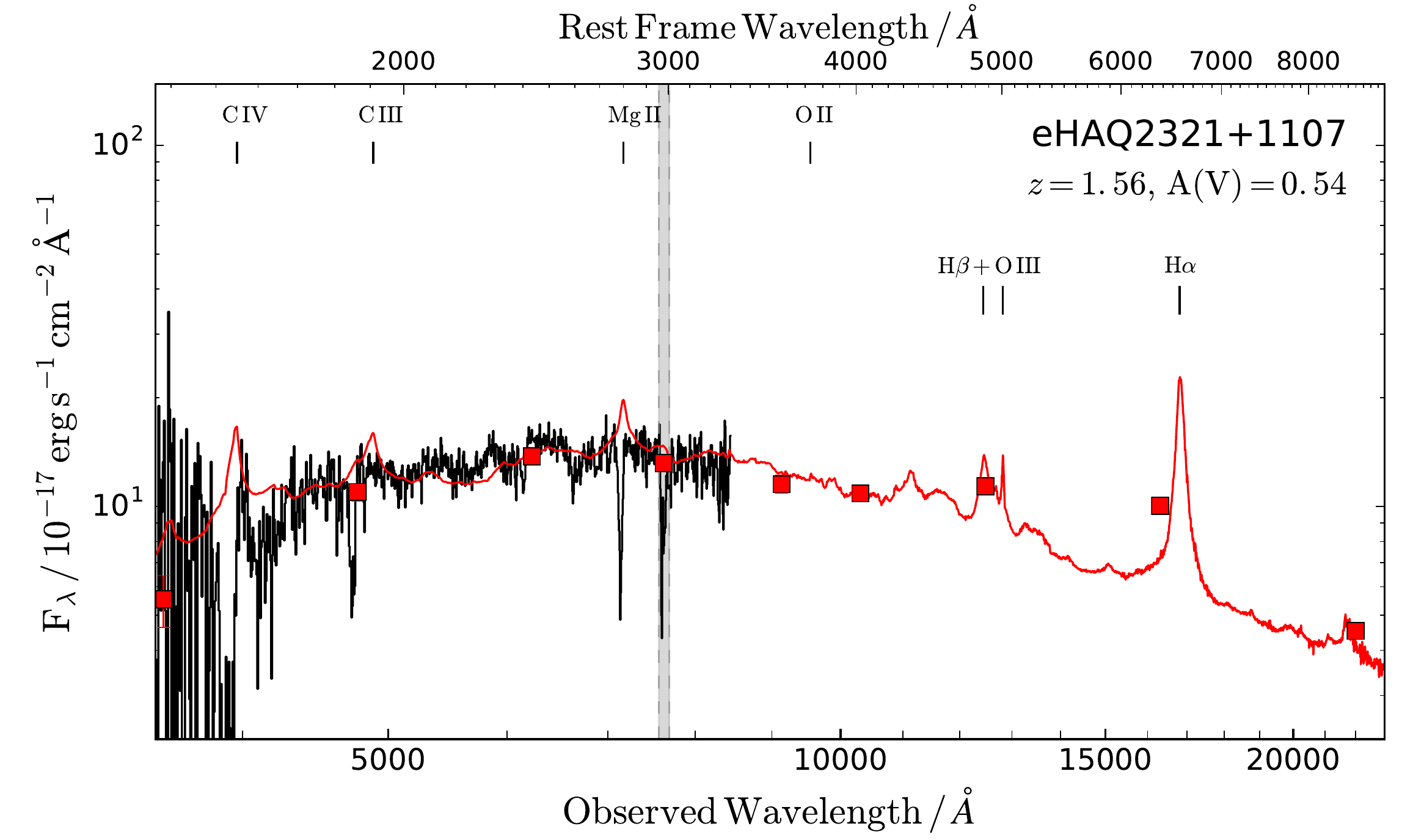}
    \includegraphics[width=0.49\textwidth]{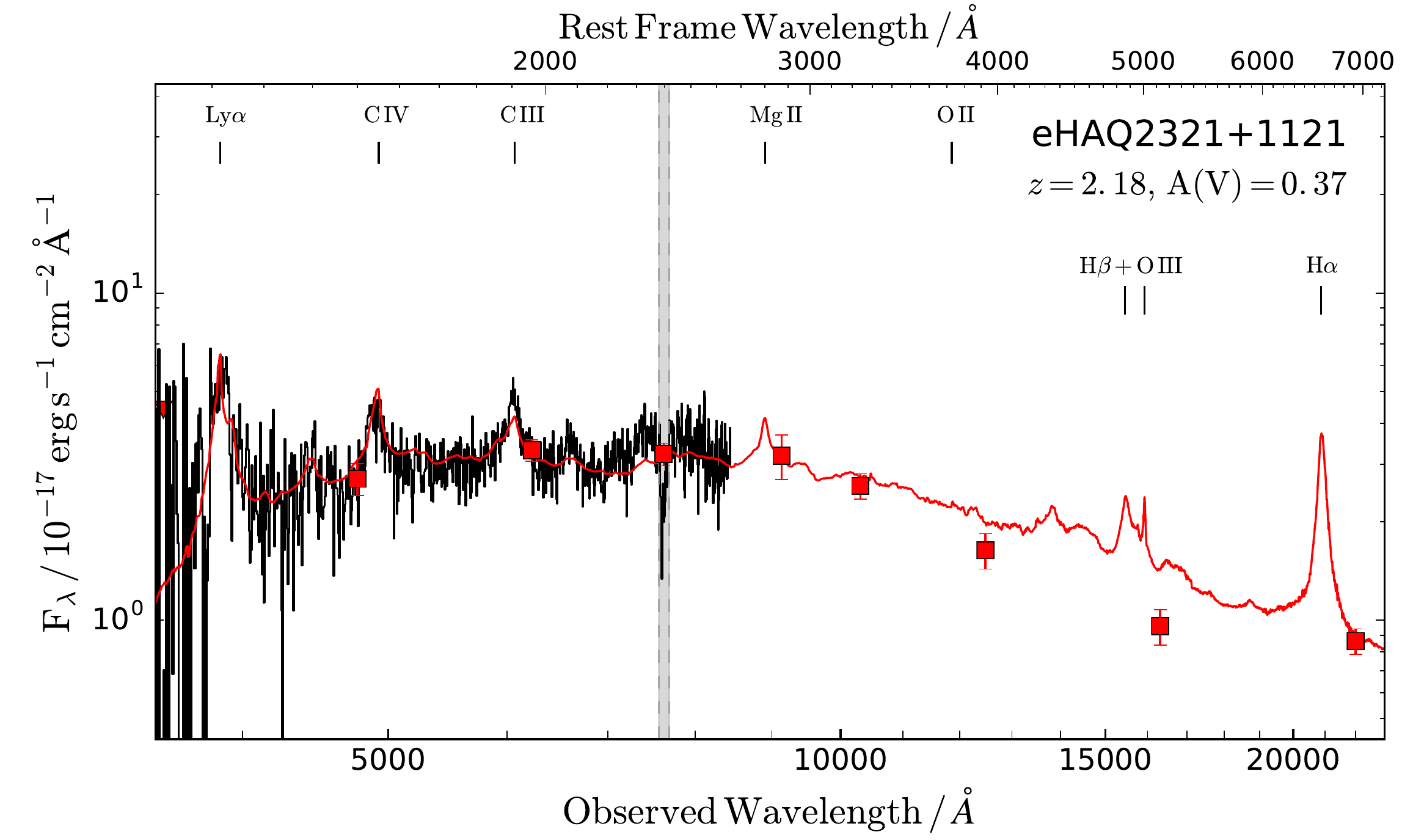}
    \vspace{4mm}

    \includegraphics[width=0.49\textwidth]{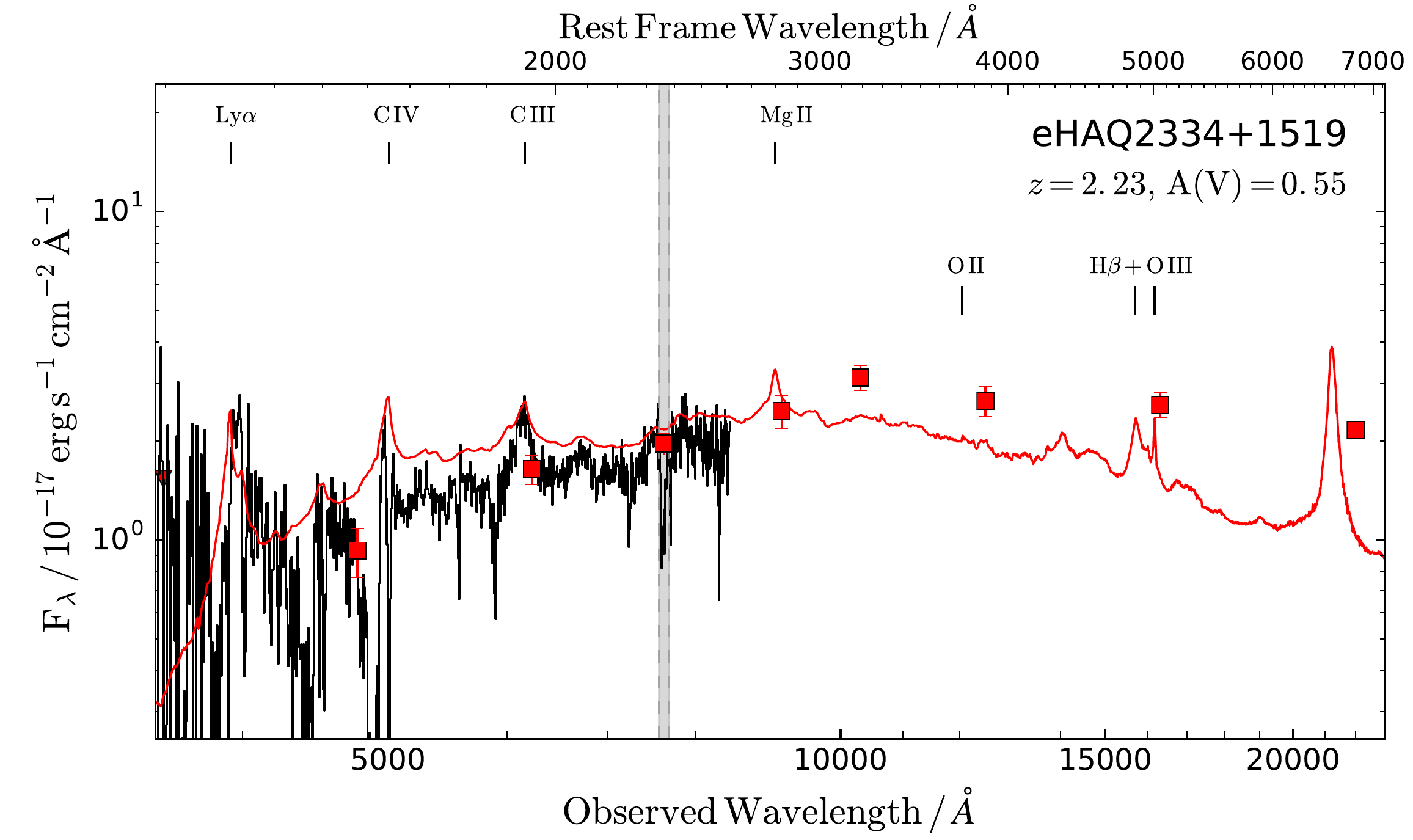}
    \includegraphics[width=0.49\textwidth]{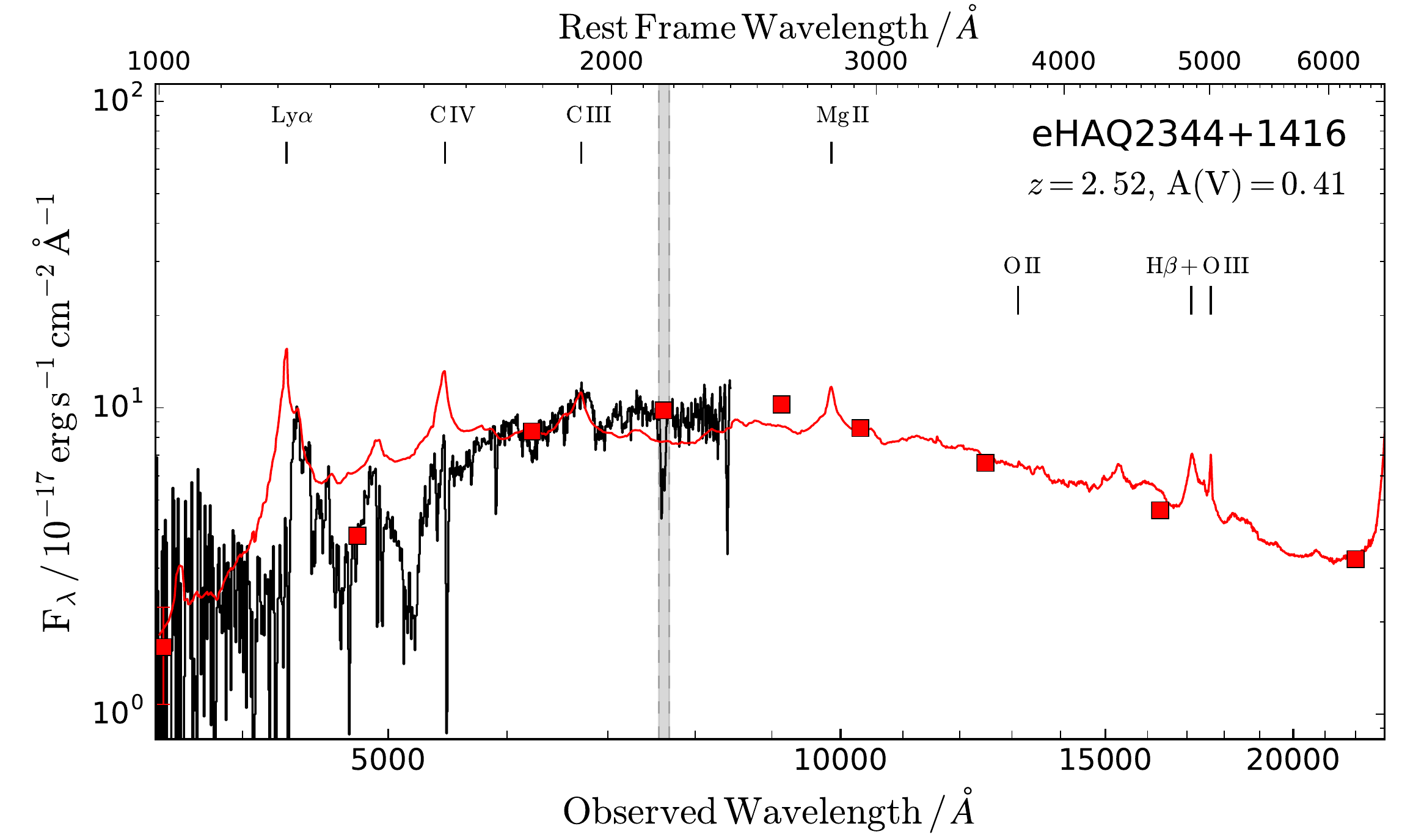}
    \vspace{4mm}

    \includegraphics[width=0.49\textwidth]{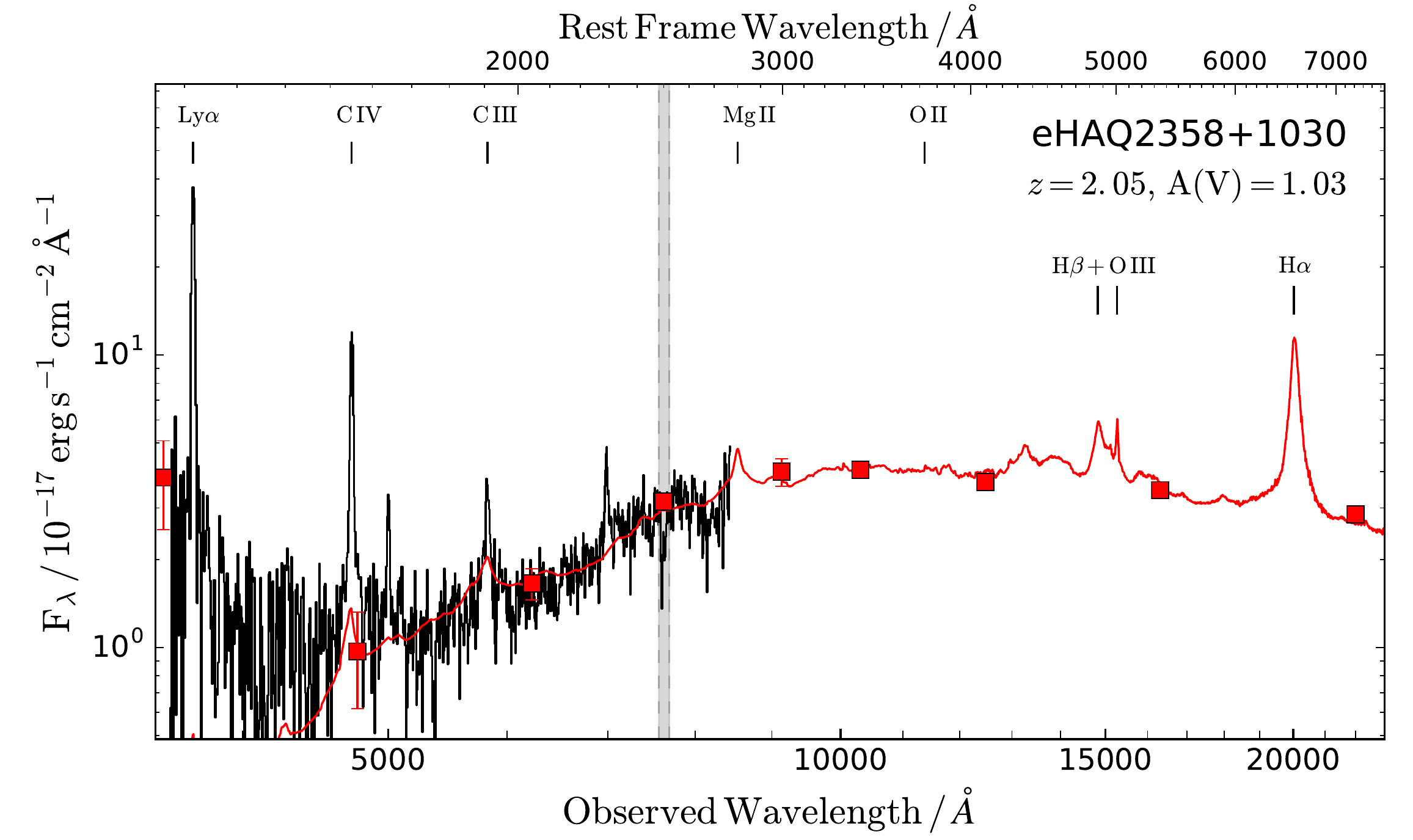}
    \includegraphics[width=0.49\textwidth]{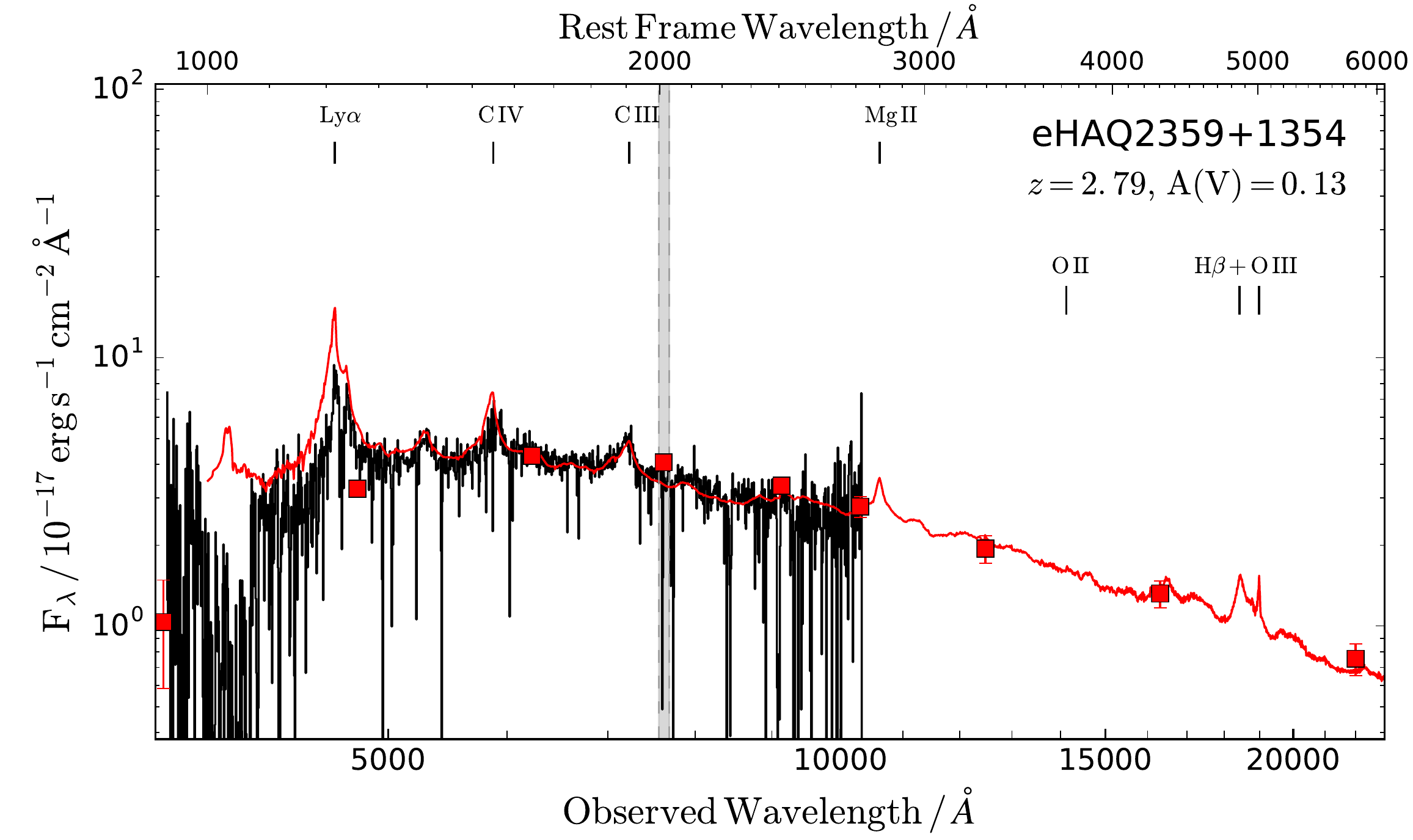}
\caption{(Continued.)}
\end{figure}

\end{document}

%% file: sample_overview.tex
\begin{deluxetable}{lllrrll}
\tabletypesize{\small}
\tablecaption{Observing log.
		\label{tab:overview}}
\tablehead{ \colhead{Target} & \colhead{R.A. (J\,2000)} & \colhead{Decl. (J\,2000)} &
\colhead{$J_{\rm AB}$} & \colhead{Exp. Time} & \colhead{Prog. ID\tablenotemark{$a$}} & \colhead{Notes} \\
  &  &  & (mag) & ($s$) &  &  }
\startdata
eHAQ0001+0431  &  00 01 042.85  &  +04 31 39.05  &  18.16  &  $2\times600$  &  P51-014 &  \\
eHAQ0007+1445  &  00 07 042.10  &  +14 45 28.63  &  17.93  &  $2\times900$  &  P51-802 &  \\
eHAQ0019+0657  &  00 19 057.28  &  +06 57 45.92  &  18.31  &  $2\times900$  &  P51-802 &  \\
eHAQ0026+0708  &  00 26 011.40  &  +07 08 41.32  &  18.08  &  $2\times600$  &  P51-014 &  \\
eHAQ0044+1321  &  00 44 056.43  &  +13 21 48.46  &  17.96  &  $2\times600$  &  P51-014 &  \\
eHAQ0102+1159  &  01 02 052.91  &  +11 59 48.40  &  18.51  &  $1\times900$  &  P51-802 & BOSS overlap \\
eHAQ0104+0756  &  01 04 017.06  &  +07 56 35.20  &  17.65  &  $2\times600$  &  P51-014 &  \\
eHAQ0104+0912  &  01 04 011.72  &  +09 12 38.49  &  18.38  &  $2\times600$  &  P51-014 &  \\
eHAQ0104+1506  &  01 04 041.53  &  +15 06 26.40  &  17.91  &  $2\times1000$  &  INT-C165 & Isaac Newton Telescope \\
eHAQ0109+0435  &  01 09 011.23  &  +04 35 44.22  &  17.65  &  $2\times900$  &  P51-014 &  \\
eHAQ0111+0641  &  01 11 034.71  &  +06 41 19.22  &  18.57  &  $2\times900$  &  P51-014 &  \\
eHAQ0113+0804  &  01 13 055.39  &  +08 04 25.68  &  17.30  &  $2\times600$  &  P51-014 &  \\
eHAQ0121+1028  &  01 21 034.45  &  +10 28 34.18  &  19.28  &  $2\times1200$  &  P51-014 &  \\
eHAQ0129+0638  &  01 29 050.61  &  +06 38 46.00  &  18.02  &  $2\times1300$  &  P51-014 &  \\
eHAQ0129+1039  &  01 29 018.99  &  +10 39 43.16  &  19.07  &  $2\times1200$  &  P51-014 &  \\
eHAQ0138+0742  &  01 38 014.45  &  +07 42 35.49  &  18.35  &  $2\times600$  &  P51-014 &  \\
eHAQ0142+0257  &  01 42 006.88  &  +02 57 13.06  &  16.68  &  $2\times600$  &  P51-014 &  \\
eHAQ0147+0411  &  01 47 032.03  &  +04 11 13.41  &  17.42  &  $2\times700$  &  INT-C165 & Isaac Newton Telescope \\
eHAQ0147+1549  &  01 47 016.89  &  +15 49 43.99  &  17.20  &  $2\times700$  &  P51-014 &  \\
eHAQ0157+1321  &  01 57 001.54  &  +13 21 18.83  &  17.89  &  $2\times1200$  &  INT-C165 & Isaac Newton Telescope \\
eHAQ0216+0426  &  02 16 043.27  &  +04 26 28.75  &  18.81  &  $2\times1200$  &  P51-014 &  \\
eHAQ0227+0521  &  02 27 057.93  &  +05 21 41.88  &  18.40  &  $1\times600$  &  P51-014 &  \\
eHAQ0300+0440  &  03 00 002.47  &  +04 40 04.89  &  18.54  &  $2\times600$  &  P51-014 &  \\
eHAQ0321+0523  &  03 21 052.20  &  +05 23 37.40  &  17.91  &  $1\times900$  &  P51-014 & BOSS overlap \\
eHAQ0347+0348  &  03 47 034.17  &  +03 48 36.88  &  18.41  &  $2\times600$  &  P51-014 &  \\
eHAQ0828+0313  &  08 28 013.98  &  +03 13 54.35  &  17.21  &  $3\times200$  &  P50-802 &  \\
eHAQ0835+0127  &  08 35 015.82  &  +01 27 31.42  &  18.00  &  $3\times600$  &  P50-802 & $J-K<-0.05$ \\
eHAQ0835+0830  &  08 35 042.19  &  +08 30 10.37  &  16.75  &  $3\times200$  &  P50-802 & $J-K<-0.05$ \\
eHAQ0839+0556  &  08 39 002.87  &  +05 56 27.77  &  18.03  &  $3\times500$  &  P50-802 &  \\
eHAQ0852+0204  &  08 52 003.84  &  +02 04 37.79  &  18.41  &  $3\times900$  &  P52-017 &  \\
eHAQ0856+0643  &  08 56 028.31  &  +06 43 54.18  &  18.67  &  $3\times900$  &  P50-802 &  \\
eHAQ0913+0910  &  09 13 004.43  &  +09 10 44.24  &  17.59  &  $3\times250$  &  P50-802 &  \\
eHAQ0915+0115  &  09 15 035.57  &  +01 15 15.43  &  17.38  &  $3\times400$  &  P50-802 & $J-K<-0.05$ \\
eHAQ0919+0843  &  09 19 007.11  &  +08 43 05.04  &  18.38  &  $1\times900$  &  P52-017 &  \\
eHAQ0921+0149  &  09 21 042.04  &  +01 49 59.75  &  17.70  &  $3\times900$  &  P50-802 &  \\
eHAQ0923+0520  &  09 23 045.76  &  +05 20 26.41  &  17.79  &  $3\times600$  &  P50-802 &  \\
eHAQ0927$-$0233  &  09 27 040.04  &  $-$02 33 47.47  &  18.95  &  $3\times900$  &  P52-017 &  \\
eHAQ0930+0148  &  09 30 045.09  &  +01 48 46.69  &  18.69  &  $3\times900$  &  P52-017 &  \\
eHAQ0940+0532  &  09 40 051.36  &  +05 32 13.29  &  17.67  &  $3\times500$  &  P50-802 &  \\
eHAQ0943+0954  &  09 43 049.65  &  +09 54 00.93  &  17.88  &  $3\times600$  &  P50-802 & BOSS overlap \\
eHAQ0943+1300  &  09 43 032.99  &  +13 00 55.74  &  17.91  &  $3\times600$  &  P52-017 &  \\
eHAQ0949+1207  &  09 49 046.98  &  +12 07 56.16  &  18.53  &  $2\times600$  &  P52-017 &  \\
eHAQ0950+0440  &  09 50 004.09  &  +04 40 40.35  &  18.60  &  $3\times600$  &  P52-017 &  \\
eHAQ0952+0835  &  09 52 002.99  &  +08 35 55.84  &  18.12  &  $3\times900$  &  P52-017 &  \\
eHAQ1002+0406  &  10 02 051.28  &  +04 06 53.95  &  16.42  &  $3\times400$  &  P50-802 & $J-K<-0.05$ \\
eHAQ1005+0602  &  10 05 001.82  &  +06 02 19.88  &  18.41  &  $2\times900$  &  P52-017 &  \\
eHAQ1010+1158  &  10 10 054.03  &  +11 58 05.70  &  18.10  &  $3\times600$  &  P50-802 &  \\
eHAQ1025+1324  &  10 25 018.62  &  +13 24 12.05  &  17.53  &  $3\times500$  &  P50-802 &  \\
eHAQ1026$-$0241  &  10 26 000.91  &  $-$02 41 17.08  &  18.34  &  $2\times900$  &  P52-017 &  \\
eHAQ1030+1040  &  10 30 032.55  &  +10 40 51.34  &  17.25  &  $3\times200$  &  P50-802 &  \\
eHAQ1106+0844  &  11 06 035.85  &  +08 44 02.35  &  18.44  &  $3\times900$  &  P50-802 & BOSS overlap \\
eHAQ1109+0135  &  11 09 002.31  &  +01 35 38.81  &  16.62  &  $3\times400$  &  P50-802 & $J-K<-0.05$ \\
eHAQ1109+1058  &  11 09 024.83  &  +10 58 25.43  &  18.82  &  $2\times900$  &  P52-017 &  \\
eHAQ1111+0151  &  11 11 037.69  &  +01 51 47.27  &  17.07  &  $3\times360$  &  P50-802 & $J-K<-0.05$ \\
eHAQ1119+1430  &  11 19 039.72  &  +14 30 00.76  &  18.43  &  $3\times600$  &  P52-017 &  \\
eHAQ1120+0812  &  11 20 003.66  &  +08 12 12.02  &  15.57  &  $3\times100$  &  P52-017 &  \\
eHAQ1132+1243  &  11 32 007.19  &  +12 43 42.09  &  17.32  &  $3\times600$  &  P50-802 &  \\
eHAQ1136+0027  &  11 36 044.12  &  +00 27 00.30  &  18.12  &  $3\times900$  &  P52-017 &  \\
eHAQ1144+0902  &  11 44 045.44  &  +09 02 26.70  &  17.92  &  $3\times600$  &  P50-802 & BOSS overlap \\
eHAQ1202+0423  &  12 02 004.85  &  +04 23 41.00  &  18.57  &  $2\times600$  &  P52-017 &  \\
eHAQ1203+0652  &  12 03 036.93  &  +06 52 34.47  &  18.91  &  $3\times900$  &  P52-017 &  \\
eHAQ1203+1118  &  12 03 033.96  &  +11 18 26.02  &  18.00  &  $3\times900$  &  P50-802 &  \\
eHAQ1210+1429  &  12 10 020.99  &  +14 29 58.53  &  16.10  &  $3\times200$  &  P50-802 &  \\
eHAQ1222+0826  &  12 22 014.97  &  +08 26 10.15  &  18.69  &  $2\times900$  &  P52-017 &  \\
eHAQ1226$-$0236  &  12 26 055.31  &  $-$02 36 57.90  &  18.63  &  $3\times900$  &  P52-017 &  \\
eHAQ1237+1233  &  12 37 026.47  &  +12 33 50.23  &  18.71  &  $2\times900$  &  P52-017 &  \\
eHAQ1244+0841  &  12 44 018.82  &  +08 41 36.10  &  18.32  &  $3\times800$  &  P50-802 &  \\
eHAQ1252+0842  &  12 52 057.73  &  +08 42 06.06  &  17.93  &  $3\times600$  &  P50-802 &  \\
eHAQ1312+1431  &  13 12 014.18  &  +14 31 11.28  &  18.53  &  $2\times900$  &  P52-017 &  \\
eHAQ1326+1317  &  13 26 024.77  &  +13 17 27.52  &  17.64  &  $6\times300$  &  P50-802 & BOSS overlap \\
eHAQ1331+1304  &  13 31 029.90  &  +13 04 20.90  &  17.11  &  $3\times600$  &  P50-802 & $J-K<-0.05$ \\
eHAQ1340+0151  &  13 40 044.56  &  +01 51 41.44  &  18.16  &  $4\times300$  &  P52-017 &  \\
eHAQ1340+1458  &  13 40 053.91  &  +14 58 53.81  &  17.80  &  $2\times900$  &  P52-017 &  \\
eHAQ1346+0114  &  13 46 032.37  &  +01 14 08.38  &  18.27  &  $3\times900$  &  P50-802 &  \\
eHAQ1357$-$0051  &  13 57 030.57  &  $-$00 51 41.09  &  16.60  &  $3\times300$  &  P52-017 &  \\
eHAQ1400+0720  &  14 00 039.19  &  +07 20 11.99  &  18.38  &  $2\times900$  &  P52-017 &  \\
eHAQ1447+0521  &  14 47 021.61  &  +05 21 41.70  &  18.54  &  $3\times900$  &  P52-017 &  \\
eHAQ1450+1002  &  14 50 043.69  &  +10 02 38.79  &  18.38  &  $2\times900$  &  P52-017 &  \\
eHAQ1455+0705  &  14 55 010.53  &  +07 05 25.50  &  16.39  &  $3\times200$  &  P50-802 &  \\
eHAQ1514$-$0002  &  15 14 001.87  &  $-$00 02 59.66  &  18.31  &  $3\times600$  &  P52-017 &  \\
eHAQ1525+0155  &  15 25 055.01  &  +01 55 13.54  &  18.25  &  $3\times900$  &  P52-017 &  \\
eHAQ1528+0546  &  15 28 053.64  &  +05 46 57.42  &  17.99  &  $6\times360$  &  P50-802 & BOSS overlap \\
eHAQ1539+0351  &  15 39 035.25  &  +03 51 25.57  &  17.00  &  $3\times360$  &  P50-802 &  \\
eHAQ1543+0447  &  15 43 027.29  &  +04 47 17.75  &  17.96  &  $2\times900$  &  P52-017 &  \\
eHAQ1549+0501  &  15 49 036.57  &  +05 01 28.04  &  18.11  &  $3\times900$  &  P52-017 &  \\
eHAQ2209+0304  &  22 09 020.36  &  +03 04 36.02  &  18.24  &  $2\times900$  &  P51-802 &  \\
eHAQ2235+0635  &  22 35 012.40  &  +06 35 45.49  &  18.11  &  $2\times400$  &  P51-014 &  \\
eHAQ2236+0731  &  22 36 009.38  &  +07 31 08.20  &  19.08  &  $2\times900$  &  P51-014 &  \\
eHAQ2247+0922  &  22 47 009.08  &  +09 22 33.46  &  18.46  &  $2\times900$  &  P51-802 &  \\
eHAQ2255+1213  &  22 55 058.75  &  +12 13 10.91  &  18.73  &  $2\times600$  &  P51-014 &  \\
eHAQ2256+0531  &  22 56 043.90  &  +05 31 15.07  &  18.65  &  $2\times1500$  &  P51-014 &  \\
eHAQ2258+0251  &  22 58 043.01  &  +02 51 07.21  &  19.42  &  $2\times1500$  &  P51-014 &  \\
eHAQ2259+0208  &  22 59 055.76  &  +02 08 25.39  &  18.10  &  $2\times400$  &  P51-014 &  \\
eHAQ2259+0256  &  22 59 020.36  &  +02 56 43.76  &  17.80  &  $2\times600$  &  P51-014 &  \\
eHAQ2259+0736  &  22 59 017.85  &  +07 36 33.47  &  18.30  &  $2\times1000$  &  P51-802 &  \\
eHAQ2301+0752  &  23 01 059.41  &  +07 52 38.95  &  18.78  &  $2\times900$  &  P51-014 &  \\
eHAQ2303+0747  &  23 03 000.30  &  +07 47 36.78  &  18.34  &  $2\times1080$  &  INT-C165 & Isaac Newton Telescope \\
eHAQ2309+1159  &  23 09 014.78  &  +11 59 55.99  &  18.25  &  $2\times900$  &  P51-014 &  \\
eHAQ2310+0447  &  23 10 043.96  &  +04 47 46.77  &  18.94  &  $2\times900$  &  P51-802 &  \\
eHAQ2313+1259  &  23 13 024.39  &  +12 59 39.01  &  18.32  &  $2\times900$  &  P51-014 &  \\
eHAQ2314+0552  &  23 14 053.38  &  +05 52 33.11  &  18.95  &  $2\times1000$  &  P51-014 &  \\
eHAQ2316+0651  &  23 16 007.06  &  +06 51 45.09  &  18.90  &  $2\times1000$  &  P51-014 &  \\
eHAQ2321+1107  &  23 21 051.29  &  +11 07 20.67  &  17.01  &  $2\times300$  &  P51-014 &  \\
eHAQ2321+1121  &  23 21 052.27  &  +11 21 28.50  &  19.11  &  $2\times1000$  &  P51-014 &  \\
eHAQ2334+1519  &  23 34 021.51  &  +15 19 47.56  &  18.60  &  $2\times1200$  &  P51-014 &  \\
eHAQ2344+1416  &  23 44 051.94  &  +14 16 40.46  &  17.59  &  $2\times500$  &  P51-014 &  \\
eHAQ2358+1030  &  23 58 040.47  &  +10 30 39.94  &  18.28  &  $2\times1200$  &  P51-014 &  \\
eHAQ2359+1354  &  23 59 016.50  &  +13 54 43.35  &  18.92  &  $2\times800$  &  P51-802 & Grism \#6, BOSS overlap \\
\enddata
\tablenotetext{a}{Observing programme identifier. If nothing else is stated, the observations were carried out at the Nordic Optical Telescope.}
\end{deluxetable}

%% file: sample_data.tex
\begin{deluxetable}{lccl}
\tabletypesize{\small}
\tablecaption{Quasar Classification
		\label{tab:sample}}
\tablehead{ \colhead{Target} & \colhead{$z$} & \colhead{A(V)} &
\colhead{Classification} \\
  &  & (mag) &  }
\startdata
eHAQ0001+0431  &  1.96  &  $0.52\pm0.01$  &  Quasar  \\
eHAQ0007+1445  &  1.37  &  $0.68\pm0.01$  &  Broad absorption line quasar  \\
eHAQ0019+0657  &  3.37  &  $0.17\pm0.01$  &  Quasar  \\
eHAQ0026+0708  &  1.78  &  $0.50\pm0.01$  &  Quasar  \\
eHAQ0044+1321\tablenotemark{$a$}  &  1.70  &  $0.16\pm0.01$  &  Broad absorption line quasar  \\
eHAQ0102+1159  &  4.25  &  $0.15\pm0.03$  &  Broad absorption line quasar  \\
eHAQ0104+0756\tablenotemark{$a$}  &  2.21  &  $0.47\pm0.01$  &  Broad absorption line quasar  \\
eHAQ0104+0912  &  2.01  &  $0.22\pm0.01$  &  Quasar  \\
eHAQ0104+1506\tablenotemark{$a$}  &  1.42  &  0.3--0.6  &  Quasar  \\
eHAQ0109+0435  &  3.47  &  $0.20\pm0.05$  &  Broad absorption line quasar  \\
eHAQ0111+0641  &  3.23  &  $0.17\pm0.01$  &  Quasar  \\
eHAQ0113+0804  &  1.97  &  $0.60\pm0.01$  &  Broad absorption line quasar  \\
eHAQ0121+1028  &  2.10  &  $0.63\pm0.03$  &  Quasar  \\
eHAQ0129+0638  &  2.40  &  $0.52\pm0.01$  &  Broad absorption line quasar  \\
eHAQ0129+1039  &  0.57  &  $1.23\pm0.02$  &  Weak line quasar  \\
eHAQ0138+0742  &  3.21  &  $0.14\pm0.03$  &  Broad absorption line quasar  \\
eHAQ0142+0257\tablenotemark{$a$}  &  2.30  &  $0.38\pm0.01$  &  Broad absorption line quasar  \\
eHAQ0147+0411  &  1.77  &  $0.37\pm0.01$  &  Quasar  \\
eHAQ0147+1549  &   --   &   --   &  BL Lacertae Object  \\
eHAQ0157+1321  &  1.73  &  $0.44\pm0.02$  &  Quasar  \\
eHAQ0216+0426  &  1.58  &  $0.60\pm0.01$  &  Quasar  \\
eHAQ0227+0521  &  2.68  &  $0.14\pm0.02$  &  Broad absorption line quasar  \\
eHAQ0300+0440\tablenotemark{$a$}  &  1.88  &  $0.30\pm0.01\tablenotemark{$b$}$  &  Quasar  \\
eHAQ0321+0523  &  2.97  &  $0.20\pm0.01$  &  Broad absorption line quasar  \\
eHAQ0347+0348  &  1.85  &  $0.19\pm0.01$  &  Quasar  \\
eHAQ0828+0313\tablenotemark{$a$}  &  1.99  &  0--0.27  &  Weak line quasar  \\
eHAQ0835+0127  &  3.30  &  $0.14\pm0.01$  &  Quasar  \\
eHAQ0839+0556\tablenotemark{$a$}  &  1.49  &  $0.64\pm0.10\tablenotemark{$b$}$  &  Weak line quasar  \\
eHAQ0852+0204  &  2.20  &  $0.58\pm0.01$  &  Broad absorption line quasar  \\
eHAQ0856+0643\tablenotemark{$a$}  &  2.49  &  0--0.44  &  Quasar  \\
eHAQ0913+0910\tablenotemark{$a$}  &  1.95  &  $0.00\pm0.01$  &  Broad absorption line quasar  \\
eHAQ0919+0843  &  2.43  &  $0.35\pm0.05$  &  Broad absorption line quasar  \\
eHAQ0921+0149  &  0.99  &  $1.06\pm0.01$  &  Quasar  \\
eHAQ0923+0520  &  3.39  &  $0.12\pm0.06$  &  Broad absorption line quasar  \\
eHAQ0927$-$0233  &  1.24  &  $0.72\pm0.02$  &  Quasar  \\
eHAQ0930+0148  &  2.95  &  $0.24\pm0.01$  &  Quasar  \\
eHAQ0940+0532  &  2.32  &  $0.19\pm0.05$  &  Weak line quasar  \\
eHAQ0943+0954  &  4.18  &  $0.00\pm0.01$  &  Broad absorption line quasar  \\
eHAQ0943+1300\tablenotemark{$a$}  &  2.68  &  $0.18\pm0.11$  &  Broad absorption line quasar  \\
eHAQ0949+1207  &  3.36  &  $0.03\pm0.01$  &  Broad absorption line quasar  \\
eHAQ0950+0440\tablenotemark{$a$}  &  2.30  &  0--0.53  &  Broad absorption line quasar  \\
eHAQ0952+0835  &  1.86  &  $0.64\pm0.01$  &  Broad absorption line quasar  \\
eHAQ1005+0602  &  2.57  &  $0.35\pm0.01$  &  Broad absorption line quasar  \\
eHAQ1010+1158\tablenotemark{$a$}  &  2.24  &  0--0.40  &  Broad absorption line quasar  \\
eHAQ1025+1324\tablenotemark{$a$}  &  2.57  &  0.15--0.78  &  Broad absorption line quasar  \\
eHAQ1026$-$0241  &  3.23  &  $0.24\pm0.01$  &  Quasar  \\
eHAQ1030+1040\tablenotemark{$a$}  &  1.58  &  0--0.50  &  Broad absorption line quasar  \\
eHAQ1106+0844  &  1.86  &  $0.37\pm0.01$  &  Broad absorption line quasar  \\
eHAQ1109+1058  &  1.90  &  $0.43\pm0.01$  &  Quasar  \\
eHAQ1119+1430  &  1.88  &  $0.31\pm0.01$  &  Broad absorption line quasar  \\
eHAQ1136+0027  &  1.12  &  $0.95\pm0.02$  &  Quasar  \\
eHAQ1144+0902  &  1.78  &  $0.33\pm0.01$  &  Broad absorption line quasar  \\
eHAQ1202+0423  &  2.32  &  $0.02\pm0.01$  &  Broad absorption line quasar  \\
eHAQ1203+0652  &  2.40  &  $0.26\pm0.03$  &  Broad absorption line quasar  \\
eHAQ1203+1118  &  3.62  &  $0.15\pm0.02$  &  Broad absorption line quasar  \\
eHAQ1210+1429\tablenotemark{$a$}  &  1.48  &  0.36--1.10  &  Weak line quasar  \\
eHAQ1222+0826  &  2.90  &  $0.22\pm0.02$  &  Broad absorption line quasar  \\
eHAQ1226$-$0236  &  1.25  &  $0.77\pm0.01$  &  Quasar  \\
eHAQ1237+1233\tablenotemark{$a$}  &  2.31  &  $0.12\pm0.03$  &  Broad absorption line quasar  \\
eHAQ1244+0841\tablenotemark{$a$}  &  1.87  &  0.25--0.45  &  Quasar  \\
eHAQ1252+0842  &  3.53  &  $0.07\pm0.01$  &  Broad absorption line quasar  \\
eHAQ1312+1431  &  1.64  &  $0.51\pm0.01$  &  Quasar  \\
eHAQ1326+1317\tablenotemark{$a$}  &  2.00  &  $0.00\pm0.01$  &  Broad absorption line quasar  \\
eHAQ1340+1458\tablenotemark{$a$}  &  1.73  &  0.3--0.61  &  Quasar  \\
eHAQ1340+0151\tablenotemark{$a$}  &  1.94  &  0--0.94  &  Tentative identification; no spectral features  \\
eHAQ1346+0114\tablenotemark{$a$}  &  1.73  &  0.3--1.10  &  Quasar  \\
eHAQ1357$-$0051  &  1.80  &  $0.71\pm0.01$  &  Broad absorption line quasar  \\
eHAQ1400+0720  &  2.40  &  $0.38\pm0.02$  &  Broad absorption line quasar  \\
eHAQ1447+0521  &  2.35  &  $0.51\pm0.02$  &  Broad absorption line quasar  \\
eHAQ1450+1002  &  1.58  &  $0.36\pm0.01$  &  Broad absorption line quasar  \\
eHAQ1455+0705  &  1.90  &  $0.52\pm0.01$  &  Quasar  \\
eHAQ1514$-$0002\tablenotemark{$a$}  &  1.70  &  $0.00\pm0.01$  &  Broad absorption line quasar  \\
eHAQ1525+0155  &  1.62  &  $0.74\pm0.01$  &  Broad absorption line quasar  \\
eHAQ1528+0546  &  2.12  &  $0.38\pm0.01$  &  Quasar  \\
eHAQ1539+0351  &  1.63  &  $0.29\pm0.01$  &  Broad absorption line quasar  \\
eHAQ1543+0447  &  2.05  &  $0.71\pm0.01$  &  Quasar  \\
eHAQ1549+0501  &  1.46  &  $0.74\pm0.01$  &  Quasar  \\
eHAQ2209+0304\tablenotemark{$a$}  &  2.33  &  $0.35\pm0.01$  &  Quasar  \\
eHAQ2235+0635  &  1.13  &  $0.81\pm0.01$  &  Quasar  \\
eHAQ2236+0731  &  1.72  &  $0.26\pm0.01$  &  Quasar  \\
eHAQ2247+0922  &  1.63  &  $0.56\pm0.02$  &  Quasar  \\
eHAQ2255+1213  &  2.05  &  $0.14\pm0.01$  &  Quasar  \\
eHAQ2256+0531\tablenotemark{$a$}  &  1.35  &  $0.88\pm0.03$  &  Quasar  \\
eHAQ2258+0251  &  2.43  &  $0.40\pm0.03$  &  Broad absorption line quasar  \\
eHAQ2259+0208  &  2.05  &  $0.36\pm0.01$  &  Quasar  \\
eHAQ2259+0256  &  1.83  &  $0.59\pm0.01$  &  Quasar  \\
eHAQ2259+0736  &  0.78  &  $1.23\pm0.02$  &  Quasar  \\
eHAQ2301+0752  &  2.62  &  $0.27\pm0.02$  &  Broad absorption line quasar  \\
eHAQ2303+0747  &  1.31  &  $0.57\pm0.04$  &  Quasar  \\
eHAQ2309+1159  &  2.00  &  $0.54\pm0.01$  &  Quasar  \\
eHAQ2310+0447  &  1.42  &  $0.42\pm0.01$  &  Broad absorption line quasar  \\
eHAQ2313+1259  &  2.11  &  $0.57\pm0.01$  &  Quasar  \\
eHAQ2314+0552  &  2.03  &  $0.27\pm0.01$  &  Broad absorption line quasar  \\
eHAQ2316+0651  &  1.37  &  $0.53\pm0.01$  &  Broad absorption line quasar  \\
eHAQ2321+1107  &  1.56  &  $0.54\pm0.01$  &  Broad absorption line quasar  \\
eHAQ2321+1121  &  2.18  &  $0.37\pm0.01$  &  Quasar  \\
eHAQ2334+1519  &  2.23  &  $0.55\pm0.01$  &  Broad absorption line quasar  \\
eHAQ2344+1416  &  2.52  &  $0.41\pm0.02$  &  Broad absorption line quasar  \\
eHAQ2358+1030\tablenotemark{$a$}  &  2.05  &  $1.03\pm0.02$  &  Quasar  \\
eHAQ2359+1354  &  2.79  &  $0.13\pm0.03\tablenotemark{$c$}$  &  Quasar  \\
\enddata
\tablenotetext{a}{Target is classified as `peculiar' quasar, see text.}
\tablenotetext{b}{Assuming extinction curve from \citet{Zafar2015}.}
\tablenotetext{c}{Determined from the BOSS spectrum.}
\tablecomments{Uncertainties given on A(V) are only statistical errors from the fit.\\
Where we give a range in A(V), the first and last numbers indicate the amount of reddening inferred from the near-infrared and optical data, respectively.\\
The classification of broad absorption lines is done by eye and does not follow a rigorous scheme.}
\end{deluxetable}

%% file: individual_notes.tex
\subsection{eHAQ0001+0431 ($z=1.96$)}
This is a reddened quasar with weak but broad absorption blueward of \ion{C}{3}].

\subsection{eHAQ0007+1445 ($z=1.37$)}
This is a reddened quasar with narrow associated absorption at the redshift of the quasar and broad absorption blueward of \ion{C}{3}] and \ion{Mg}{2} with varying relative velocity offsets.

\subsection{eHAQ0019+0657 ($z=3.37$)}
This is a reddened quasar with an intervening \ion{Mg}{2} absorption system at $z_{\rm abs}=1.672$.

\subsection{eHAQ0026+0708 ($z=1.78$)}
This is a reddened quasar with weak absorption around \ion{C}{4}.

\subsection{eHAQ0044+1321 ($z=1.70$)}
This is a highly absorbed BAL quasar with weak emission lines. Due to the large amount of intrinsic absorption, the template is a very poor match to the observed optical spectrum. The reddening is inferred from the infrared photometry only.

\subsection{eHAQ0102+1159 ($z=4.25$)}
This is a quasar with broad absorption blueward of \ion{Si}{4}. The Lyman-break from the associated absorption at the redshift of the quasar is observed around $\lambda_{\rm obs} \approx 4800$~\AA.

\subsection{eHAQ0104+0756 ($z=2.21$)}
This is a FeLoBAL quasar. The reddening is roughly estimated from the photometry. However, due to the high level of absorption, it is impossible to determine the intrinsic continuum level.

\subsection{eHAQ0104+0912 ($z=2.01$)}
This is a reddened quasar with an intervening \ion{Mg}{2} absorption system at $z_{\rm abs}=1.487$. There is absorption around \ion{C}{4} and \ion{C}{3}], which is consistent with the presence of the 2175~\AA\ bump at the redshift of the absorber.

\subsection{eHAQ0104+1506 ($z=1.42$)}
Insecure dust fitting due to flux calibration issues of the INT spectrum. The \Mgii\ is partly absorbed by associated absorption. The \Feii\ emission around the \Mgii\ emission line is possibly very strong causing the \Mgii\ line to appear weaker. The optical data favor a solution with high dust extinction (${\rm A(V)} \approx 0.6$) whereas the near-infrared data indicate somewhat lower extinction (${\rm A(V)} \approx 0.3$). The optical and near-infrared data might be reconciled by a steeper extinction curve or a steeper intrinsic power-law slope, or a combination of both.

\subsection{eHAQ0109+0435 ($z=3.47$)}
This is a reddened BAL quasar with very weak emission lines and a very complex absorption line structure. The redshift is determined from absorption lines and from the small emission line around $\lambda_{\rm obs} \approx 4800$~\AA, which is interpreted as Ly$\beta$. The absolute flux calibration is insecure due to the scaling to the $r$-band, which is highly affected by the BAL features. The reddening is therefore only estimated form the photometry. There is possibly an excess of \ion{Fe}{2} emission causing the observed excess in the $z$ and $Y$-bands compared to the best-fit template.

\subsection{eHAQ0111+0641 ($z=3.23$)}
This is a reddened quasar with an intervening absorption system at $z_{\rm abs}=2.027$ determined from \ion{Fe}{2} lines since \ion{Mg}{2} is just outside the spectral coverage.

\subsection{eHAQ0113+0804 ($z=1.97$)}
This is a reddened BAL quasar with associated metal absorption from \ion{Fe}{2} and \ion{Mg}{2} at $z=1.938$. Furthermore, there is an intervening absorption system at $z_{\rm abs}=1.589$.

\subsection{eHAQ0121+1028 ($z=2.10$)}
This is a reddened quasar.

\subsection{eHAQ0129+0638 ($z=2.40$)}
This is a reddened BAL quasar with associated iron absorption.

\subsection{eHAQ0129+1039 ($z=0.567$)}
This is a highly reddened low-redshift AGN with weak broad lines. The infrared excess in the observed $H$ and $K$-bands is most likely caused by a combination of host galaxy emission and hot dust.

\subsection{eHAQ0138+0742 ($z=3.21$)}
This is a reddened quasar with narrow associated absorption. The Lyman-break from the associated absorber is visible in the blue edge of the spectrum.

\subsection{eHAQ0142+0257 ($z=2.30$)}
This is a reddened quasar with narrow associated absorption especially in \ion{C}{4}, \ion{Si}{4}, and \ion{N}{5}. The strong absorption around \Civ\ and \Ciii] is possibly caused by absorption of the iron `pseudo-continuum'. The object is detected in both FIRST and the NRAO VLA Sky Survey \citep[NVSS][]{Condon1998} at 1.4~GHz.

\subsection{eHAQ0147+0411 ($z=1.77$)}
This is a reddened quasar. Due to possible issues with the flux calibration of the INT spectrum, we have only estimated reddening from the photometry.
The object has a detection by NASA's Galaxy Evolution Explorer (GALEX) in the two bands (near-UV and far-UV): ${\rm NUV} = 20.59 \pm 0.10$~AB~mag and ${\rm FUV} = 20.73 \pm 0.18$~AB~mag.

\subsection{eHAQ0147+1549 (BL Lacertae Object)}
There are no spectral features that can hint at a redshift. The object is detected in radio by NVSS at 1.4~GHz, $F_{1.4~{\rm GHz}} = 17$~mJy. Given the optical flux, $i=18.3$~mag, this object is classified as {\it radio loud} \citep[following][]{Ivezic2002}. The object exhibits variability between the different epochs of observation; The photometry from SDSS and UKIDSS seems to be offset with respect to one another, and the spectrum does not seem to match the SDSS photometry in the $u$ and $g$ bands. These pieces of evidence hint at this object being a BL Lac object. Furthermore, this object appears in the ``Catalog of Candidate $\gamma$-ray Blazars'' of \citet{DAbrusco2014}.

\subsection{eHAQ0157+1321 ($z=1.73$)}
This is a reddened quasar. The spectral shape might be influenced by flux calibration issues for the INT spectra. This object also appears in the KX-catalog of \citet{Maddox2012}; however, these authors assign a redshift of $z=0.8524$. This redshift is inconsistent with the data presented here, since we see no evidence of the H$\beta$ and \ion{O}{3} emission lines, which should be observable at the red end of our spectrum at this redshift. Moreover, our template fitting using only the photometry (excluding the $W_3$ and $W_4$ bands) yields $z_{\rm phot}=1.75 \pm 0.10$.

\subsection{eHAQ0216+0426 ($z=1.58$)}
This is a reddened quasar with associated narrow absorption at $z_{\rm abs}=1.550$.

\subsection{eHAQ0227+0521 ($z=2.68$)}
This is a reddened BAL quasar.

\subsection{eHAQ0300+0440 ($z=1.88$)}
This is a reddened quasar with narrow associated absorption. The spectrum and photometry require a steeper extinction curve than the SMC-like curve from \citet{Gordon2003}. We therefore use the extinction curve derived for similar quasars from \citet{Zafar2015} to obtain a better fit.

\subsection{eHAQ0321+0523 ($z=2.97$)}
This is a reddened BAL quasar. The quasar has also been observed in SDSS from which a consistent redshift ($z_{\rm SDSS}=2.9811$) is derived.
The quasar also appears in the BALQSO catalog of \citet{Allen2011}, and in the catalog of unusual quasars from \citet{Meusinger2012}. 

\subsection{eHAQ0347+0348 ($z=1.85$)}
This is a reddened quasar.

\subsection{eHAQ0828+0313 ($z=1.99$)}
This is a weak line quasar with weak BAL features blueward of \Civ. We are not able to match the spectrum and photometry simultaneously. However, we note that the template is not suitable for weak line quasars, since it was derived for `regular' quasars. The NIR photometry is well matched by quasar template with no dust reddening applied. The quasar appears in the photometric catalogs of \citet{Richards2004, Richards2009} who estimate photometric redshifts of $z_{\rm phot}=0.275$ and $z_{\rm phot}=0.225$, respectively.

\subsection{eHAQ0835+0127 ($z=3.30$)}
This is a reddened quasar with a tentative detection of a \Feii\ absorption system at $z_{\rm abs}=1.717$, for which the \Mgii\ line falls directly on top of the telluric A-band.

\subsection{eHAQ0835+0830 (Star)}
This is a star of spectral type: K4\, {\sc v}.

\subsection{eHAQ0839+0556 ($z=1.49$)}
There are no strong emission lines for this object, however, we observe small absorption features that match up with \Civ, \Ciii], and \Mgii. This redshift is furthermore consistent with the photometric redshift of $z_{\rm phot}=1.5 \pm 0.1$. The overall SED is well-fitted by a continuum template with an intrinsically steep slope with no emission lines and reddened by a steep extinction curve. See Sect.~\ref{WLQ:phot_fit} for details about the fit.

\subsection{eHAQ0852+0204 ($z=2.20$)}
This is a reddened BAL quasar. The object appears in the catalog of photometric quasar candidates from \citet{Richards2009} who report a photometric redshift of $z_{\rm phot}=4.255$.

\subsection{eHAQ0856+0643 ($z=2.49$)}
This is a `peculiar' quasar with narrow associated absorption. The optical spectrum is well described by dust reddening ${\rm A(V)}=0.44$, however, the infrared photometry requires no reddening. The steeper extinction curve of \citet{Zafar2015} does not provide a satisfactory solution either.

\subsection{eHAQ0913+0910 ($z=1.95$)}
This is a FeLoBAL quasar with no apparent reddening due to dust. The large number of blended BAL features in the observed optical spectrum makes it impossible to identify the quasar continuum. This leads to a strong evident mismatch between the optical data and the template. The object appears in the catalog of photometric quasar candidates from \citet{Richards2009} who report a photometric redshift of $z_{\rm phot}=0.225$.

\subsection{eHAQ0915+0115 (Star)}
This is a star of spectral type: K5\, {\sc v}.

\subsection{eHAQ0919+0843 ($z=2.43$)}
This is a reddened BAL quasar. The redshift is determined from the \Ciii] emission line.

\subsection{eHAQ0921+0149 ($z=0.99$)}
This is a highly reddened quasar with narrow emission lines.

\subsection{eHAQ0923+0520 ($z=3.40$)}
This is a reddened BAL quasar. The reddening is determined from fitting the template to the near-infrared photometry since the intrinsic continuum level is hard to estimate.

\subsection{eHAQ0927$-$0233 ($z=1.24$)}
This is a reddened quasar possibly with BAL features blueward of \Ciii]. There is a small deficit in the $Y$ and $J$ bands relative to the best-fit template. This is possibly due to variations in the \Feii emission around H$\beta$. The object appears in the catalog of photometric quasar candidates from \citet{Richards2009} who report a photometric redshift of $z_{\rm phot}=3.875$. 

\subsection{eHAQ0930+0148 ($z=2.95$)}
This is a reddened quasar with narrow associated absorption. We see tentative evidence for a DLA at $z_{\rm DLA}=2.720$ which matches with metal absorption lines.

\subsection{eHAQ0940+0532 ($z=2.32$)}
This is a reddened quasar with very weak emission lines and high ionization BAL.

\subsection{eHAQ0943+0954 ($z=4.25$)}
This is a BAL quasar with no evidence for reddening due to dust. The quasar is observed in SDSS with a spectroscopic redshift of $z_{\rm SDSS}=4.27$ and appears in the catalog of BAL quasars \citep{Allen2011} and unusual quasars \citep{Meusinger2012}.

\subsection{eHAQ0943+1300 ($z=2.68$)}
This is a highly absorbed BAL quasar with mild reddening and very weak emission lines. The absorption exhibits various velocity profiles which makes an exact redshift determination difficult. The estimate of reddening is insecure and only based on the near-infrared photometry.

\subsection{eHAQ0949+1207 ($z=3.36$)}
This is a BAL quasar consistent with no dust reddening.

\subsection{eHAQ0950+0440 ($z=2.30$)}
This is a `peculiar' BAL quasar. The optical spectrum is well described by the quasar template reddened by ${\rm A(V)} = 0.53$, however, the near-infrared photometry does not match neither the reddened or unreddened template.

\subsection{eHAQ0952+0835 ($z=1.86$)}
This is a reddened BAL quasar.

\subsection{eHAQ1002+0406 (Star)}
This is a star of spectral type: M2\, {\sc v}.

\subsection{eHAQ1005+0602 ($z=2.57$)}
This is a reddened BAL quasar.

\subsection{eHAQ1010+1158 ($z=2.24$)}
This is a `peculiar' BAL quasar. The optical spectrum is well described by the dust reddened quasar template, however, the near-infrared photometry is consistent with no dust reddening.

\subsection{eHAQ1025+1324 ($z=2.57$)}
This is a `peculiar' BAL quasar. The spectrum is well described by a high amount of reddening (${\rm A(V)} = 0.78$), but the near-infrared photometry requires less reddening (${\rm A(V)} = 0.15$). We note that the template may not provide a good description for the optical spectrum, due to the large amount of absorption.

\subsection{eHAQ1026$-$0241 ($z=3.23$)}
This is a reddened quasar with associated absorption (strongest in \Civ, \ion{N}{5} and Ly$\alpha$).

\subsection{eHAQ1030+1040 ($z=1.58$)}
From photometric modelling (using bands from $g$ to $W_2$), we infer a redshift of $z_{\rm phot}=4.2 \pm 0.1$, consistent with the 
estimate from \citet{Richards2009}: $z_{\rm phot}=4.215$.
However, the spectrum reveals that this object is indeed a FeLoBAL quasar at lower redshift. The near-infrared photometry is consistent with no reddening whereas the optical spectrum seems to require some reddening, however, the estimated A(V) is highly uncertain due to the strong absorption of the continuum. 

\subsection{eHAQ1106+0844 ($z=1.86$)}
This is a reddened BAL quasar. The object is also observed by SDSS with a consistent redshift estimate: $z_{\rm SDSS}=1.8645$, and appears in the catalog of unusual quasars by \citet{Meusinger2012}.

\subsection{eHAQ1109+0135 (Star)}
This is a star of spectral type: M2\, {\sc v}.

\subsection{eHAQ1109+1058 ($z=1.90$)}
This is a reddened quasar with narrow associated absorption. We identify a \Mgii\ absorption system at $z_{\rm abs}=1.667$. \citet{Richards2009} estimate a photometric redshift of $z_{\rm phot}=2.735$.

\subsection{eHAQ1111+0151 (Star)}
This is a star of spectral type: K7\, {\sc v}.

\subsection{eHAQ1119+1430 ($z=1.88$)}
This is a reddened FeLoBAL quasar with weak emission lines.

\subsection{eHAQ1120+0812 (Star)}
This is a star of spectral type: K5\, {\sc iii}.

\subsection{eHAQ1132+1243 (Star?)}
The classification of this object is insecure.
The best-fit template yields a spectral type of a metal-rich K4 giant. There is an apparent offset between the SDSS and UKIDSS photometry.

\subsection{eHAQ1136+0027 ($z=1.12$)}
This is a highly reddened quasar with associated absorption. The near-infrared photometry is offset relative to the optical data. This is possibly caused by intrinsic variability in the quasar. The object appears in the photometric catalogs of \citet{Richards2004} and \citet{Richards2009} who assign photometric redshifts of $z_{\rm phot}=0.33$ and $z_{\rm phot}=3.89$, respectively.
The spectroscopic redshift if measured accurately from [\ion{O}{2}]\,$\lambda3727$.

\subsection{eHAQ1144+0902 ($z=1.78$)}
This is a reddened BAL quasar. The object is also observed by SDSS at a consistent redshift ($z_{\rm SDSS}=1.7849$). The object appears in the BAL catalog of \citet{Allen2011}.

\subsection{eHAQ1202+0423 ($z=2.32$)}
This is a highly absorbed BAL quasar, but we observe no signs of dust reddening.

\subsection{eHAQ1203+0652 ($z=2.40$)}
This is a reddened BAL quasar with weak emission lines.

\subsection{eHAQ1203+1118 ($z=3.62$)}
This is a reddened BAL quasar.

\subsection{eHAQ1210+1429 ($z=1.48$)}
This is a highly reddened quasar with weak emission lines and very strong associated absorption at the quasar redshift $z_{\rm abs}=1.486$. The optical spectrum is well fitted by a high amount of dust (${\rm A(V)}=1.1$), however, the near-infrared photometry requires less reddening (${\rm A(V)}=0.36$; shown as the blue template in the figure). This indicates that the intrinsic power-law slope of the quasar is steeper than average or that the extinction curve is steeper than SMC.

\subsection{eHAQ1222+0826 ($z=2.90$)}
This is a reddened BAL quasar. The near-infrared photometry is slightly offset compared to the optical data. This is possibly due to intrinsic variability.

\subsection{eHAQ1226$-$0236 ($z=1.25$)}
This is a reddened ordinary quasar.

\subsection{eHAQ1237+1233 ($z=2.31$)}
This is a reddened BAL quasar with weak emission lines. The optical spectrum is better fitted by a rather high A(V) whereas a fit to the near-infrared photometry results in a significantly lower amount of reddening: ${\rm A(V)}=0.12$. We caution the reader that the template does not provide a reasonable description of the optical data, due to the strong absorption. We therefore only provide the A(V) derived from the near-infrared photometry.

\subsection{eHAQ1244+0841 ($z=1.87$)}
This is a reddened quasar with weak BAL features. We note that the optical spectrum seems to indicate a high A(V) of $0.45$~mag, but this is highly uncertain due to the possible absorption around \Civ\ and \Ciii]. The object appears in the photometric catalog of \citet{Richards2009} who infer a photometric redshift of $z_{\rm phot}=2.72$.

\subsection{eHAQ1252+0842 ($z=3.53$)}
This is a BAL quasar. The overall SED is consistent with a low amount of reddening, which could plausibly be ascribed to variations in the intrinsic quasar slope.

\subsection{eHAQ1312+1431 ($z=1.64$)}
This is a reddened quasar with weak \Civ\ and \Ciii] lines. There appears to be absorption from \Feii\ lines. There is narrow absorption on top of \Civ.
The object appears in the photometric catalog of \citet{Richards2009} who infer a photometric redshift of $z_{\rm phot}=3.54$.

\subsection{eHAQ1326+1317 ($z=2.00$)}
This is a strongly absorbed BAL quasar. We see no signs of dust reddening. The object is observed by SDSS ($z_{\rm SDSS}=1.9869$) and is included in the BAL catalog of \citet{Allen2011}. 

\subsection{eHAQ1331+1304 (Star)}
This is a star of spectral type: M1\, {\sc v}.

\subsection{eHAQ1340+0151 ($z=1.94?$)}
There are no strong spectral features to securely identify this target. There is a sharp break in the SED after the $z$-band. We tentatively identify the emission feature in the blue edge of the spectrum as Ly$\alpha$ at redshift $z=1.94$, which matches with the weak absorption feature in the $g$-band. The reddening required for this redshift solution is quite high (${\rm A(V)}\sim1$~mag), but does not match the near-infrared photometry. At the same redshift, the near-infrared photometry requires no reddening. There is no radio detection to aid an classification.
The target is marked as a quasar candidate for the Large sky Area Multi-Object Fiber Spectroscopic Telescope (LAMOST) survey \citep{Luo2015}; However, the spectrum observed in LAMOST DR1 shows no flux. The overall SED is fitted well by the continuum model of \citet{Richards2006} with variable slope and extinction curve. For details, see Sect.~\ref{WLQ:phot_fit}.

\subsection{eHAQ1340+1458 ($z=1.73$)}
This is a reddened quasar with weak BAL in \Civ\ and \ion{Si}{4}. There is possibly variations in the iron pseudo continuum around \Ciii]. The near-infrared data require less reddening (${\rm A(V)}\sim0.3$) than the optical data. This indicates that the extinction curve is steeper than SMC or that the intrinsic slope of the quasar power-law is steeper than average.

\subsection{eHAQ1346+0114 ($z=1.73$)}
This is a highly reddened quasar with weak \Civ\ emission and blueshifted absorption in both \Civ\ and \Ciii]. There is possibly variations in the iron pseudo continuum around \Ciii]. The near-infrared data require less reddening (${\rm A(V)}\sim0.3$) than the optical data. This indicates that the extinction curve is steeper than SMC or that the intrinsic slope of the quasar power-law is steeper than average.

\subsection{eHAQ1357$-$0051 ($z=1.80$)}
This is a reddened BAL quasar with strong associated low-ionization metal absorption ($z_{\rm abs}=1.792$).

\subsection{eHAQ1400+0720 ($z=2.40$)}
This is a reddened BAL quasar with associated low-ionization metal absorption at the redshift of the quasar. Possibly also weak \Feii\ BAL features.

\subsection{eHAQ1447+0521 ($z=2.35$)}
This is a reddened BAL quasar. The object appears in the photometric catalog of \citet{Richards2009} who infer a photometric redshift of $z_{\rm phot}=4.26$.

\subsection{eHAQ1450+1002 ($z=1.58$)}
This is a reddened FeLoBAL quasar.

\subsection{eHAQ1455+0705 ($z=1.90$)}
This is a reddened quasar with associated absorption in \Civ\ and \Ciii]. We see tentative evidence of an intervening \Mgii\ absorption system at $z={\rm abs}=1.650$.
\citet{Richards2009} infer a photometric redshift of $z_{\rm phot}=4.26$. The object is observed by \citet{Maddox2012} who also classify the object as a quasar at redshift $z=1.9094$.

\subsection{eHAQ1514$-$0002 ($z=1.70$)}
This is a FeLoBAL quasar with no evidence of dust reddening from the near-infrared photometry.

\subsection{eHAQ1525+0155 ($z=1.62$)}
This is a highly reddened quasar with broad absorption around the \Ciii] emission line, possibly caused by absorption from \ion{Fe}{3}, and narrow associated absorption on top of \Mgii. We observe blueshifted \Civ\ absorption.

\subsection{eHAQ1528+0546 ($z=2.12$)}
This is a reddened quasar with broad absorption around the \Civ\ emission line. The object is furthermore observed by SDSS ($z_{\rm SDSS}=2.1051$) and by \citet{Maddox2012} who infer a consistent redshift $z=2.11$. Moreover, \citet{Richards2009} infer a photometric redshift of $z_{\rm phot}=0.41$.

\subsection{eHAQ1539+0351 ($z=1.63$)}
This is a reddened BAL quasar. The higher ionization lines (\Civ\ and \ion{Si}{4}) exhibit stronger BAL features, whereas \Ciii] and \Mgii\ show progressively weaker lines.

\subsection{eHAQ1543+0447 ($z=2.05$)}
This is a highly reddened quasar with strong excess emission from \Feii\ and \ion{Fe}{3}.

\subsection{eHAQ1549+0501 ($z=1.46$)}
This is a highly reddened quasar with weak emission lines and narrow associated absorption. \citet{Richards2009} infer a photometric redshift of $z_{\rm phot}=3.88$.

\subsection{eHAQ2209+0304 ($z=2.33$)}
This is a very `peculiar' quasar. The spectrum reveals very weak emission lines and narrow associated absorption, however, the continuum shape is not matched by the template at all neither in the optical nor in the near-infrared. 

\subsection{eHAQ2235+0635 ($z=1.13$)}
This is a highly reddened quasar with associated low-ionization metal absorption. The object is detected in the FIRST survey at 1.4~GHz, $F_{\rm 1.4~GHz}=0.62 \pm 0.14$~mJy.

\subsection{eHAQ2236+0731 ($z=1.72$)}
This is a reddened quasar categorized as an optical variable in the Palomar-QUEST survey \citep{Bauer2009}.

\subsection{eHAQ2247+0922 ($z=1.63$)}
This is a reddened quasar.

\subsection{eHAQ2255+1213 ($z=2.05$)}
This is a reddened quasar with broad absorption around \Civ\ and \Ciii] possibly caused by a combination of weak high-ionization BAL and FeLoBAL features. The iron absorption would also explain the deficit observed in the $Y$ and $J$ bands. The object is detected in the FIRST survey at 1.4~GHz, $F_{\rm 1.4~GHz}=4.19 \pm 0.16$~mJy.

\subsection{eHAQ2256+0531 ($z=1.35$)}
This is a highly reddened quasar with weak emission lines an narrow associated absorption. We observe a large excess in the $H$ and $K$ bands, most likely due to the H$\alpha$ line in the $H$ band and emission from the host-galaxy and hot dust that start to become important in the $K$-band (rest-frame $\sim1\mu$m).
The object is a known radio source detected at several frequencies. It is detected in FIRST and NVSS at 1.4~GHz with a flux of $F_{\rm FIRST}=11.99 \pm 0.13$ and $F_{\rm NVSS}=14.8 \pm 0.6$~mJy, respectively. \citet{Jackson2007} report a flux at 8.4~GHz of $F_{\rm 8.4~GHz}=10.7\pm0.2$~mJy and \citet{Coble2007} report a flux at 28.5~GHz of $F_{\rm 28.5~GHz}=8.15\pm0.15$~mJy.

\subsection{eHAQ2258+0251 ($z=2.43$)}
This is a reddened BAL quasar.

\subsection{eHAQ2259+0208 ($z=1.83$)}
This is a reddened quasar.

\subsection{eHAQ2259+0256 ($z=1.83$)}
This is a reddened quasar with broad absorption around the \Civ\ and \Ciii] emission lines. The object is detected in the FIRST survey at 1.4~GHz, $F_{\rm 1.4~GHz}=0.72 \pm 0.13$~mJy.

\subsection{eHAQ2259+0736 ($z=0.78$)}
This is a highly reddened quasar with narrow emission lines an associated narrow absorption. We can accurately measure the redshift from [\ion{O}{2}]\,$\lambda3727$.

\subsection{eHAQ2301+0752 ($z=2.62$)}
This is a reddened BAL quasar.

\subsection{eHAQ2303+0747 ($z=1.31$)}
The spectrum from INT does not match the photometry due to issues in the flux calibration. We therefore only use the spectrum to securely classify this object as a quasar on the basis of the broad emission line (\Mgii\ at $z=1.31$). We estimate the reddening using only the photometry.
The object is detected in the two bands of GALEX (near-UV and far-UV): ${\rm NUV} = 22.7 \pm 0.2$~AB~mag and ${\rm FUV} > 23.2$~AB~mag.

\subsection{eHAQ2309+1159 ($z=2.00$)}
This is a reddened quasar with absorption in the red wing of \Ciii] possibly due to \ion{Fe}{3} absorption.

\subsection{eHAQ2310+0447 ($z=1.42$)}
This is a reddened BAL quasar.

\subsection{eHAQ2313+1259 ($z=2.11$)}
This is a reddened quasar with narrow associated absorption on top of the \Civ\ emission line. The reddening estimate is uncertain due to noise in the blue edge of the spectrum.

\subsection{eHAQ2314+0552 ($z=2.03$)}
This is a reddened quasar with narrow emission lines and broad emission around the \Civ\ and \Ciii] emission lines possibly due to a blend of weak high-ionization BAL and low-ionization iron absorption.

\subsection{eHAQ2316+0651 ($z=1.37$)}
This is a reddened quasar with associated absorption at the quasar redshift. We observe strong absorption bluewards of \Ciii], possibly due to BAL features.

\subsection{eHAQ2321+1107 ($z=1.56$)}
This is a reddened BAL quasar.

\subsection{eHAQ2321+1121 ($z=2.18$)}
This is a reddened quasar. We observe a deficit in the $J$ and $H$ bands compared to the best-fit template. This can plausibly be explained by variations in the iron pseudo continuum. This would also explain the small excess in the spectrum around the $z$-band.

\subsection{eHAQ2334+1519 ($z=2.23$)}
This is a reddened BAL quasar with weak absorption from low-ionization iron lines. The near-infrared photometry is offset from the optical data possibly due to intrinsic variability.

\subsection{eHAQ2344+1416 ($z=2.52$)}
This is a reddened BAL quasar possibly with variations in the iron pseudo continuum around \Civ\ and \Ciii].
The object is detected in the FIRST survey at 1.4~GHz, $F_{\rm 1.4~GHz}=0.97 \pm 0.16$~mJy.

\subsection{eHAQ2358+1030 ($z=2.05$)}
This is a highly reddened quasar with very strong and narrow emission lines. We clearly detect the \ion{He}{2} emission line redwards of \Civ. We furthermore observe an emission line at $\lambda_{\rm obs} = 6988.8$~\AA, which is caused by second order contamination from Ly$\alpha$. This quasar is described in detail in \citep{Heintz2016a}.

\subsection{eHAQ2359+1354 ($z=2.79$)}
We observed this target with grism \#6 to look for a DLA around $z_{\rm abs}= 2.248$. We see evidence for metal absorption from \ion{Fe}{2} and \ion{Si}{2} and a sharp drop around $4000$~\AA, which corresponds to the location of the Ly$\alpha$ line from the DLA. We only use the spectrum to fix the redshift and subsequently determine the reddening by fitting the BOSS spectrum.